\documentclass[prb,twocolumn,floatfix,aps,showpacs]{revtex4-2}
\usepackage{graphicx,amsmath,amssymb,color,nicefrac,multirow, makecell, ragged2e,float}
\usepackage[unicode=true,colorlinks=true,linkcolor=blue,citecolor=blue,urlcolor=blue]{hyperref}
\usepackage[titletoc,title]{appendix}

\newcommand{\be}{\begin{equation}}
\newcommand{\ee}{\end{equation}}

\newcommand{\ba}{\begin{eqnarray}}
\newcommand{\ea}{\end{eqnarray}}

\renewcommand{\vec}[1]{\mbox{\boldmath$#1$}}

\def\beq{\begin{eqnarray}}
\def\eeq{\end{eqnarray}}

\newcommand{\etal}{{\it et al.}}

\newcommand*{\rom}[1]{\expandafter\@slowromancap\romannumeral #1@}

\begin{document}

\title{Revisiting excitation gaps in the fractional quantum Hall effect}
\author{Tongzhou Zhao$^{1}$, Koji Kudo$^{1}$, W. N. Faugno$^{1}$, Ajit C. Balram$^{2,3}$,  and J. K. Jain$^{1}$}
\affiliation{$^{1}$Department of Physics, 104 Davey Lab, Pennsylvania State University, University Park, Pennsylvania 16802, USA}
\affiliation{$^{2}$Institute of Mathematical Sciences, CIT Campus, Chennai 600113, India}
\affiliation{$^{3}$Homi Bhabha National Institute, Training School Complex, Anushaktinagar, Mumbai 400094, India}

\date{\today}

\begin{abstract}
Recent systematic measurements of the quantum well width dependence of the excitation gaps of fractional quantum Hall states in high mobility samples [Villegas Rosales {\it et al.}, Phys. Rev. Lett. {\bf 127}, 056801 (2021)] open the possibility of a better quantitative understanding of this important issue. We present what we believe to be accurate theoretical gaps including the effects of finite width and Landau level (LL) mixing. While theory captures the width dependence, there still remains a deviation between the calculated and the measured gaps, presumably caused by disorder. It is customary to model the experimental gaps of the $n/(2n\pm 1)$ states as $\Delta_{n/(2n\pm 1)} = Ce^2/[(2n\pm 1)\varepsilon  l]-\Gamma$, where $\varepsilon$ is the dielectric constant of the background semiconductor and $l$ is the magnetic length; the first term is interpreted as the cyclotron energy of composite fermions and $\Gamma$ as a disorder-induced broadening of composite-fermion LLs. Fitting the gaps for various fractional quantum Hall states, we find that $\Gamma$ can be nonzero even in the absence of disorder. 

\end{abstract}
\maketitle

\section{Introduction}
It has been appreciated since the very beginning that the existence of a gap is a fundamental property of a fractional quantum Hall effect (FQHE) state~\cite{Tsui82, Laughlin83}. Despite the passage of almost four decades, the quantitative agreement between the experimentally measured~\cite{Boebinger85, Willett88, Shayegan90, Du93, Pan20, Rosales21} and the theoretically predicted~\cite{Jain07, Haldane85a, Melik-Alaverdian95, Ortalano97, Jain97, Park99b, Scarola02, Morf02, Balram16d, Balram20b} values of the gaps is not as good as one might have hoped. In the idealized limit where electrons are in a strict two-dimensional (2D) layer and Landau level (LL) mixing and disorder are absent, comparisons with computer calculations show that the zeroth-order composite fermion (CF) theory predicts gaps of the $n/(2n\pm 1)$ FQHE states to within 10\%, and the agreement can be further improved by allowing for CF-LL mixing~\cite{Jain07, Balram13}. The discrepancy between the theoretical and the experimental gaps must, therefore, originate from features not included in the idealized model.

A recent article by Villegas Rosales {\em et al.} has reported on systematic measurements of the gaps and their quantum well width dependence in the highest mobility samples available to date~\cite{Rosales21}.  The modest aim of our work below is to determine the theoretical values of gaps incorporating, to the best extent we currently know, the effects of finite width and LL mixing (LLM), hoping to gain a better quantitative understanding of this important issue. Our strategy below is first to accurately calculate the thermodynamic limits of the variational gaps with the effect of finite width treated through a local density approximation (LDA); finite width is responsible for the most significant reduction in the gap for typical experimental parameters. We then estimate the thermodynamic limits for the deviation between the variational and the exact gaps in the lowest LL (LLL) and also for correction due to LLM. The final theoretical gaps along with the experimental gaps are shown in Fig.~\ref{Gap_theory_experiment_n11}. We find that while theory nicely captures the behavior of the gaps as a function of the quantum well width, a quantitative discrepancy remains, which is most likely due to disorder, not included in our calculations.

We mention here some of the previous theoretical studies of gaps in the FQHE regime. Zhang and Das Sarma~\cite{Zhang86} calculated finite width correction for the 1/3 gap modeling the interaction as $1/(\sqrt{r^2+d^2})$, where $r$ is the inter-particle spacing and $d$ is related to the width.  Park~\etal~\cite{Park99b} evaluated gaps for fractions along the $n/(2n+1)$ sequence with a variational Monte Carlo (VMC) method using wave functions from the CF theory, treating finite width in a LDA; they did not go to large widths that have been studied in Ref.~\cite{Rosales21}. Morf~\etal~\cite{Morf02} evaluated gaps by performing exact diagonalization (ED); they used a Gaussian model for the transverse wave function to simulate the LDA wave function.  As for all ED studies, this work is restricted to small systems. Yoshioka~\cite{Yoshioka84} calculated the effect of LLM on the 1/3 gap by performing ED in the Hilbert space of the two lowest LLs. Melik-Alaverdian and Bonesteel~\cite{Melik-Alaverdian95} studied the effect of LL mixing on the energy gap of the 1/3 state. They evaluated the quasiparticle energy by diagonalizing the Coulomb interaction in a 2$\times$2 basis of the projected and unprojected Jain quasiparticle wave functions; within this approximation, the energies of the ground state and the quasihole are not modified by LLM.

\section{Calculational details}
\label{sec: background}

We work with the spherical geometry~\cite{Haldane83}. In this geometry, a magnetic monopole placed at the center of the sphere generates a uniform radial magnetic flux of strength $2Q hc/e$ ($2Q$ is an integer) through the spherical surface, on which $N$ electrons reside. Owing to the rotational symmetry, states can be characterized by their total orbital angular momentum quantum number $L$. Incompressible quantum Hall ground states are uniform, i.e., have $L=0$ while their excitations, in general, have $L>0$.  Compared with the planar geometry, the flux-particle relationship on the sphere is given by $2Q=N/\nu-\mathcal{S}$, where $\mathcal{S}$ is a topological quantum number called the shift~\cite{Wen92}. The planar momentum $k$ is related to $L$ as $k=L/R$, where $R=\sqrt{Q}l$ is the radius of the sphere and $l=\sqrt{\hbar c/eB}$ is the magnetic length at the magnetic field $B$. 

The FQHE of electrons at $\nu=n/(2pn\pm 1)$ is a manifestation of the integer quantum Hall effect (IQHE) of CFs~\cite{Jain89, Jain07}, which are bound states of electrons and an even number ($2p$) of vortices. The ground-state wave function at these fractions is known to be accurately given by $\Psi_{n/(2n+1)}^\text{Jain} = \mathcal{P}_{\rm LLL} \Phi_{n}\Phi^{2}_{1}$, where $\Phi_{n}$ is the IQHE wave function of electrons at filling factor $n$ and $\mathcal{P}_{\rm LLL}$ implements projection to the LLL as is appropriate in the limit that $B\rightarrow \infty$. Throughout this work, we carry out the LLL projection using the Jain-Kamilla method, details of which can be found in the literature~\cite{Jain97, Jain97b, Jain07}. The lowest-energy neutral excitation is obtained by promoting a CF from the highest occupied $\Lambda$L to the lowest unoccupied $\Lambda$L. The wave function for this state termed the CF exciton (CFE), is given by
$\Psi_{n/(2n+1)}^\text{CFE} = \mathcal{P}_{\rm LLL} \Phi_{n}^\text{ex}\Phi^{2}_{1}$, 
where $\Phi_{n}^\text{ex.}$ is the IQHE wave function of an exciton with a hole in the LL indexed by $n-1$ and a particle in the LL indexed by $n$.  The constituent quasiparticle and quasihole of the CFE are referred to as the CF particle (CFP) and CF hole (CFH). The gap measured in transport corresponds to the energy of a far separated CFP-CFH pair. In the spherical geometry, at $\nu=n/(2n+1)$ this state is obtained by placing the CFH and the CFP at the north and south poles, respectively, which corresponds to the CFE with the largest $L_{\rm max}=(N-n^2)/{n}+(2n-1)$~\cite{Balram16d}. The detailed form of the above wave functions in the spherical geometry is given in the Supplemental Material (SM)~\cite{SM}. 

To calculate the transport gap accurately, the attractive interaction between the CFH and the CFP that exists in any finite system needs to be accounted for.  The CFP and CFH have an extent of only a few magnetic lengths~\cite{Balram13b}, so in the simplest approximation, we treat them as point particles with charge $(\pm e)/(2n+1)$~\cite{Jain07} separated by a distance of $2\sqrt{Q} l$, which is the diameter of the sphere. The resulting Coulomb attraction between the CFP and the CFH is thus given by
\begin{equation}
V_\text{CFH-CFP}= -\frac{1}{\left(2n+1\right)^2(2\sqrt{Q})} \frac{e^2}{\varepsilon l}.
\label{eq: qp_qh_interaction}
\end{equation}
Taking this attractive interaction into account, we define the transport gap as:
\begin{equation}
\Delta =\sqrt{\frac{2Q\nu}{N}} \left(E^{L_{\rm max}}_\text{CFE}-E_\text{gs}-V_\text{CFH-CFP}\right),
\label{eq: gap}
\end{equation}
where $E^{L_{\rm max}}_\text{CFE}$ and $E_\text{gs}$ are the expectation values of the Coulomb energies of the wave functions corresponding to the largest-$L$ CFE and ground state, respectively. The expectation values for the variational wave functions are evaluated using the Metropolis Monte Carlo method. In Eq.~\eqref{eq: gap} the factor of $\sqrt{2Q\nu/N}$ corrects for the density difference between a finite system on the sphere and that in the thermodynamic limit, and thereby reduces the $N$-dependence of the gaps~\cite{Morf86}. 

To compare the theoretical gaps against the experimental values, we need to carefully incorporate the effect of the finite width of the quantum well. To do so we consider the effective interaction given by
\begin{equation}
\begin{aligned}\label{eqn:V_eff}
V_{\text{eff}}(r)=\frac{e^2}{\varepsilon}\int d\xi_1 \int d\xi_2\frac{\left|\psi(\xi_1)\right|^2\left|\psi(\xi_2)\right|^2}{\sqrt{r^2+(\xi_1-\xi_2)^2}},
\end{aligned}
\end{equation}
where $r$ is the in-plane distance between two particles, $\xi$ is the transverse coordinate, and $\psi\left(\xi\right)$ is the transverse wave function which is obtained from a separate LDA calculation~\cite{Martin20}. 

\section{LLM: perturbative approach}

One of our objectives is to determine the modification of the gap due to LLM. The LLM parameter $\kappa$, defined as $\kappa=\left(e^2/\varepsilon l\right)/\left(\hbar \omega_c\right)$, characterizes the strength of the Coulomb interaction relative to the cyclotron energy, where $\omega_c=eB/m_b c$ is the cyclotron frequency of the band electron, where $m_b$ is the band mass. To study the effect of LLM, we carry out ED using a perturbative method that incorporates LLM through a correction to the interaction (including a three-body interaction)~\cite{Bishara09, Peterson13, Peterson14, Sodemann13}. 

The effective Hamiltonian we use is given by 
$\hat{V}_c(W)+\kappa[\hat{V}_2(W)+\hat{V}_3(W)]$, where $\hat{V}_c(W)$ is the Coulomb interaction for a quantum well of width $W$, $\kappa \hat{V}_2(W)$ is the correction to the two-body interaction due to LLM, and $\kappa \hat{V}_3(W)$ is the three-body interaction term generated by LLM. In the disk geometry, these have the form
\begin{align}
 \label{eq: LLM}
 \hat{V}_c(W)
 &=\sum_{m=0}^\infty V_c(W,m)\sum_{i<j}P_{ij}(m),\\
 \hat{V}_2(W)
 &=\sum_{m=0}^{m_{2,{\rm max}}} V_2(W,m)\sum_{i<j}P_{ij}(m), \nonumber \\
 \hat{V}_3(W)
 &=\sum_{m=0}^{m_{3,{\rm max}}} V_3(W,m)\sum_{i<j<k}P_{ijk}(m), \nonumber
\end{align}
where $P_{ij}(m)$ and $P_{ijk}(m)$ are the projection operators onto a pair 
or a triplet of electrons, respectively, with relative angular momentum $m$. The pseudopotential $V_c(W,m)$ is given by~\cite{DasSarma85}.
\begin{align}
 	V_c(W,m)
	&=\int_0^\infty dk k\left[L_0(k^2/2)\right]^2L_m(k^2)e^{-k^2}V(k)\\
	V(k)
	&=\frac{e^2l}{\epsilon}\frac{1}{k}
	\frac{3kW+\frac{8\pi^2}{kW}
	-\frac{32\pi^4(1-e^{-kW})}{(kW)^2\left[(kW)^2+4\pi^2\right]}}
	{(kW)^2+4\pi^2},
\end{align}
where $k$ is quoted in units of $1/l$. As for $V_2(W,m)$ and $V_3(W,m)$, we use the values quoted in Tables I and II of Ref.~\cite{Peterson13} \{the corrections up to $m_{2,{\rm max}}=8$ and $m_{3,{\rm max}}=8$ are given in Ref.~\cite{Peterson13} and thus we truncate the sums in Eq.~\eqref{eq: LLM} to these values\}. (It is noted that these values are obtained for a transverse wave function of the cosine form. We make the simplifying assumption below that the LLM reduction factor is not strongly sensitive to the form of the transverse wave function.) Using these planar disk pseudopotentials in the spherical geometry, we compute the energy gap of a far separated CFP-CFH pair as defined in Eq.~\eqref{eq: gap} using ED.  In the following calculations, we set $\kappa=0.70$ and $0.76$ at $\nu=1/3$ and $2/5$, respectively,  as appropriate for the experiments of Ref.~\cite{Rosales21}, with electron density $\rho=1.1{\times}10^{11}\text{cm}^{-2}$.

\begin{figure}[t]
\includegraphics[width=0.9 \columnwidth]{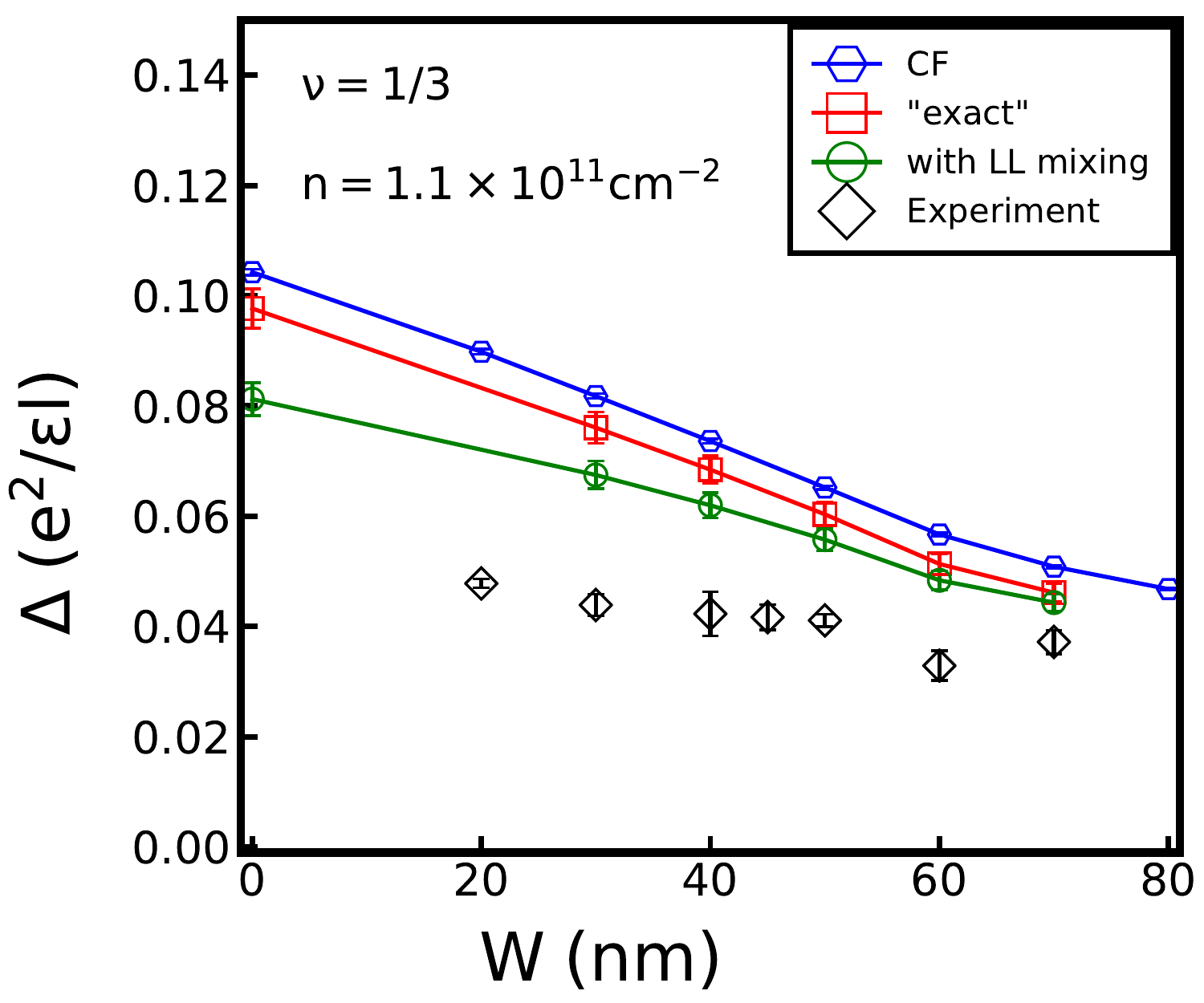}
\includegraphics[width=0.9 \columnwidth]{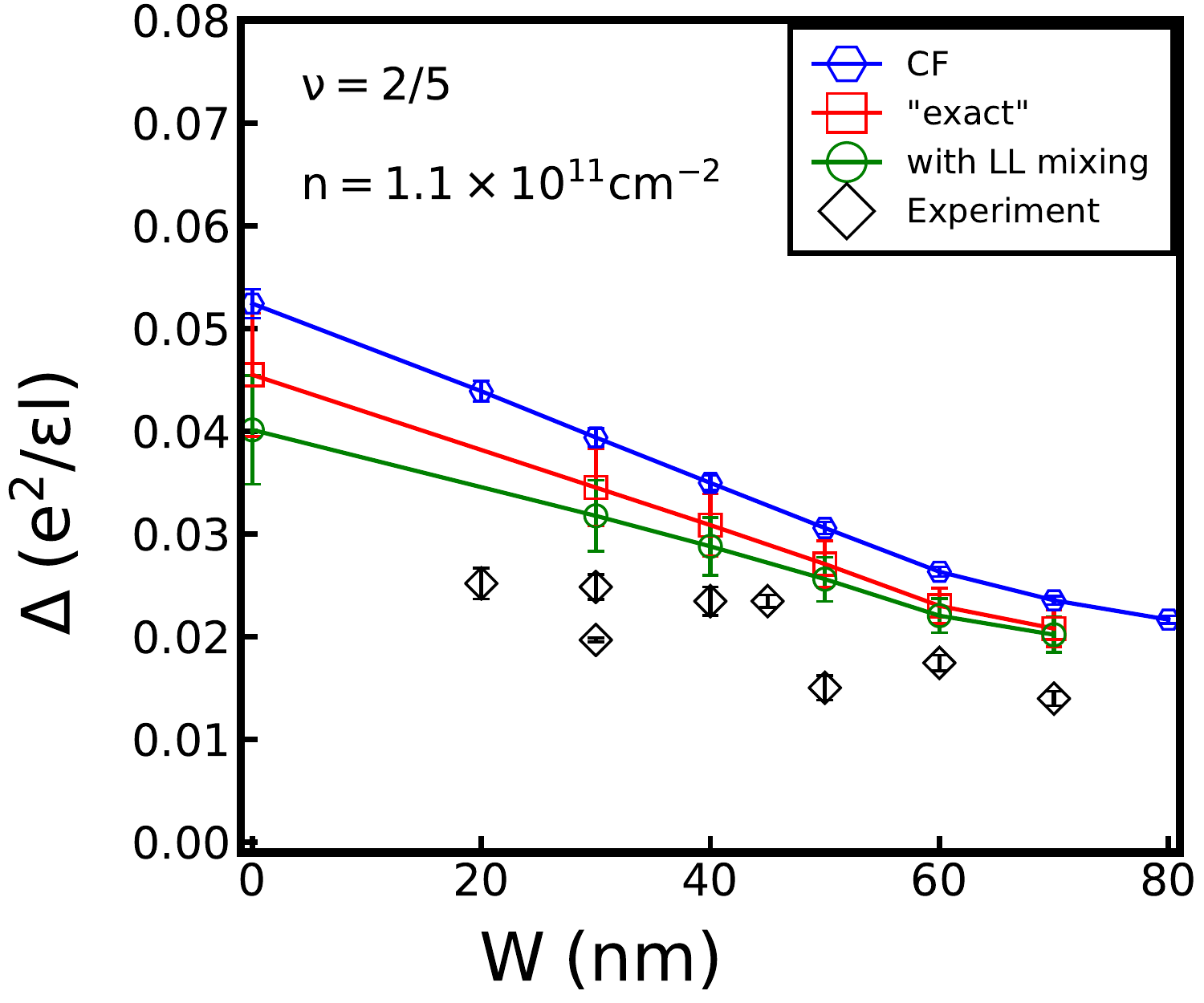}
\caption{Comparison between the theoretical and the experimental gaps for $\rho=1.1\times10^{11} \mathrm{cm}^{-2}$ at $\nu=1/3$ (top) and $2/5$ (bottom). Several different methods are used for the calculation: VMC with Jain wavefunctions; VMC gaps corrected for variational error (``exact"), as explained in the main text; LLM corrected gaps (``with LLM"), as explained in the text. Note that an insulating phase is observed in the experiment at $1/3$ for $W=80\,\mathrm{nm}$. 
}\label{Gap_theory_experiment_n11}
\end{figure}

\begin{figure}[t!]
 \begin{center}
  \includegraphics[width=\columnwidth]{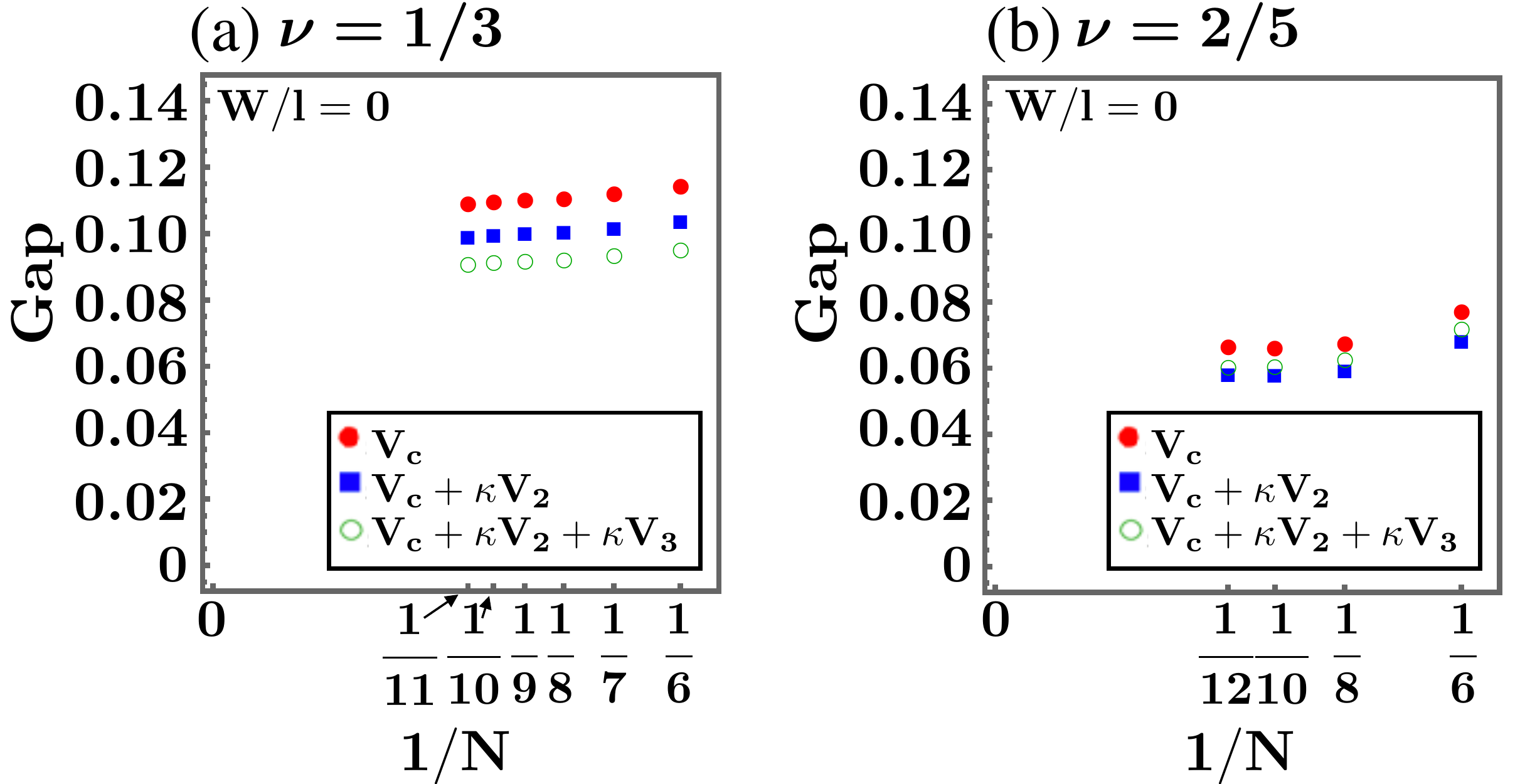} 
 \end{center}
 \caption{Energy gap [defined in Eq.~\eqref{eq: gap}] incorporating the effect of LLM [see Eq.~\eqref{eq: LLM}] as a function of the inverse of the particle number at (a) $\nu=1/3$ and (b) $\nu=2/5$ obtained from ED in the spherical geometry.}
 \label{fig:gap_ED}
\end{figure}

\begin{figure}[t!]
 \begin{center}
  \includegraphics[width=\columnwidth]{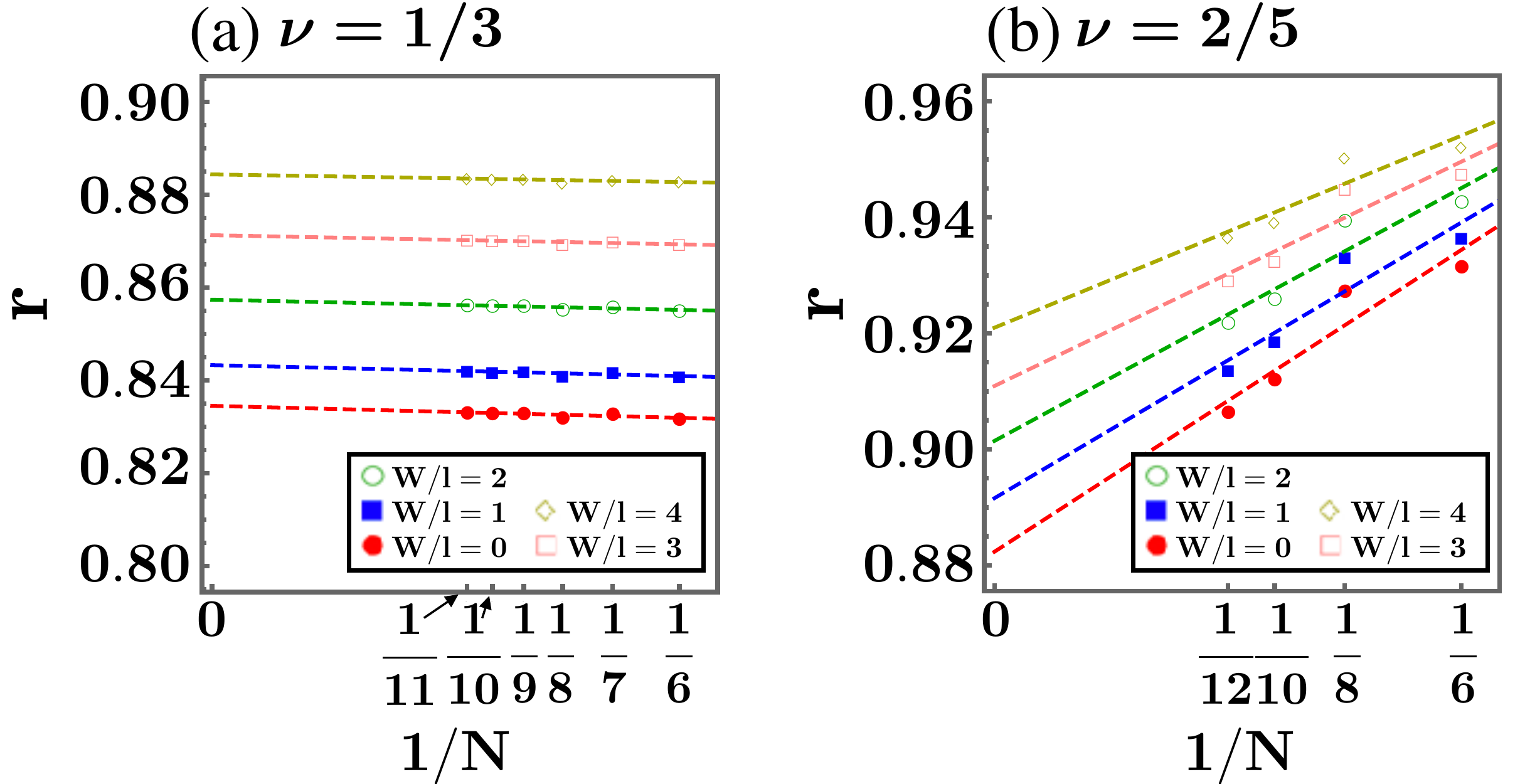} 
 \end{center}
 \caption{The reduction factor $r$ as a function of $1/N$ for (a) $\nu=1/3$ and (b) $\nu=2/5$. The reduction factor $r$ is defined as the ratio of the gap for $\hat{V}_c+\kappa[\hat{V}_2+\hat{V}_3]$ (which includes LLM) to the gap for the bare Coulomb interaction $\hat{V}_c$. The dashed 
 lines represent a linear approximation.}
 \label{fig:r_1_ED}
\end{figure}

\begin{figure}[t!]
 \begin{center}
  \includegraphics[width=\columnwidth]{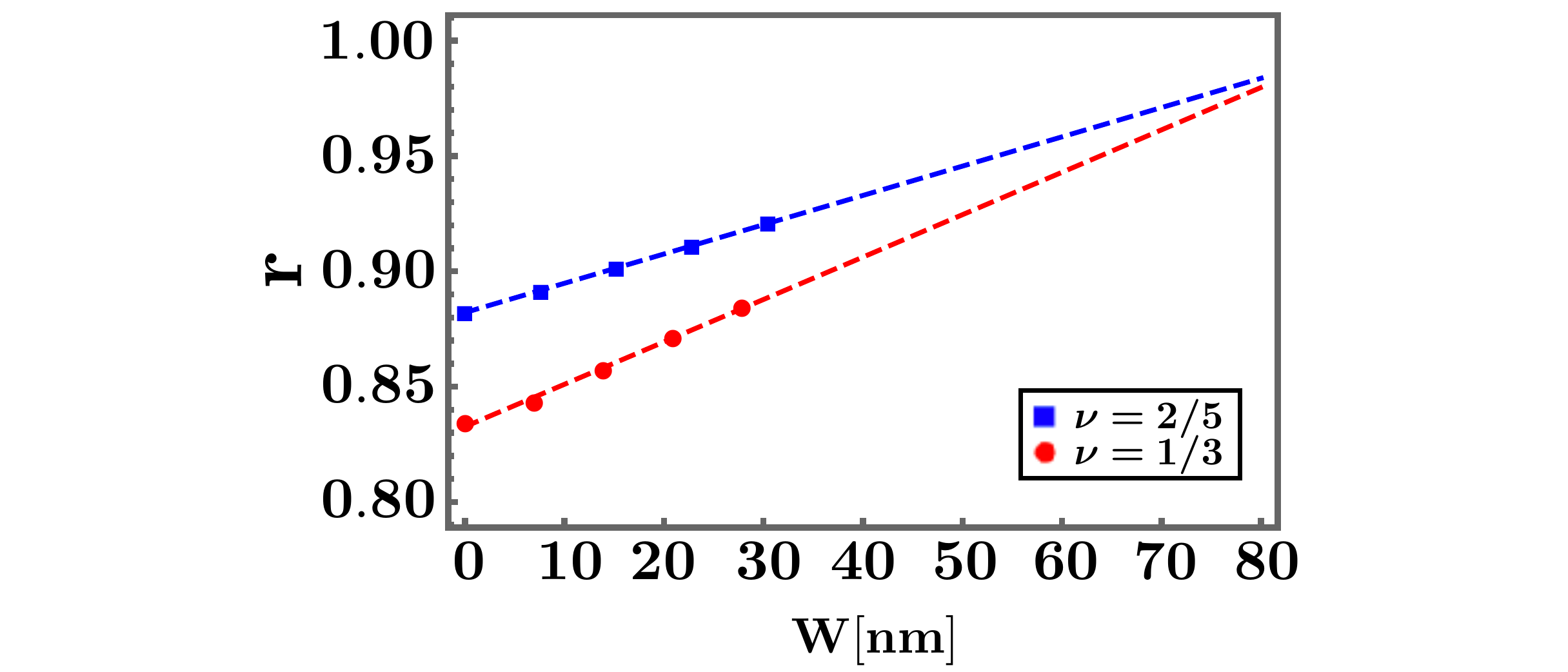} 
 \end{center}
 \caption{The reduction factor $r$ as a function of $W$, where $r$ is the factor by which the gap is reduced due to LLM. The dashed 
 lines represent a linear approximation.}
 \label{fig:r_2_ED}
\end{figure}

In Fig.~\ref{fig:gap_ED}, we plot the energy gaps at $\nu=1/3$ and $\nu=2/5$ for $W=0$ as functions of the inverse of the particle number. The energy gaps for the interaction $\hat{V}_c+\kappa[\hat{V}_2+\hat{V}_3]$ are always smaller than those for $\hat{V}_c$ at any $N$ since LLM screens the interaction. Figure~\ref{fig:r_1_ED} shows LLM reduction factor $r\equiv\Delta_{V_c+\kappa V_2+\kappa V_3}/\Delta_{V_c}$, which is the ratio of the gaps with and without LLM, as a function of $1/N$ for several quantum well widths. We deduce the value in the thermodynamic limit for each $W/l$ by linear extrapolation.  Using these data, we generate Fig.~\eqref{fig:r_2_ED} that plots the reduction factor $r$ as a 
function of $W$. Here, we set the magnetic length as $l=\sqrt{\nu/(2\pi\rho)}$ with $\rho=1.1\times10^{11}\text{cm}^{-2}$. Because the inter-electron interaction weakens with increasing width, we expect the effect of LLM to become less prominent, which is consistent with the finding that the reduction in gap decreases with increasing quantum well width. 

In the SM~\cite{SM}, we discuss an alternative approach for treating LLM, namely the fixed-phase diffusion Monte Carlo method, which has proved to be effective in dealing with the effect of LLM in the context of competition between FQHE states with different spins and also between liquid and crystal states \cite{Ortiz93, Zhao18, Zhang16, Zhao20, Hossain20a}. We believe that this method may underestimate LLM corrections to gaps because fixing the phase of the wave function limits the flexibility of the CFP and CFH wave functions. Also, this method allows a determination of LLM corrections only for zero-width; for finite widths, the thermodynamic extrapolations are not reliable.

\section{Results and discussion}
\label{sec: results}

We evaluate the gaps as follows. First, we determine the thermodynamic limits for the gaps at filling fractions $1/3$, $2/5$, $3/7$, $4/9$, and $5/11$ from the Jain wave functions for the ground states and far separated quasiparticle-quasihole pair. We then estimate the thermodynamic limit of the ``variational error," namely, the discrepancy between the gaps from trial wave functions of the CF theory and ED; this can be accomplished reliably for 1/3 and 2/5 but not for the other fractions for which the number of systems on which ED can be performed is not sufficient for a thermodynamic extrapolation. Finally, we multiply the gap by the reduction factor obtained above to include the effect of LLM.

\begin{figure}
\includegraphics[width=0.8 \columnwidth]{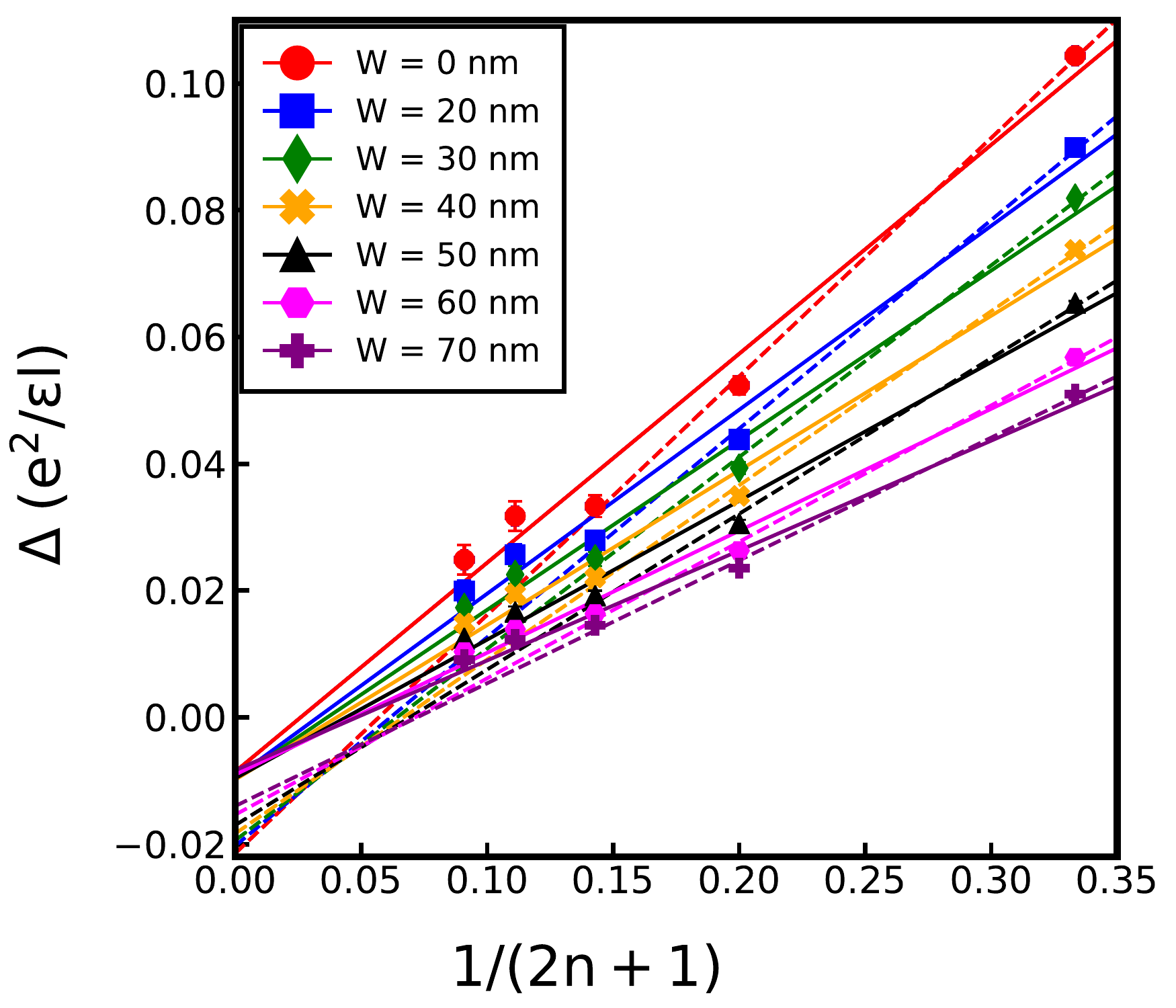}
\includegraphics[width=0.8 \columnwidth]{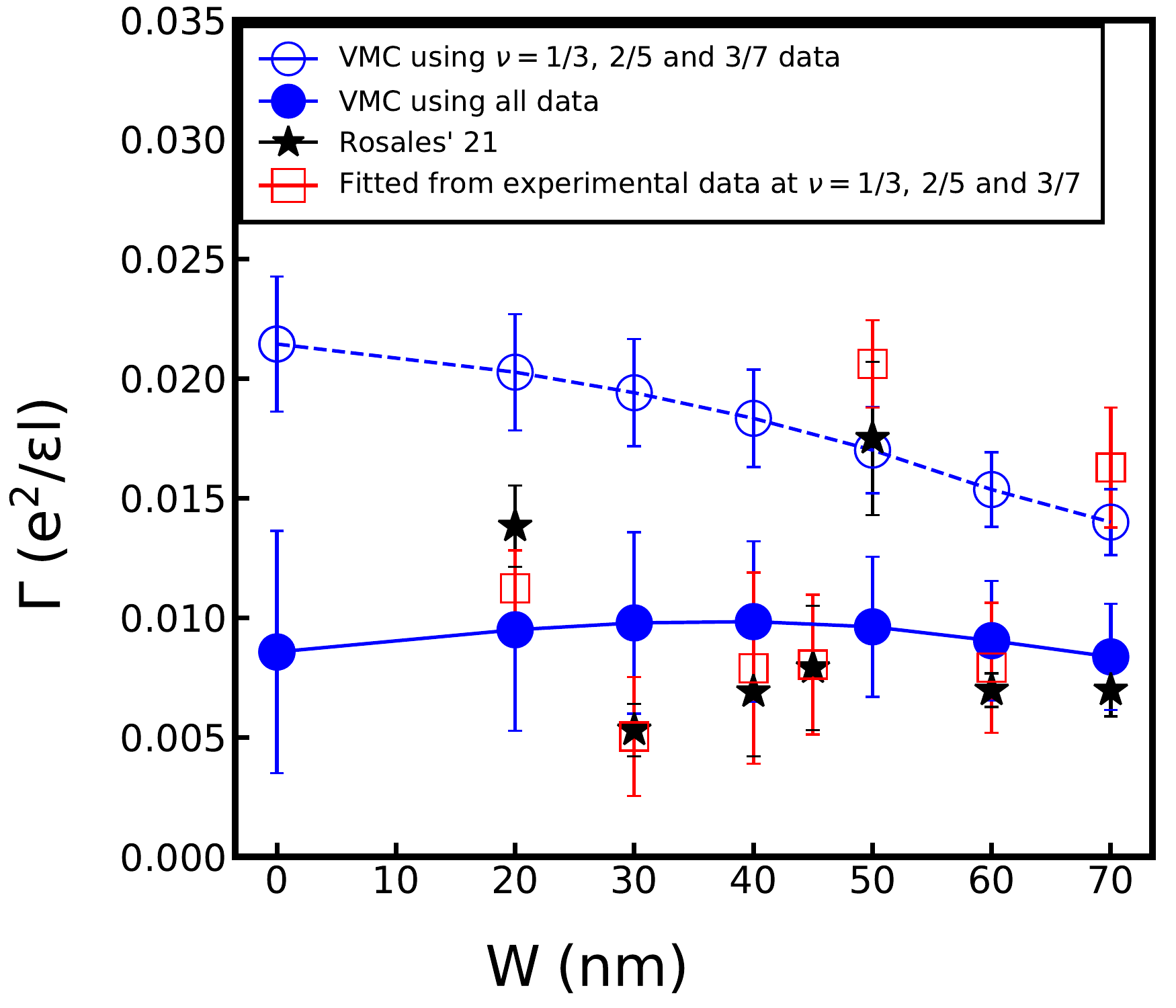}
\caption{Top panel: Theoretical transport gap $\Delta$ as a function of $1/(2n+1)$ for state in the $n/(2n+1)$ sequence in GaAs quantum wells at different widths $W$ with density $\rho = 1.1\times 10^{11}\mathrm{cm}^{-2}$. The gap is calculated using the VMC method. Different markers correspond to different well-widths. The solid lines are fits using data points at all filling factors while the dashed lines are fits of the gaps at only $\nu=1/3, 2/5$ and $3/7$. Bottom panel: Comparison between $\Gamma$ from the VMC gaps with its experimentally measured value as a function of the well width for density $\rho=1.1\times10^{11}\mathrm{cm}^{-2}$. The blue solid circles are obtained by linear regression of data points at all filling factors ($\nu=1/3, 2/5, 3/7, 4/9$ and $5/11$), which correspond to solid lines in the top plot, while the blue hollow circles are obtained by linear regression of gaps only for $\nu=1/3, 2/5$ and $3/7$, which correspond to dashed lines in the top plot. The black stars are directly taken from Fig.~5 of Ref.~\cite{Rosales21} and the red squares are obtained from a linear fit of the experimental data at $\nu=1/3, 2/5, 3/7$ in Ref.~\cite{Rosales21}.}\label{Gamma_n11}
\end{figure}

The resulting gaps for 1/3 and 2/5 states are shown in Fig.~\ref{Gap_theory_experiment_n11} as a function of the quantum well width $W$ for density $\rho = 1.1\times 10^{11}\mathrm{cm}^{-2}$. The blue symbols show the thermodynamic limits of the VMC gaps (labeled VMC).  Evidently, finite width causes the largest correction to the gap.  The red symbols are gaps corrected for the variational error. Comparisons with ED studies have shown (see overlaps shown in Refs.~\cite{Balram20b, Zhao20, SM}) that the Laughlin and Jain trial wave functions for the 1/3, 2/5, and 3/7 ground states remain very accurate even for finite widths; we find that the use of these trial wave functions overestimates the gaps by $\sim 10$\% (the estimation of this error is primarily responsible for the uncertainty in the theoretical gaps).  The green dashed line is obtained by multiplying the gaps by the reduction factor $r$ given in Fig.~\eqref{fig:r_2_ED} to include corrections due to LLM. This is the primary result of our calculation. We note that the theoretical and experimental gaps behave qualitatively similarly, and the discrepancy between them, presumably attributable to disorder, is only weakly dependent on the quantum well width. For $\nu=2/5$ the deviation between theory and experiment is of the same order as the scatter in the experimental gaps. Given various approximations in the model and the calculations, we find this level of agreement to be satisfactory.

We next come to the behavior of gaps as a function of the filling factor. In the zeroth-order approximation of noninteracting CFs, it is natural to interpret the gaps in terms of the CF cyclotron energy $\hbar  eB^*/m^*c$, where $B^*$ is the effective magnetic field sensed by CFs and $m^*$ is their mass. This suggests that the gap is proportional to $[1/( 2n\pm 1)] (e^2 / \varepsilon  l)$, where $\varepsilon$ is the dielectric constant of the background semiconductor and $ l=\sqrt{\hbar c/eB}$ is the magnetic length. This follows from the observations that for the $\nu=n/(2n\pm 1)$ state we have $B^*=B/(2n\pm 1)$, and that we must have $m^*\propto \sqrt{B}$ for the gap to be proportional to the Coulomb energy $e^2/\varepsilon  l$, the only energy scale in the absence of LLM. Experimental gaps can be fitted, approximately, to
\be
\Delta_{n/(2n\pm 1)} = {C\over 2n\pm 1} {e^2 \over \varepsilon  l}-\Gamma,
\label{eq: gap_gamma_definition}
\ee
where $C$ and $\Gamma$ are constants that are determined by the fitting. The quantity $\Gamma$ is often interpreted as a disorder-induced broadening of CF-LLs, known as $\Lambda$ levels ($\Lambda$Ls). Many experiments have reported values for $\Gamma$~\cite{Du93, Rosales21}. 

\begin{figure}
\includegraphics[width=0.9 \linewidth]{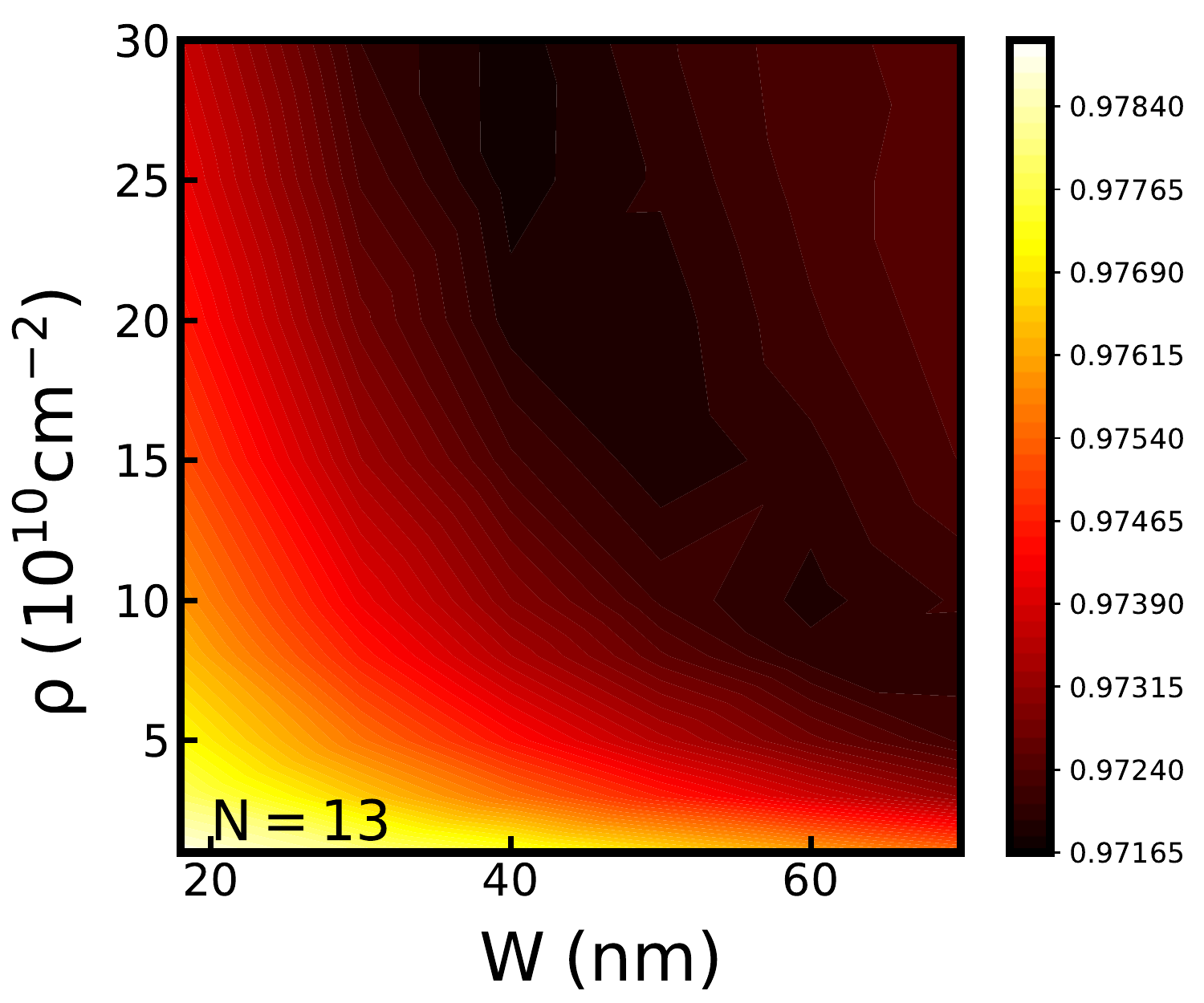}
\includegraphics[width=0.9 \linewidth]{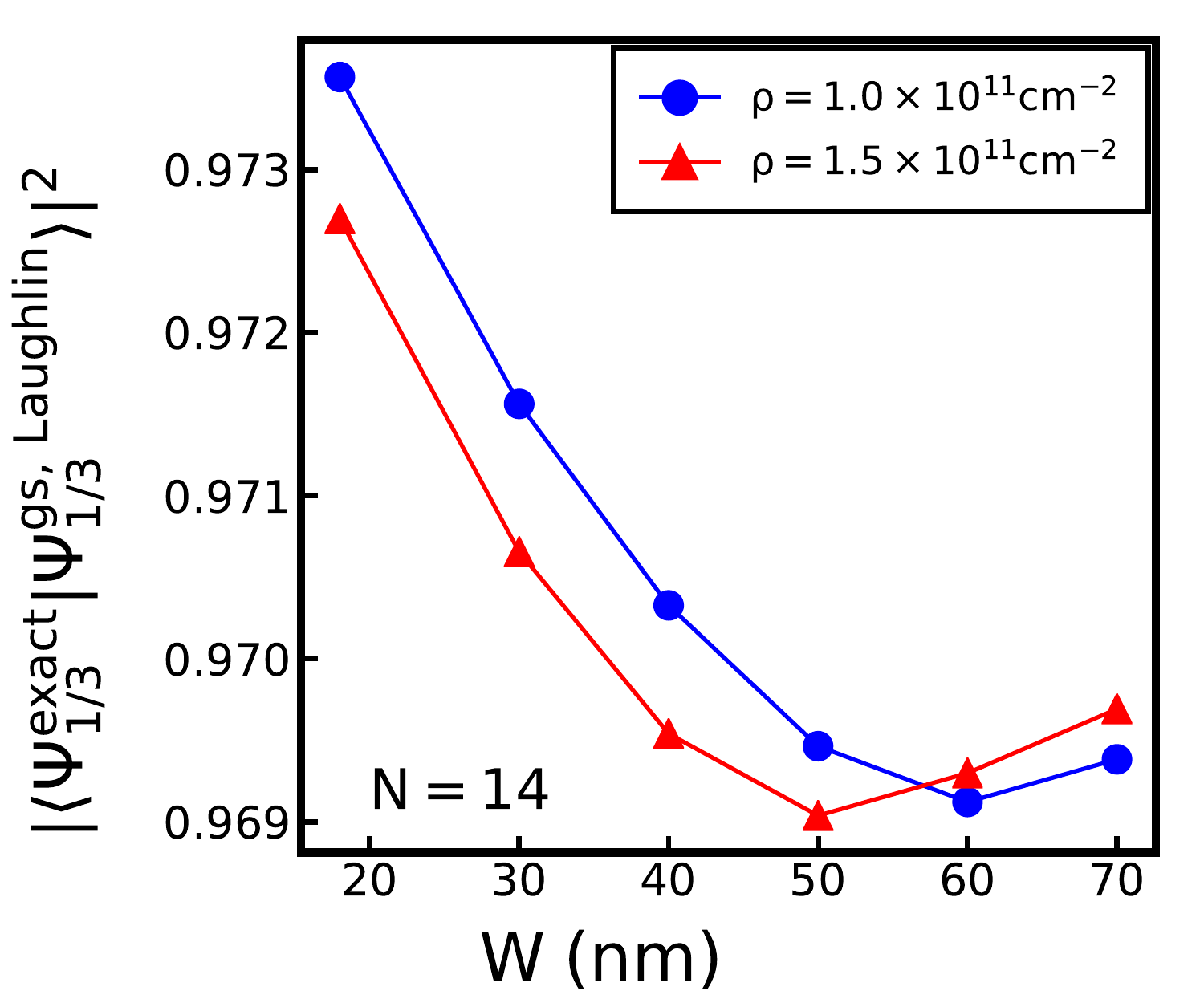}
\caption{Squared overlaps of the exact ground states with the Laughlin state at $\nu=1/3$ for $N=13$ (upper panel) and $N=14$ (lower panel) electrons in the spherical geometry. The plot for $N=14$ only includes data at $\rho=1.0\times10^{11} \text{cm}^{-2}$ and $\rho=1.5\times10^{11} \text{cm}^{-2}$. The exact ground states are evaluated using the pseudopotentials of the finite-width interaction. }
\label{fig: overlap}
\end{figure}

Villegas Rosales {\it et al.} estimate $\Gamma$ in Eq.~\eqref{eq: gap_gamma_definition} to be in the range $0.005 - 0.02~e^2/\varepsilon l$~\cite{Rosales21}, with the precise value depending on the quantum well width. Traditionally,  a nonzero $\Gamma$ has been attributed to the disorder-induced broadening of the $\Lambda$ levels~\cite{Du93}.  In Fig.~\ref{Gamma_n11} we plot the theoretical VMC gaps for $1/3$, $2/5$, $3/7$, $4/9$, and $5/11$ states without including the effects of the disorder, LLM, or variational error (which are difficult to estimate for the higher-order fractions). We find that if we attempt a linear fit, the gaps are approximately consistent with Eq.~\eqref{eq: gap_gamma_definition} with a ~\emph{nonzero} $\Gamma$ for typical widths of the quantum well (a similar behavior was noted in Ref.~\cite{Park99b}, but with a less realistic treatment of finite width). This value of $\Gamma$ depends on the range of fractions used for the fit; we show fits using the gaps at 1/3, 2/5, and 3/7 (which are known more precisely), as well as fits using all of the gaps. The resulting $\Gamma$ is in the same range as that seen experimentally~\cite{Rosales21}.  Moreover, in the experiments of Rosales {\it et al.}~\cite{Rosales21}, there is no clear correlation between the measured $\Gamma$ and the sample mobility which characterizes the disorder strength further suggesting that $\Gamma$ may not be attributable solely to disorder.  The implications of these results to the CF mass, which is related to the gaps, are discussed in the SM~\cite{SM}, which also includes many other details.

One may also ask how robust the FQHE states are as the width is increased. The ED study in Ref.~\cite{Zhao20} has shown that the overlap of the $N=12$ exact state with the Laughlin wave function at $1/3$ is very close to unity even in wide quantum wells. In this paper, we extend their result by calculating the overlap between the exact ground state and the Laughlin state for $N=13, 14$ at $\nu=1/3$ for quantum well widths ranging from $0$ to $70$ nm and carrier densities ranging from $10^{10} \text{cm}^{-2}$ to $3\times 10^{11} \text{cm}^{-2}$. These overlaps are shown in Fig.~\ref{fig: overlap}.  We find that for these larger systems too, the $\nu=1/3$ Laughlin wave function provides an almost exact representation of the ground states for the entire range of widths and densities considered in our work. This indicates, theoretically, that this state survives until a first-order transition takes place into a compressible liquid or crystal bilayer state~\cite{Scarola01b, Faugno18} (experimentally to an insulating state~\cite{Suen94b}). This is consistent with the rather sudden collapse of the 1/3 state observed experimentally as a function of the quantum well width~\cite{Rosales21}. We expect that the same remains true for other prominent FQHE states.

\begin{figure}[t]
	\includegraphics[width=\columnwidth]{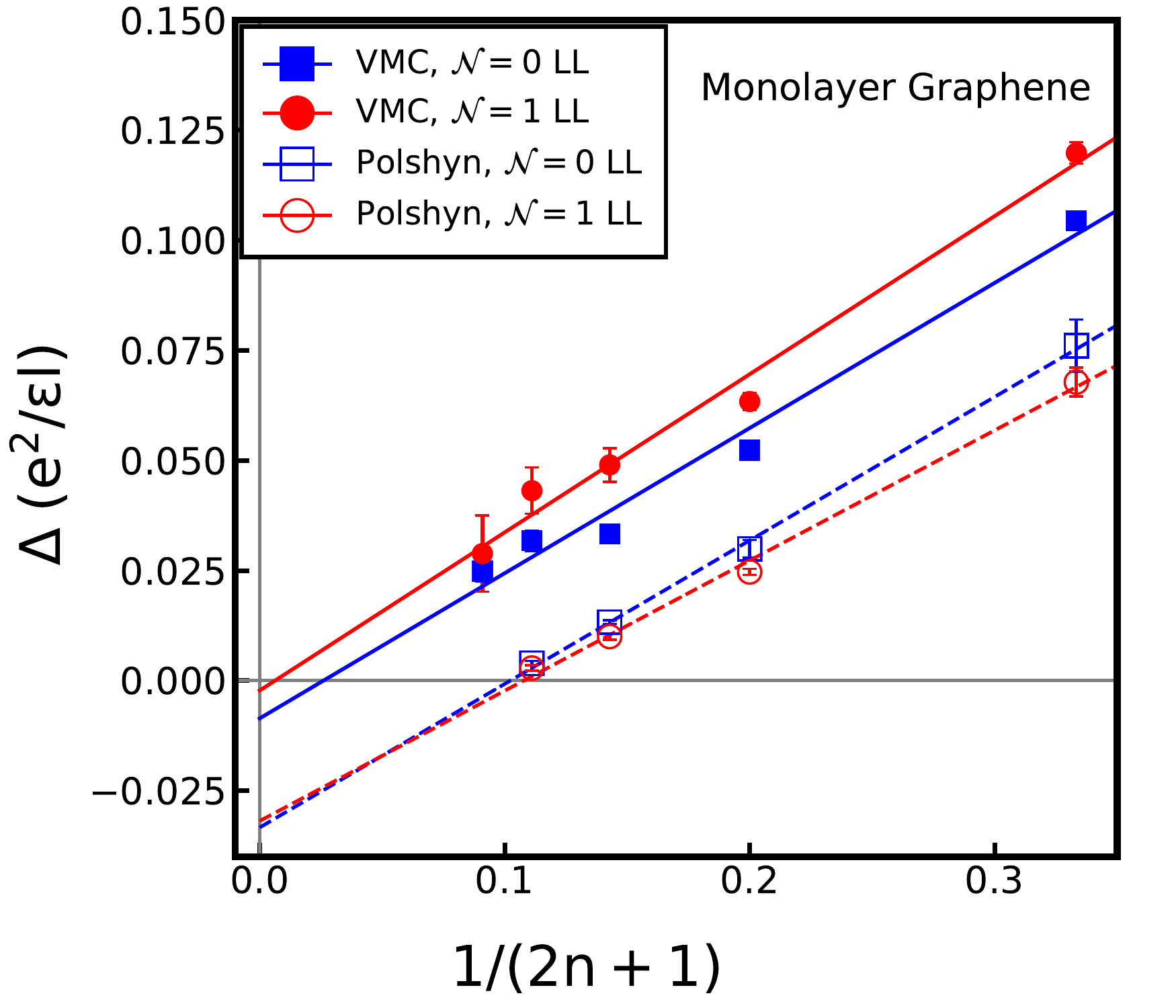}
	\caption{Theoretical transport gap $\Delta$ as a function of $1/(2n+1)$ for states at filling factors $|\nu|=n/(2n+1)$ sequence in the $\mathcal{N}=0$ and $\mathcal{N}=1$ LLs of graphene. The solid blue squares and red circles are activation gaps obtained from the VMC method without including the effects of LLM. The hollow symbols are the experimental gaps measured by Polshyn {\it et al.}~\cite{Polshyn18}.  The solid lines are fitted from the VMC gap data at all filling factors ($1/3, 2/5, 3/7, 4/9$ and $5/11$), and the dashed lines are fitted from the experimental gaps at all filling factors ($1/3, 2/5, 3/7$ and $4/9$).}
	\label{fig_graphene_Gamma}
\end{figure}

Finally, we ask how these considerations apply to FQHE in graphene.  A similar analysis has been performed for the gaps in monolayer graphene by Polshyn {\em et al.}~\cite{Polshyn18}. They find that $\Gamma$ is larger by a factor of $\sim 3$ than that seen in GaAs quantum wells. The measured gaps are shown in Fig.~\ref{fig_graphene_Gamma}. [We show the gaps at $\nu=-1+n/(2n+1)$ in the $\mathcal{N}=0$ graphene LL, and the gaps at $\nu=-3+n/(2n+1)$ in the $\mathcal{N}=1$ graphene LL, because these are the largest experimental gaps.] For the $\mathcal{N}=0$ graphene LL, the theoretical gaps are identical to those in the LLL of GaAs in the zero-width limit, when LLM is neglected~\cite{Balram15a}. Earlier studies have demonstrated that the CF theory is quantitatively accurate in the $\mathcal{N}=1$ graphene LL~\cite{Balram15c, Balram21b}. We have evaluated the VMC activation gaps for the FQHE states at $\nu=n/(2n+1)$ in the $\mathcal{N}=1$ LL of graphene using the effective interaction given in Ref.~\cite{Balram15c} (see SM). The results are also shown in Fig.~\ref{fig_graphene_Gamma}. It is not known at present how much LLM and disorder contribute to the observed value of $\Gamma$ in experiments. 

In summary, motivated by the recent experimental study of gaps of various FQHE states in extremely high mobility systems,  we have evaluated the excitation gaps including the effects of finite width and LLM. The theoretical gaps are in qualitative and semi-quantitative agreement with the experimental gaps, but some discrepancy remains, presumably due to disorder.

\textit{Acknowledgement}: We are grateful to Mansour Shayegan for many insightful discussions and for raising the questions that motivated this work and to G. J. Sreejith for help with interaction pseudopotentials. T.Z., W.N.F., and J.K.J. acknowledge financial support from the U.S. Department of Energy under Award No. DE-SC-0005042. K.K. thanks JSPS for support from Overseas Research Fellowship. A.C.B.  acknowledges the Science and Engineering Research Board (SERB) of the Department of Science and Technology (DST) for funding support via the Start-up Grant SRG/2020/000154. The numerical calculations were performed using Advanced CyberInfrastructure computational resources provided by The Institute for CyberScience at The Pennsylvania State University and the Nandadevi supercomputer, which is maintained and supported by the Institute of Mathematical Science's High-Performance Computing Center.  Some of the computations were performed using the  AQILA~\cite{Martin20} and DiagHam packages, for which we are grateful to its authors.

\pagebreak

\setcounter{figure}{0}
\setcounter{equation}{0}
\setcounter{section}{0}
\setcounter{table}{0}
\renewcommand\thesection{S\arabic{section}}
\renewcommand\thefigure{S\arabic{figure}}
\renewcommand\thetable{S\arabic{table}}
\renewcommand\theequation{S\arabic{equation}}

\begin{widetext}
{\LARGE {\bf Supplemental Material for ``Revisiting excitation gaps in the fractional quantum Hall effect"}}
\end{widetext}

In the Supplemental Material, we present details of (i) the wave functions we use to evaluate the gaps; (ii) variational Monte Carlo (VMC) results for a broad range of densities, quantum well widths and filling factors;  (iii) estimation of the corrections from the use of variational wave functions as well as from Landau level (LL) mixing (LLM); (iv) estimate of the composite fermion (CF) masses from the gaps; and (v) VMC results for gaps in the first excited LL of graphene (vi) overlaps of the 1/3 Laughlin wave function with the exact ground state for large systems.

\section{Trial wavefunctions}
\label{X_CF_WF}
For the VMC calculations, we use the Jain wave functions projected into the lowest Landau level (LLL) using the Jain-Kamilla (JK) method\cite{Jain97,Jain98,Jain07}. The ground state wave function at $\nu=n/(2n+1)$ is given by:
\begin{widetext}
\begin{equation}
\begin{aligned}
\Psi^\text{gs, Jain}_{n/(2n+1)}\left(\left\{\vec{\Omega}\right\}\right)=
\begin{vmatrix}
\hat{Y}_{q,q,-q}\left(\Omega_1\right) \mathcal{J}_1& \hat{Y}_{q,q,-q}\left(\Omega_2\right) \mathcal{J}_2& ... &\hat{Y}_{q,q,-q}\left(\Omega_N\right) \mathcal{J}_N\\
\hat{Y}_{q,q,-q+1}\left(\Omega_1\right) \mathcal{J}_1& \hat{Y}_{q,q,-q+1}\left(\Omega_2\right) \mathcal{J}_2& ...&\hat{Y}_{q,q,-q+1}\left(\Omega_N\right) \mathcal{J}_N\\
\vdots                                  &\vdots                                   &\vdots    &\vdots                                 \\
\hat{Y}_{q,q,q}\left(\Omega_1\right) \mathcal{J}_1& \hat{Y}_{q,q,q}\left(\Omega_2\right) \mathcal{J}_2& ...&\hat{Y}_{q,q,q}\left(\Omega_N\right) \mathcal{J}_N\\
\hat{Y}_{q,q+1,-(q+1)}\left(\Omega_1\right) \mathcal{J}_1& \hat{Y}_{q,q+1,-(q+1)}\left(\Omega_2\right) \mathcal{J}_2& ...&\hat{Y}_{q,q+1,-(q+1)}\left(\Omega_N\right) \mathcal{J}_N\\
\vdots                                  &\vdots                                   &\vdots    &\vdots                                 \\
\hat{Y}_{q,q+n-1,-(q+n-1)}\left(\Omega_1\right) \mathcal{J}_1& \hat{Y}_{q,q+n-1,-(q+n-1)}\left(\Omega_2\right) \mathcal{J}_2& ...&\hat{Y}_{q,q+n-1,-(q+n-1)}\left(\Omega_N\right) \mathcal{J}_N\\
\vdots                                  &\vdots                                   &\vdots    &\vdots                                 \\
\hat{Y}_{q,q+n-1,(q+n-1)}\left(\Omega_1\right) \mathcal{J}_1& \hat{Y}_{q,q+n-1,(q+n-1)}\left(\Omega_2\right) \mathcal{J}_2& ...&\hat{Y}_{q,q+n-1,(q+n-1)}\left(\Omega_N\right) \mathcal{J}_N\\
\end{vmatrix}
\end{aligned}
\end{equation}
\end{widetext}
where $\hat{Y}_{q,\ell,\ell_z}\left(\Omega\right)$ denote the operators of LLL-projected monopole harmonics  (see Appendix~J of Ref.~\cite{Jain07} for details), $q$ is the effective magnetic monopole strength of CFs that satisfies $q=Q-(N-1)$, $\Omega\equiv(\theta,\phi)$ is the angular coordinate on the sphere, $\ell$ and $\ell_z$ denote the quantum numbers of the orbital angular momentum and its $z$-component, and $\mathcal{J}_i=\sum_{j\neq i}{u_i v_j-u_j v_i}$ is the Jastrow factor, where $u=\cos(\theta/2)e^{i\phi/2}, v=\sin(\theta/2)e^{-i\phi/2}$ are spinor variables\cite{Haldane83}.
The largest-$L$ CF-exciton wave function is given by:
\begin{widetext}
\begin{equation}
\begin{aligned}
\Psi^\text{ex, Jain}_{n/(2n+1)}\left(\left\{\vec{\Omega}\right\}\right)=
\begin{vmatrix}
\hat{Y}_{q,q,-q}\left(\Omega_1\right) \mathcal{J}_1& \hat{Y}_{q,q,-q}\left(\Omega_2\right) \mathcal{J}_2& ... &\hat{Y}_{q,q,-q}\left(\Omega_N\right) \mathcal{J}_N\\
\hat{Y}_{q,q,-q+1}\left(\Omega_1\right) \mathcal{J}_1& \hat{Y}_{q,q,-q+1}\left(\Omega_2\right) \mathcal{J}_2& ...&\hat{Y}_{q,q,-q+1}\left(\Omega_N\right) \mathcal{J}_N\\
\vdots                                  &\vdots                                   &\vdots    &\vdots                                 \\
\hat{Y}_{q,q,q}\left(\Omega_1\right) \mathcal{J}_1& \hat{Y}_{q,q,q}\left(\Omega_2\right) \mathcal{J}_2& ...&\hat{Y}_{q,q,q}\left(\Omega_N\right) \mathcal{J}_N\\
\hat{Y}_{q,q+1,-(q+1)}\left(\Omega_1\right) \mathcal{J}_1& \hat{Y}_{q,q+1,-(q+1)}\left(\Omega_2\right) \mathcal{J}_2& ...&\hat{Y}_{q,q+1,-(q+1)}\left(\Omega_N\right) \mathcal{J}_N\\
\vdots                                  &\vdots                                   &\vdots    &\vdots                                 \\
\hat{Y}_{q,q+n-1,-(q+n-1)}\left(\Omega_1\right) \mathcal{J}_1& \hat{Y}_{q,q+n-1,-(q+n-1)}\left(\Omega_2\right) \mathcal{J}_2& ...&\hat{Y}_{q,q+n-1,-(q+n-1)}\left(\Omega_N\right) \mathcal{J}_N\\
\vdots                                  &\vdots                                   &\vdots    &\vdots                                 \\
\hat{Y}_{q,q+n,q+n}\left(\Omega_1\right) \mathcal{J}_1& \hat{Y}_{q,q+n,q+n}\left(\Omega_2\right) \mathcal{J}_2& ...&\hat{Y}_{q,q+n,q+n}\left(\Omega_N\right) \mathcal{J}_N\\
\end{vmatrix}
\end{aligned}
\end{equation}
\end{widetext}

\section{Variational Monte Carlo results}
\label{SM_sec:VMC}

\begin{figure*}[ht!]
	\includegraphics[width=0.32\linewidth]{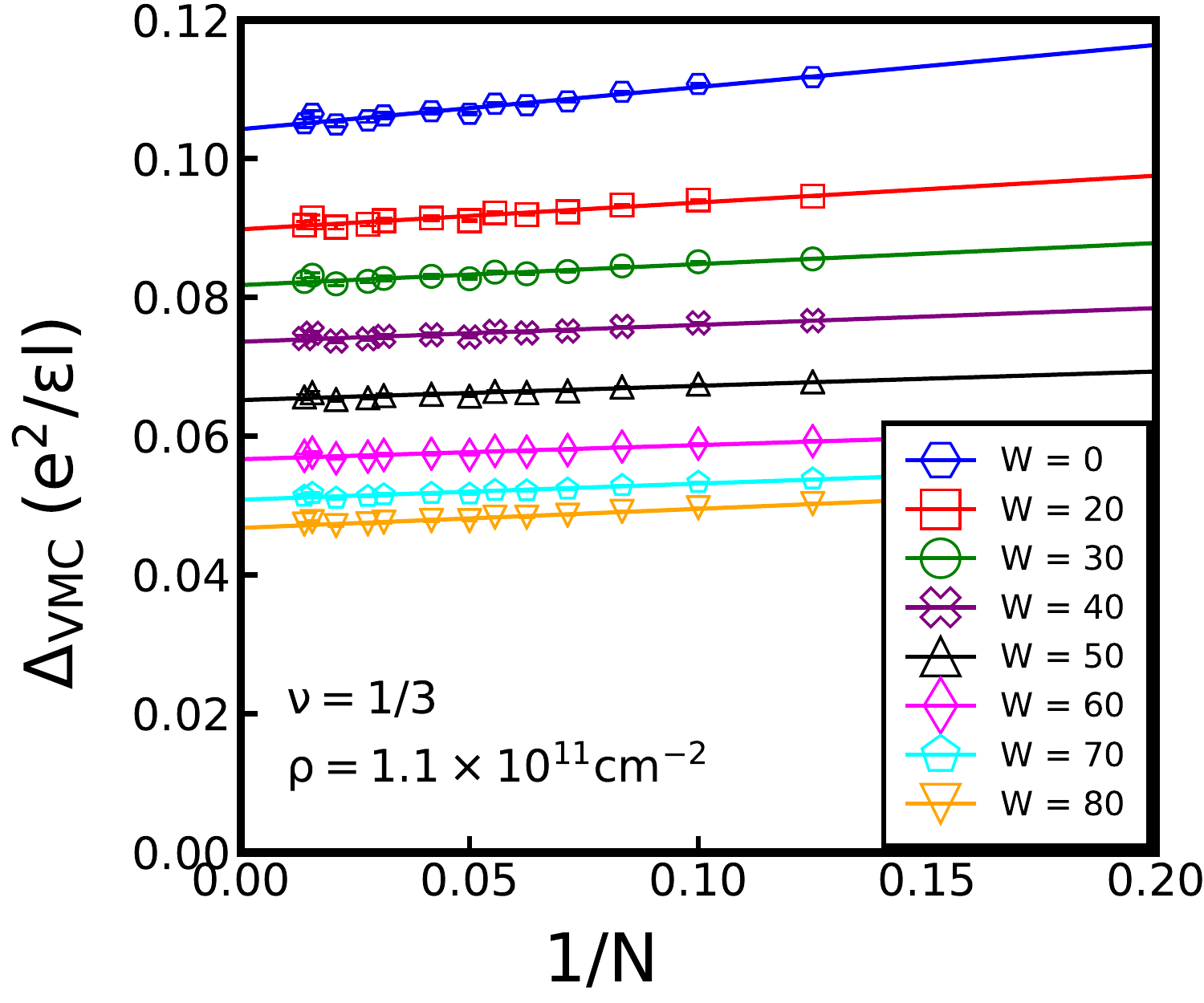}
	\includegraphics[width=0.32\linewidth]{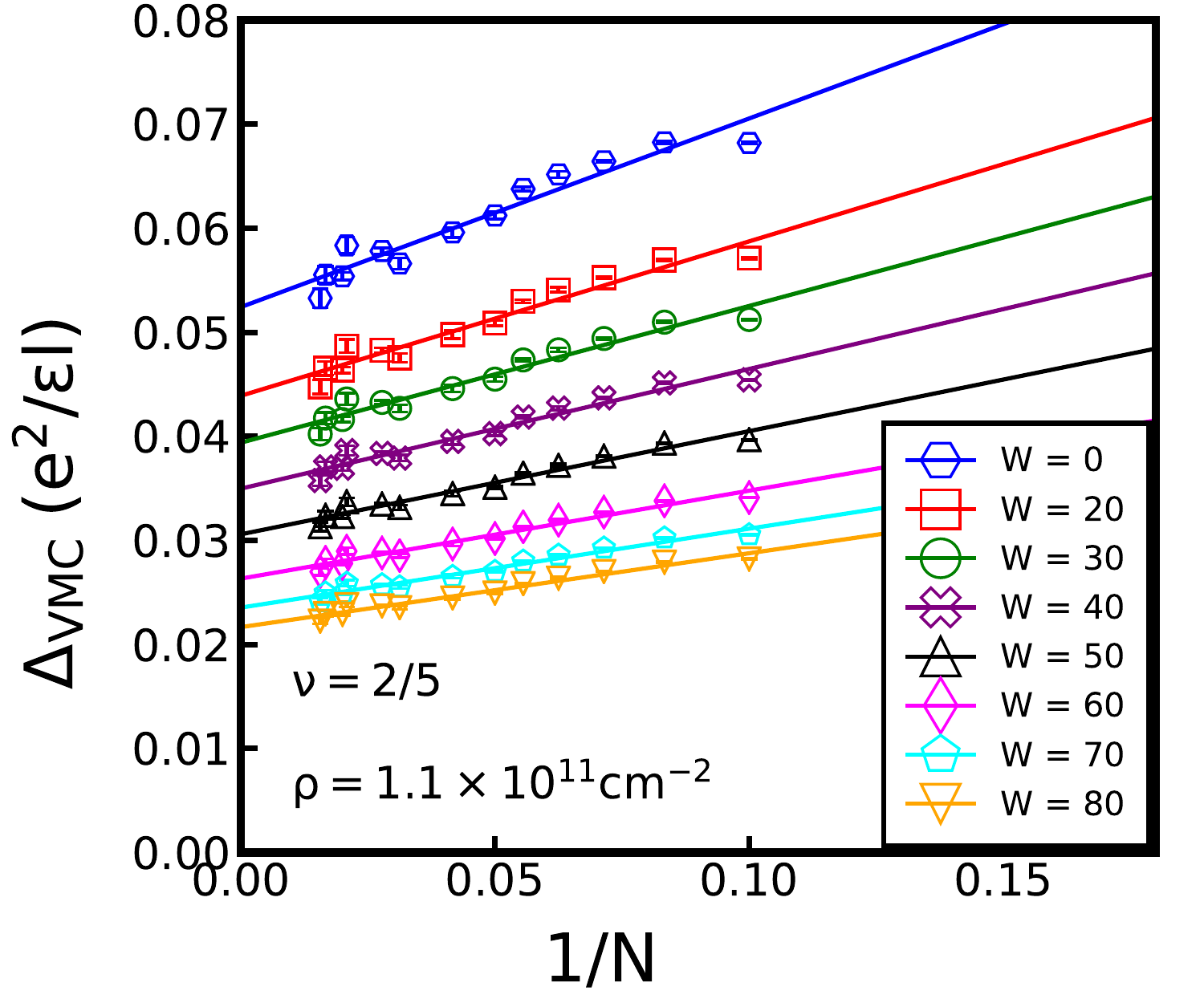}
	\includegraphics[width=0.32\linewidth]{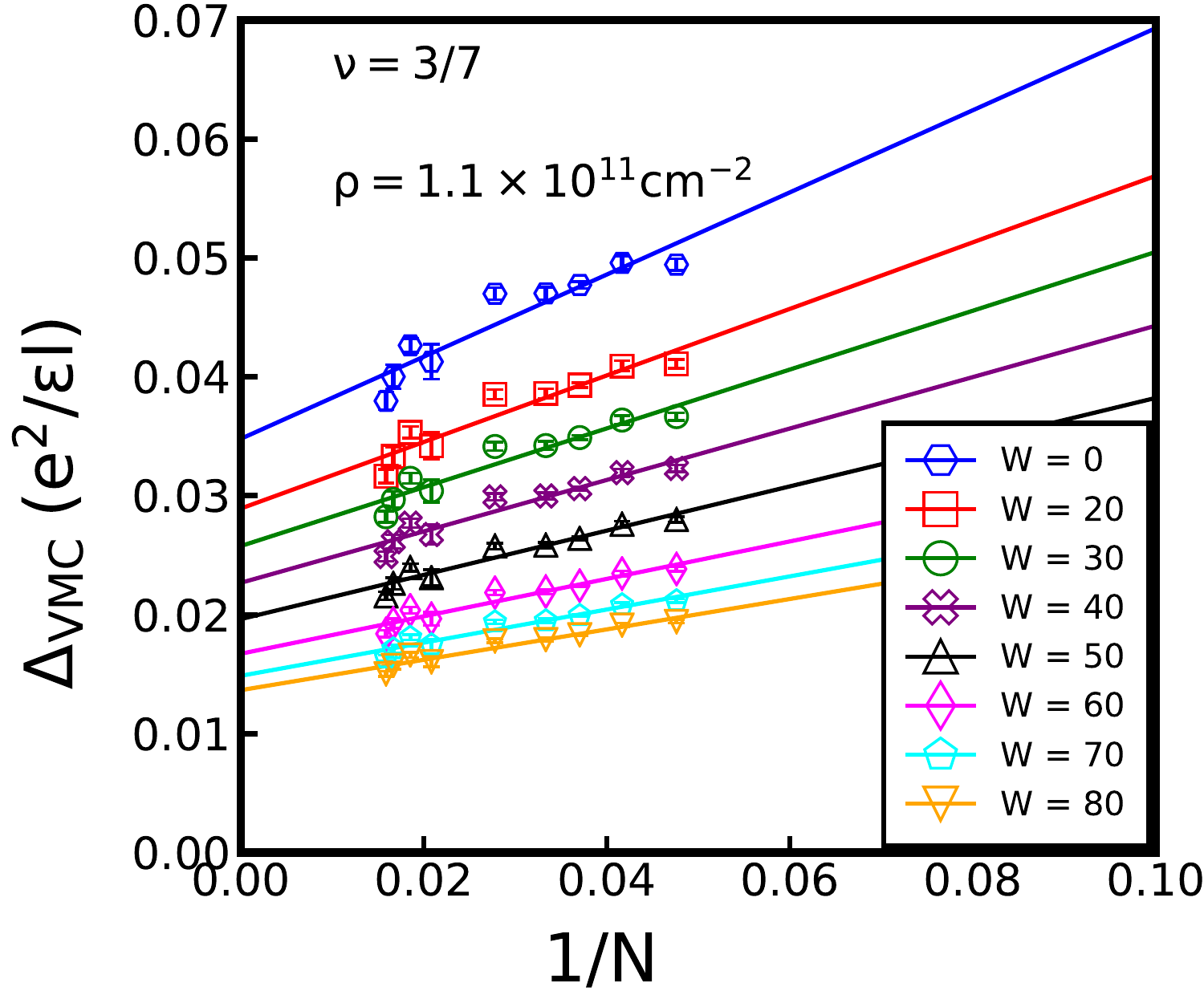}
	\caption{Thermodynamic extrapolation of the transport gap [defined in Eq.~(3) of the main text] calculated by the variational Monte Carlo (VMC) method in finite-width quantum wells at the density $\rho=1.1\times10^{11} \mathrm{cm}^{-2}$ at filling factors $1/3$ (left), $2/5$ (middle) and $3/7$ (right). The effect of the finite thickness of the quantum well is accounted for by using the effective Coulomb interaction based on a transverse distribution calculated using a local density approximation (LDA) [see Eq.~(4) of the main text].}\label{VMC_extrap_n11}
\end{figure*}

In this section, we show our VMC results for various filling factors, densities and well-widths. We first show the finite-size results and the thermodynamic extrapolations of the gaps obtained from the VMC for $\nu=1/3$, $2/5$ and $3/7$ at the density $\rho=1.1\times10^{11}~\text{cm}^{-2}$ (same as that of the sample used in Ref.~\cite{Rosales21}) in Fig.~\ref{VMC_extrap_n11}. The largest system size for which we evaluated gaps is $N=72$ for $\nu=1/3$, $N=64$ for $\nu=2/5$,  and $N=63$ for $\nu=3/7$. The linear regression fits the data points well at both $\nu=1/3$ and $2/5$. For the same density of $\rho=1.1\times10^{11}\text{cm}^{-2}$, a comparison between the calculated and the measured gaps of the $1/3$ and $2/5$ FQHE is shown in Fig.~1 of the main text. Here we also show the results for the $3/7$ FQHE in Fig.~\ref{X_fig:VMC_ED_37_n11}. Our results are largely consistent with those of Refs.~\cite{Park99b, Balram16d} with some differences arising from different choices of system sizes and the parameters used in the LDA calculations.

Figs.~\ref{X_fig:CF13_VMC_extrap},\ref{X_fig:CF25_VMC_extrap},\ref{X_fig:CF37_VMC_extrap},\ref{X_fig:CF49_VMC_extrap},\ref{X_fig:CF511_VMC_extrap} show the thermodynamic extrapolations of VMC gaps at $\nu=1/3,2/5,3/7,4/9,5/11$ for various densities and well-widths, and Figs.~\ref{X_fig:VMC_13_other},\ref{X_fig:VMC_25_other},\ref{X_fig:VMC_37_other},\ref{X_fig:VMC_49_other},\ref{X_fig:VMC_511_other} show the theoretical gaps in the thermodynamic limit as functions of the quantum well width for $\nu=1/3, 2/5, 3/7, 4/9, 5/11$. In these figures, we have also plotted the estimates of the exact gaps (labeled as ``exact''), which account for the variational error in the Laughlin/Jain wave functions (see Sec.~\ref{SM_subsec: variational} for explanation).

We also plot $\Delta_\text{VMC}$ against $1/(2n+1)$ at various densities in Fig.~\ref{X_fig:Gamma_extrap_fig}. The $\Gamma$ fitted from Fig.~\ref{X_fig:Gamma_extrap_fig} as a function of the well-width is shown in Fig.~\ref{X_fig:Gamma_thermo_fig}.

\section{Corrections to the variational Monte Carlo results}
In this section we discuss corrections to the VMC results that arise from i) use of trial wave functions (Sec.~\ref{SM_subsec: variational}) and ii) LL mixing (Sec.~\ref{SM_subsec: LLM}). In Sec.~\ref{SM_subsec:ED} we present the transport gaps estimated from exact diagonalization (ED) studies on small-systems. 

\subsection{Variational Error}
\label{SM_subsec: variational}

Although our trial wave functions provide an extremely accurate representation of the Coulomb eigenstate, they are not exact. To estimate the variational error stemming from the use of these trial wave functions, we compare the VMC gaps with the exact gaps for system sizes that are accessible to ED. In general, the deviation between the VMC gap, $ \Delta_\text{VMC}$, and the exact gap, $\Delta_\text{ED}$, depends on the system size $N$. 

For example, at $\rho=10^{11}~\text{cm}^{-2}$, VMC overestimates the transport gaps by about $0.001~e^2/\varepsilon l$ for $N=8$ at $\nu=1/3$, but the deviation rises to about $0.002~e^2/\varepsilon l$ when $N=12$. We extrapolate the deviation to the thermodynamic limit, as seen in Fig.~\ref{X_fig:ED_correction_13}. 
To estimate $\lim\limits_{N\to\infty}(\Delta_\text{ED} -\Delta_\text{VMC})$, we take an average of linear and quadratic extrapolations in $1/N$. We call this quantity ``the variational error." 
We have obtained variational errors only for some discrete values of the density; the variational errors at other densities, e.g. at $\rho=1.1\times10^{11}\text{cm}^{-2}$, are obtained by interpolation. 

\begin{figure*}[ht!]
	\includegraphics[width=\columnwidth]{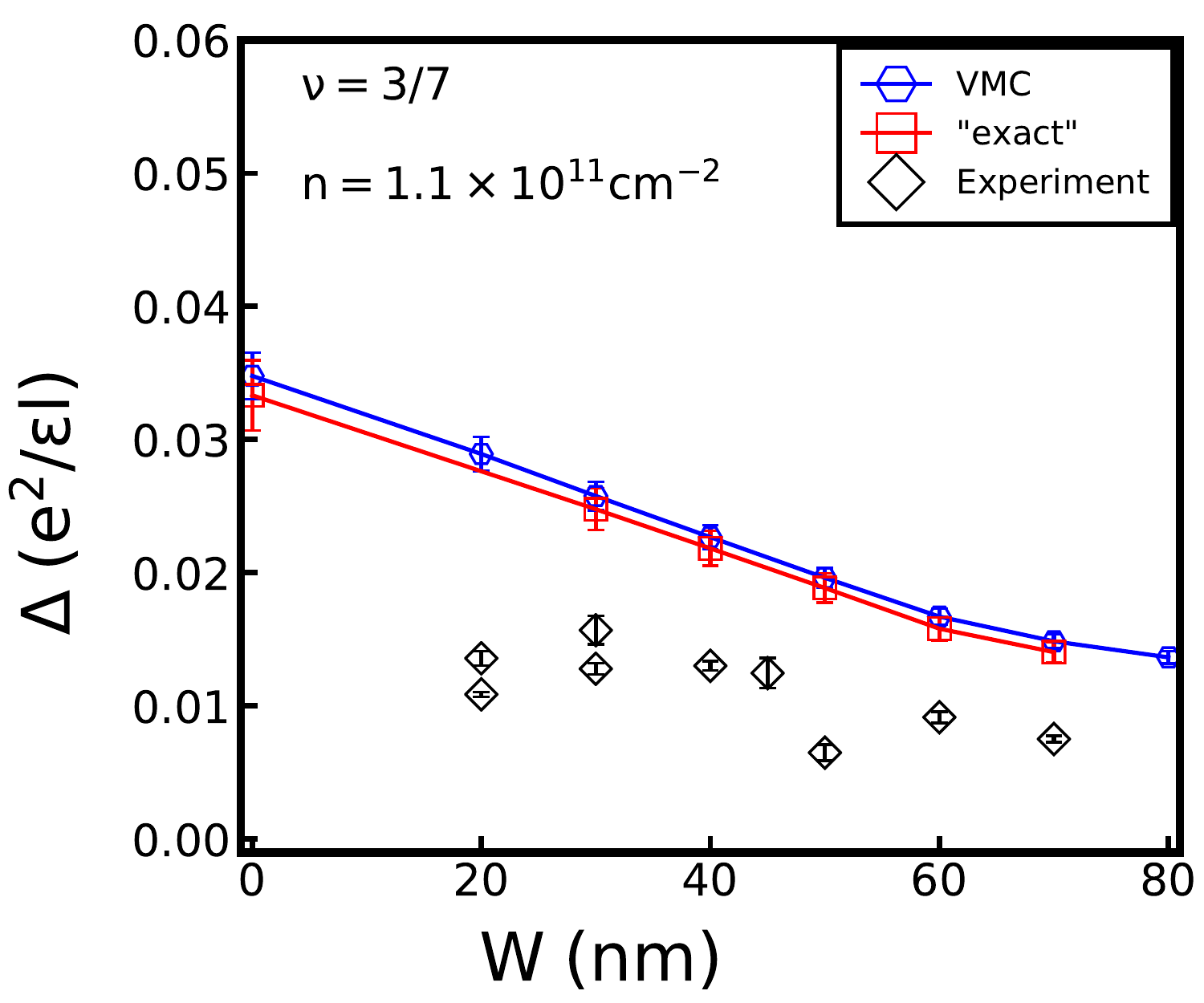}
	\caption{Comparison between the theoretical and experimental gaps at $\rho=1.1\times10^{11} \mathrm{cm}^{-2}$ for the $\nu=3/7$ FQHE. The blue symbols are variational Monte Carlo gaps (VMC) and the red symbols (``exact") include variational correction, as explained in the main text.}
	\label{X_fig:VMC_ED_37_n11}
\end{figure*}

\begin{figure*}[ht!]
	\includegraphics[width=0.32 \linewidth]{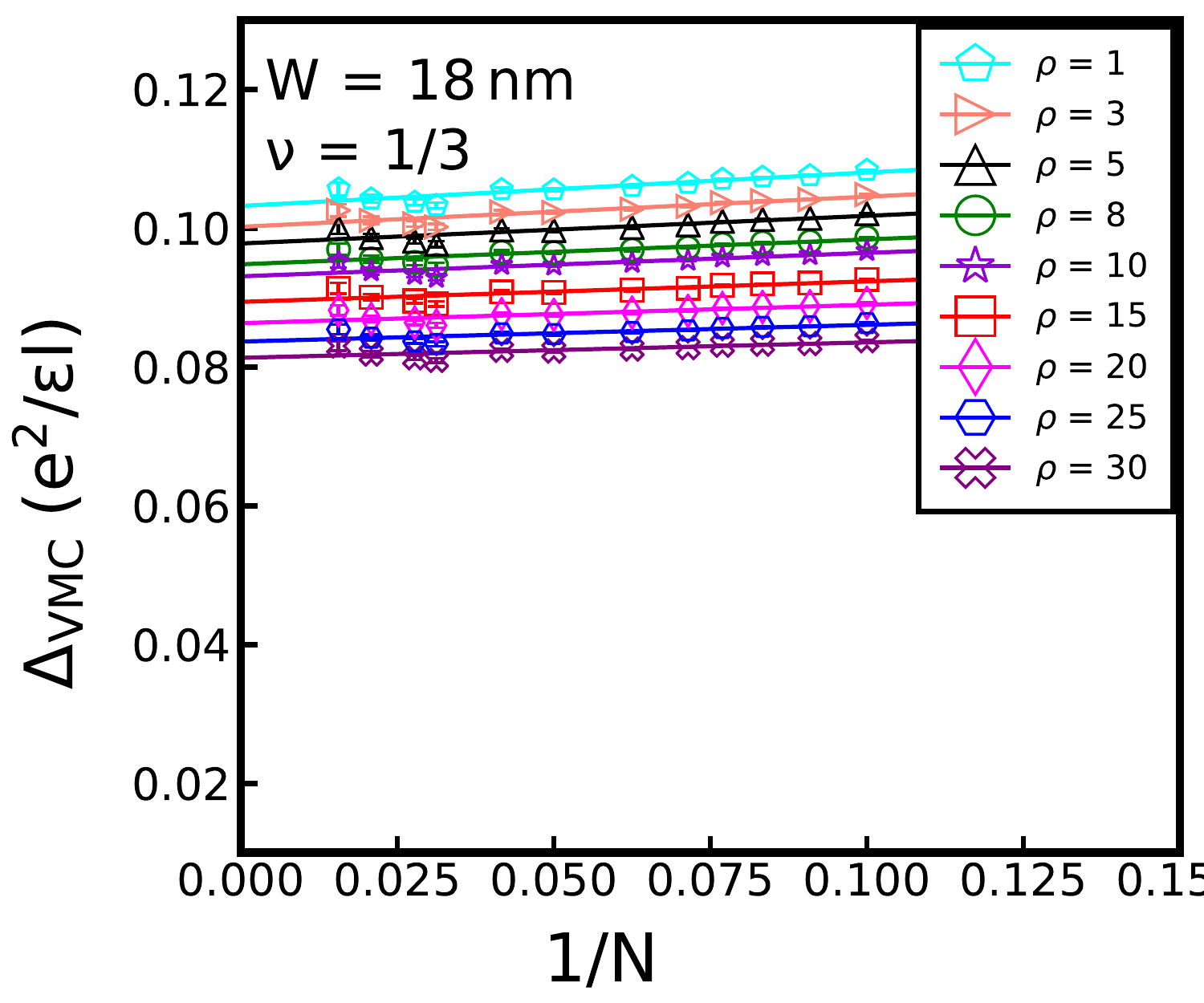}
	\includegraphics[width=0.32 \linewidth]{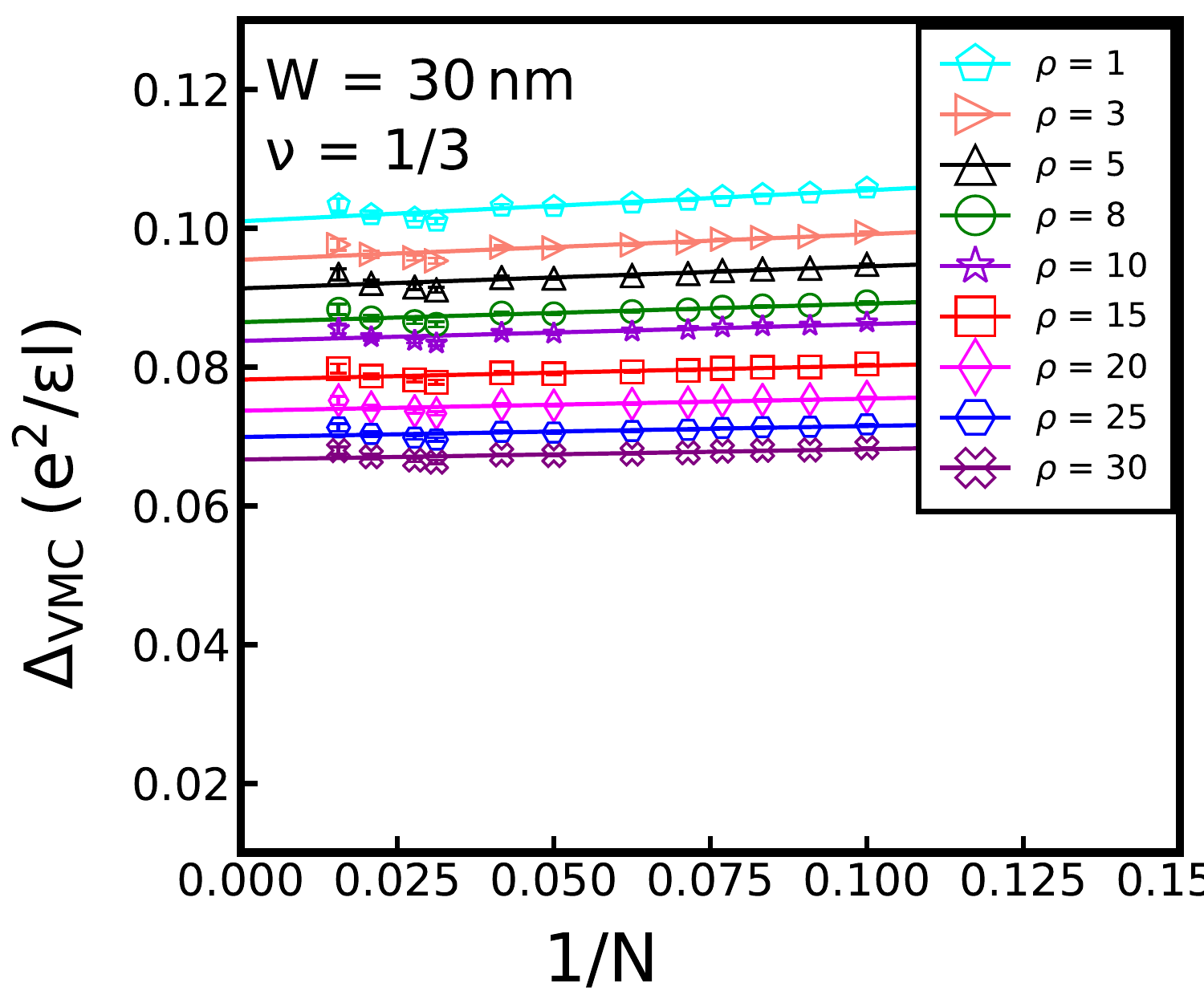}
	\includegraphics[width=0.32 \linewidth]{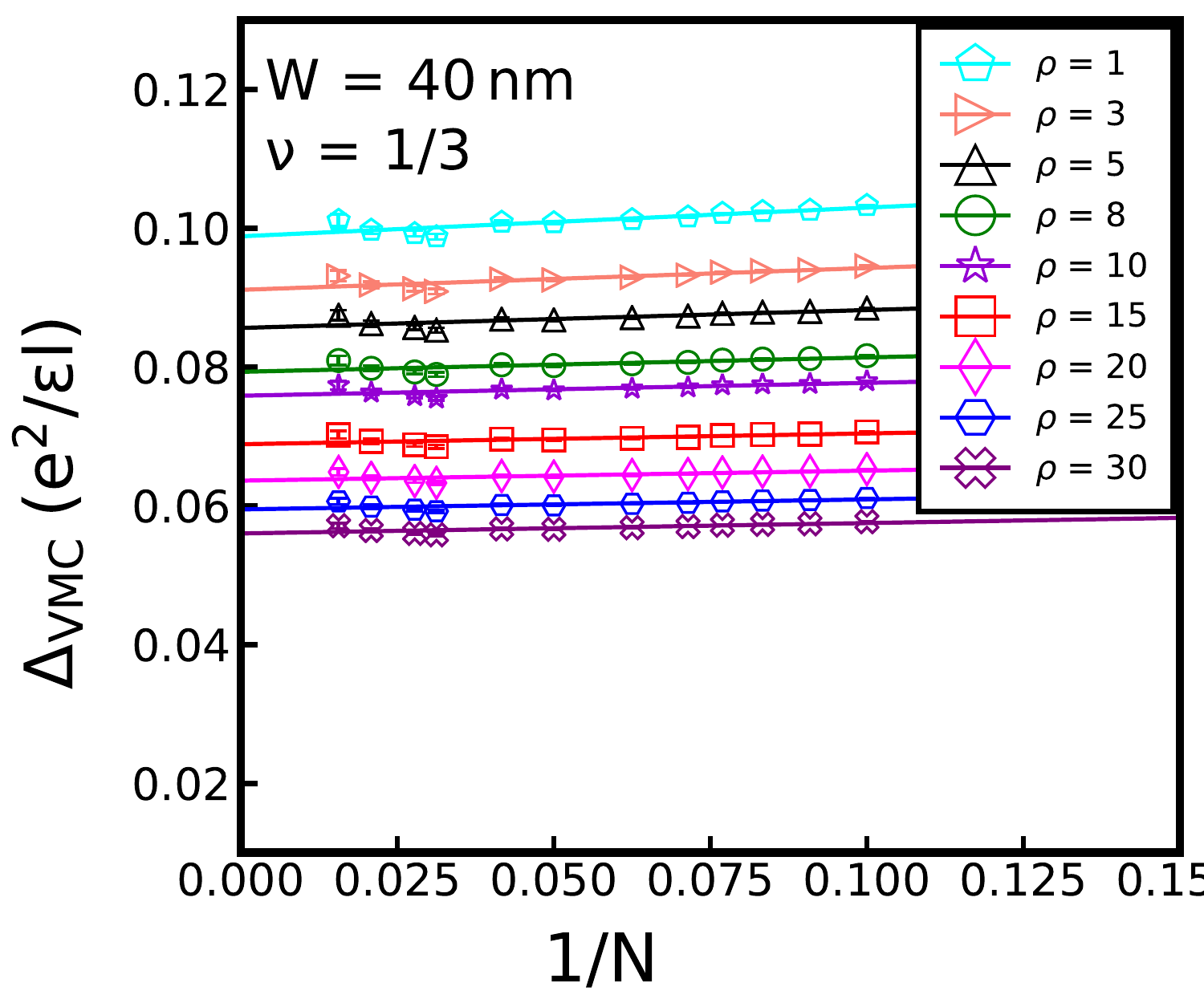}
	\includegraphics[width=0.32 \linewidth]{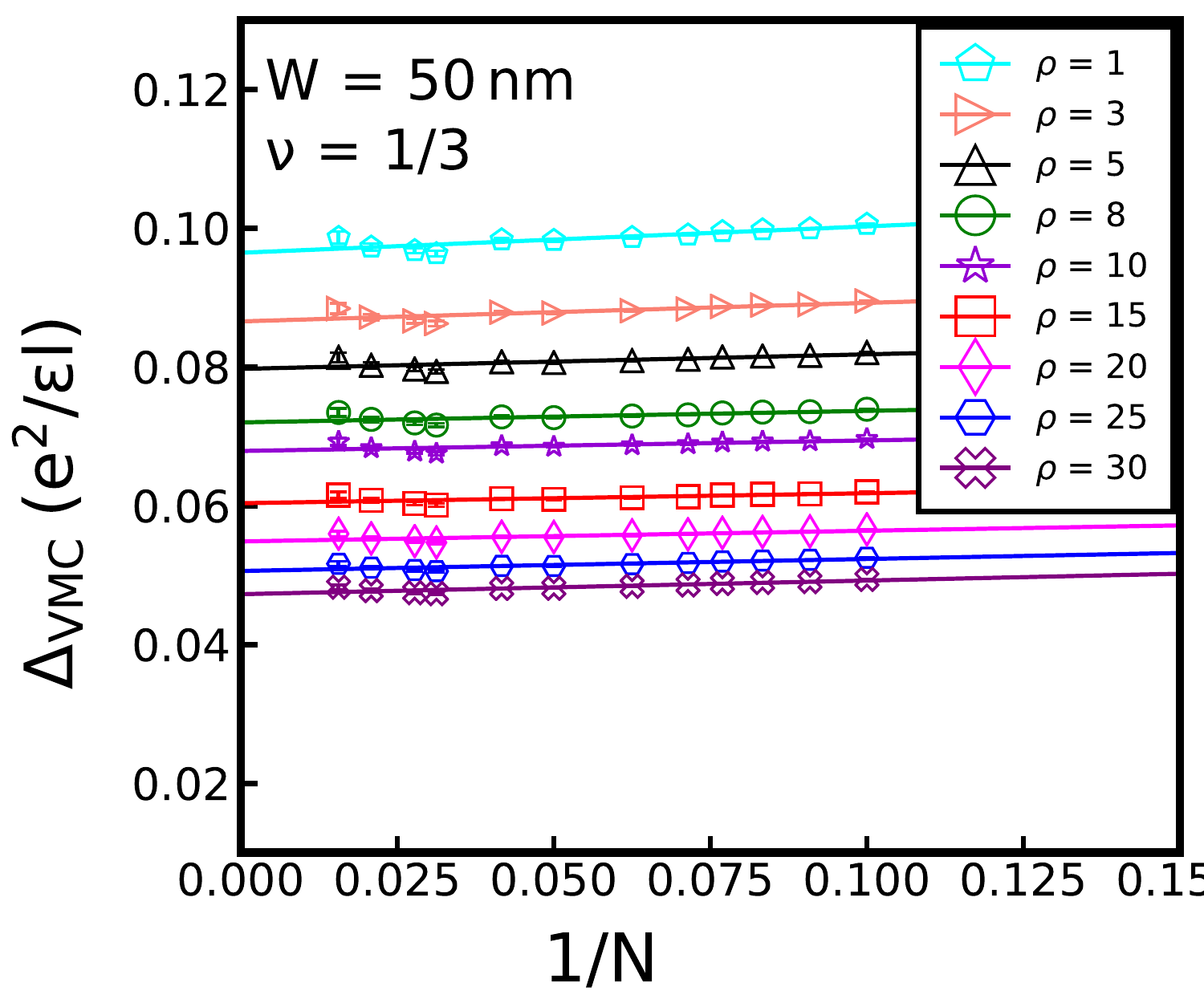}
	\includegraphics[width=0.32 \linewidth]{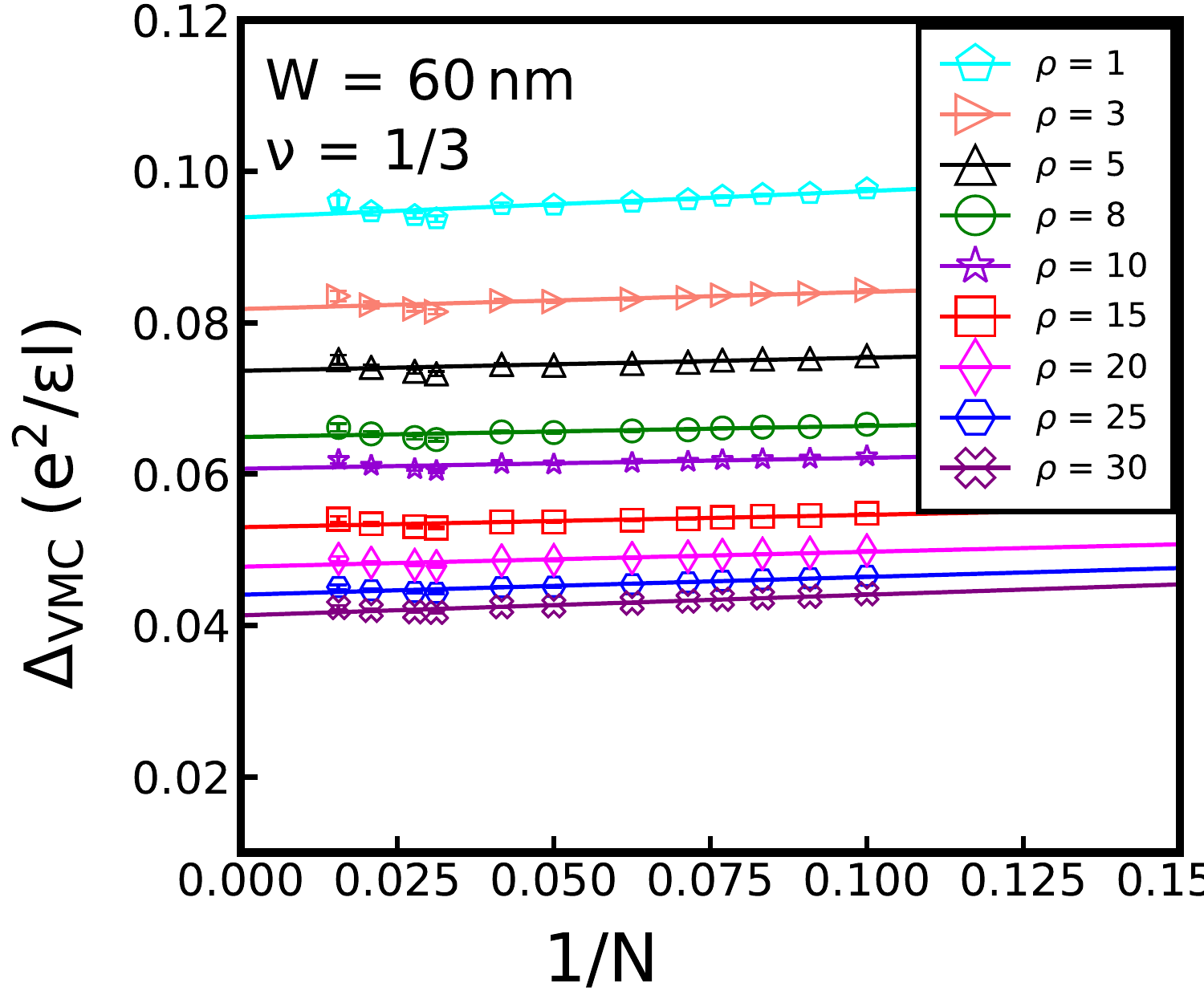}
	\includegraphics[width=0.32 \linewidth]{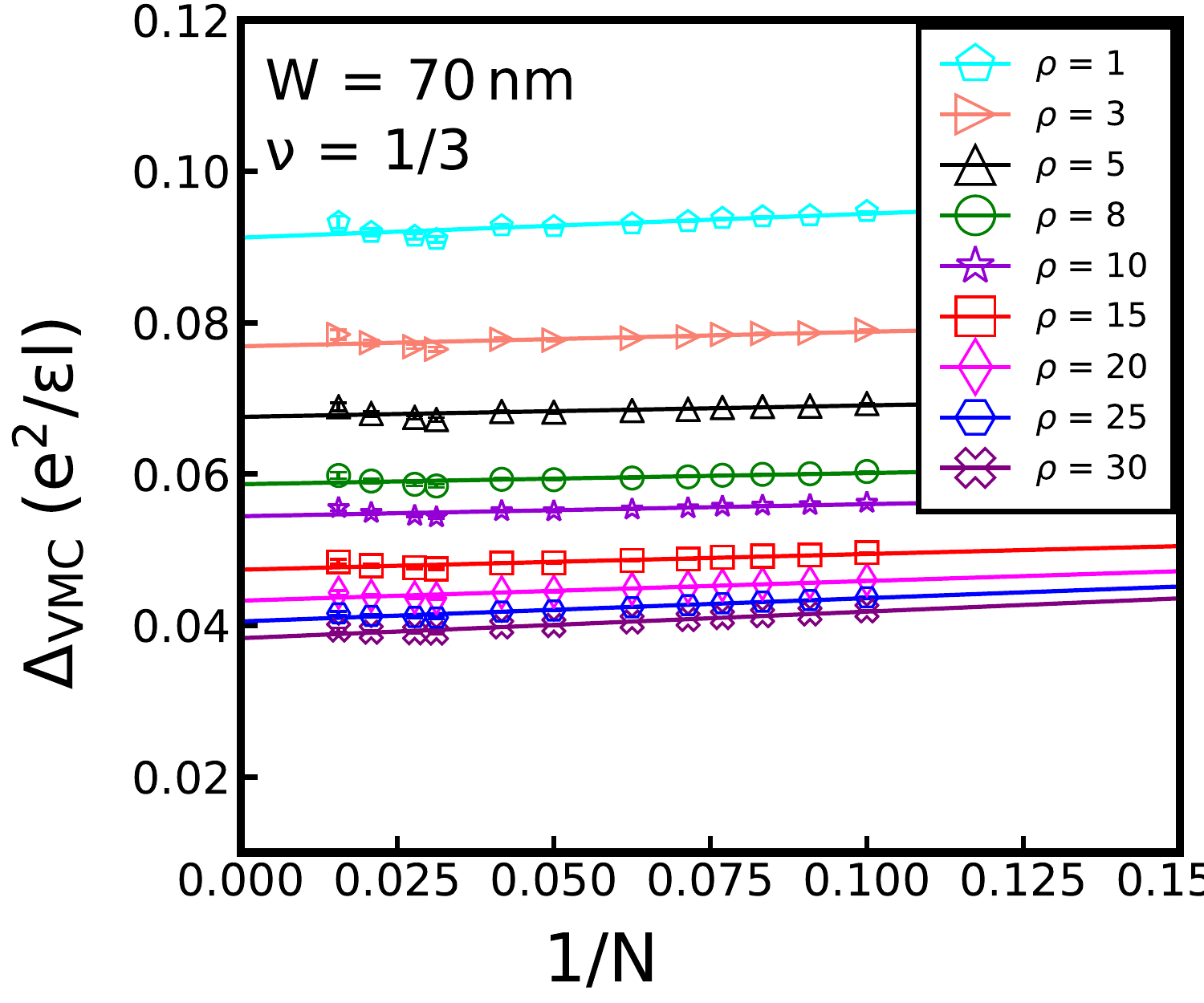}
	\caption{Thermodynamic extrapolation of the transport gap calculated by the VMC for $\nu=1/3$ at different widths and densities. Different markers in the legend label different densities in units of $10^{10}\text{cm}^{-2}$. The density correction and the CF-quasiparticle-quasihole interaction have been included (Eq. (3) of the main text). }\label{X_fig:CF13_VMC_extrap}
\end{figure*}

\begin{figure*}[ht!]
	\includegraphics[width=0.32 \linewidth]{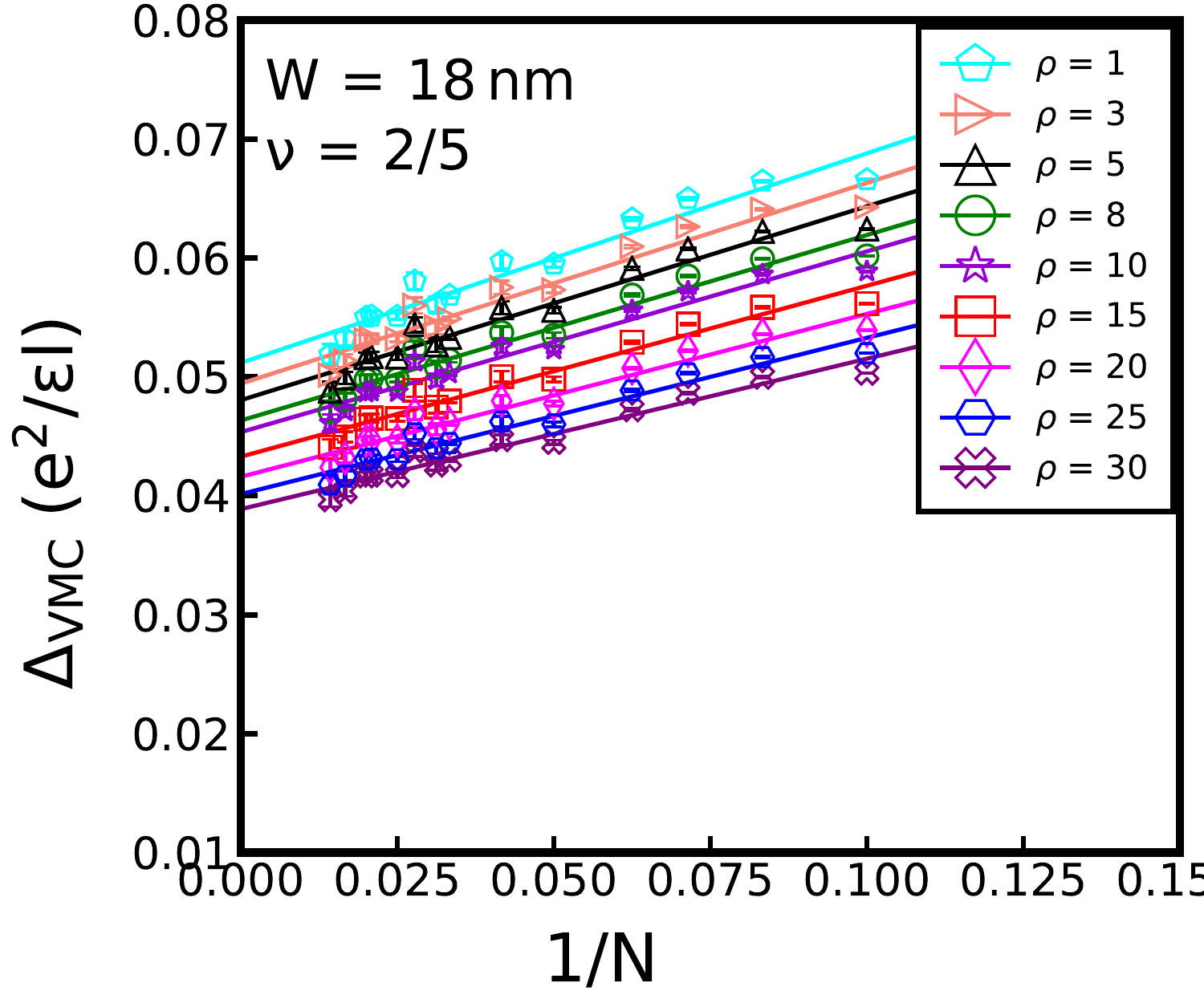}
	\includegraphics[width=0.32 \linewidth]{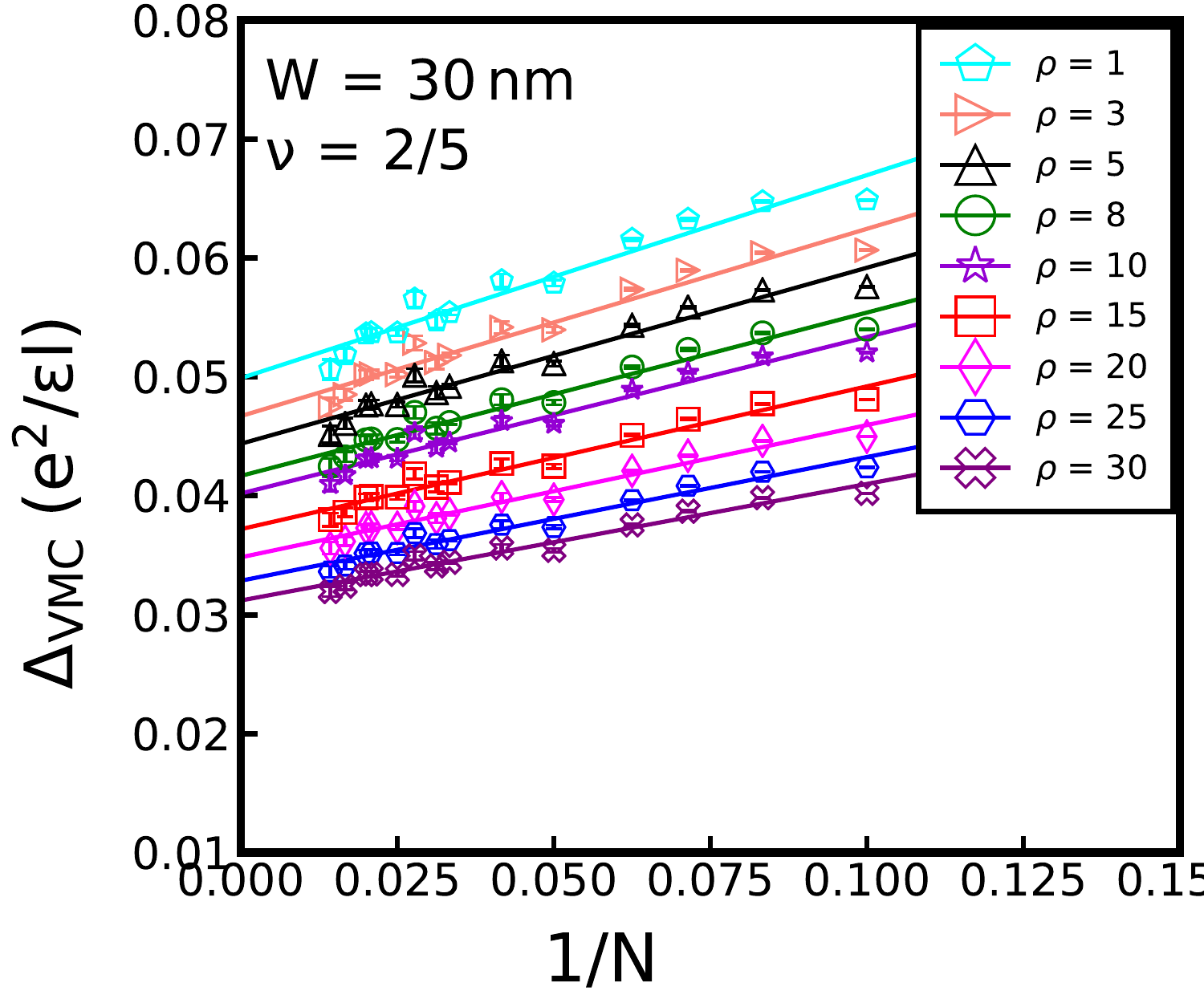}
	\includegraphics[width=0.32 \linewidth]{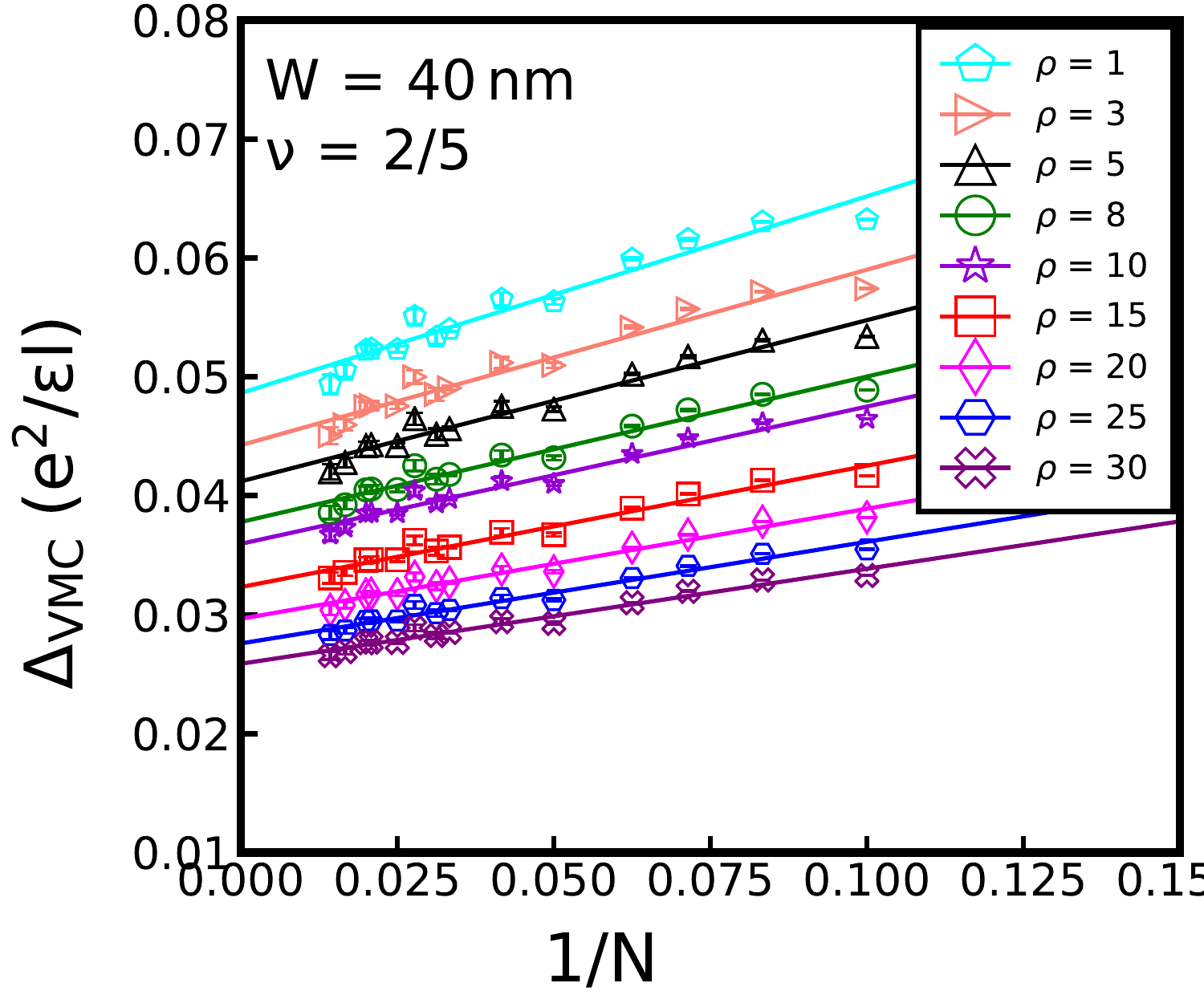}
	\includegraphics[width=0.32 \linewidth]{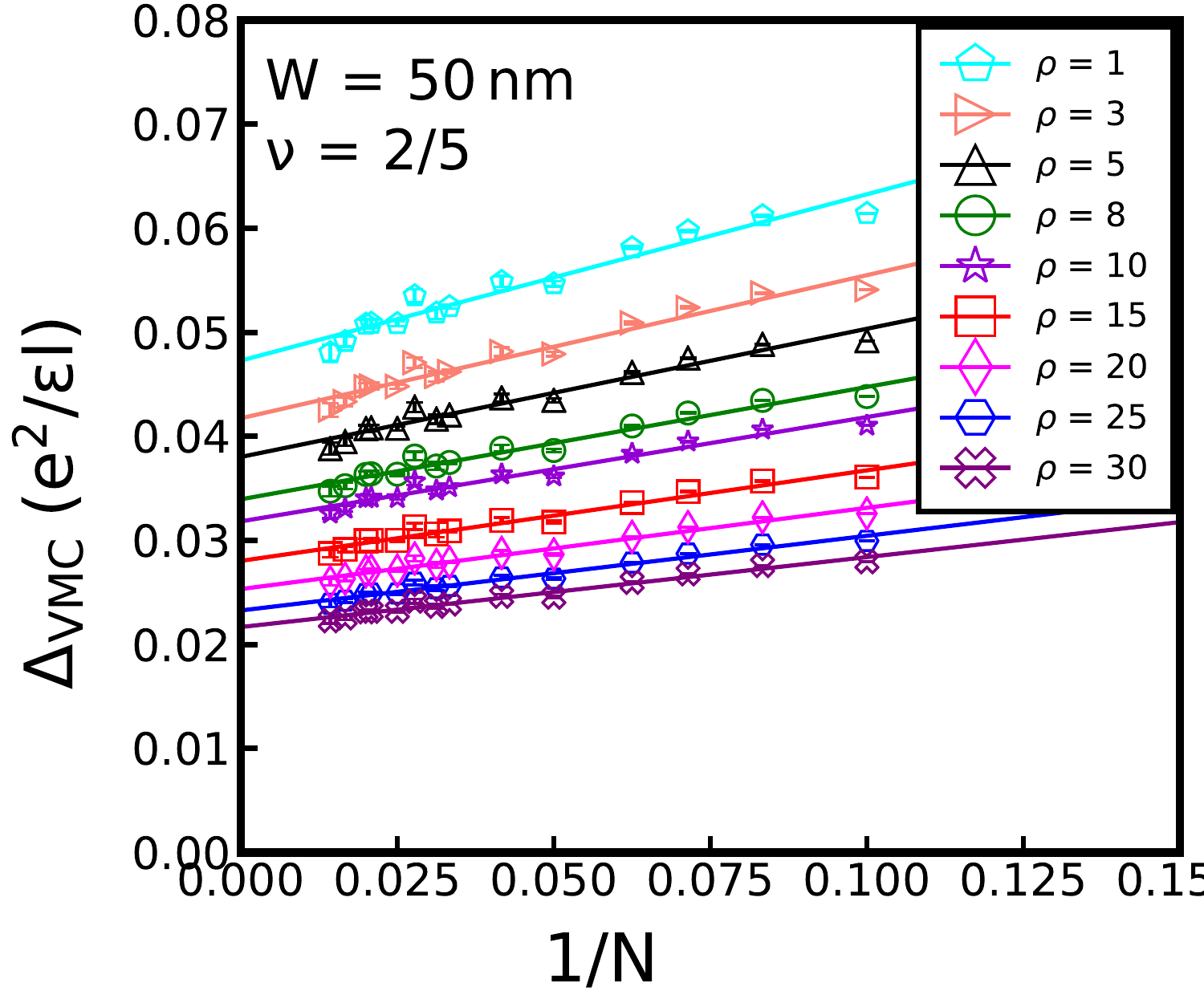}
	\includegraphics[width=0.32 \linewidth]{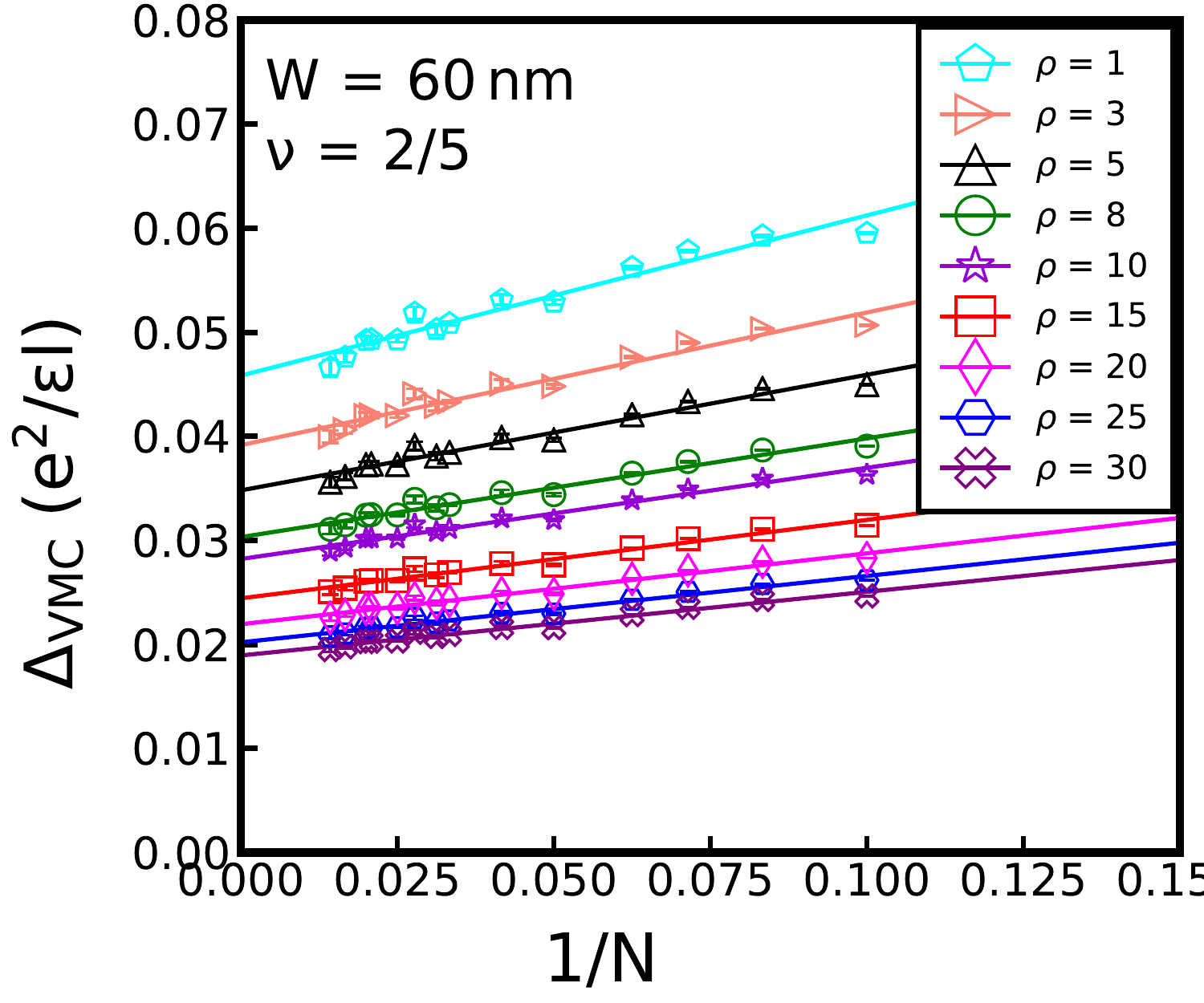}
	\includegraphics[width=0.32 \linewidth]{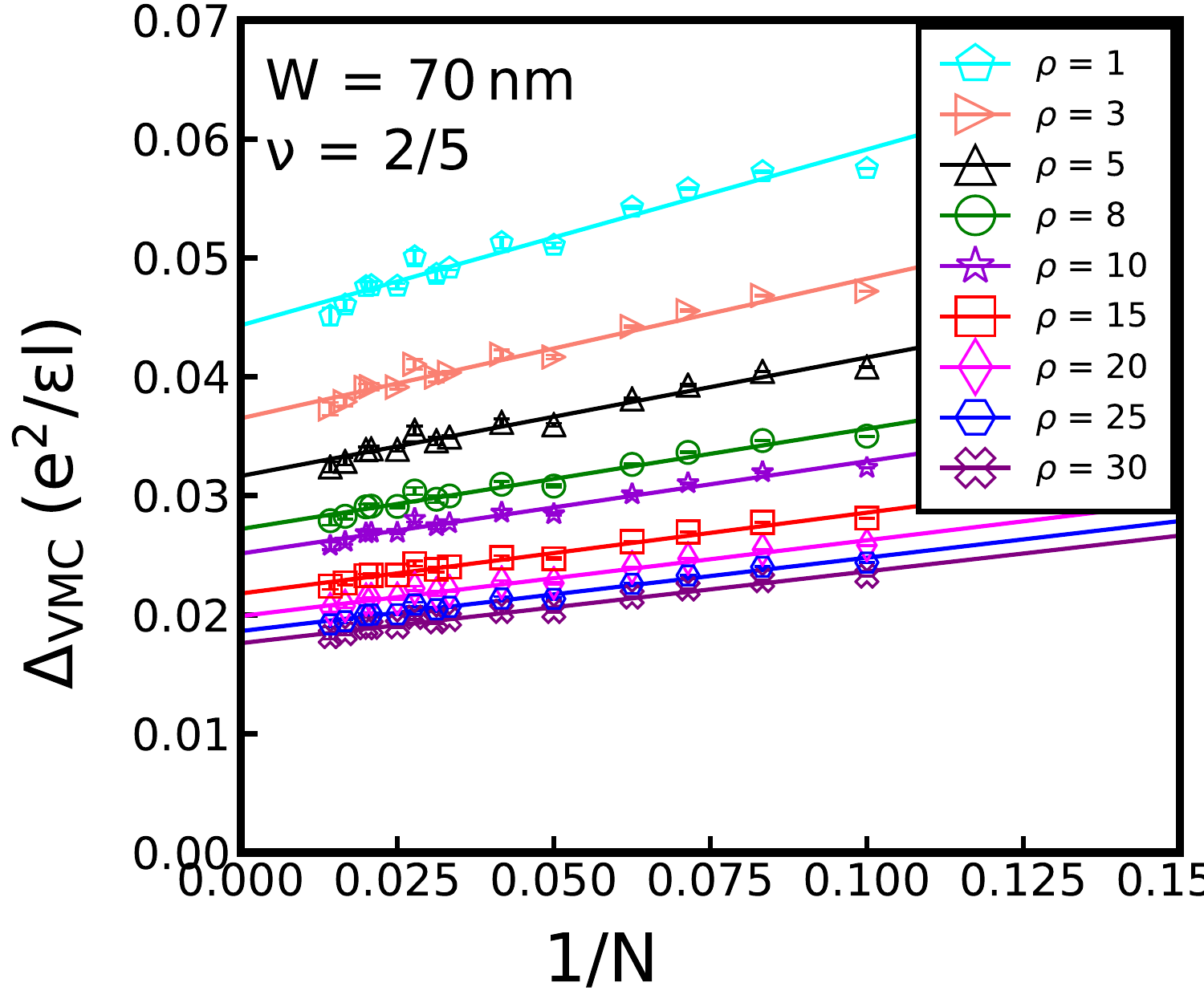}
	\caption{Thermodynamic extrapolation of the transport gap calculated by the VMC for $\nu=2/5$ at different widths and densities. Different markers in the legend label different densities in units of $10^{10}\text{cm}^{-2}$. The density correction and the CF-quasiparticle-quasihole interaction have been included (Eq. (3) of the main text). }\label{X_fig:CF25_VMC_extrap}
\end{figure*}

\begin{figure*}[ht!]
	\includegraphics[width=0.32 \linewidth]{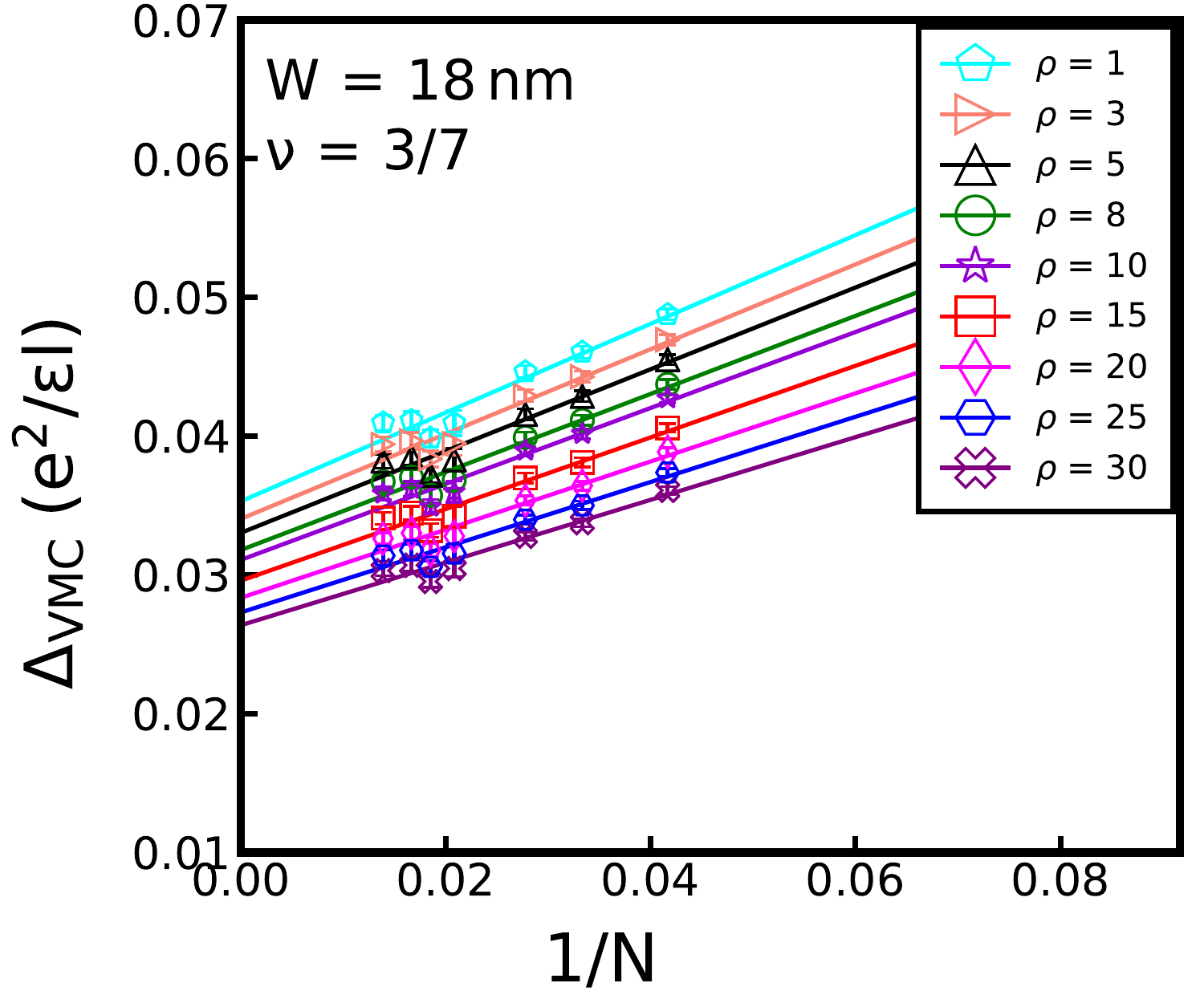}
	\includegraphics[width=0.32 \linewidth]{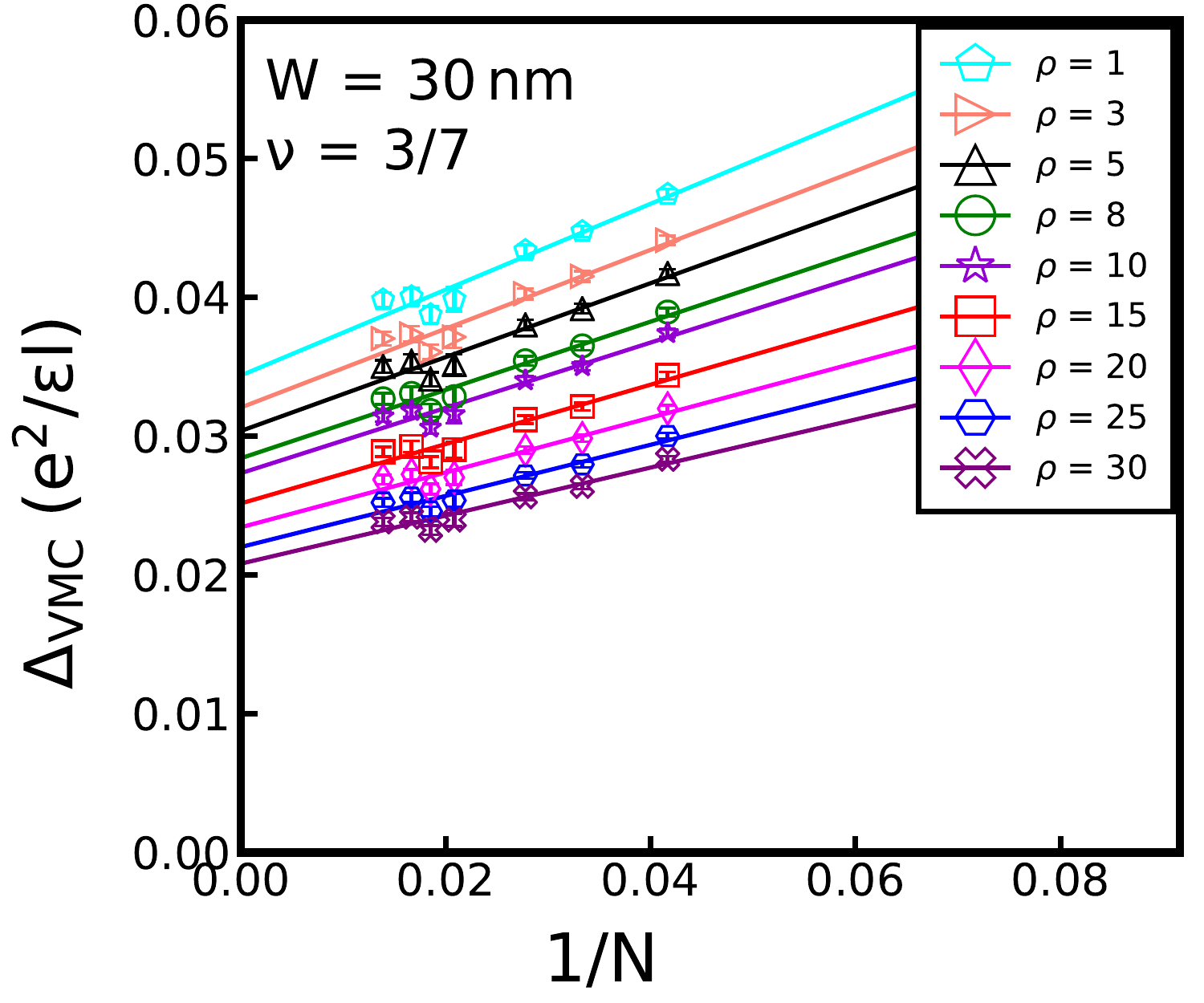}
	\includegraphics[width=0.32 \linewidth]{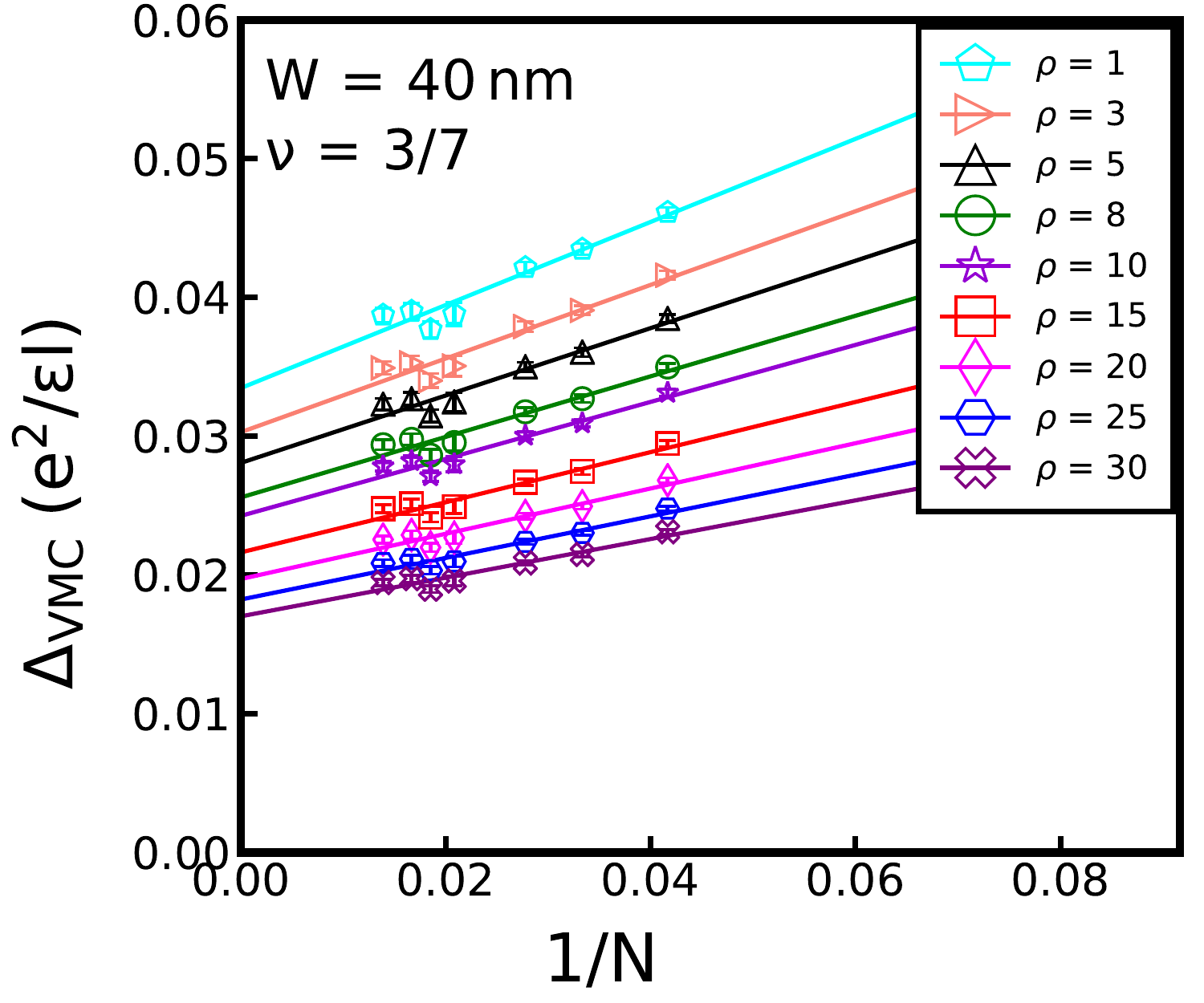}
	\includegraphics[width=0.32 \linewidth]{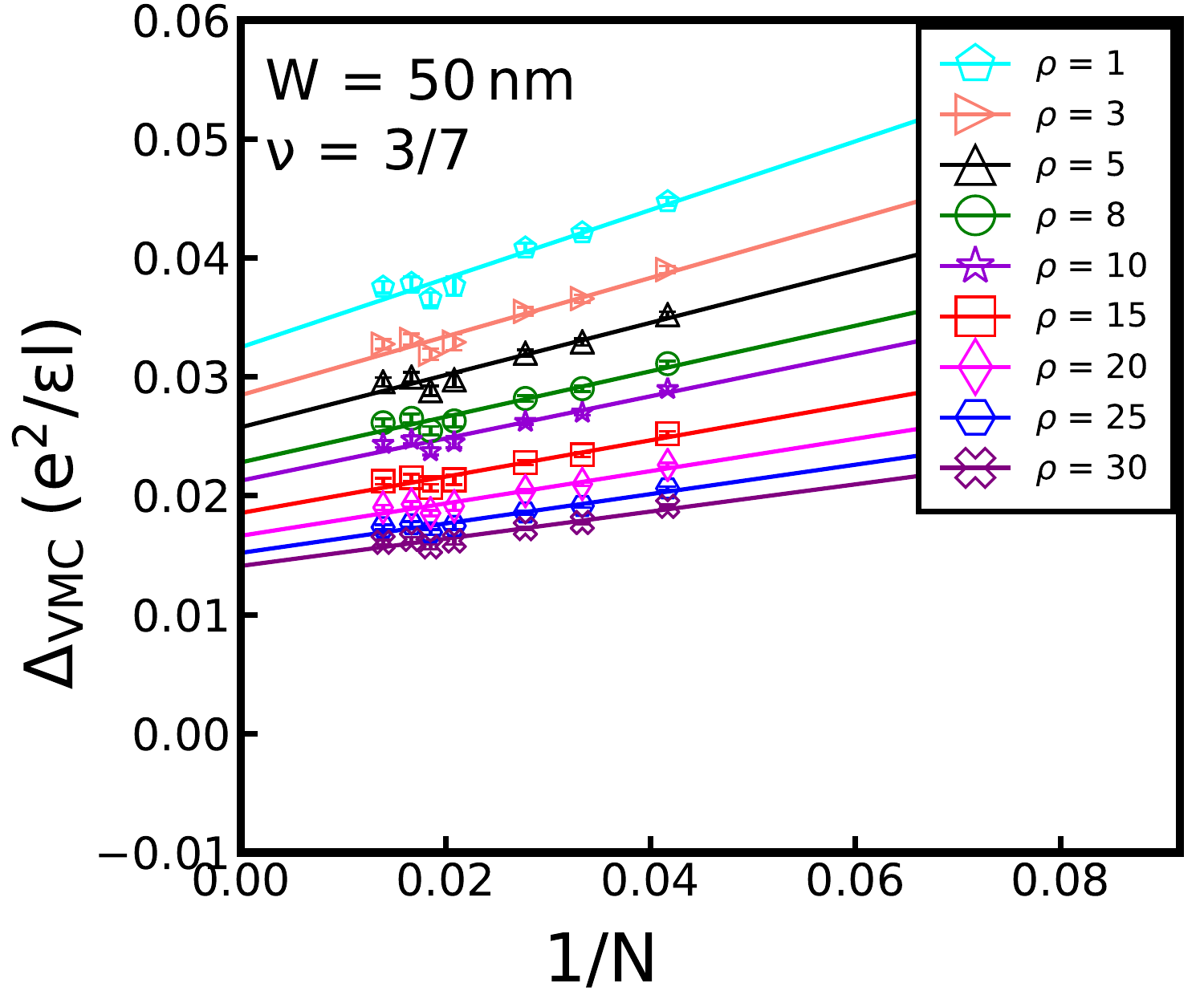}
	\includegraphics[width=0.32 \linewidth]{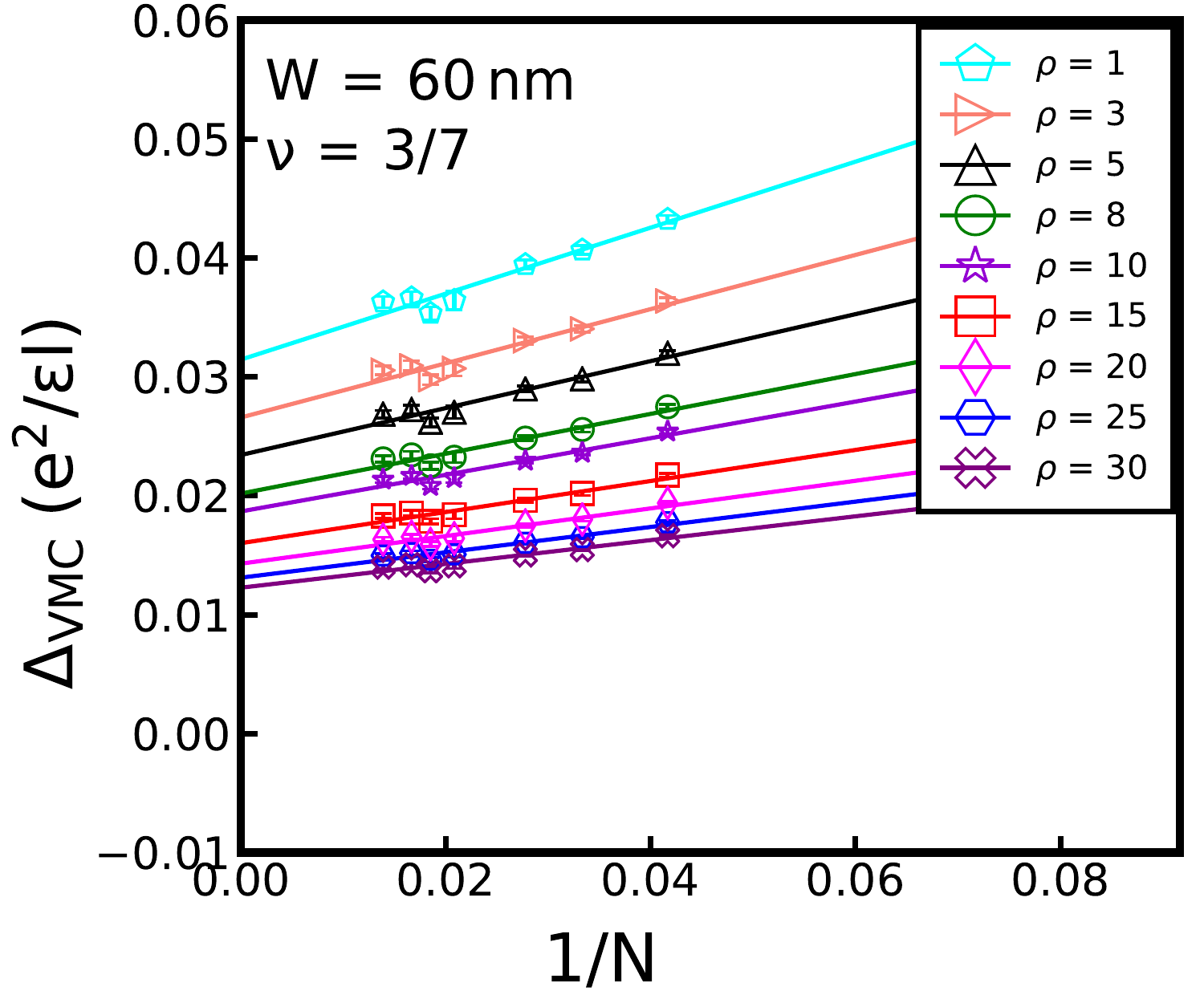}
	\includegraphics[width=0.32 \linewidth]{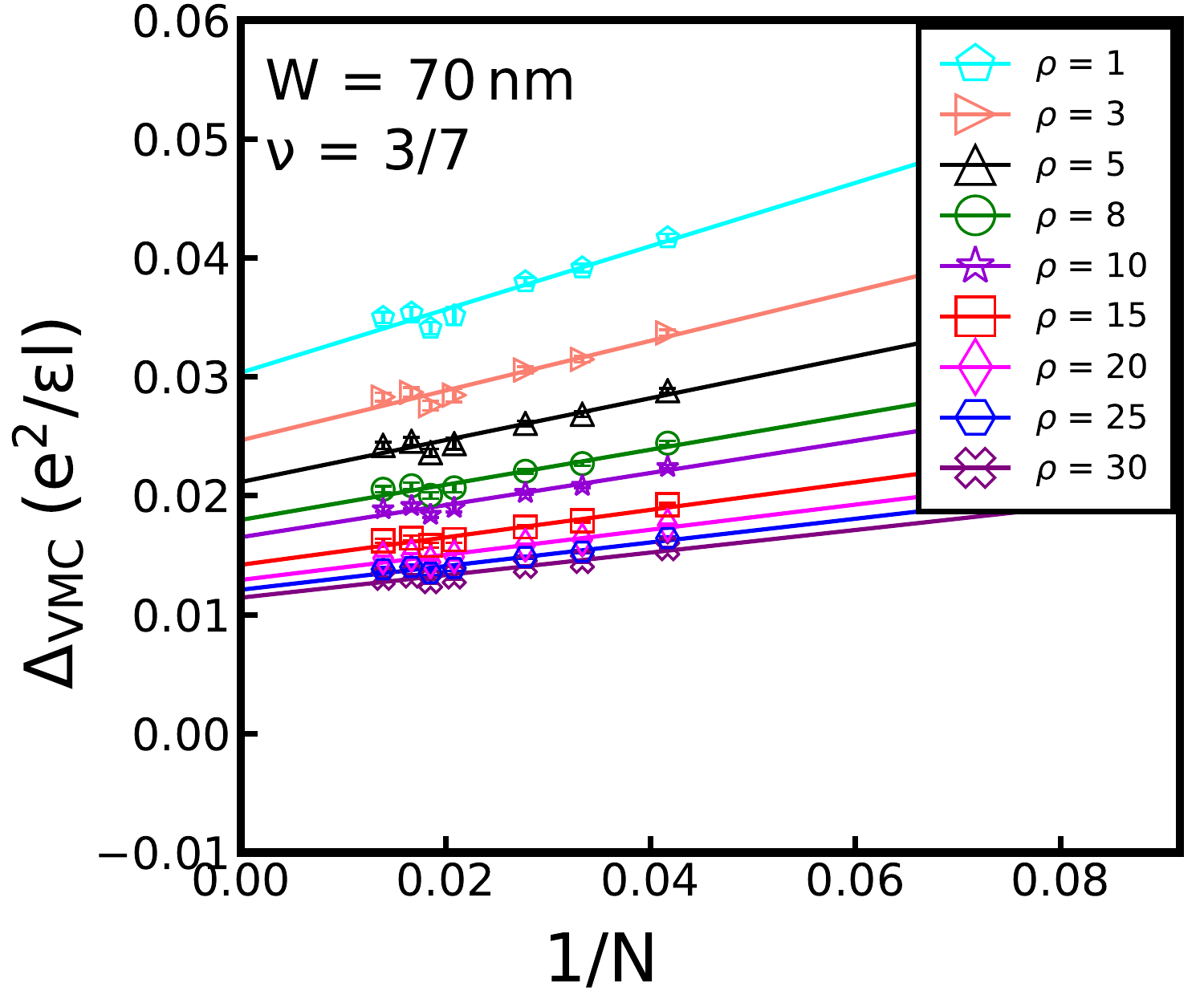}
	\caption{Thermodynamic extrapolation of the transport gap calculated by the VMC for $\nu=3/7$ at different widths and densities. Different markers in the legend label different densities in units of $10^{10}\text{cm}^{-2}$. The density correction and the CF-quasiparticle-quasihole interaction have been included (Eq. (3) of the main text). }\label{X_fig:CF37_VMC_extrap}
\end{figure*}

\begin{figure*}[ht!]
	\includegraphics[width=0.32 \linewidth]{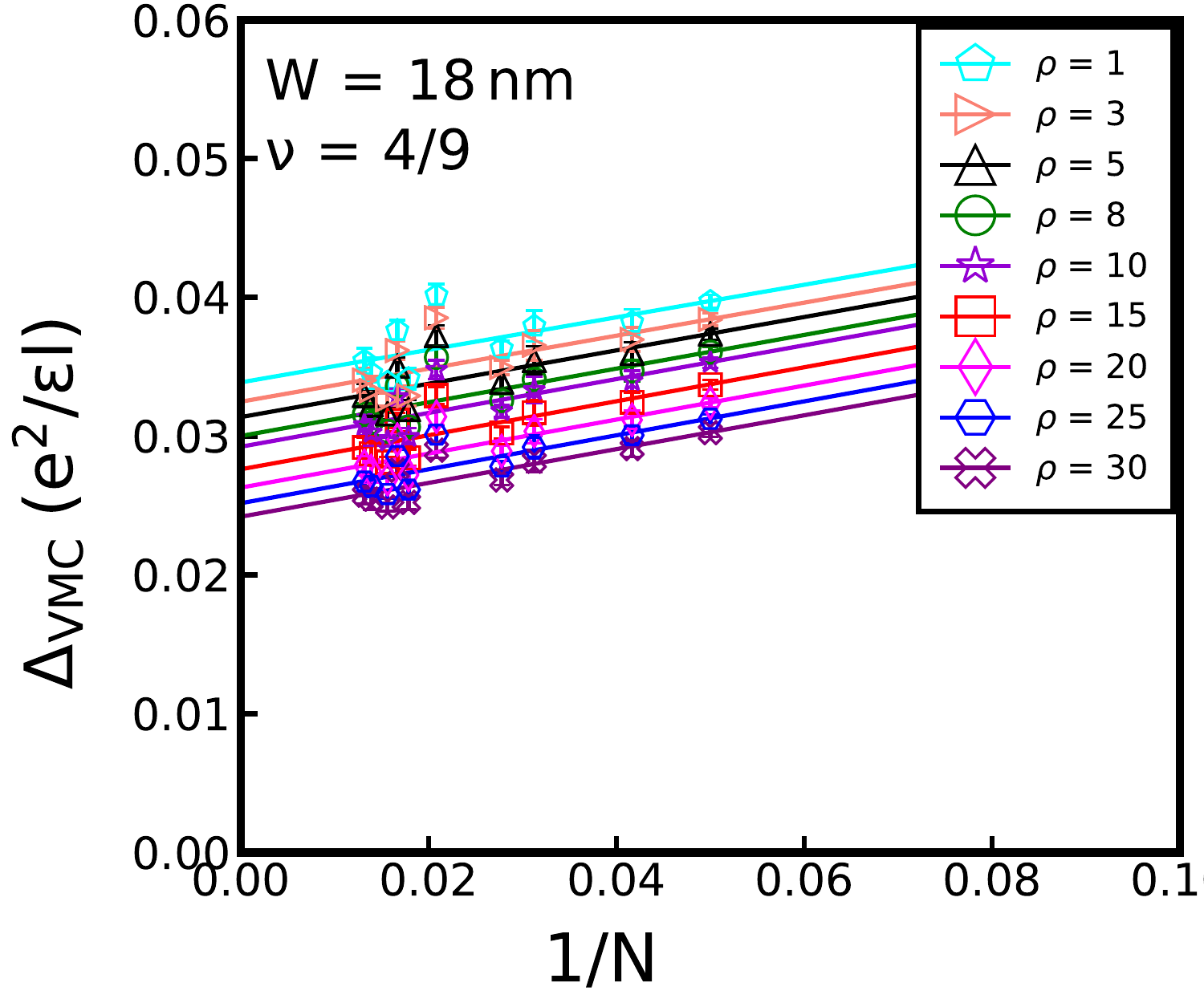}
	\includegraphics[width=0.32 \linewidth]{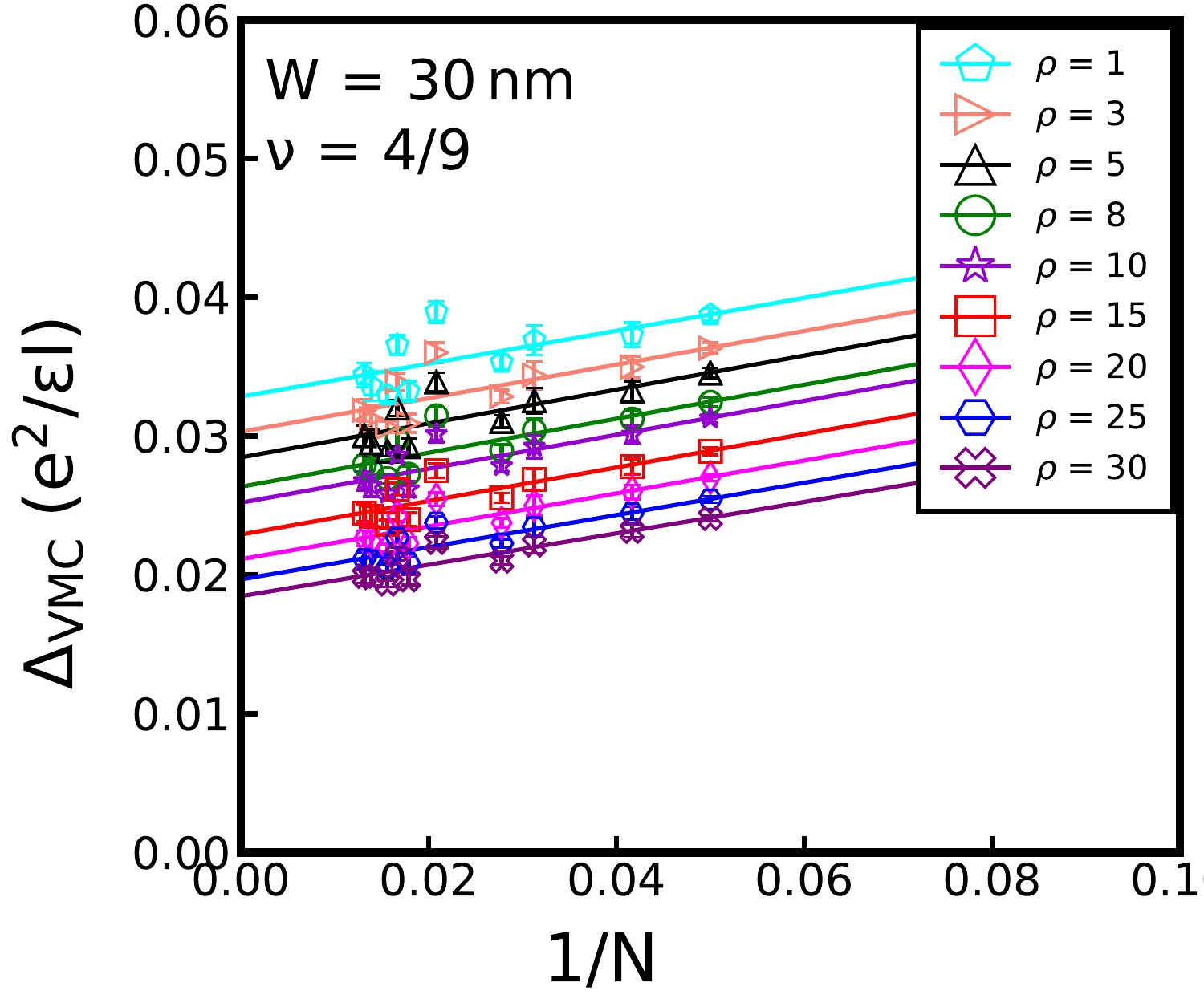}
	\includegraphics[width=0.32 \linewidth]{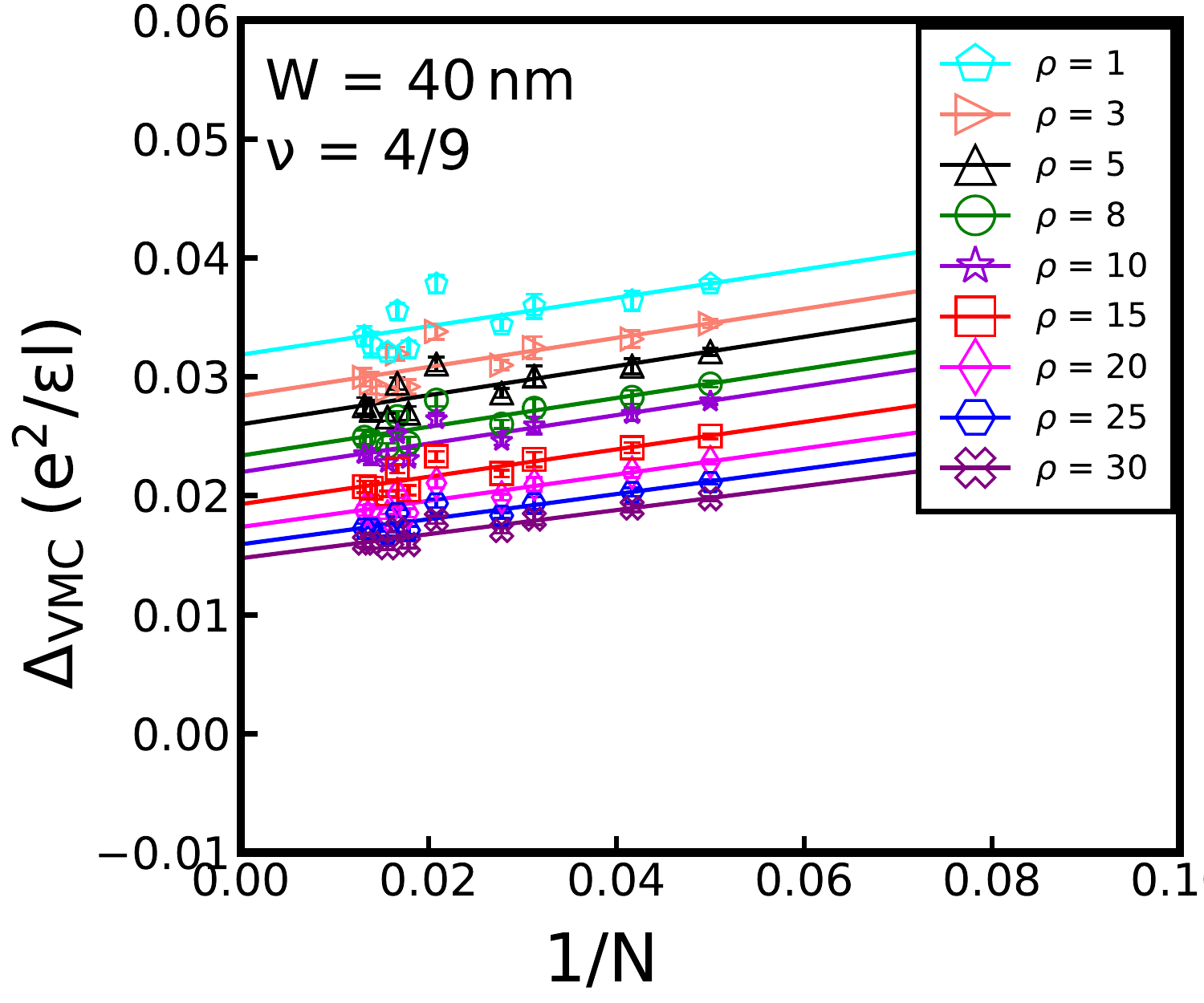}
	\includegraphics[width=0.32 \linewidth]{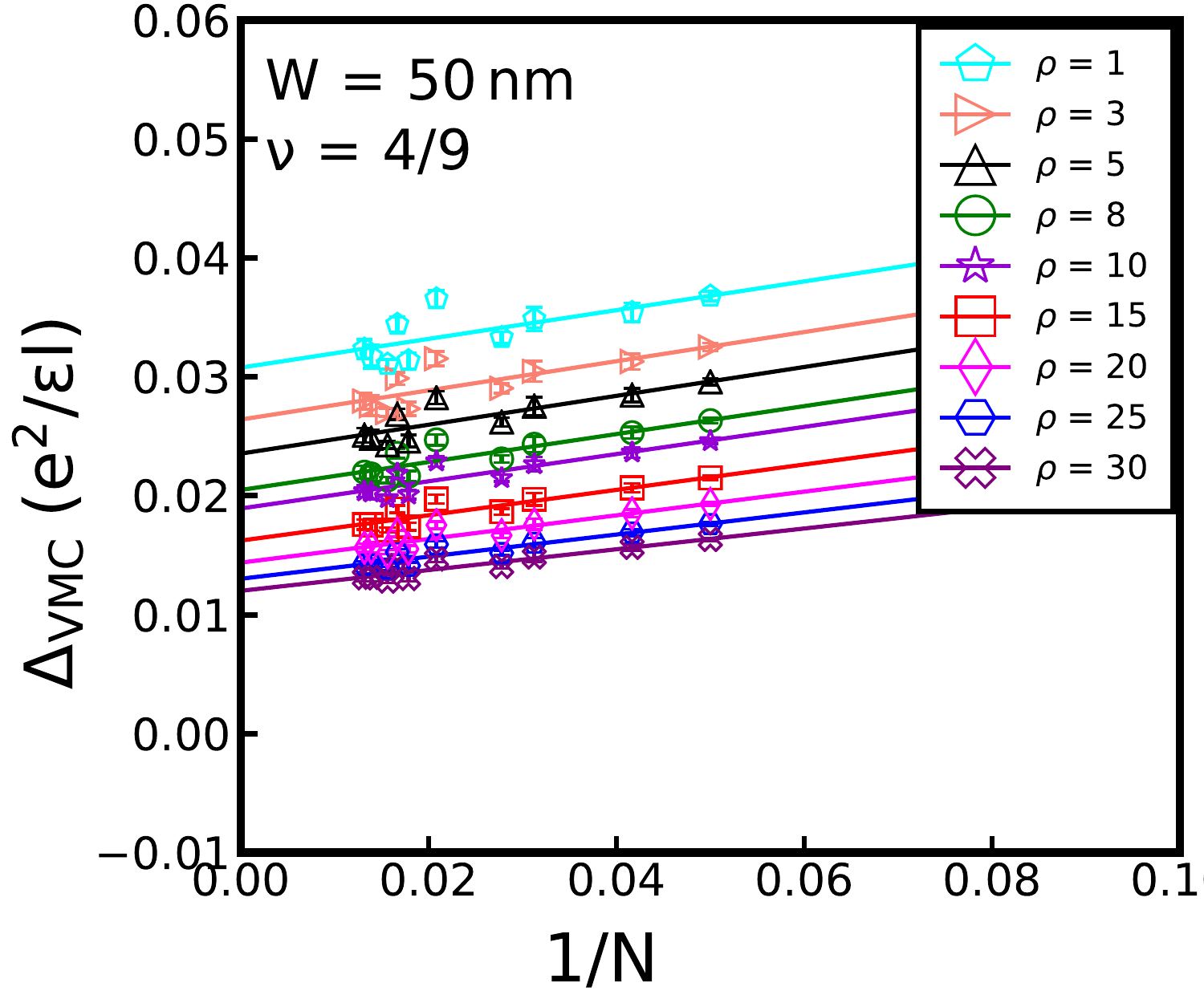}
	\includegraphics[width=0.32 \linewidth]{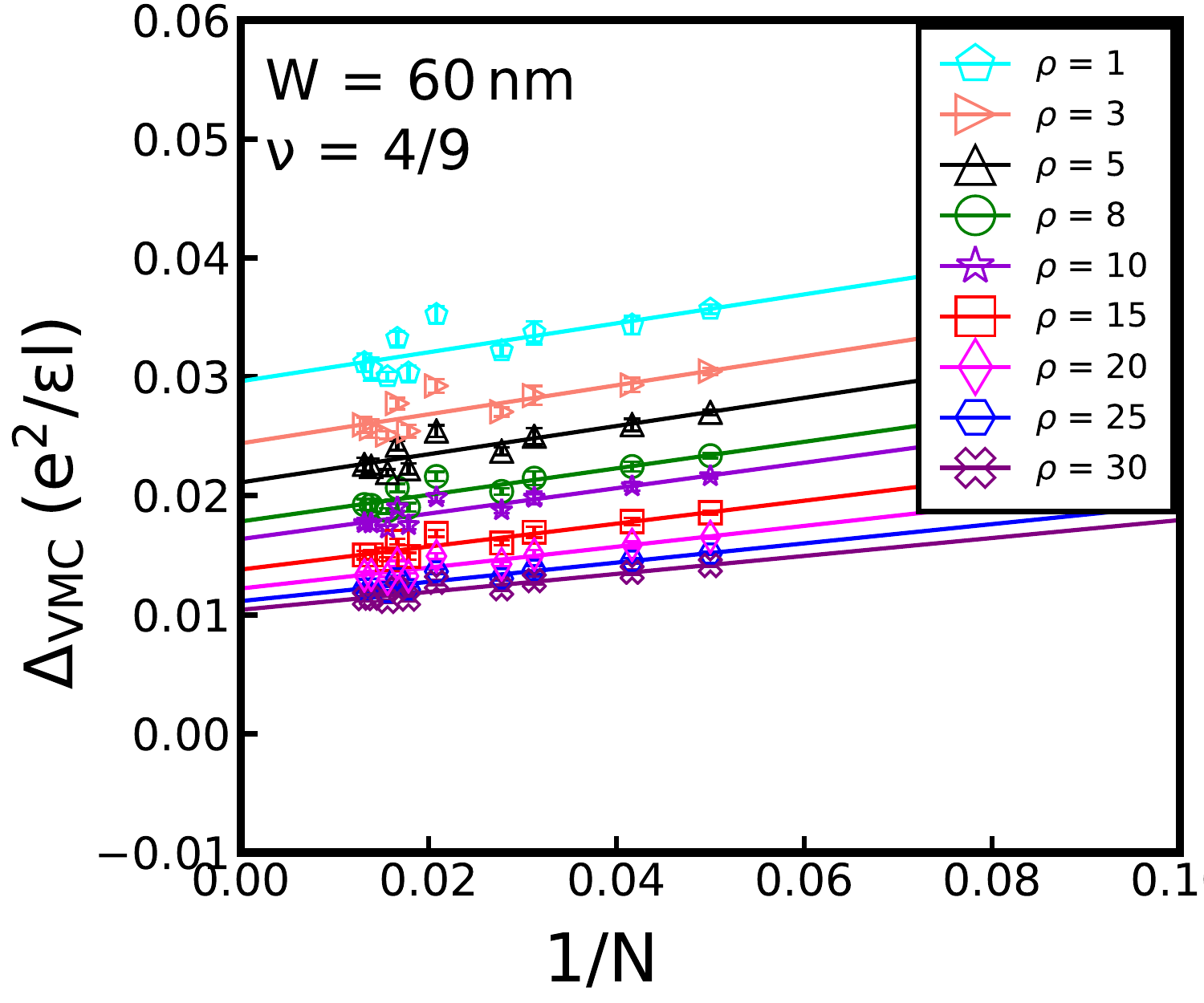}
	\includegraphics[width=0.32 \linewidth]{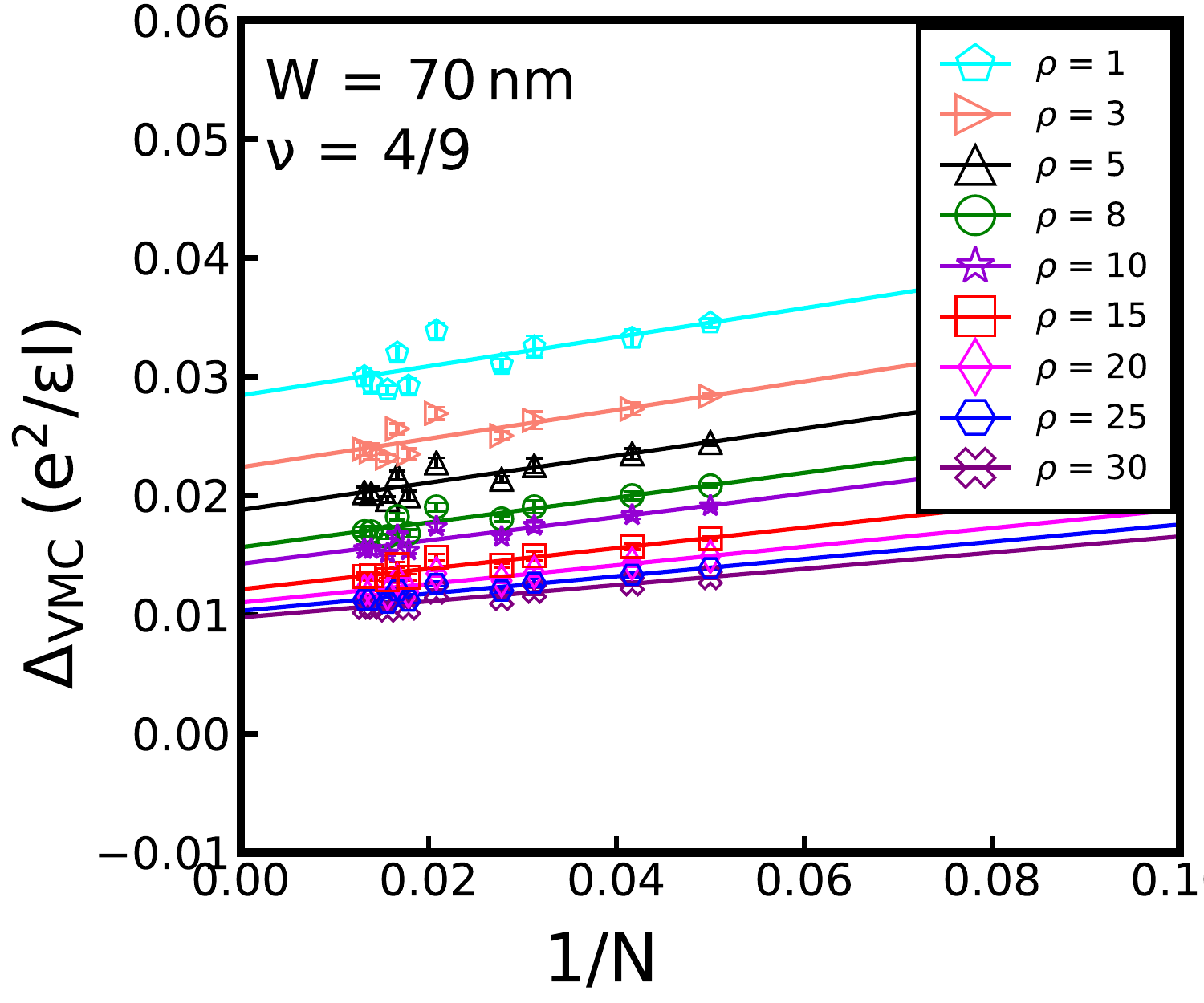}
	\caption{Thermodynamic extrapolation of the transport gap calculated by the VMC for $\nu=4/9$ at different widths and densities. Different markers in the legend label different densities in units of $10^{10}\text{cm}^{-2}$. The density correction and the CF-quasiparticle-quasihole interaction have been included (Eq. (3) of the main text). }\label{X_fig:CF49_VMC_extrap}
\end{figure*}

\begin{figure*}[ht!]
	\includegraphics[width=0.32 \linewidth]{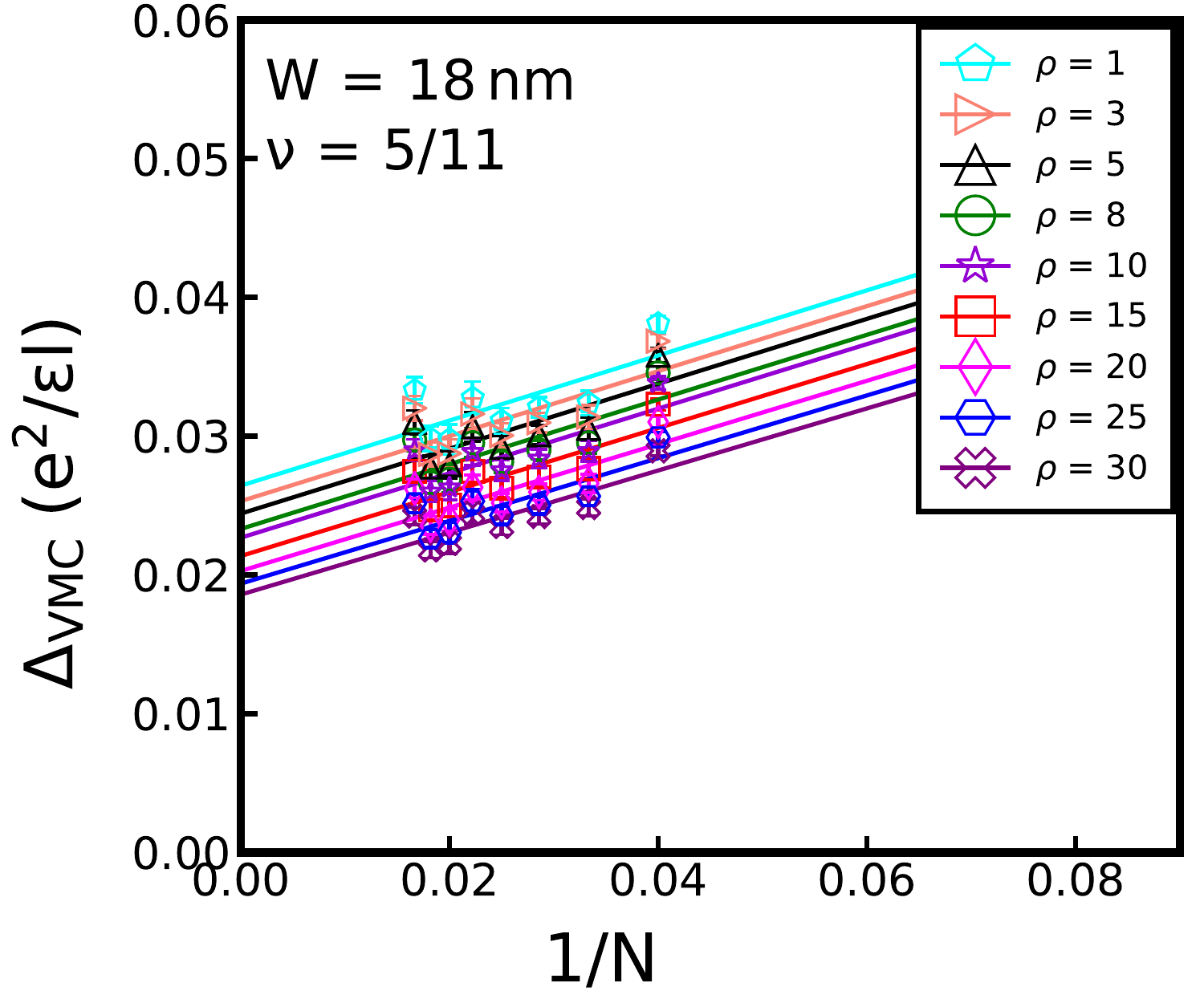}
	\includegraphics[width=0.32 \linewidth]{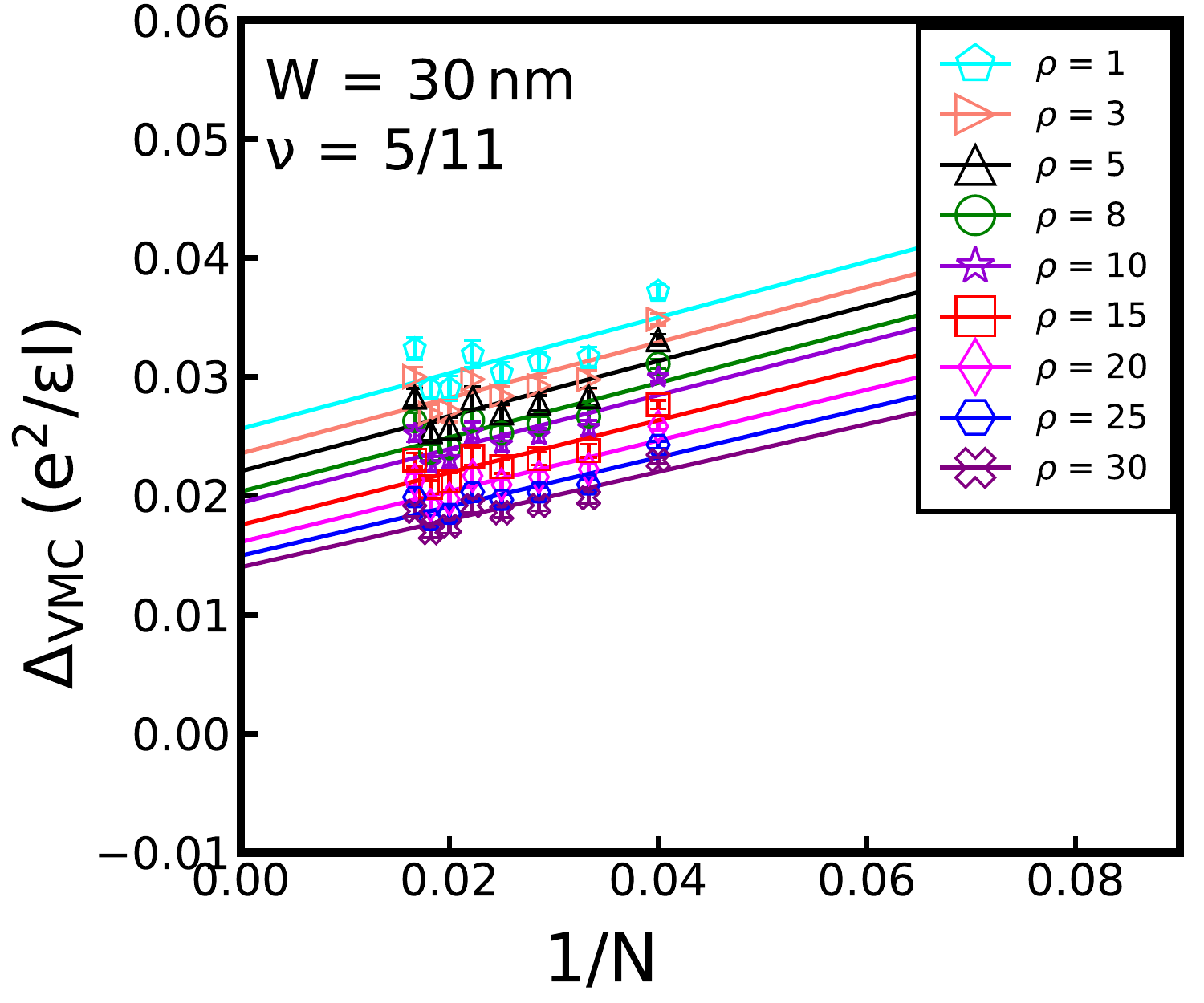}
	\includegraphics[width=0.32 \linewidth]{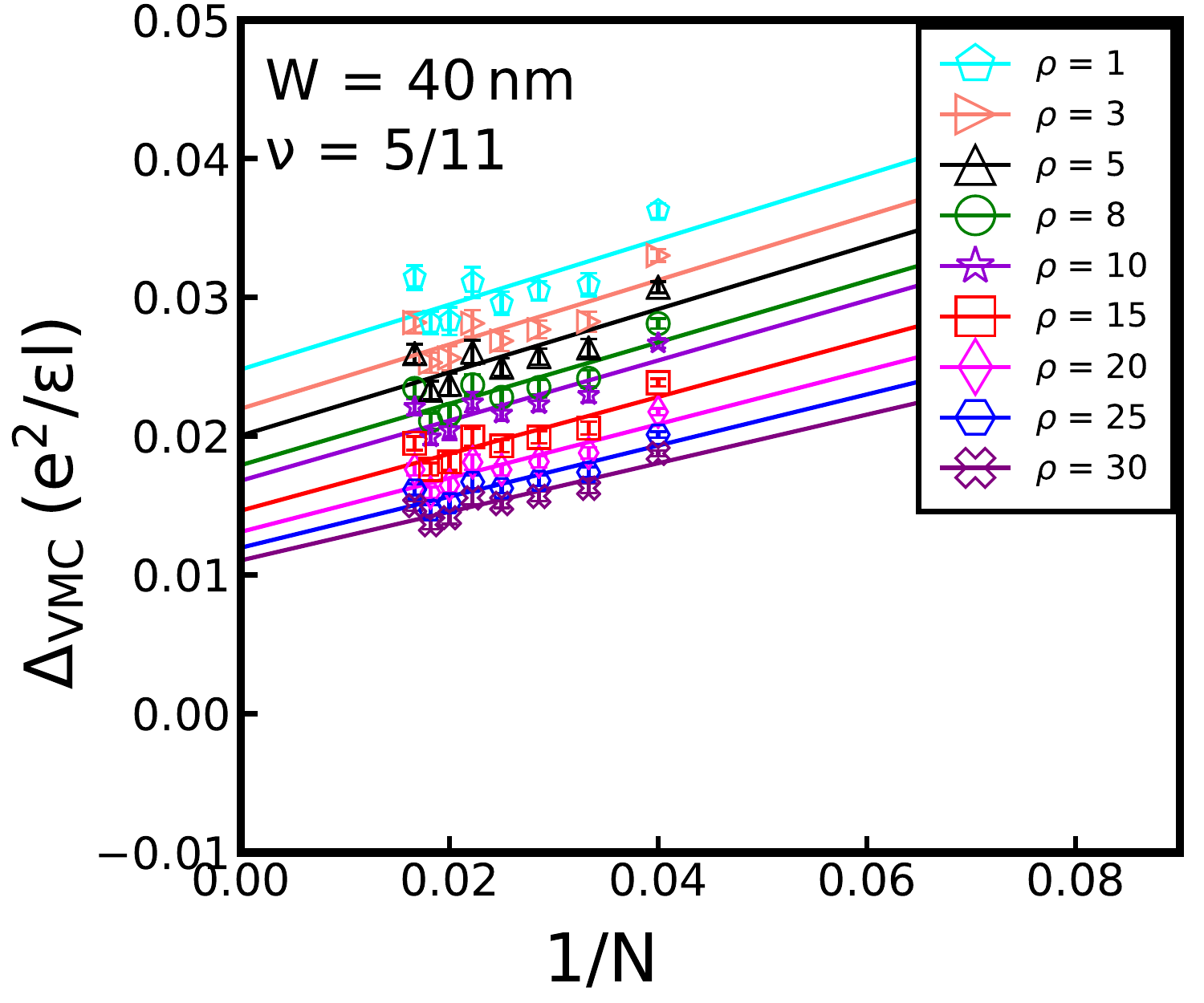}
	\includegraphics[width=0.32 \linewidth]{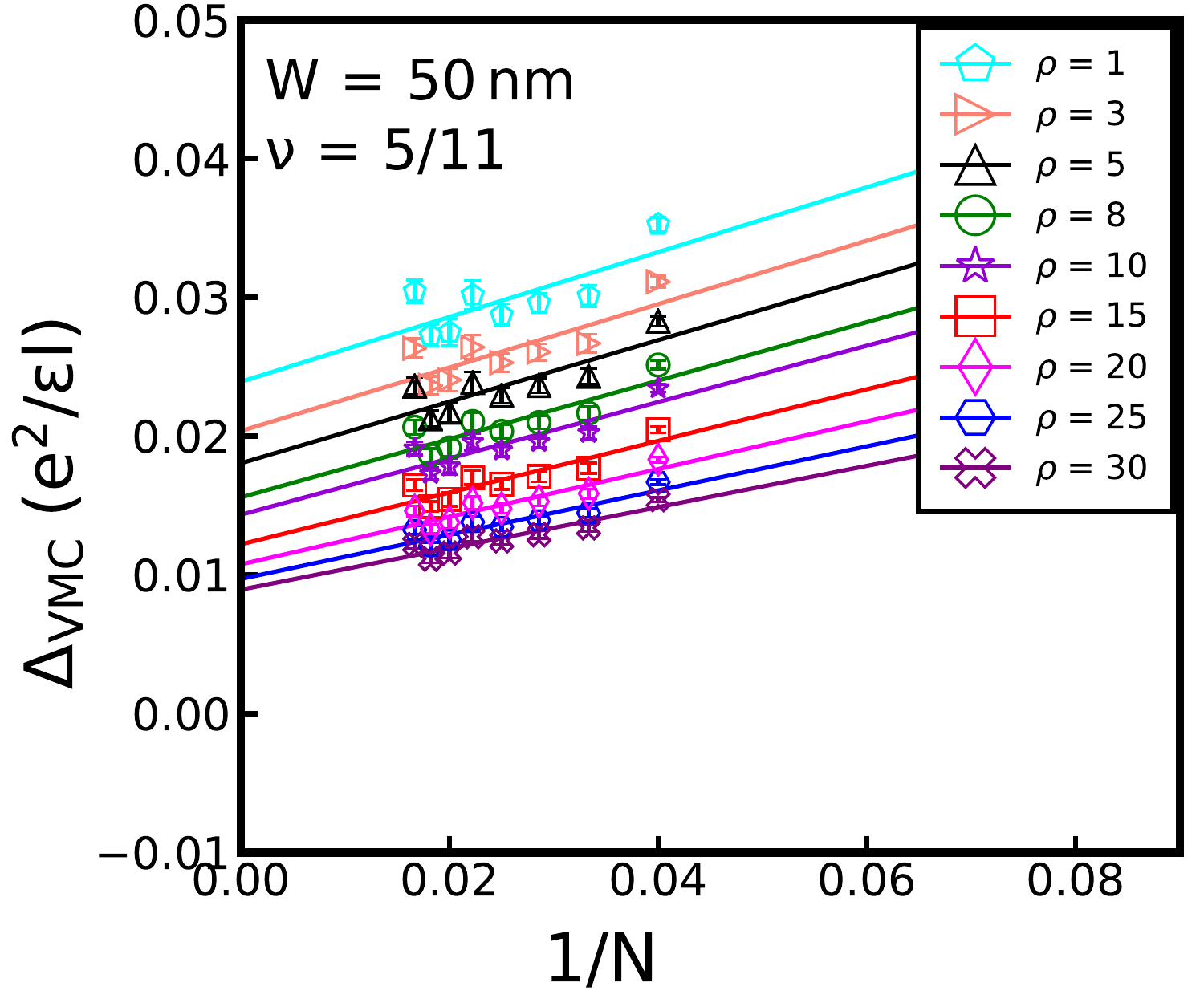}
	\includegraphics[width=0.32 \linewidth]{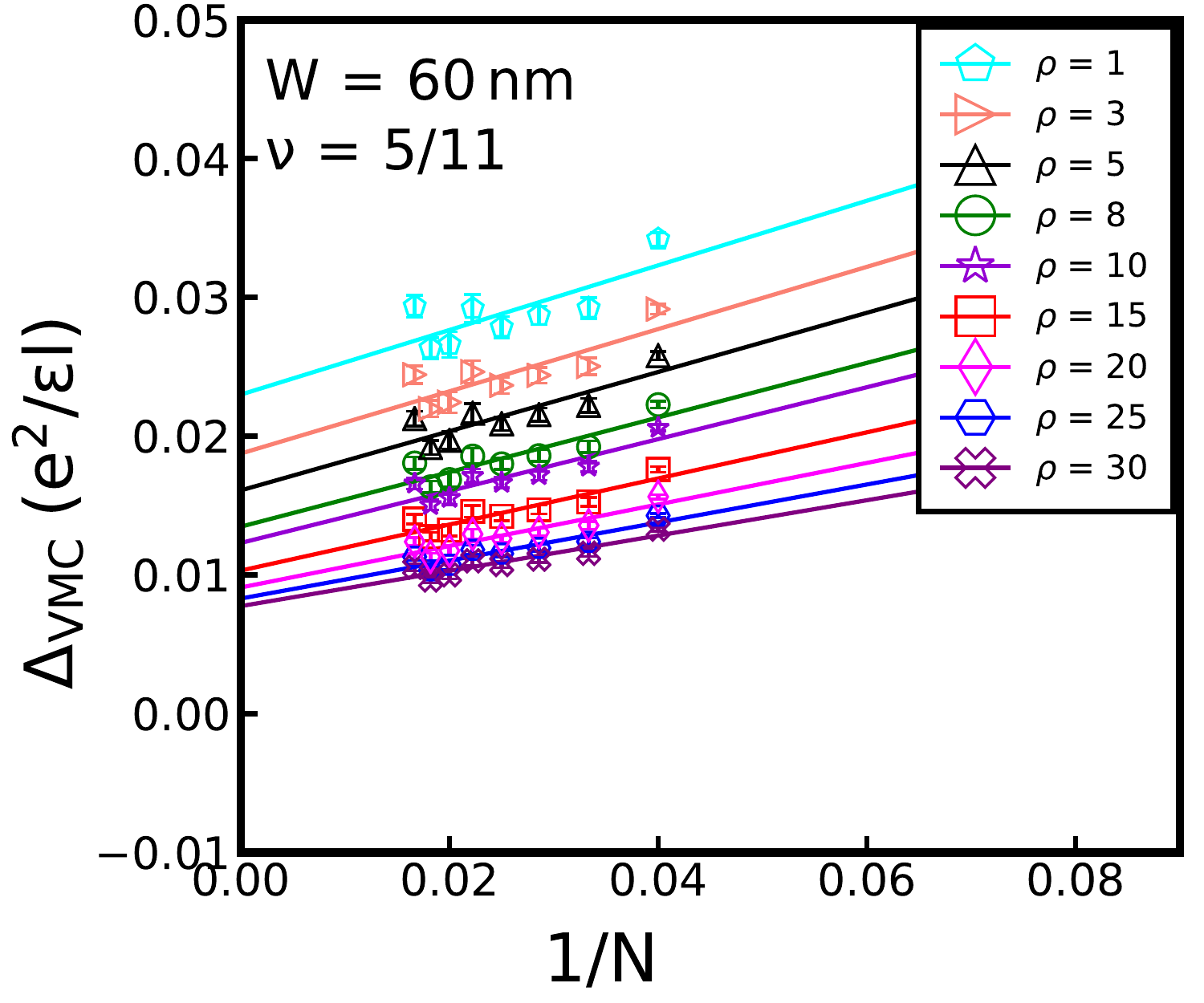}
	\includegraphics[width=0.32 \linewidth]{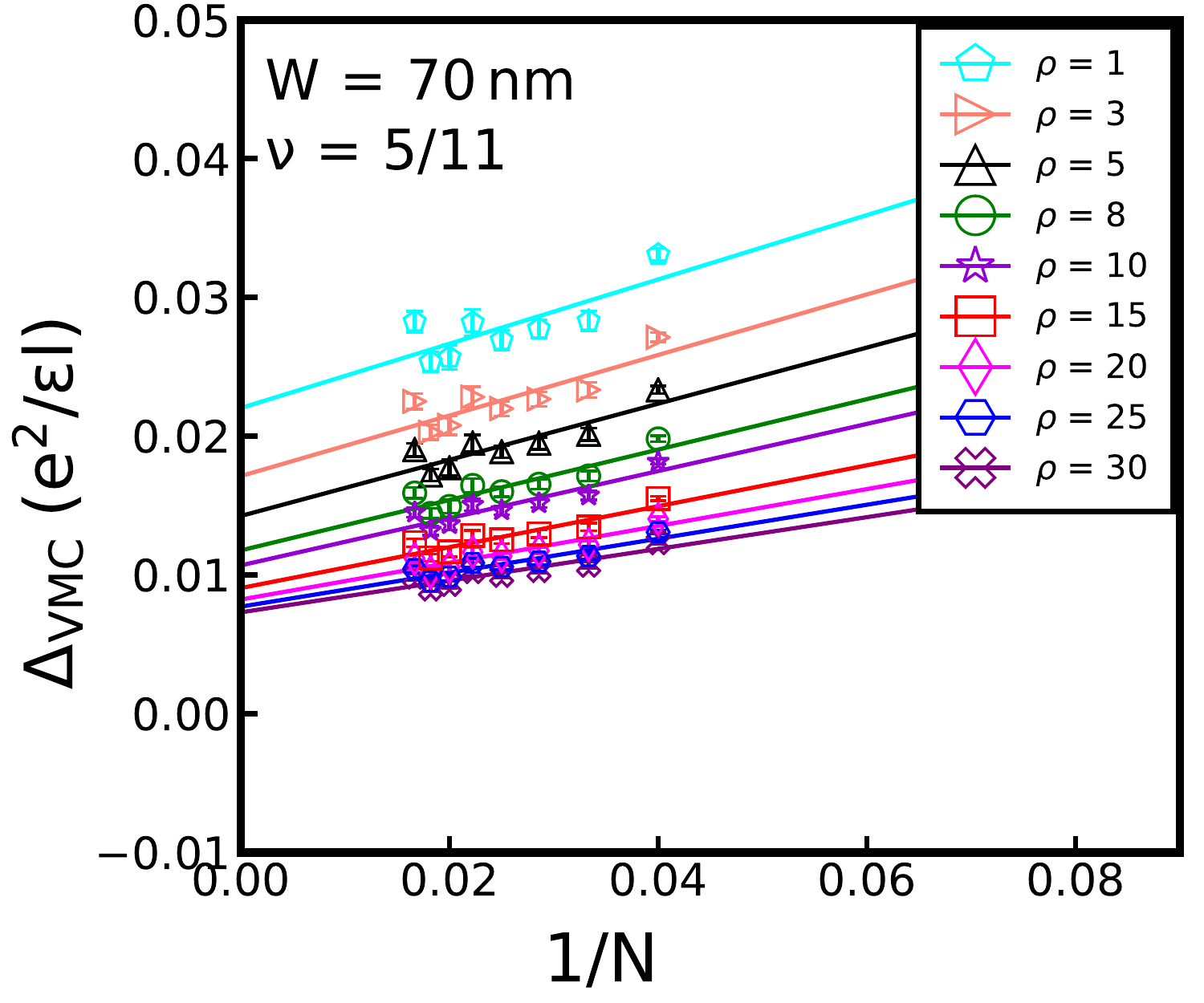}
	\caption{Thermodynamic extrapolation of the transport gap calculated by the VMC for $\nu=5/11$ at different widths and densities. Different markers in the legend label different densities in units of $10^{10}\text{cm}^{-2}$. The density correction and the CF-quasiparticle-quasihole interaction have been included (Eq. (3) of the main text). }\label{X_fig:CF511_VMC_extrap}
\end{figure*}

\begin{figure*}[ht!]
	\includegraphics[width=0.32 \linewidth]{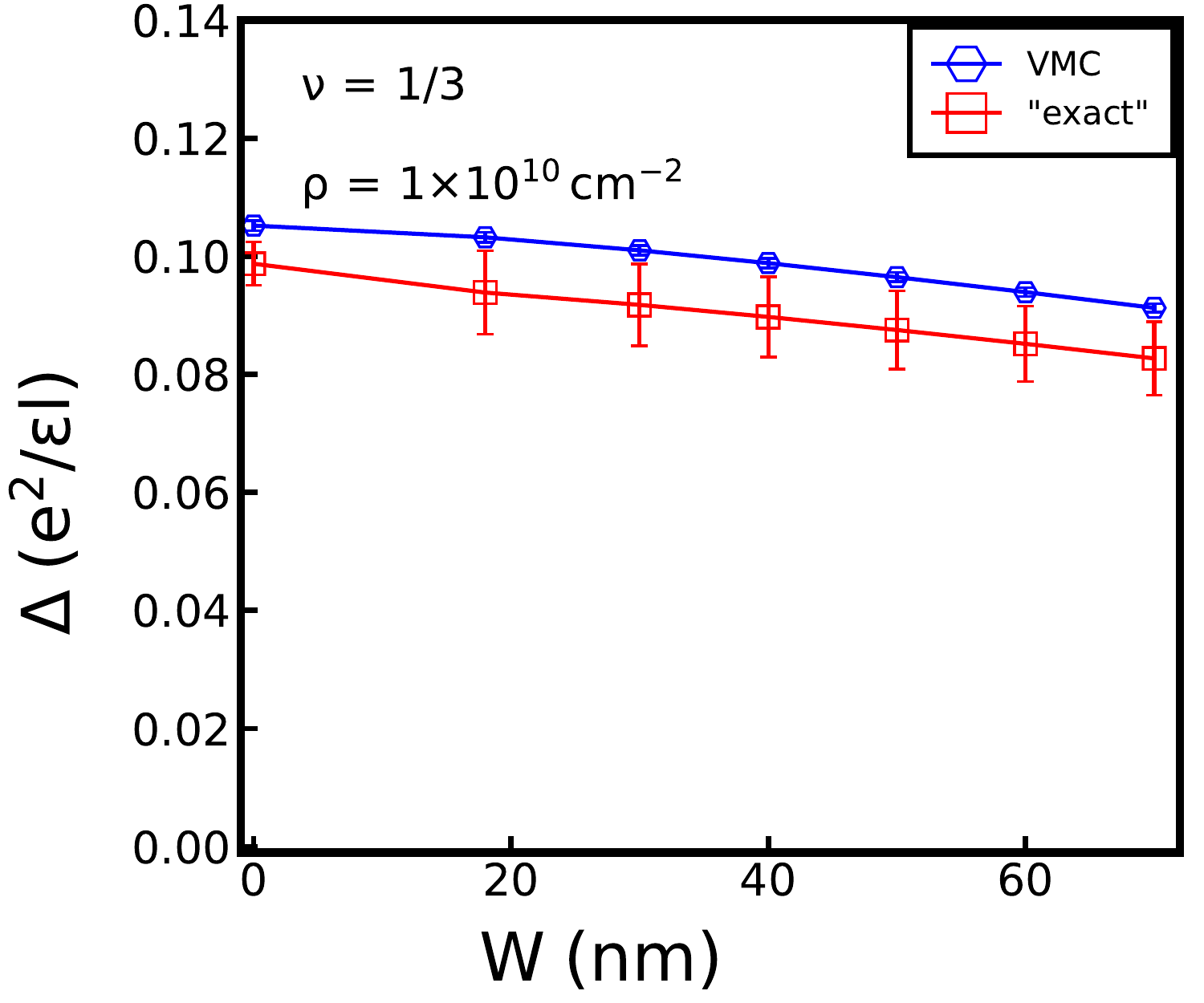}
	\includegraphics[width=0.32 \linewidth]{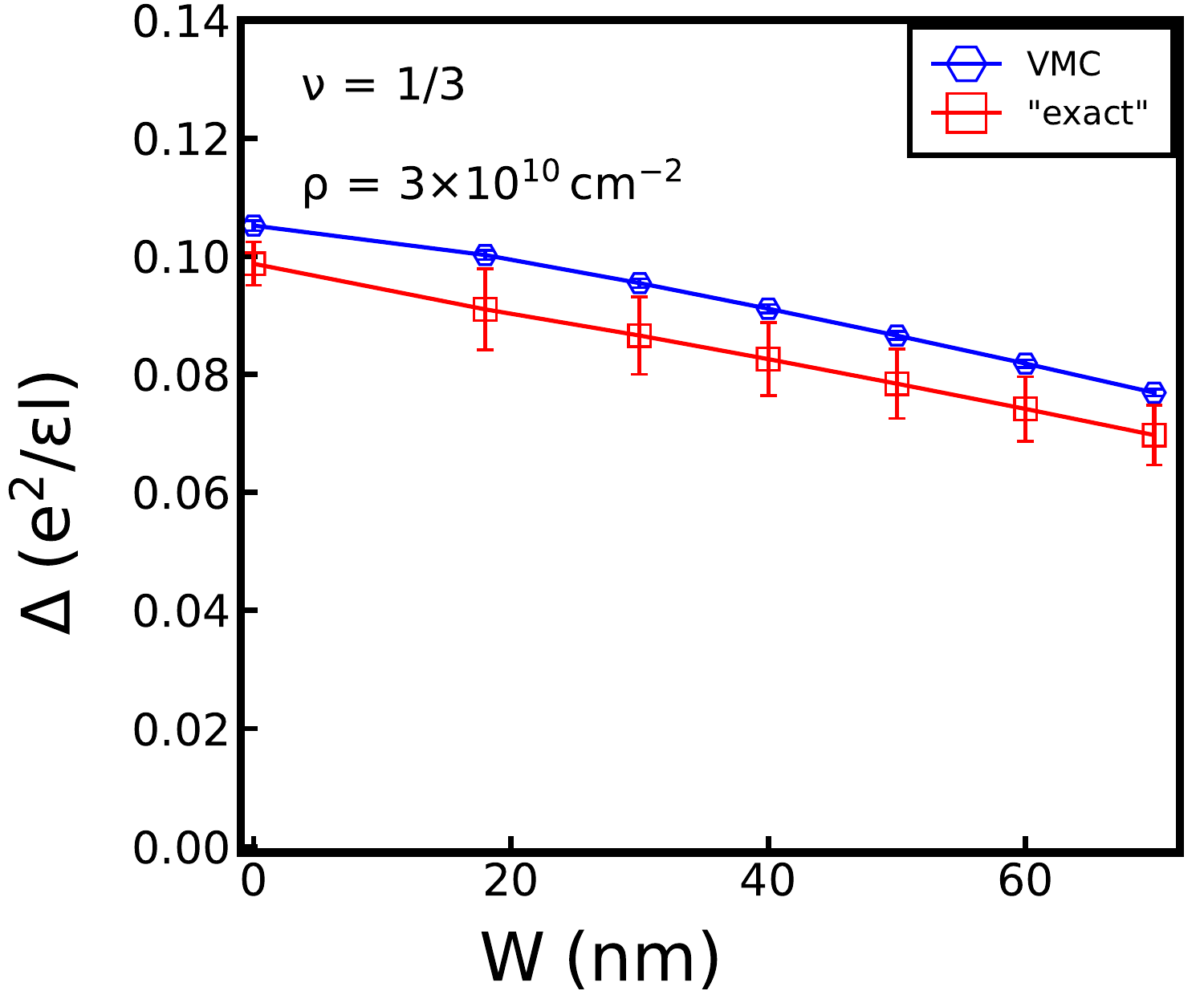}
	\includegraphics[width=0.32 \linewidth]{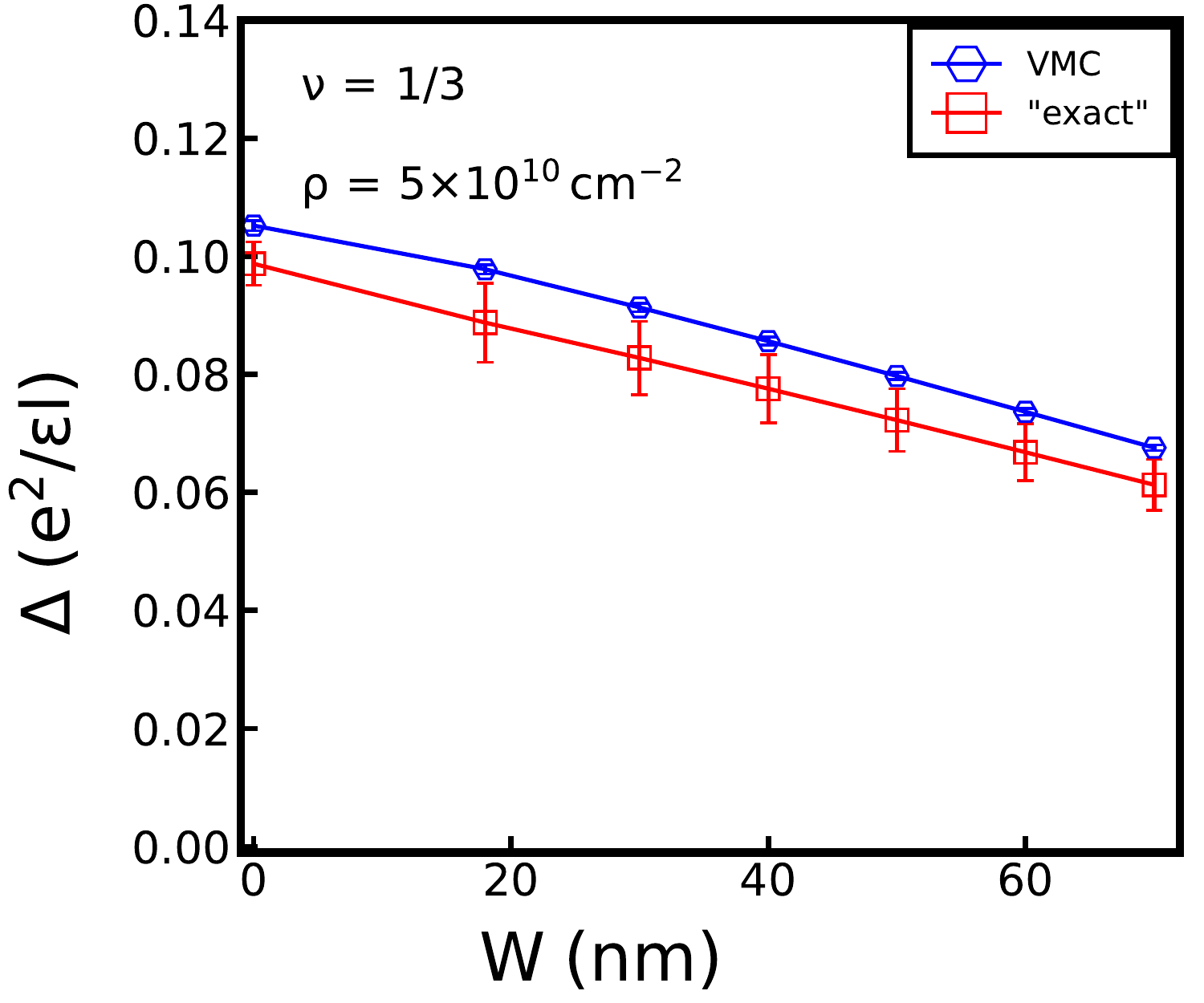}
	\includegraphics[width=0.32 \linewidth]{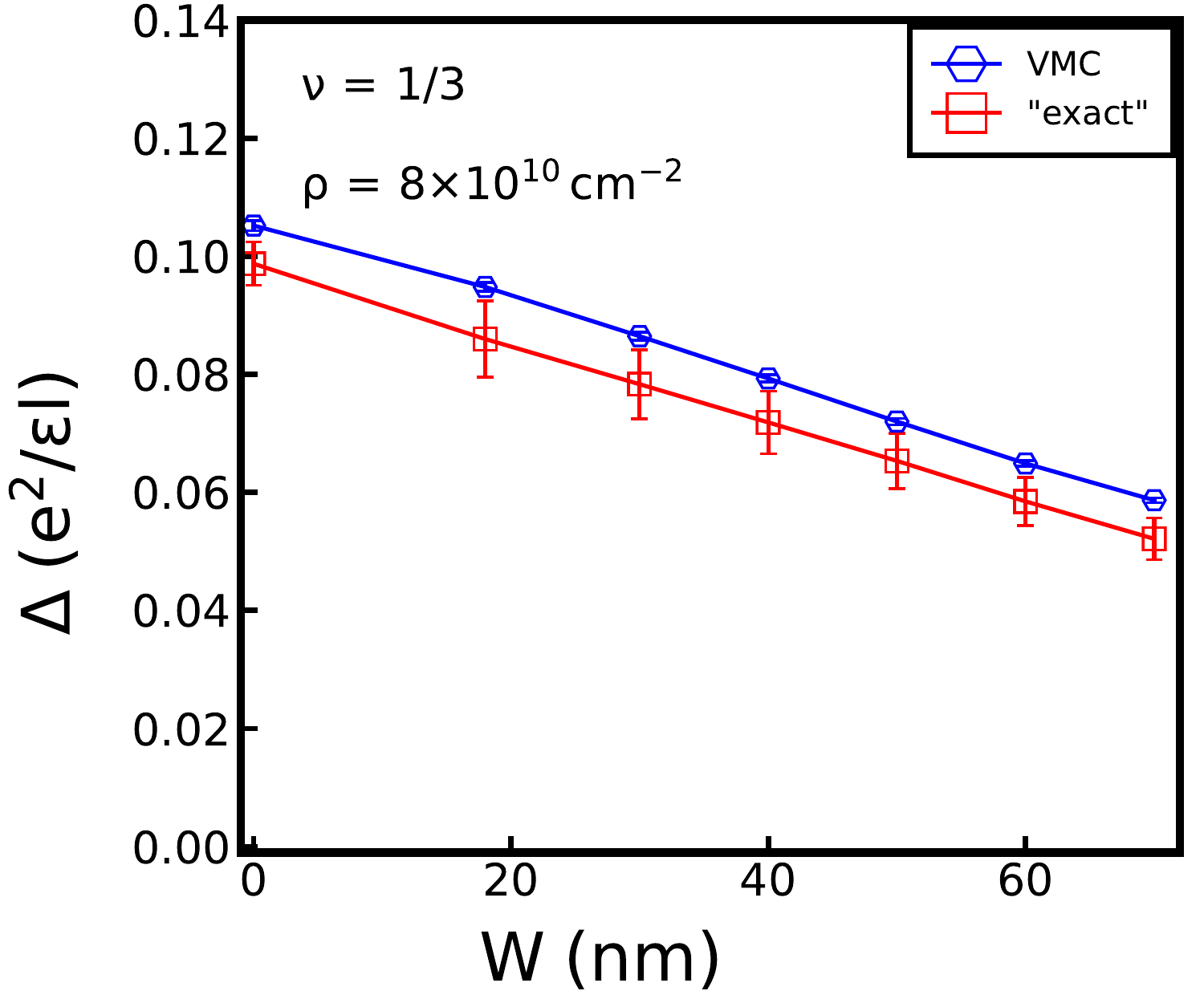}
	\includegraphics[width=0.32 \linewidth]{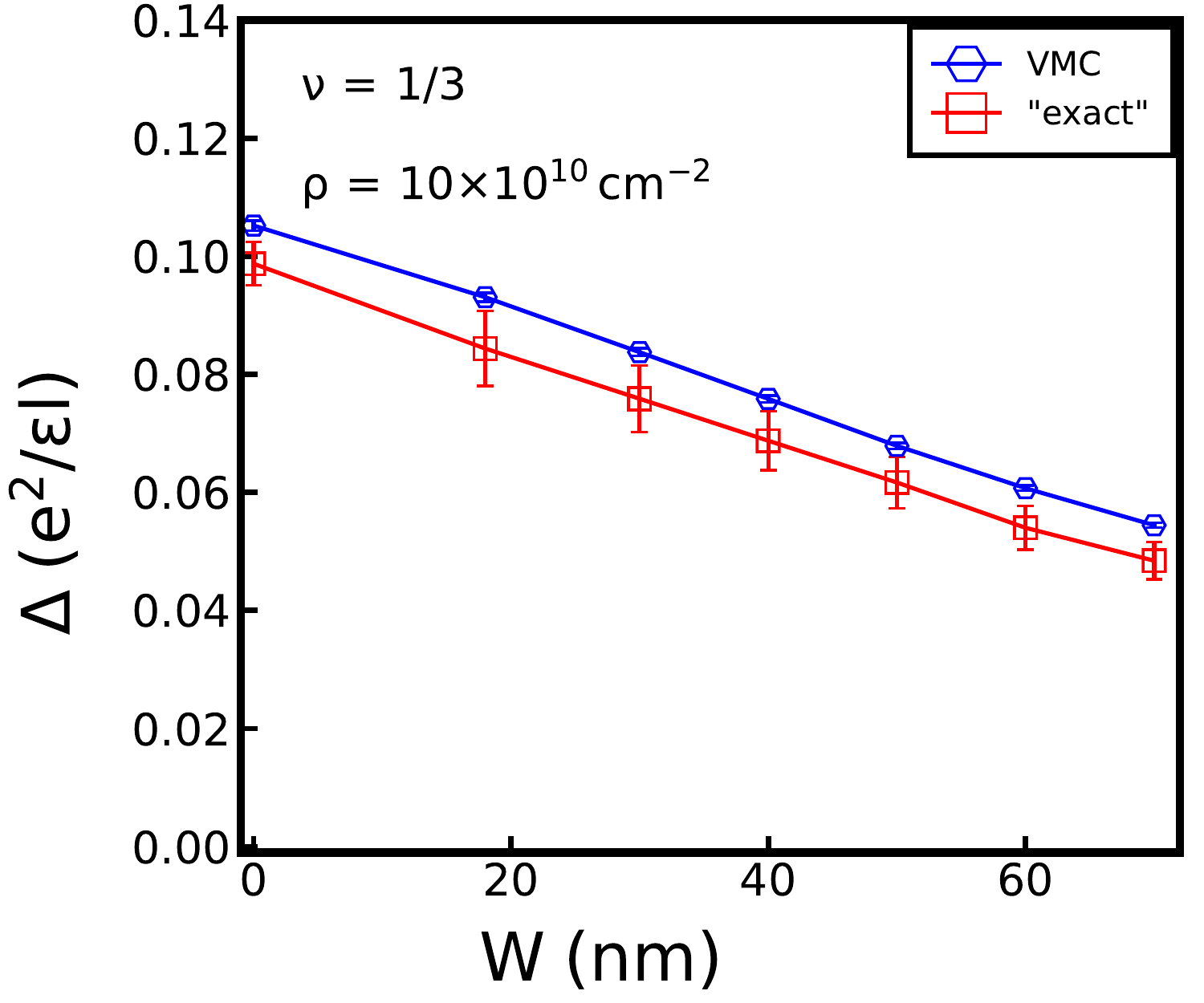}
	\includegraphics[width=0.32 \linewidth]{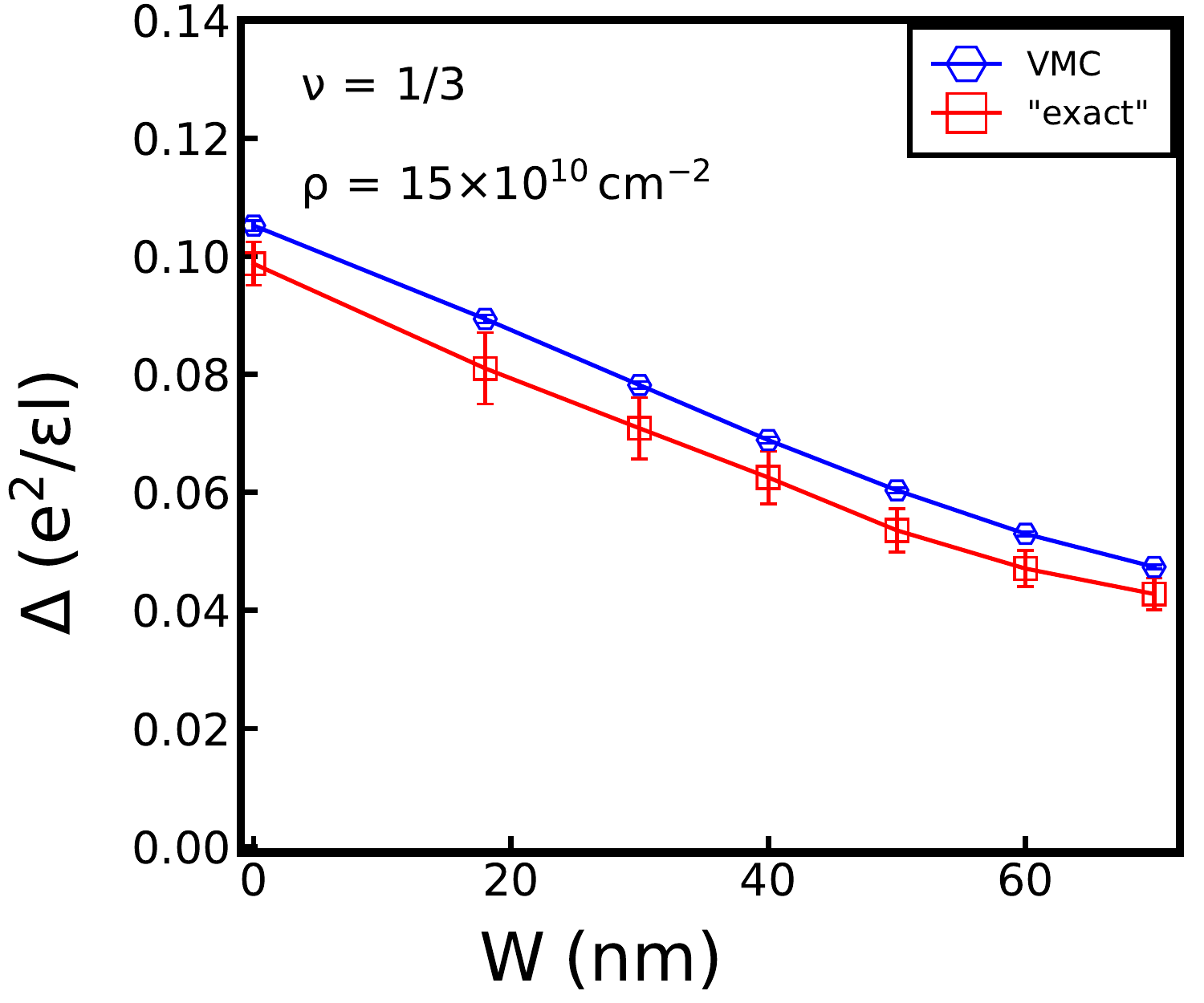}
	\includegraphics[width=0.32 \linewidth]{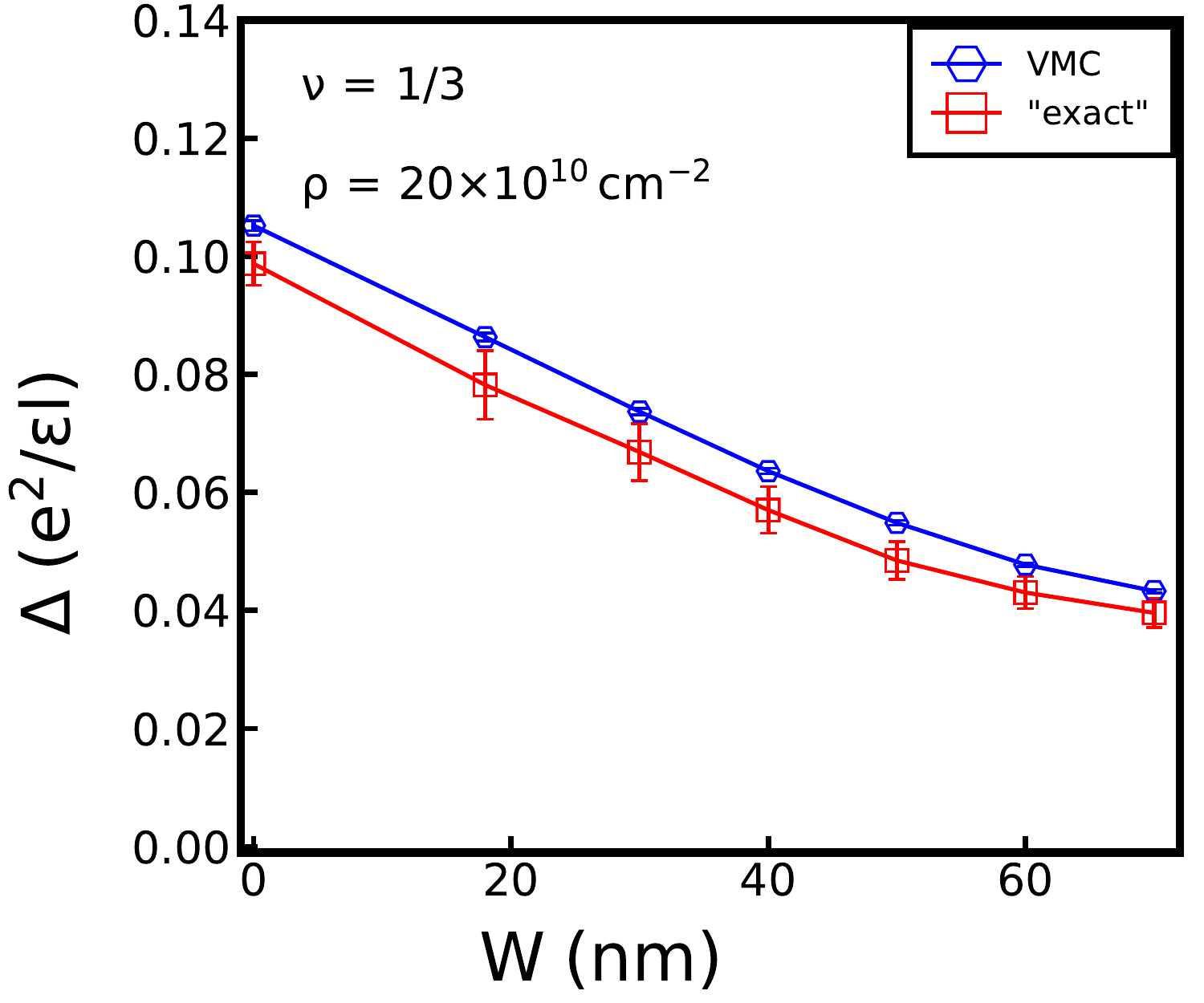}
	\includegraphics[width=0.32 \linewidth]{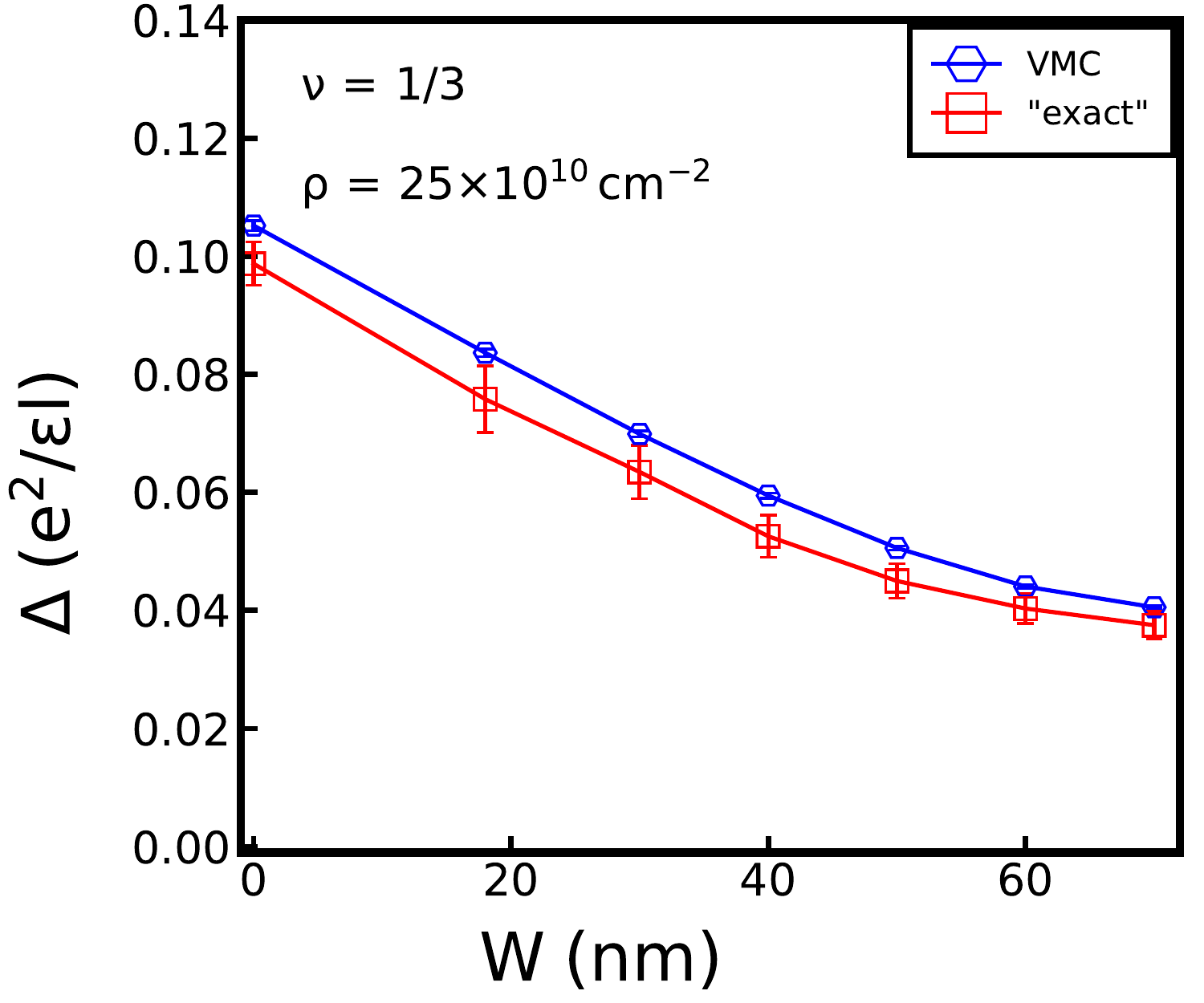}
	\includegraphics[width=0.32 \linewidth]{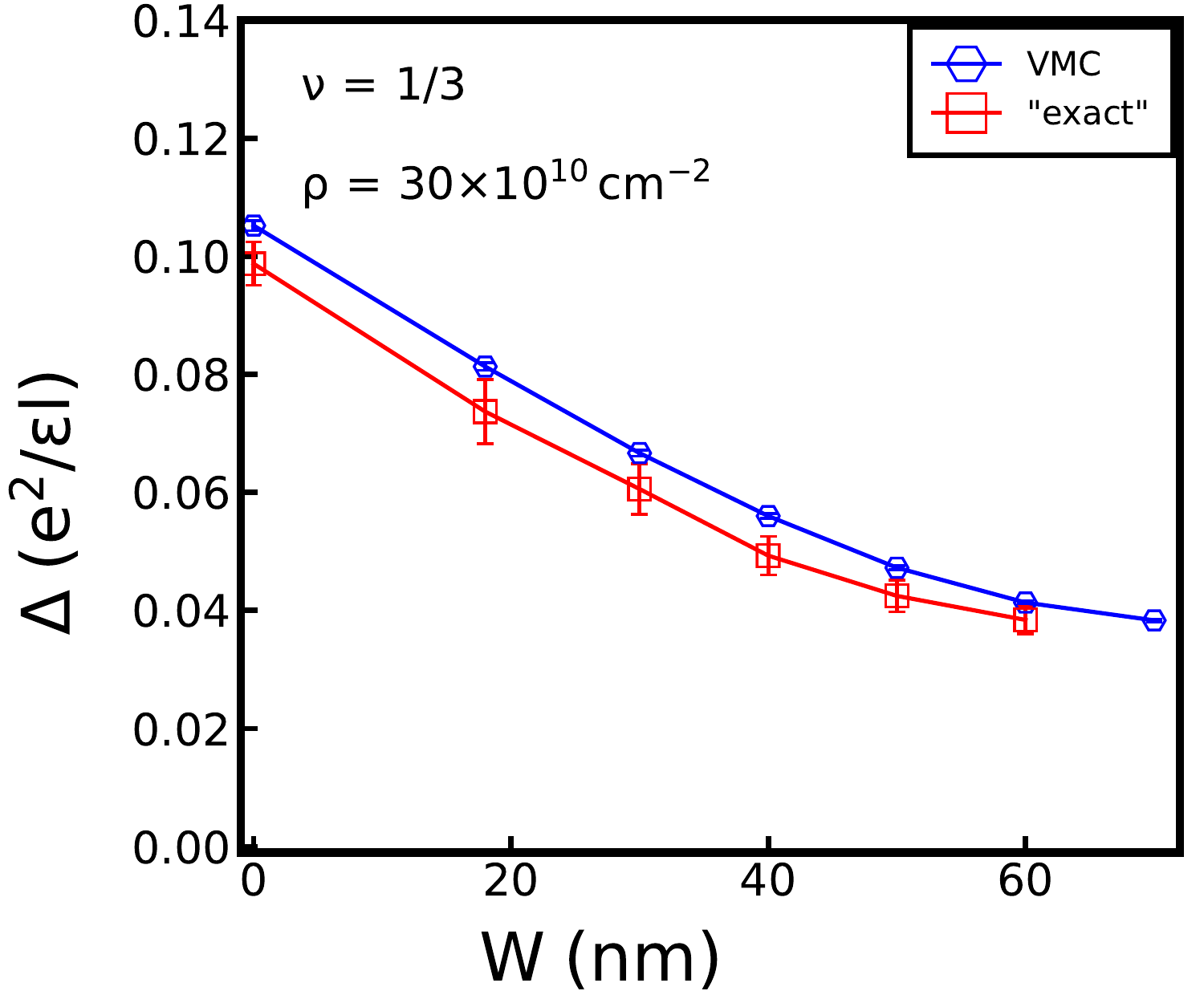}
	\caption{Transport gaps calculated by variational Monte Carlo (VMC) and corrected for variational error (``exact") at $\nu=1/3$ for several widths and densities.}\label{X_fig:VMC_13_other}
\end{figure*}

\begin{figure*}[ht!]
	\includegraphics[width=0.32 \linewidth]{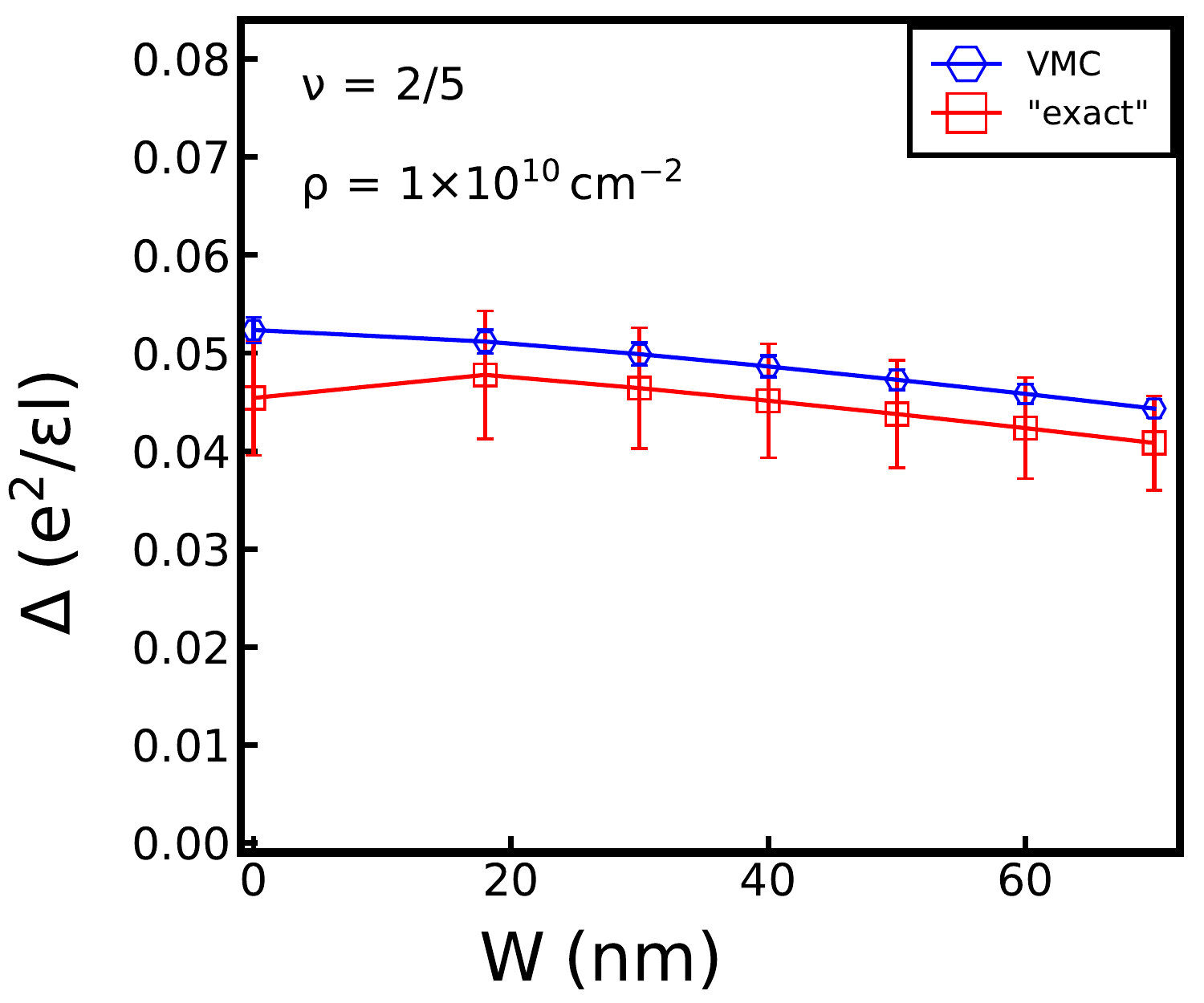}
	\includegraphics[width=0.32 \linewidth]{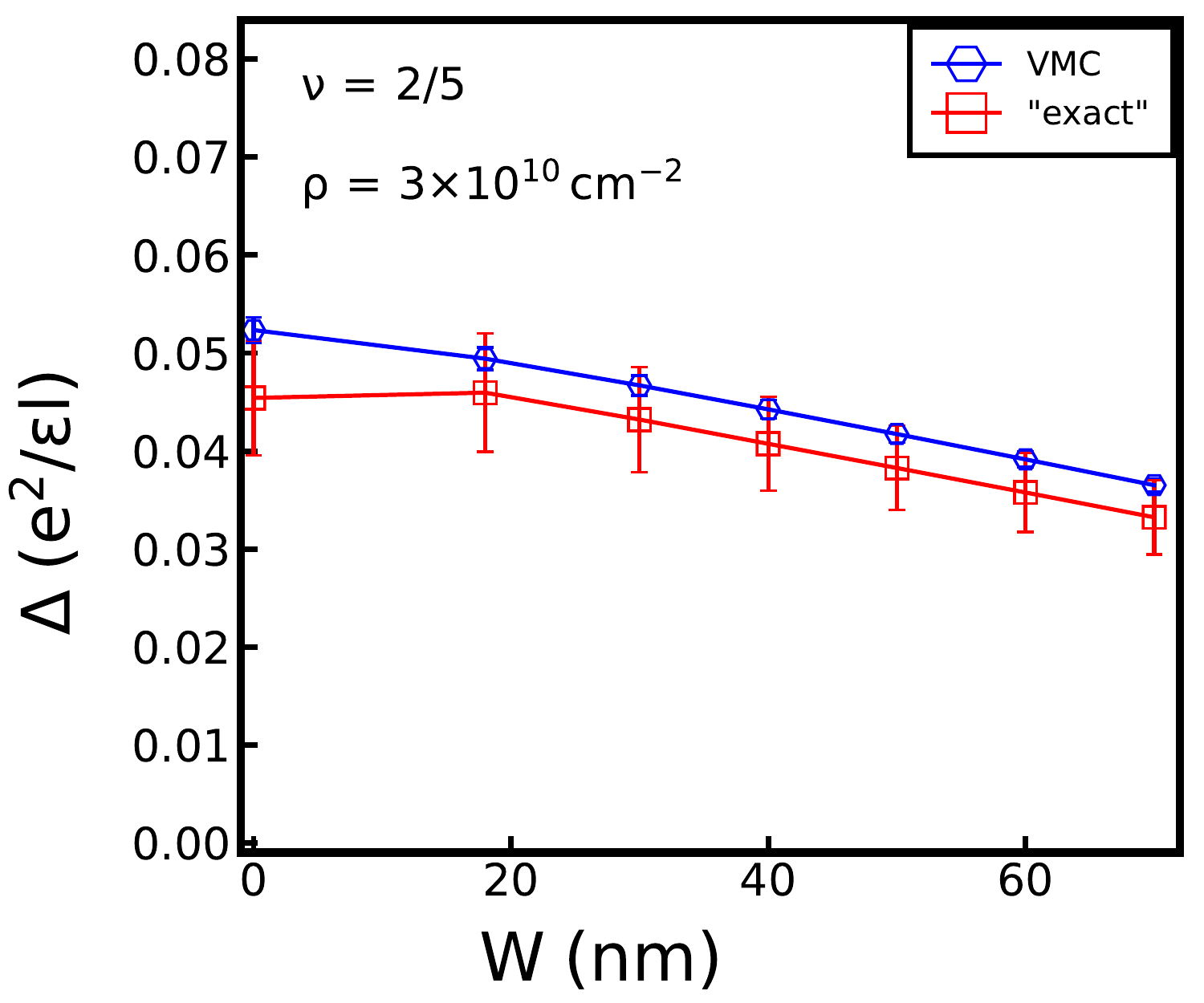}
	\includegraphics[width=0.32 \linewidth]{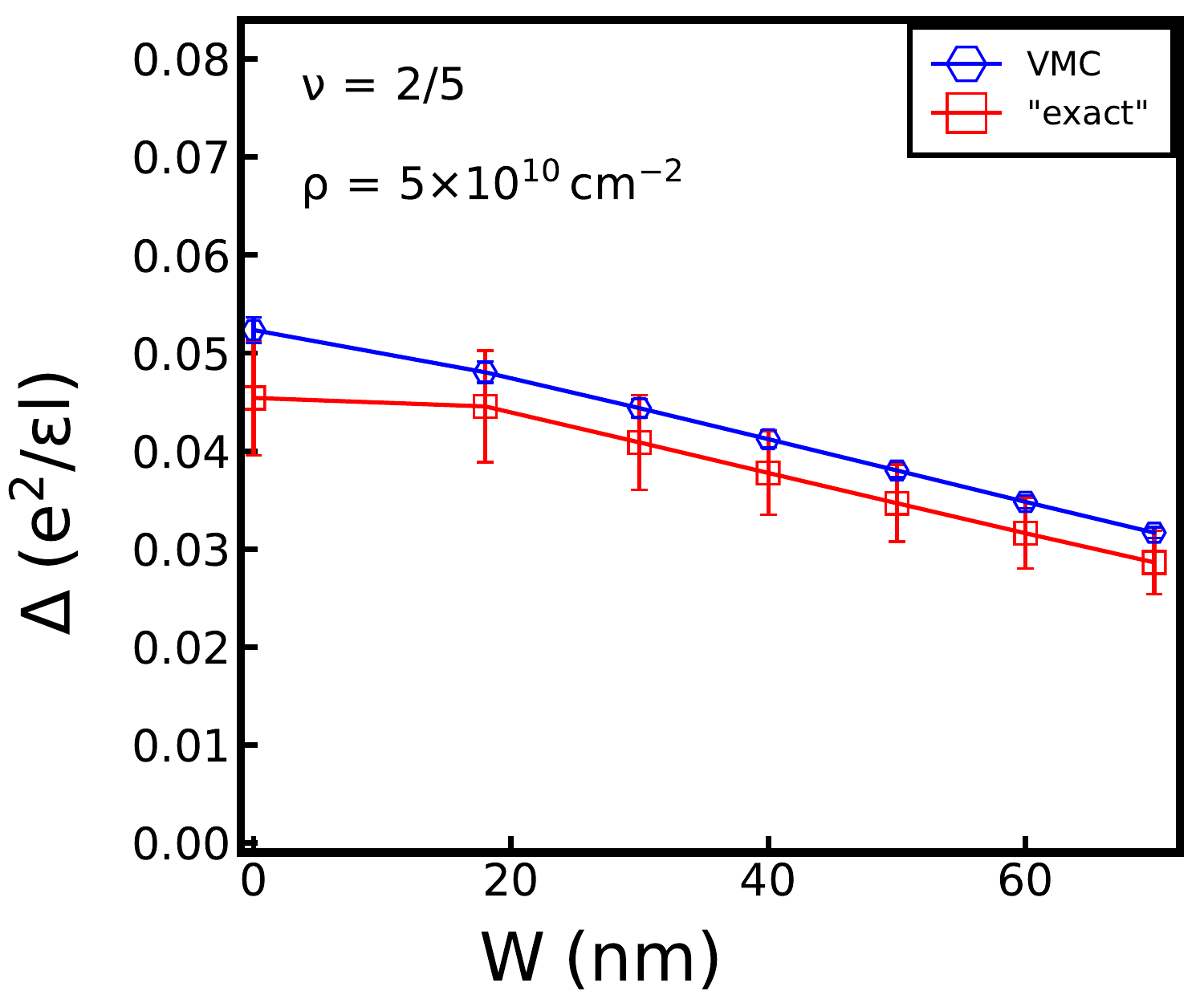}
	\includegraphics[width=0.32 \linewidth]{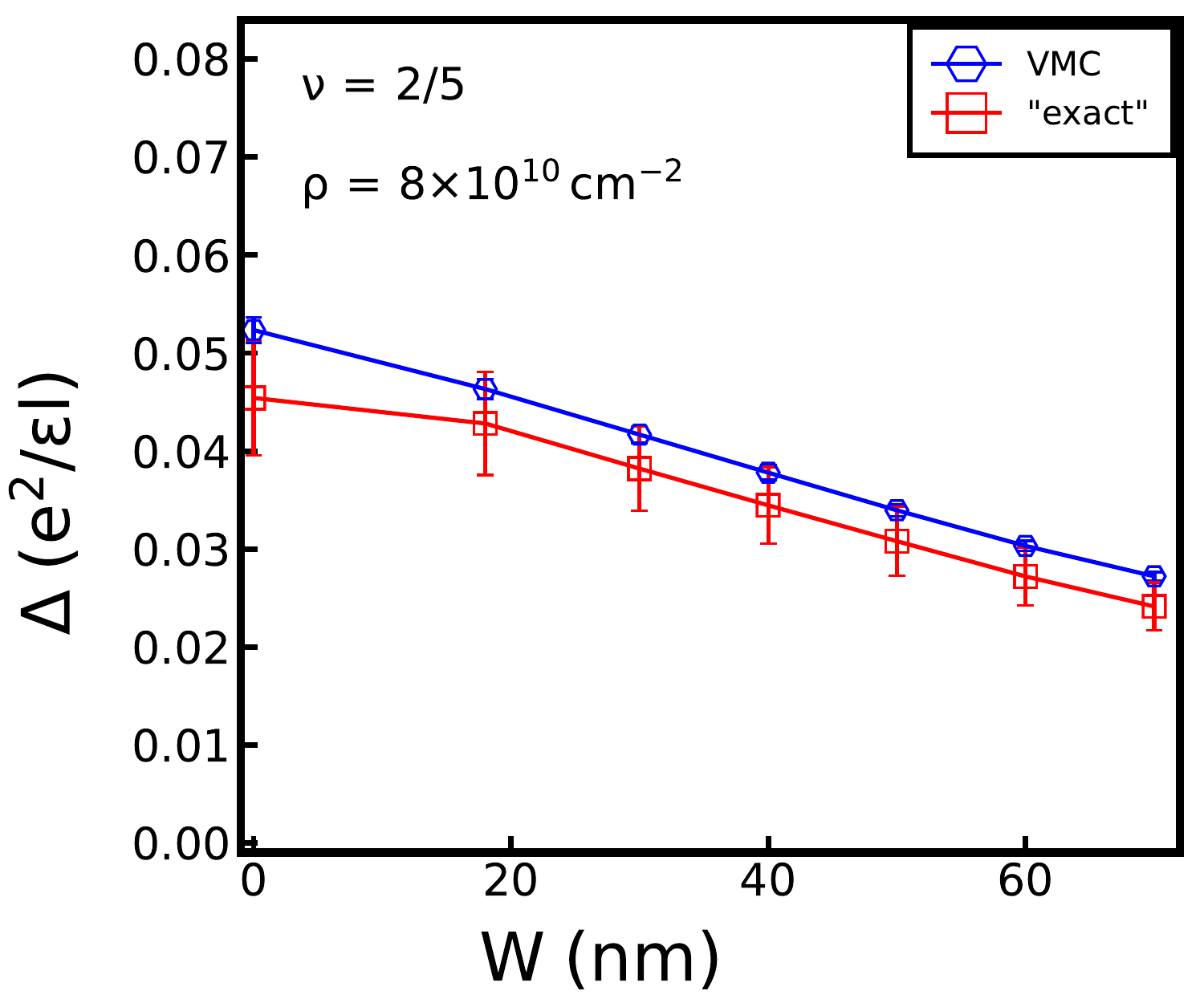}
	\includegraphics[width=0.32 \linewidth]{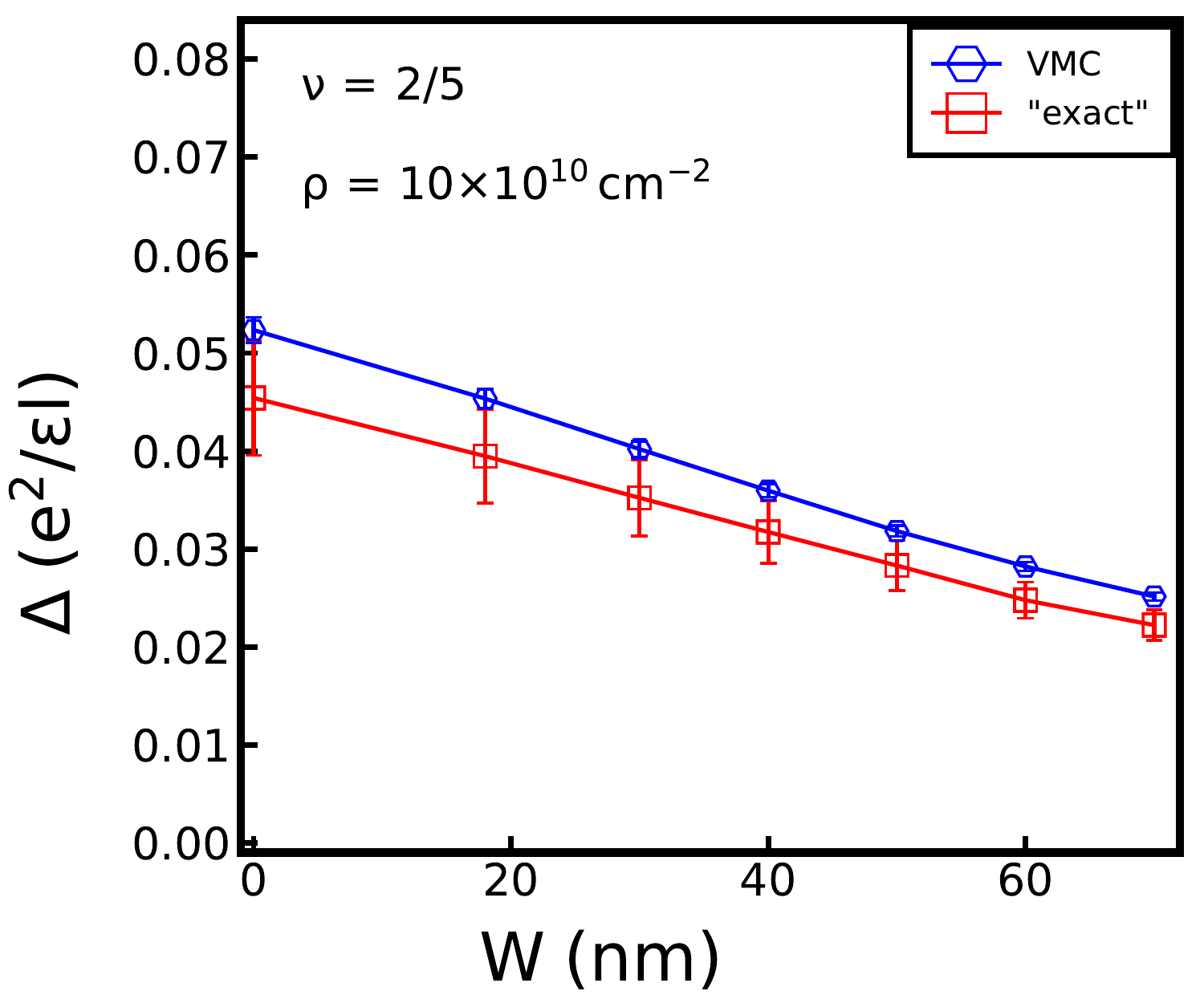}
	\includegraphics[width=0.32 \linewidth]{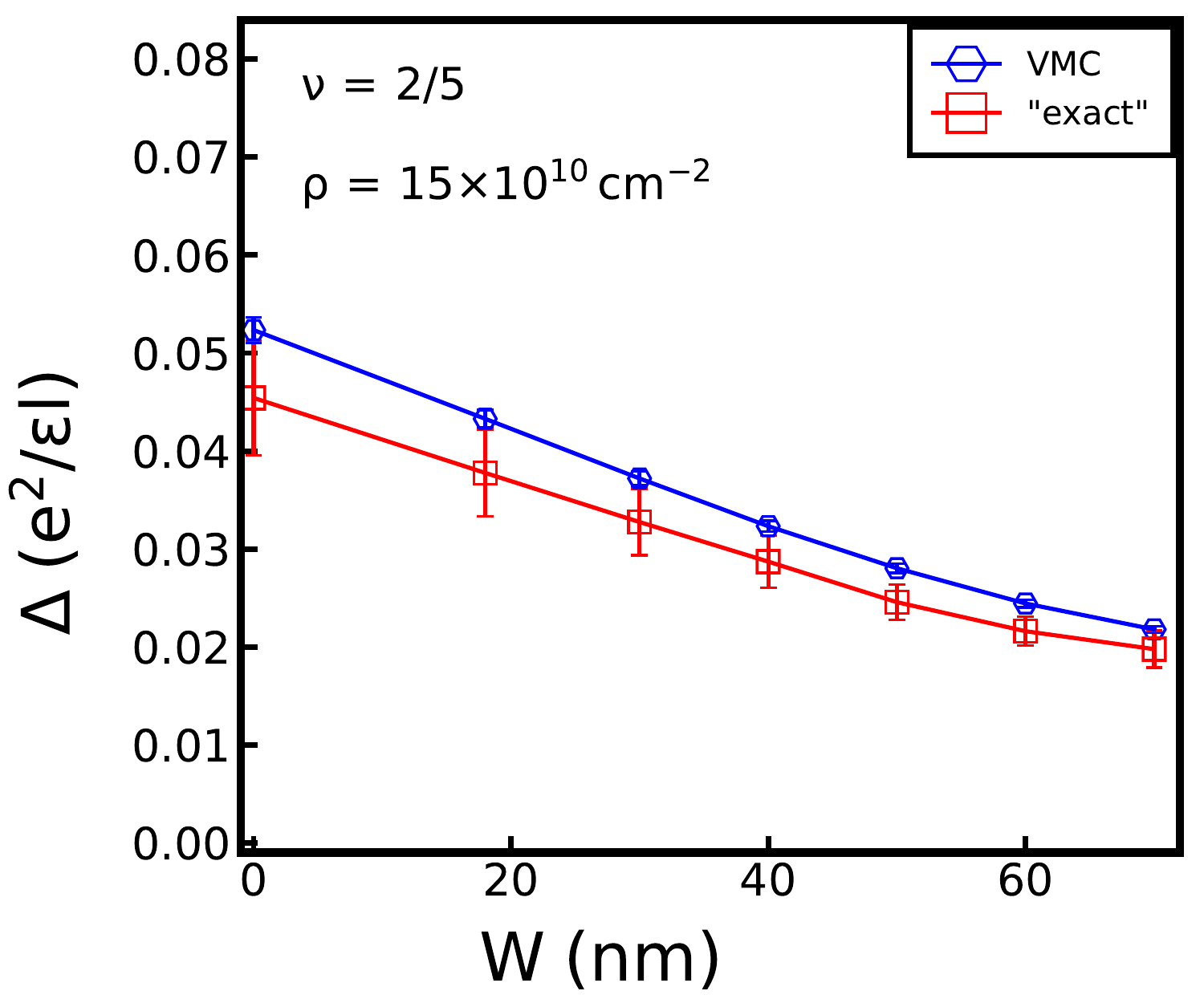}
	\includegraphics[width=0.32 \linewidth]{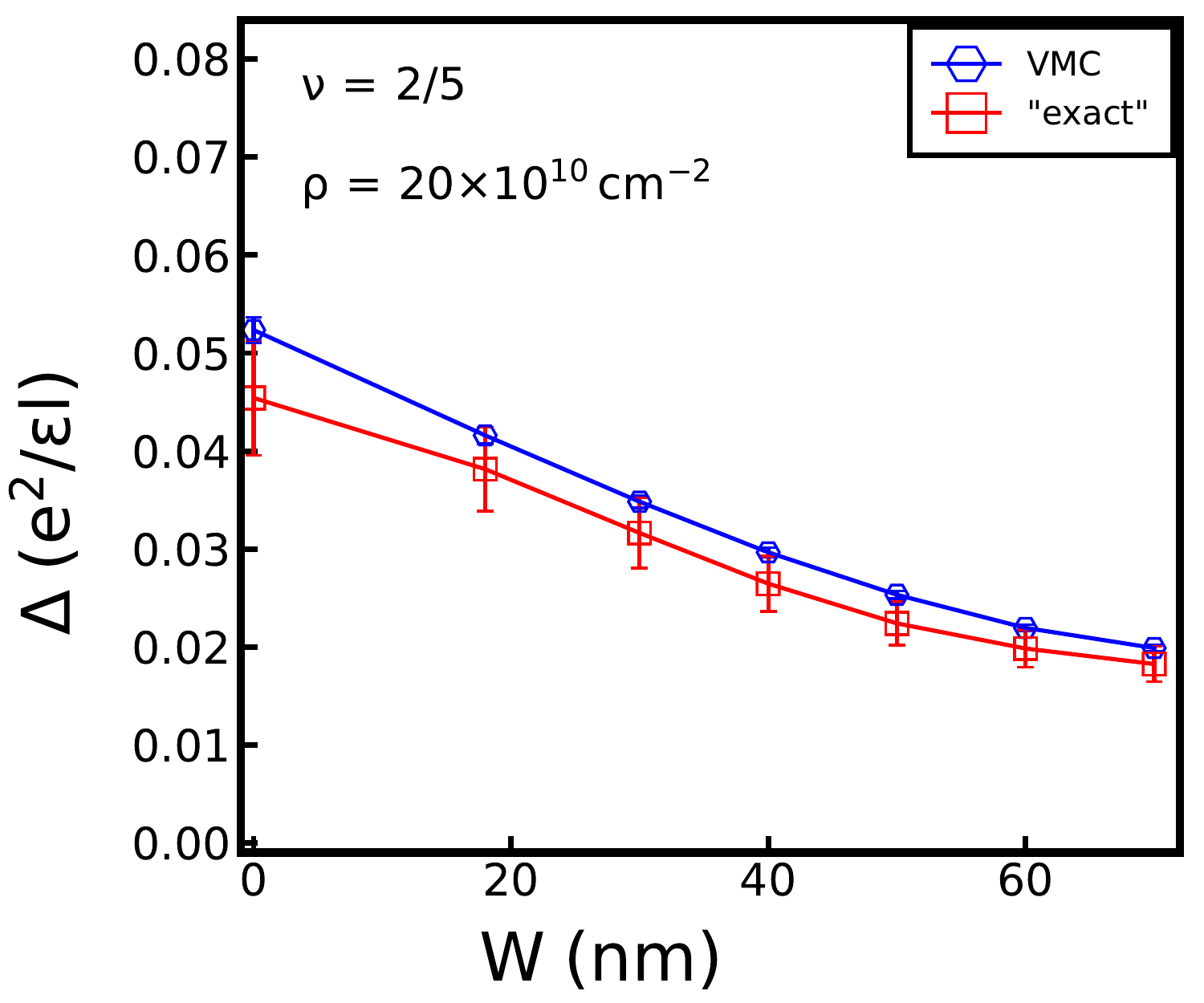}
	\includegraphics[width=0.32 \linewidth]{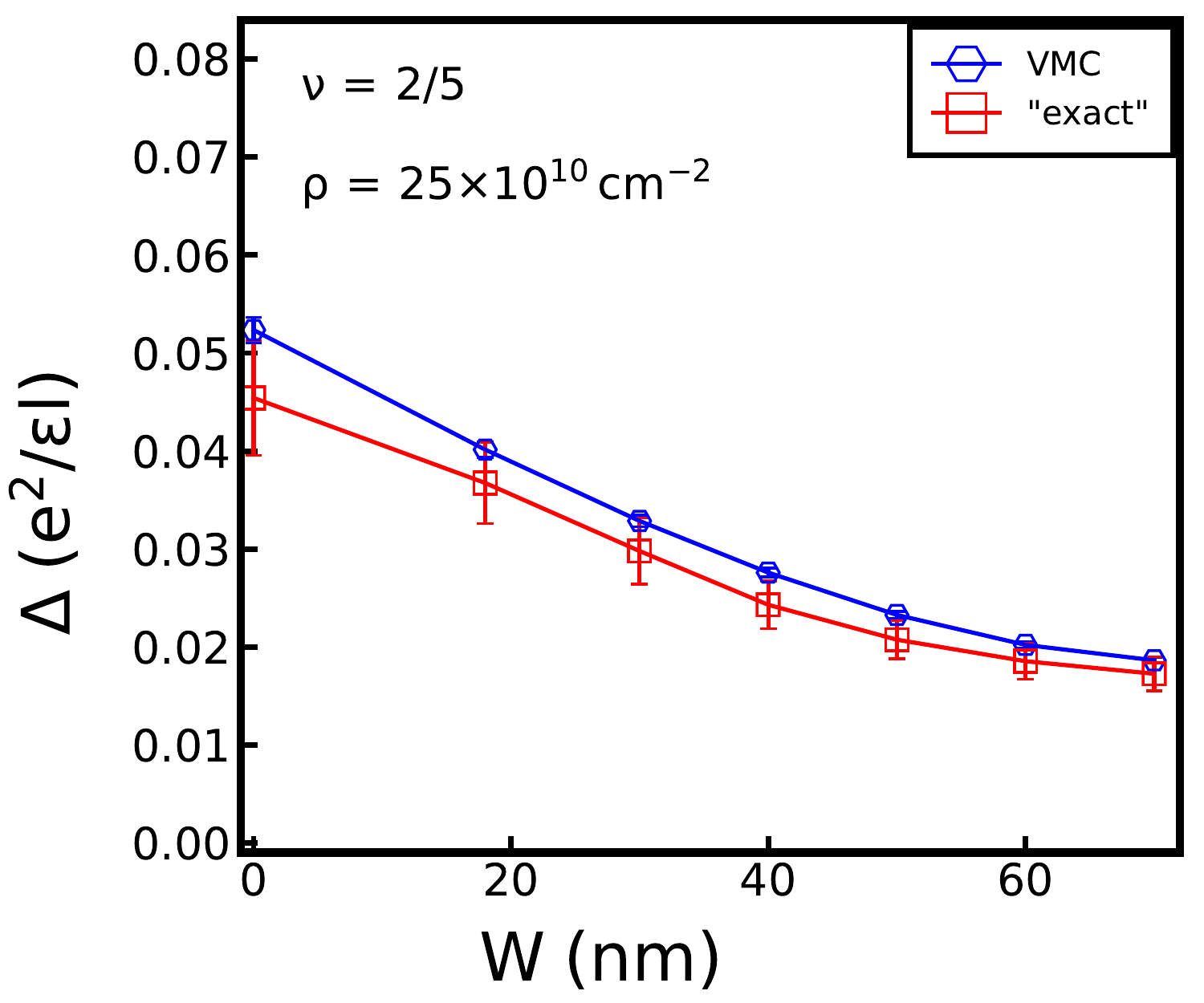}
	\includegraphics[width=0.32 \linewidth]{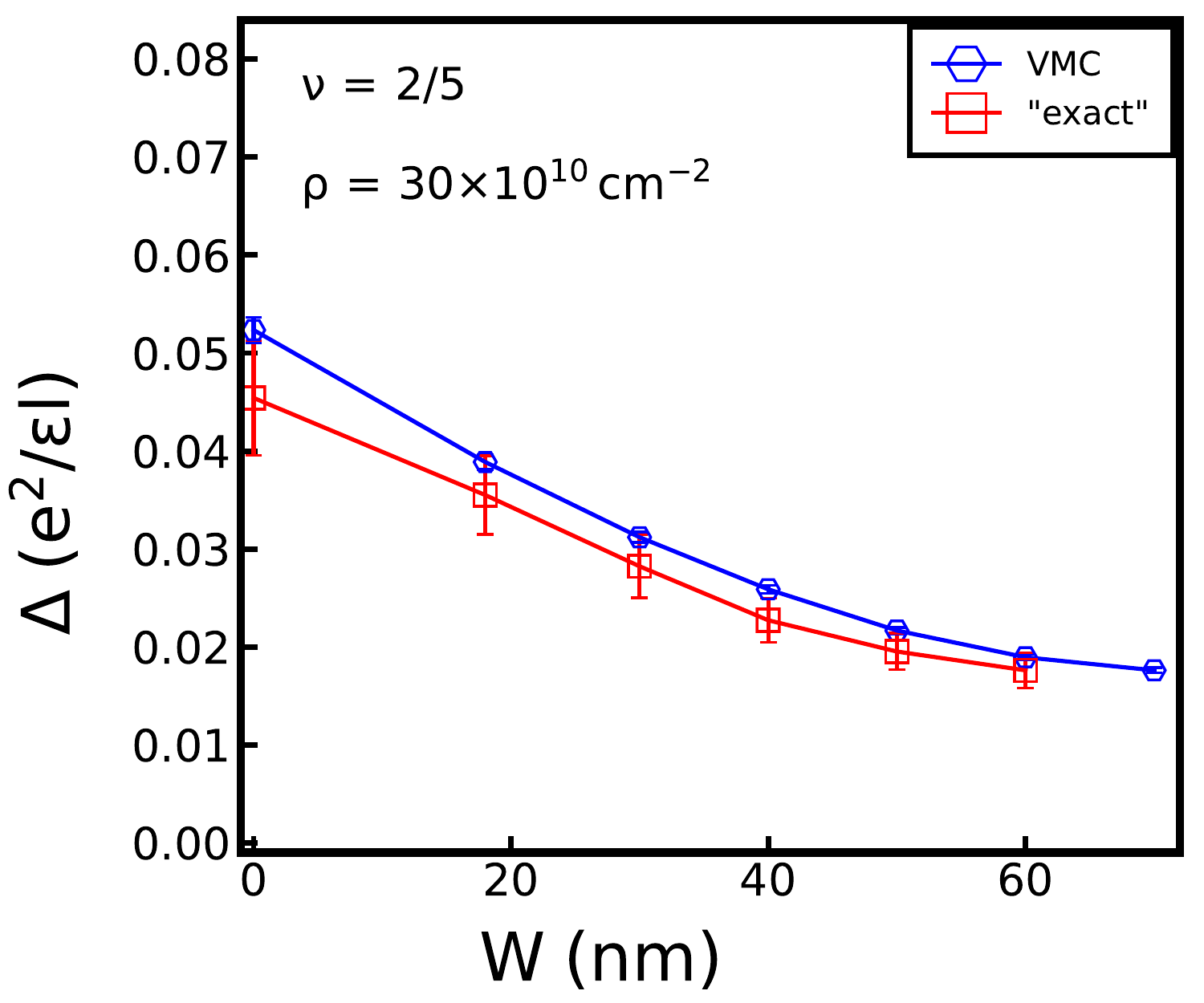}
	\caption{Transport gaps calculated by variational Monte Carlo (VMC) and corrected for variational error (``exact") at $\nu=2/5$ for several widths and densities.}\label{X_fig:VMC_25_other}
\end{figure*}

\begin{figure*}[ht!]
	\includegraphics[width=0.32 \linewidth]{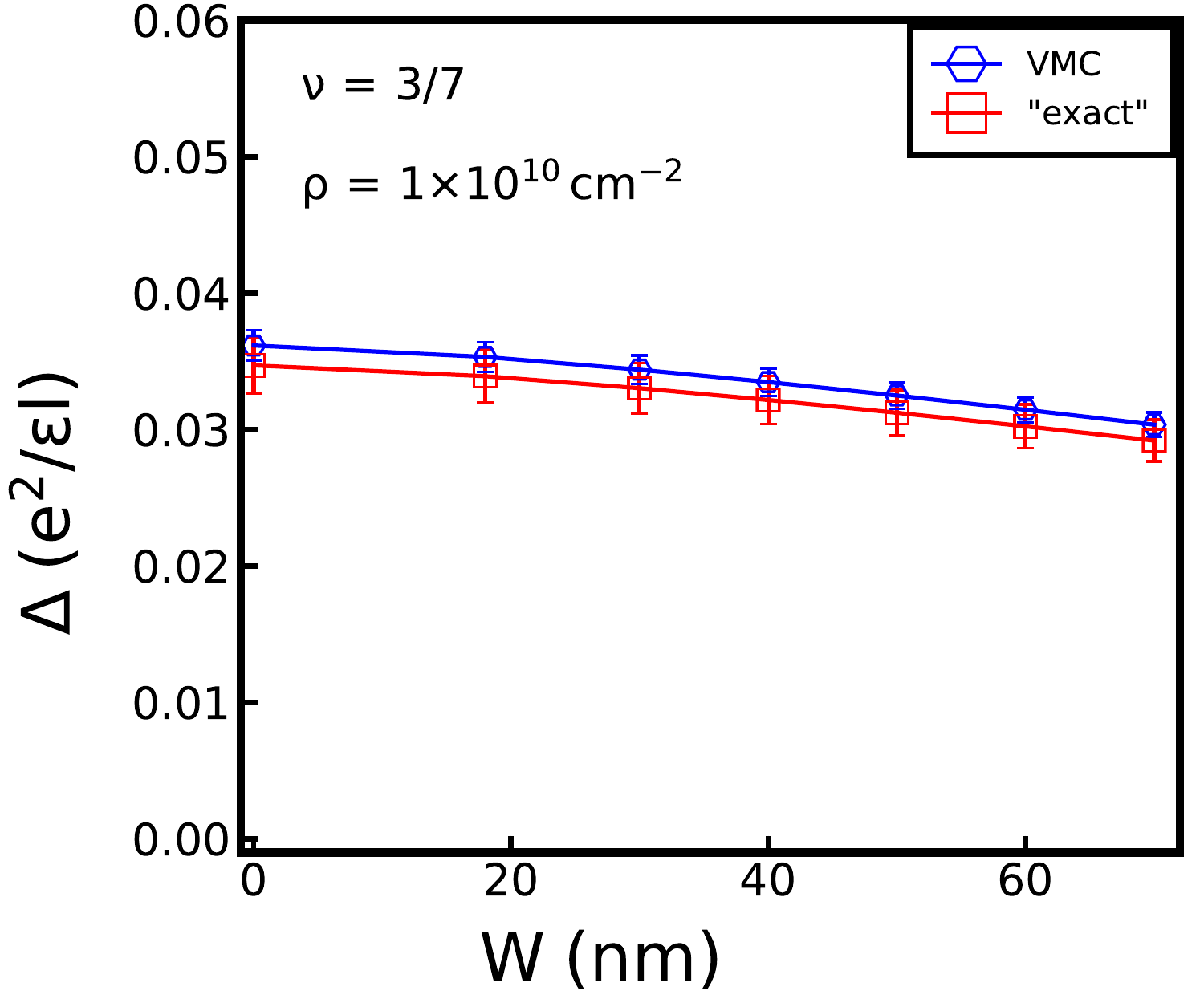}
	\includegraphics[width=0.32 \linewidth]{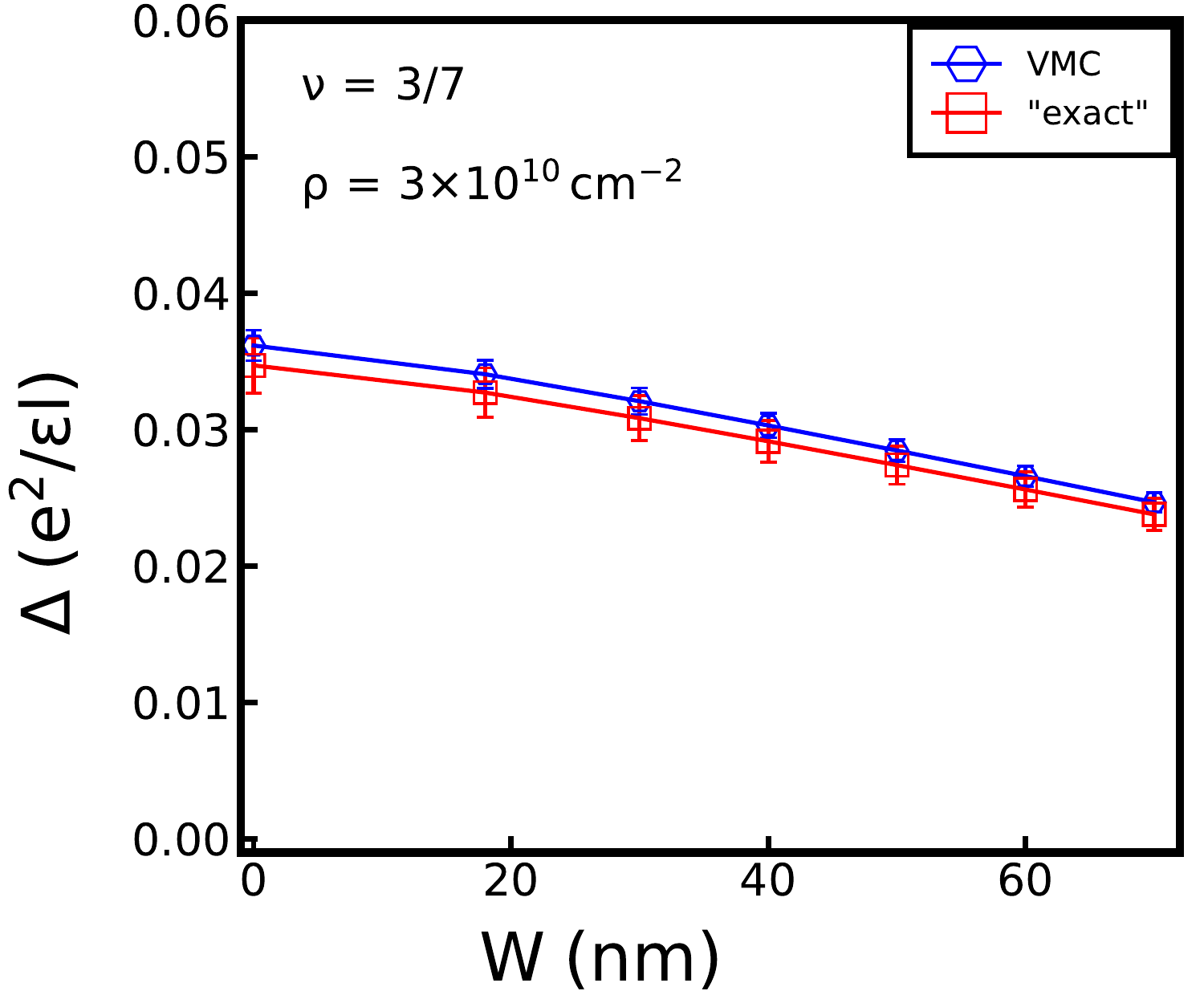}
	\includegraphics[width=0.32 \linewidth]{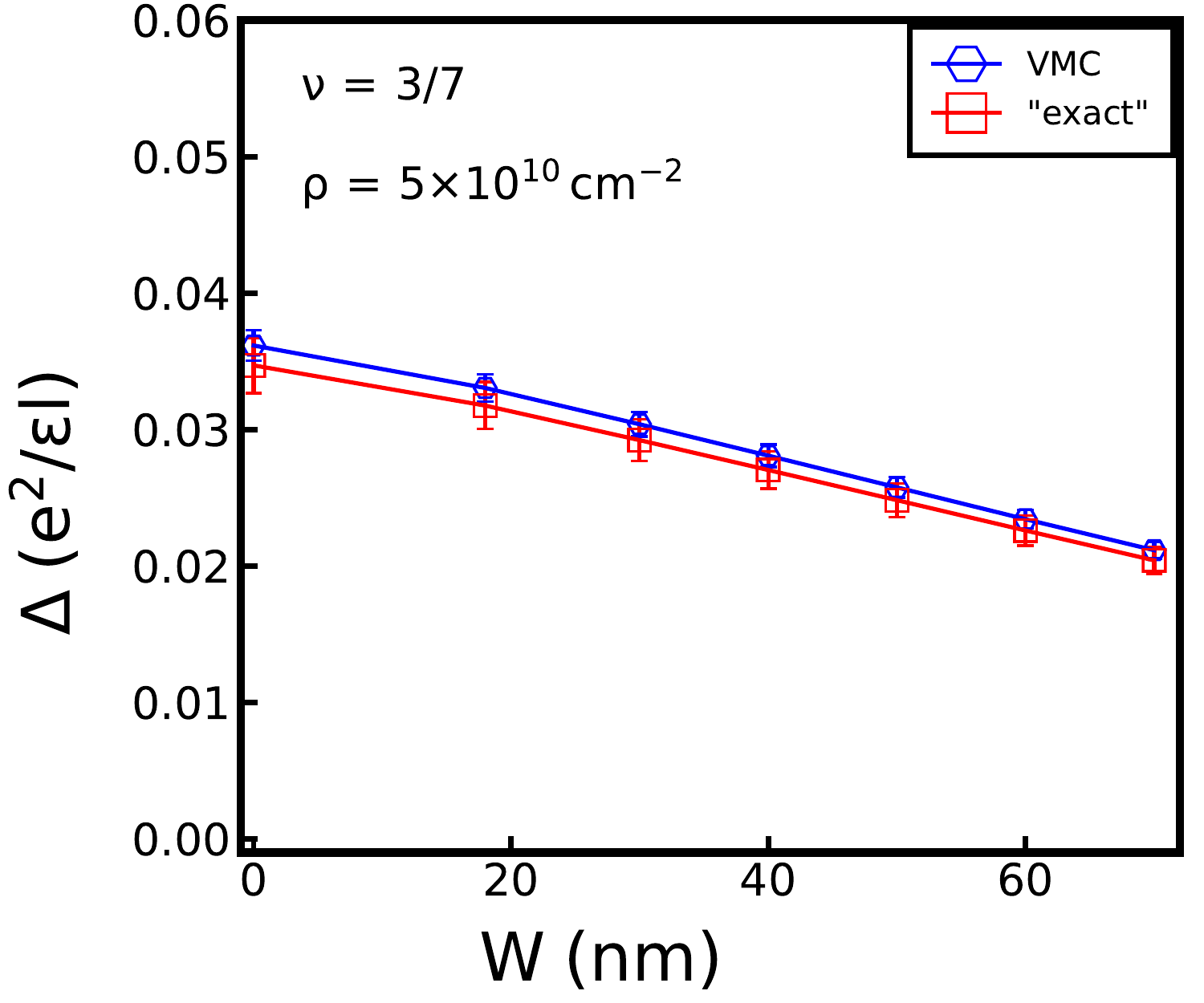}
	\includegraphics[width=0.32 \linewidth]{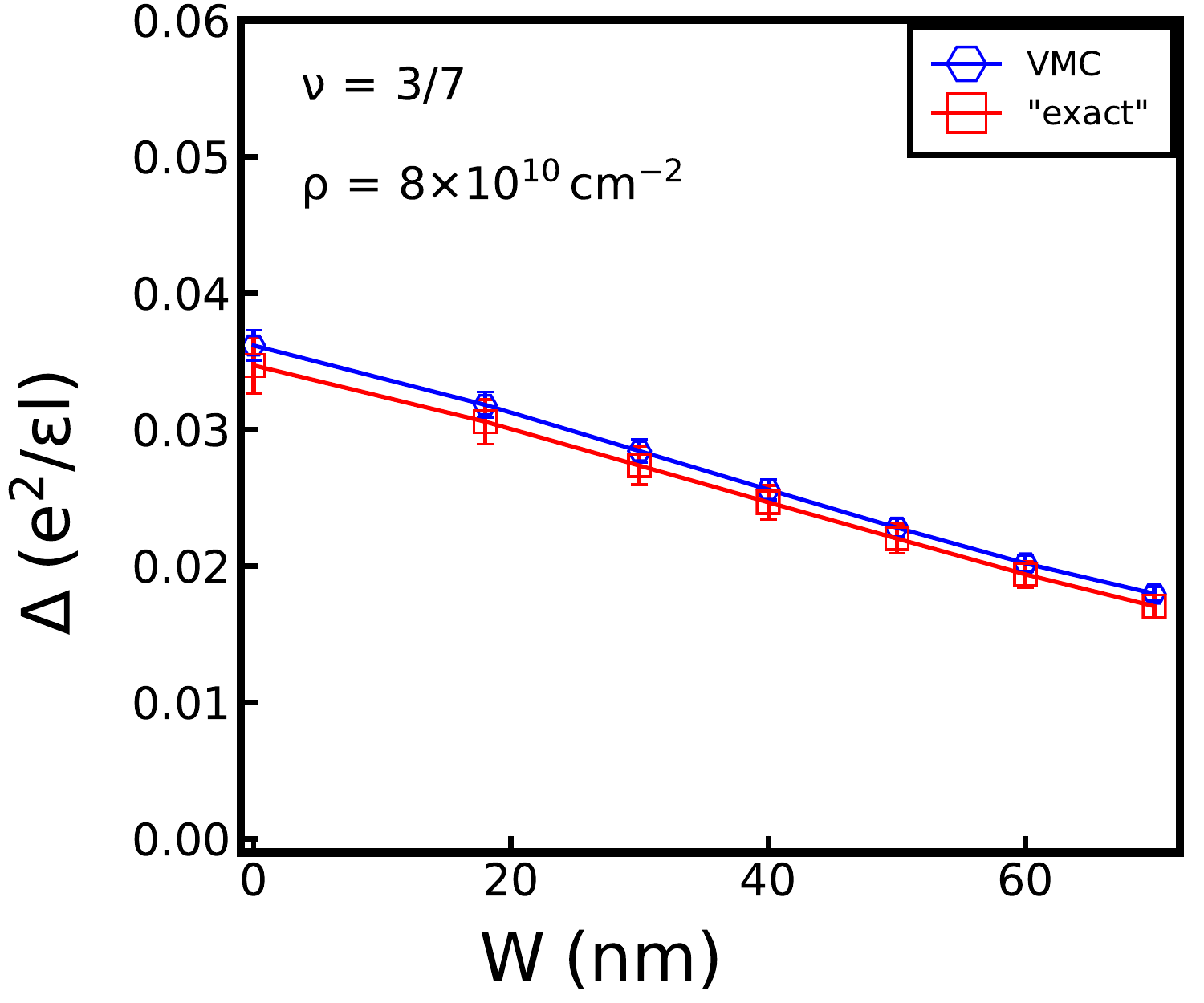}
	\includegraphics[width=0.32 \linewidth]{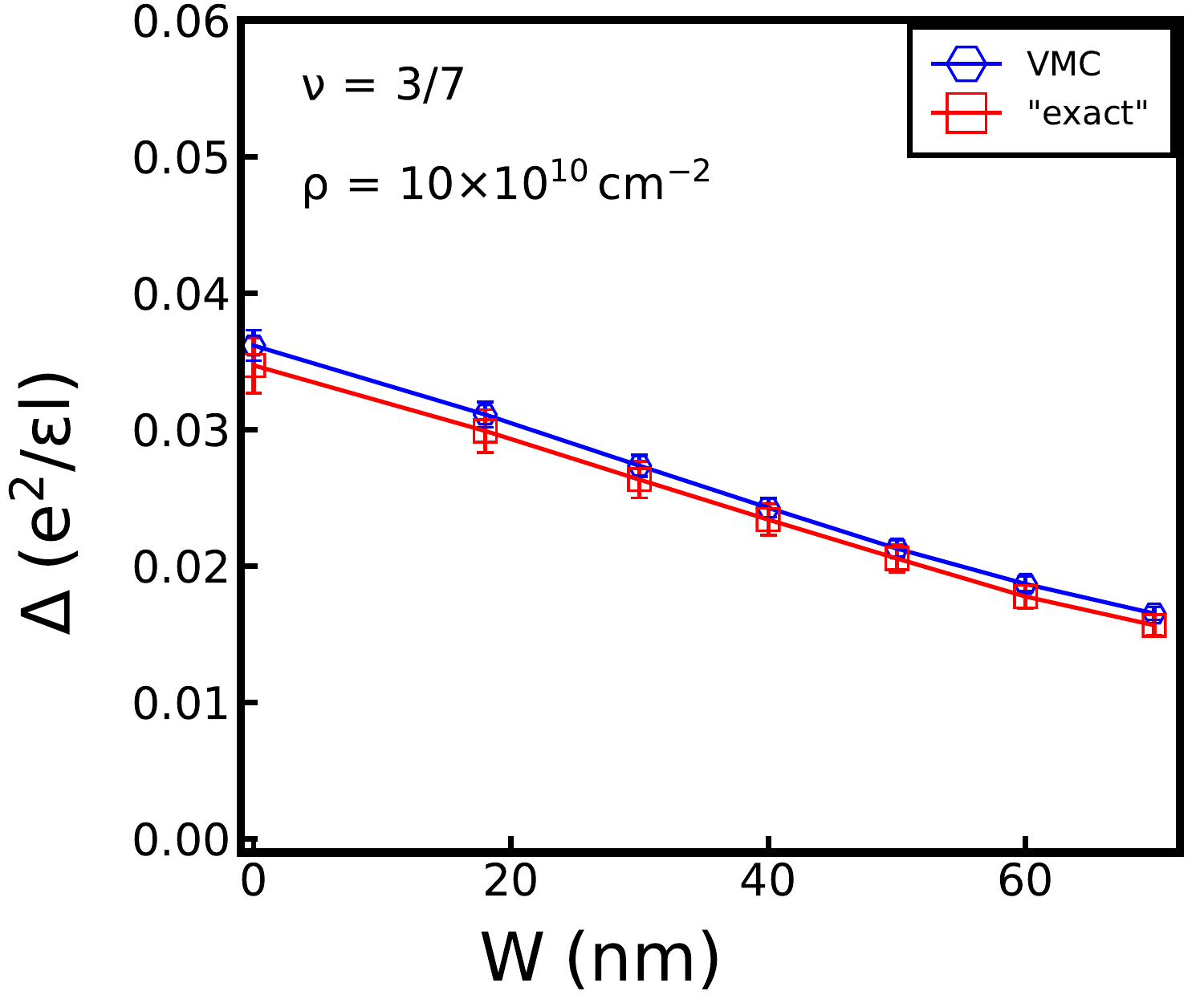}
	\includegraphics[width=0.32 \linewidth]{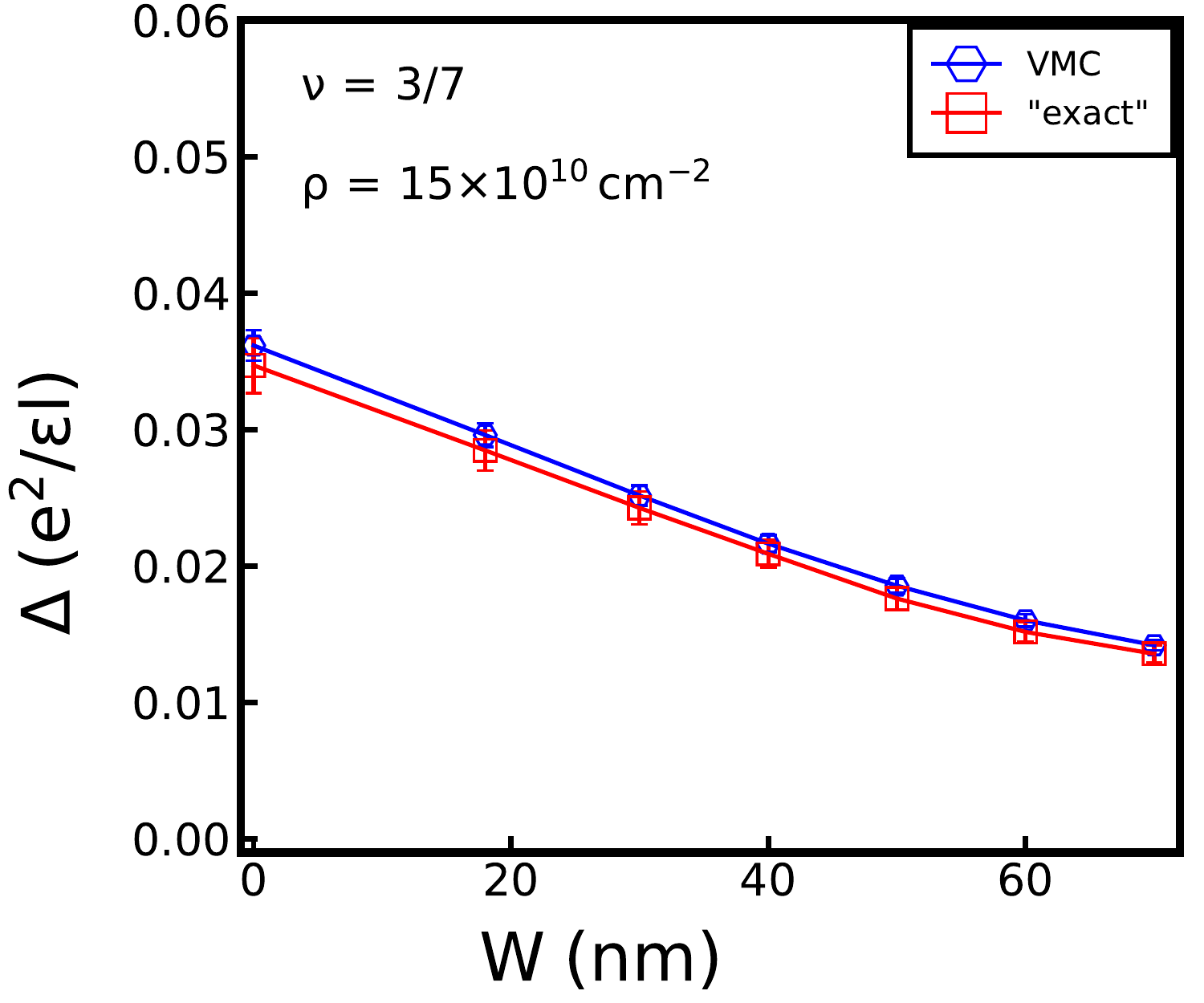}
	\includegraphics[width=0.32 \linewidth]{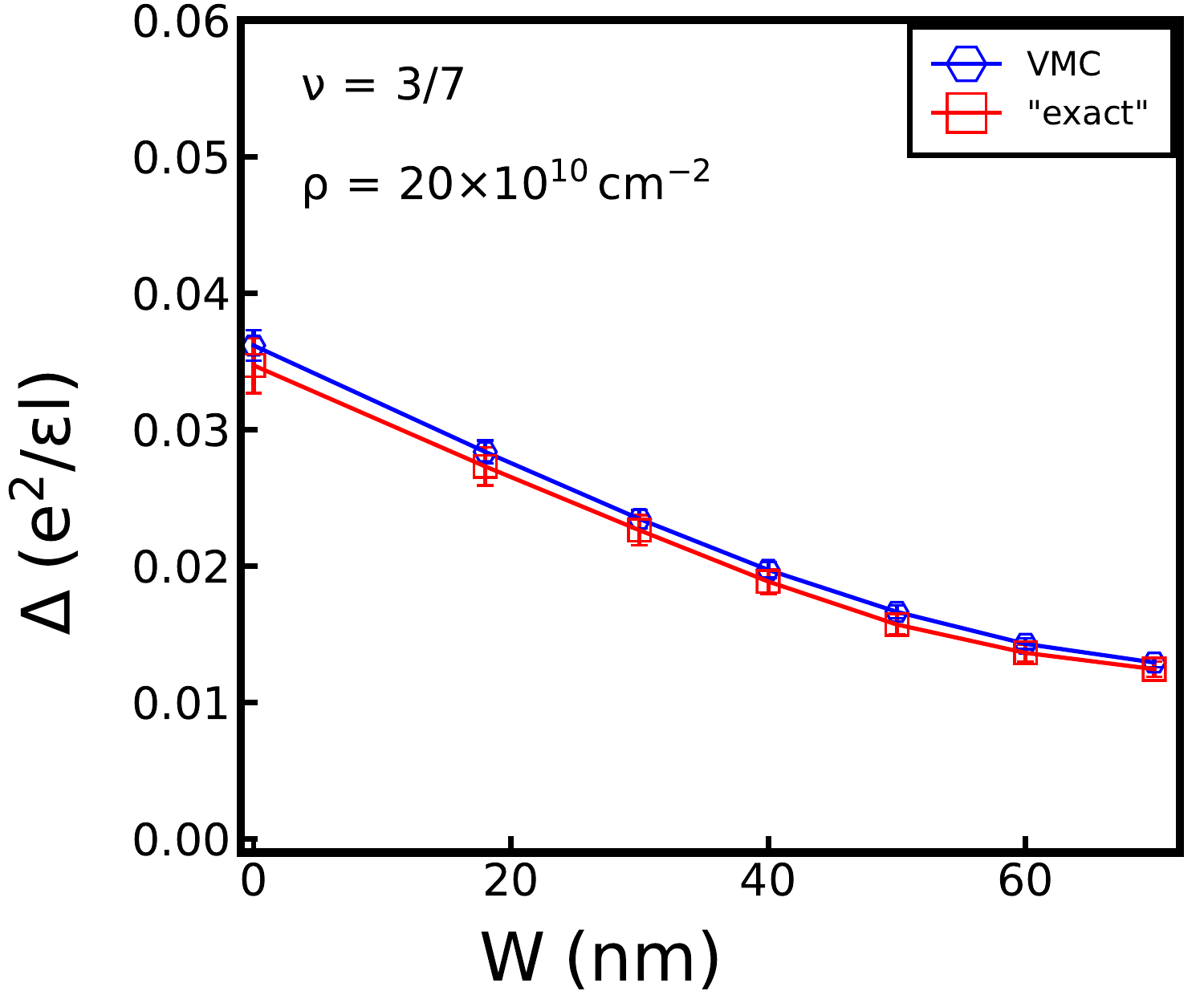}
	\includegraphics[width=0.32 \linewidth]{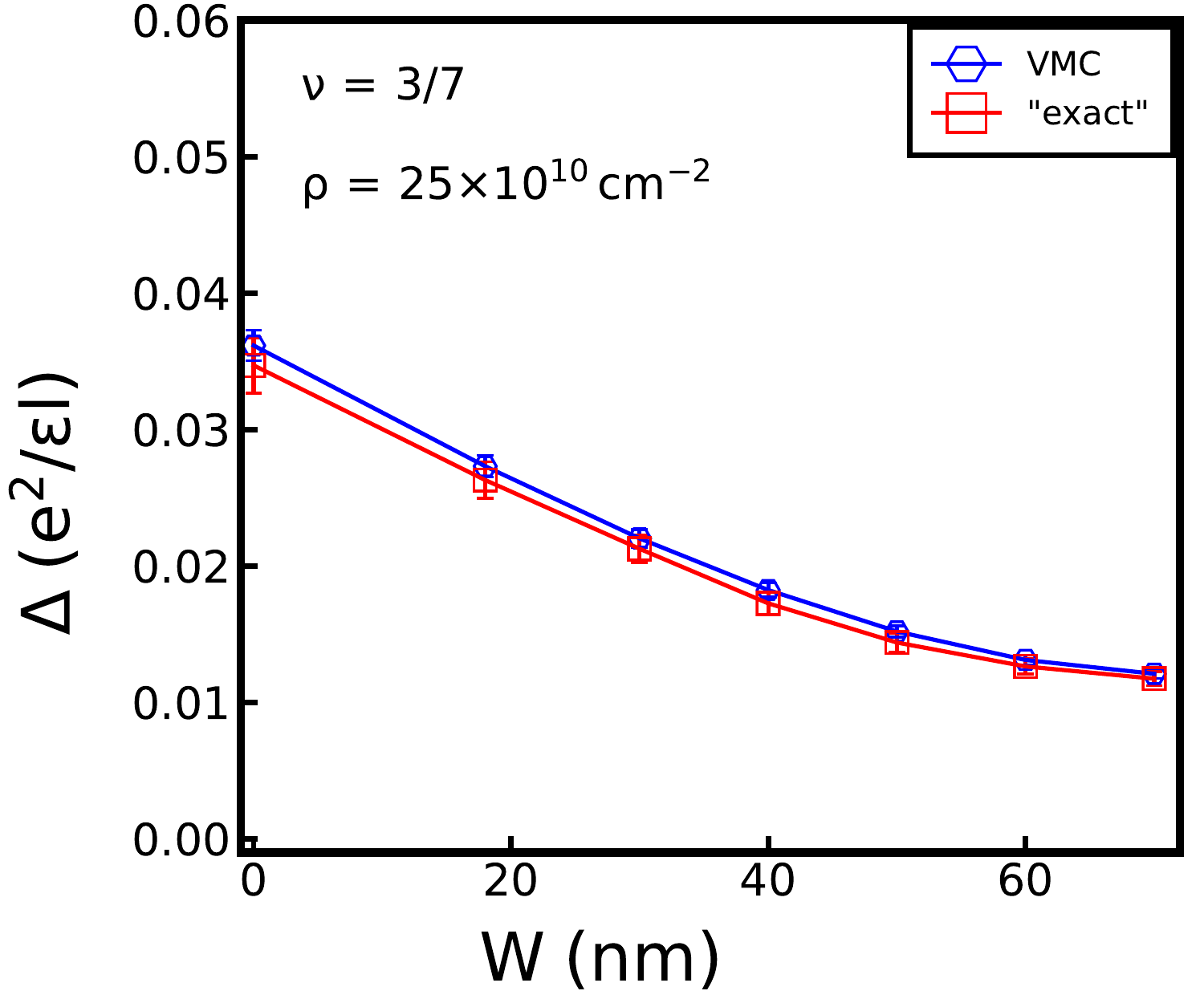}
	\includegraphics[width=0.32 \linewidth]{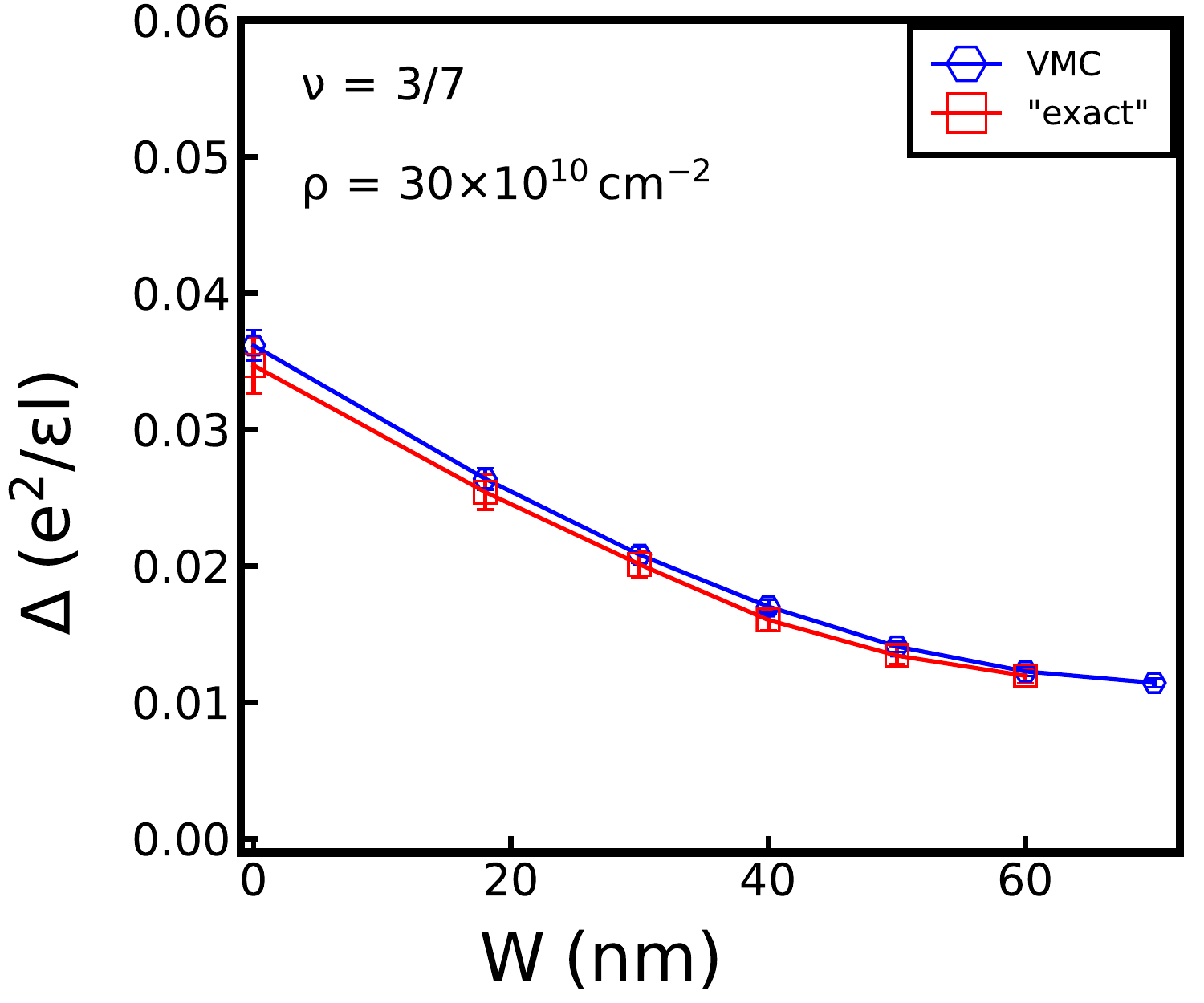}
	\caption{Transport gaps calculated by variational Monte Carlo (VMC) and corrected for variational error (``exact") at $\nu=3/7$ for several widths and densities. }\label{X_fig:VMC_37_other}
\end{figure*}

\begin{figure*}[ht!]
	\includegraphics[width=0.32 \linewidth]{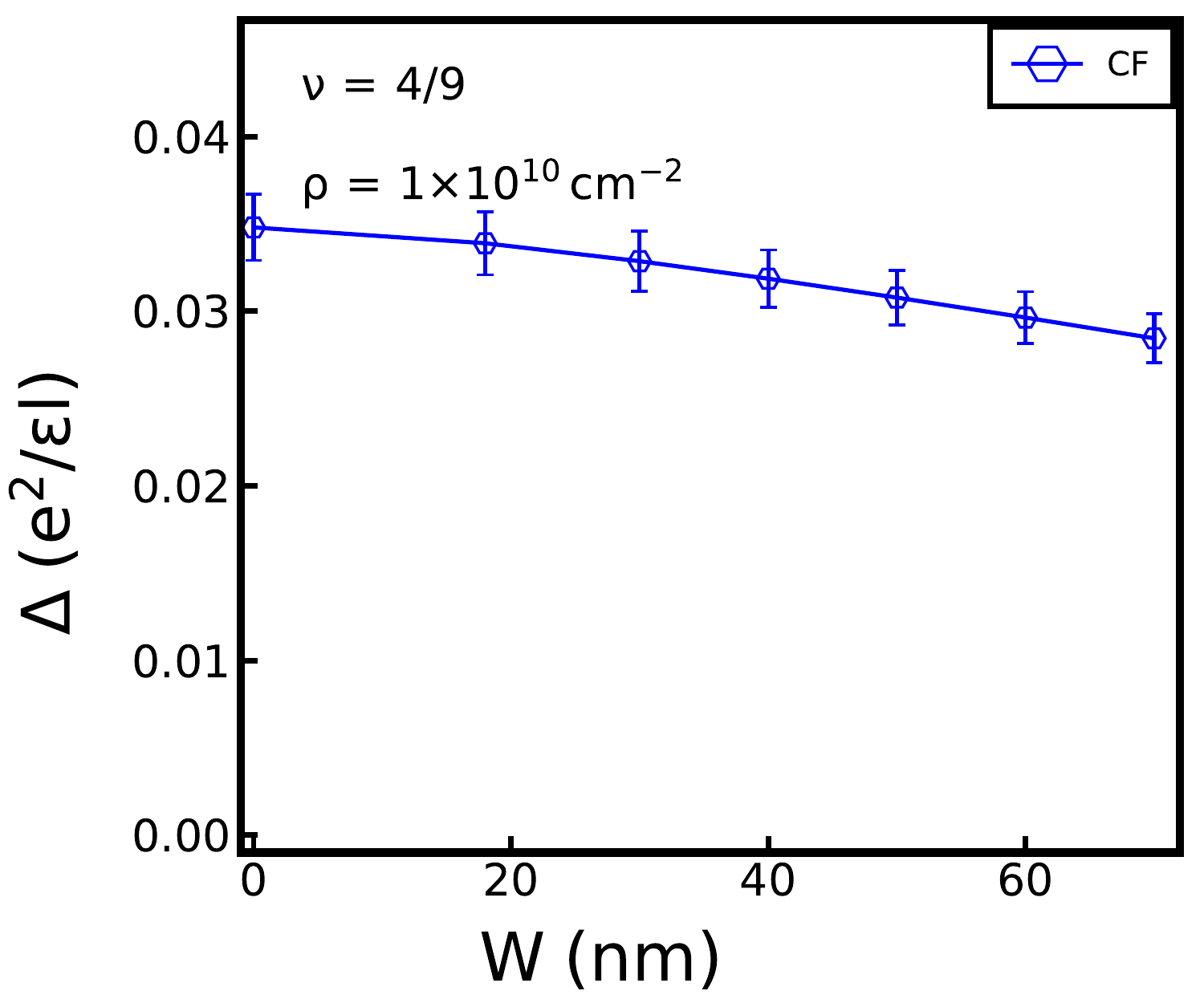}
	\includegraphics[width=0.32 \linewidth]{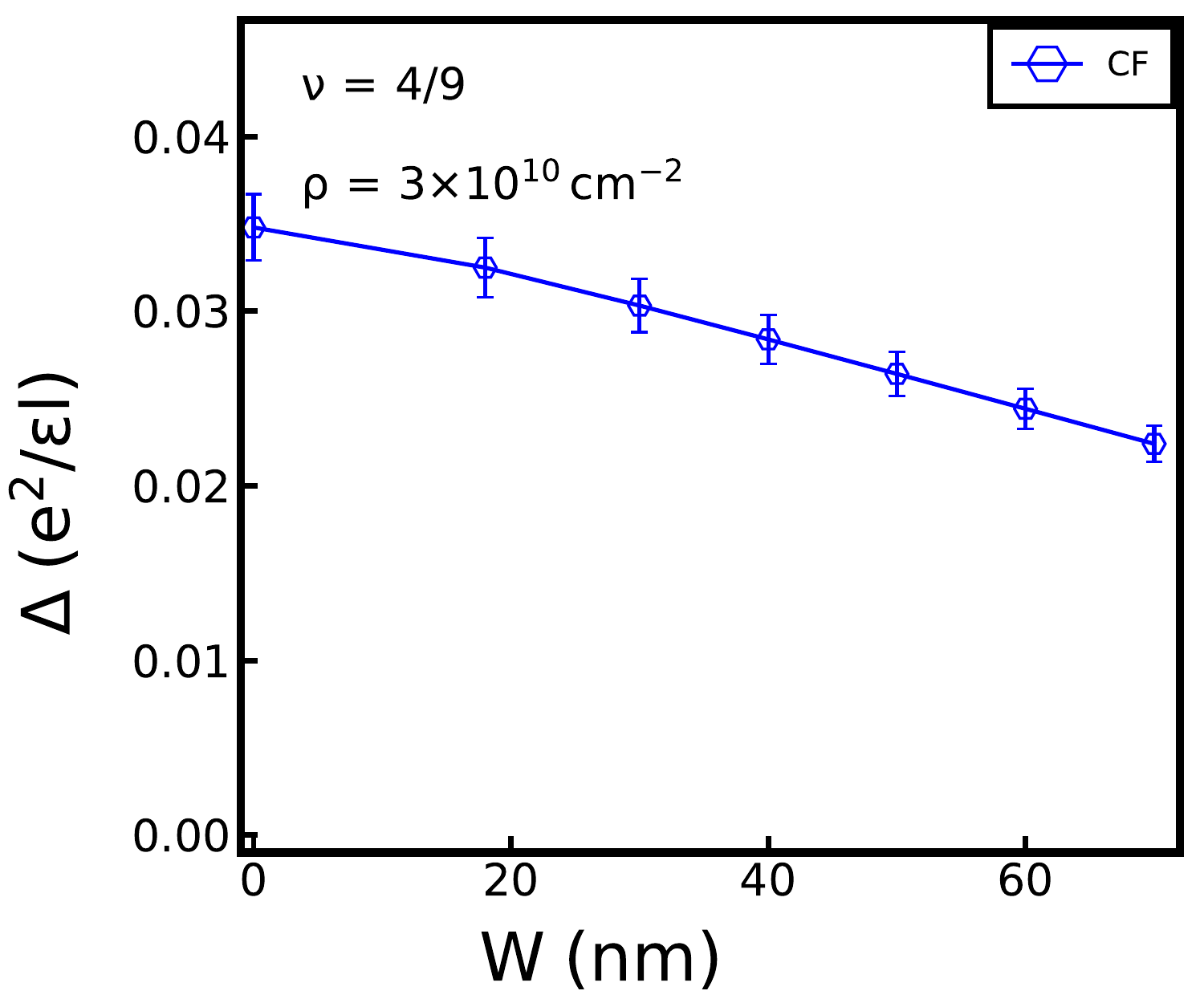}
	\includegraphics[width=0.32 \linewidth]{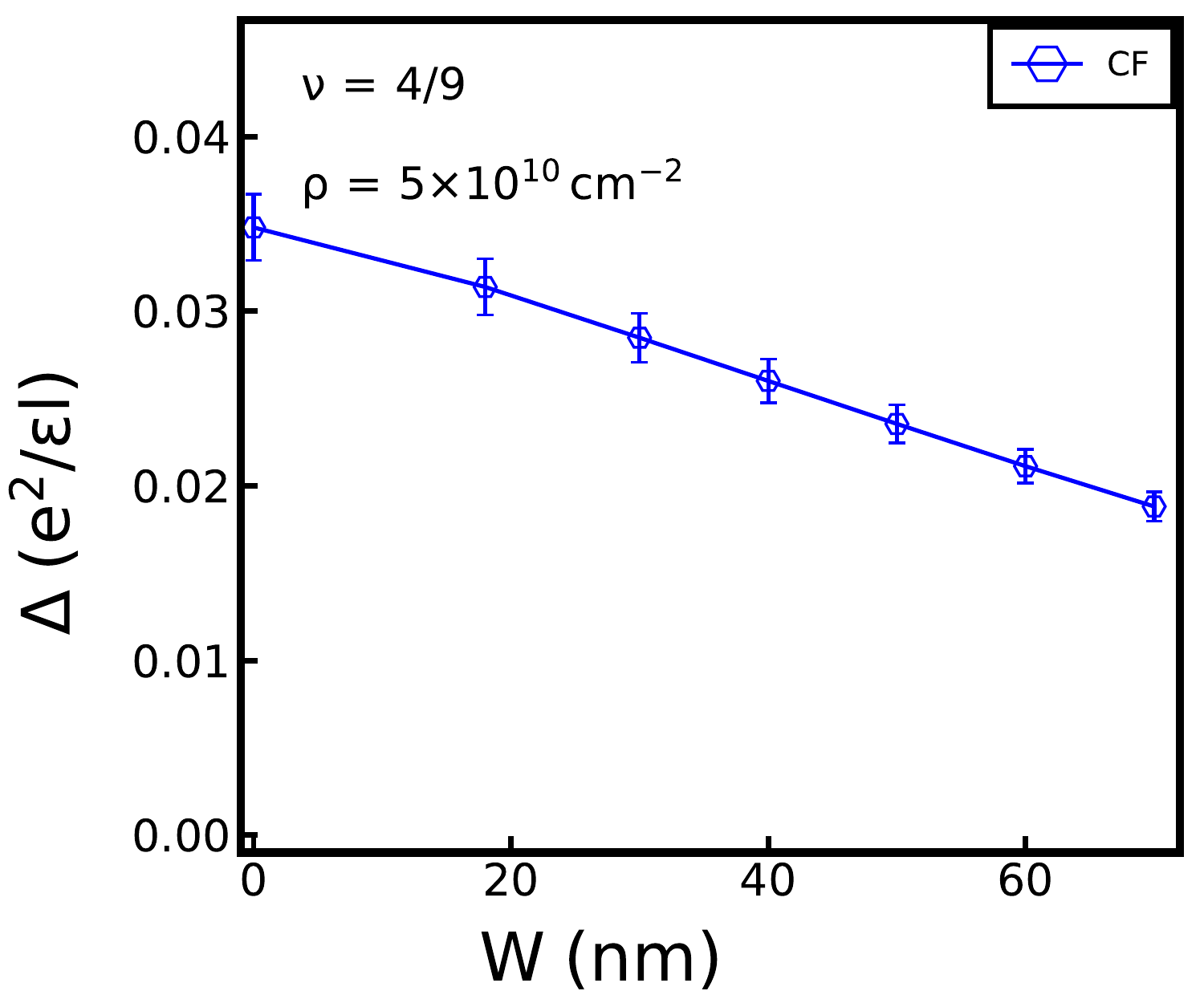}
	\includegraphics[width=0.32 \linewidth]{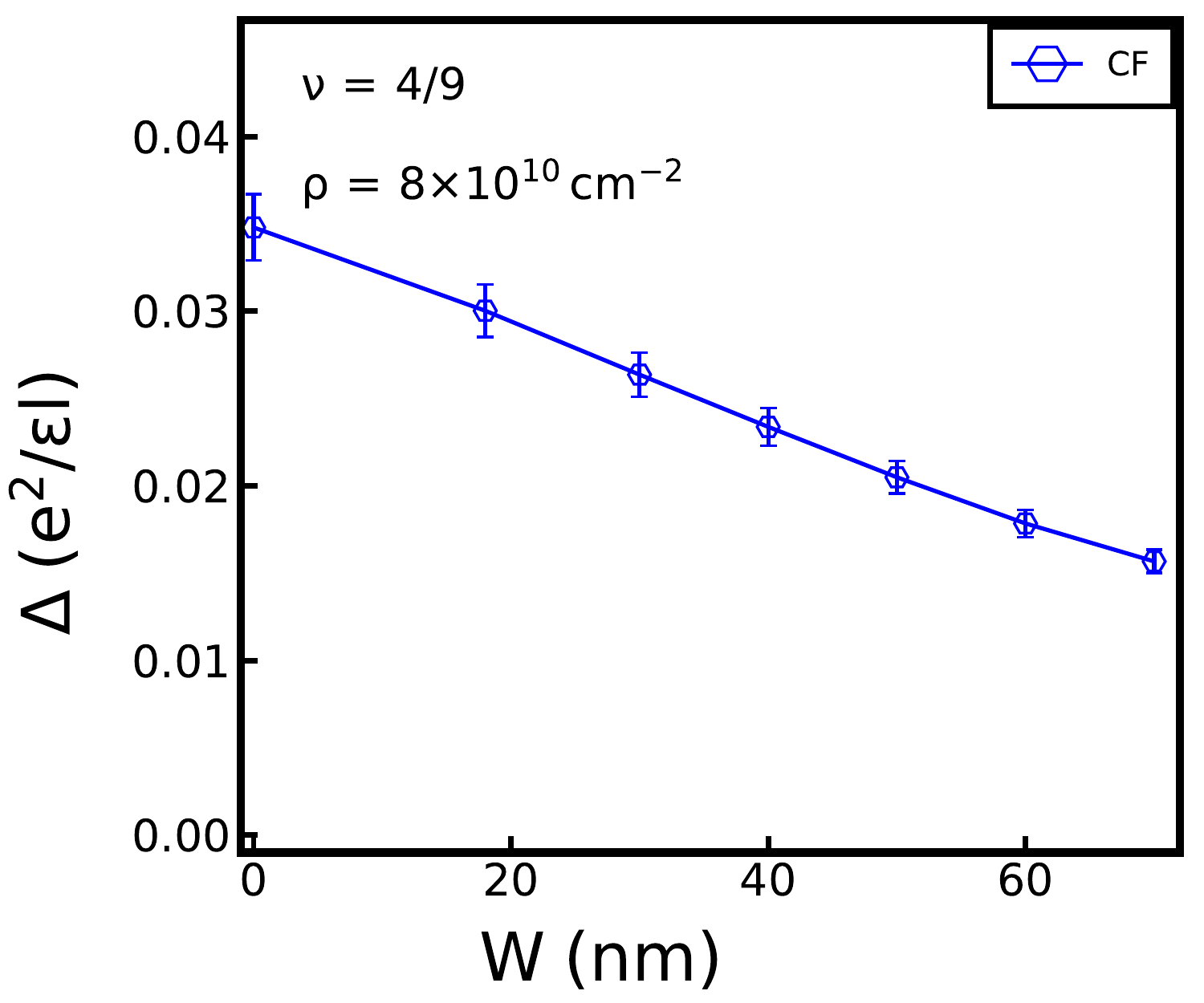}
	\includegraphics[width=0.32 \linewidth]{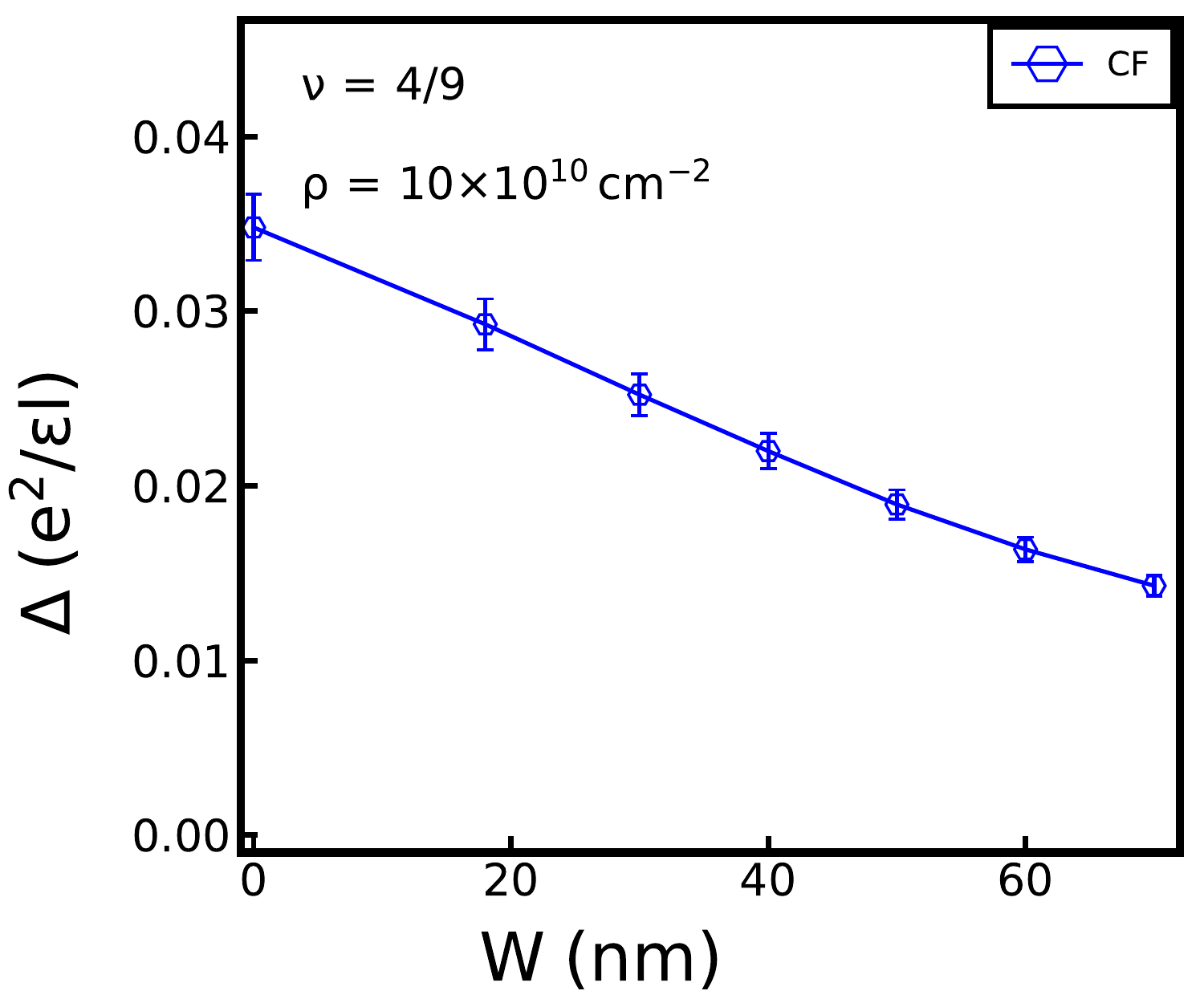}
	\includegraphics[width=0.32 \linewidth]{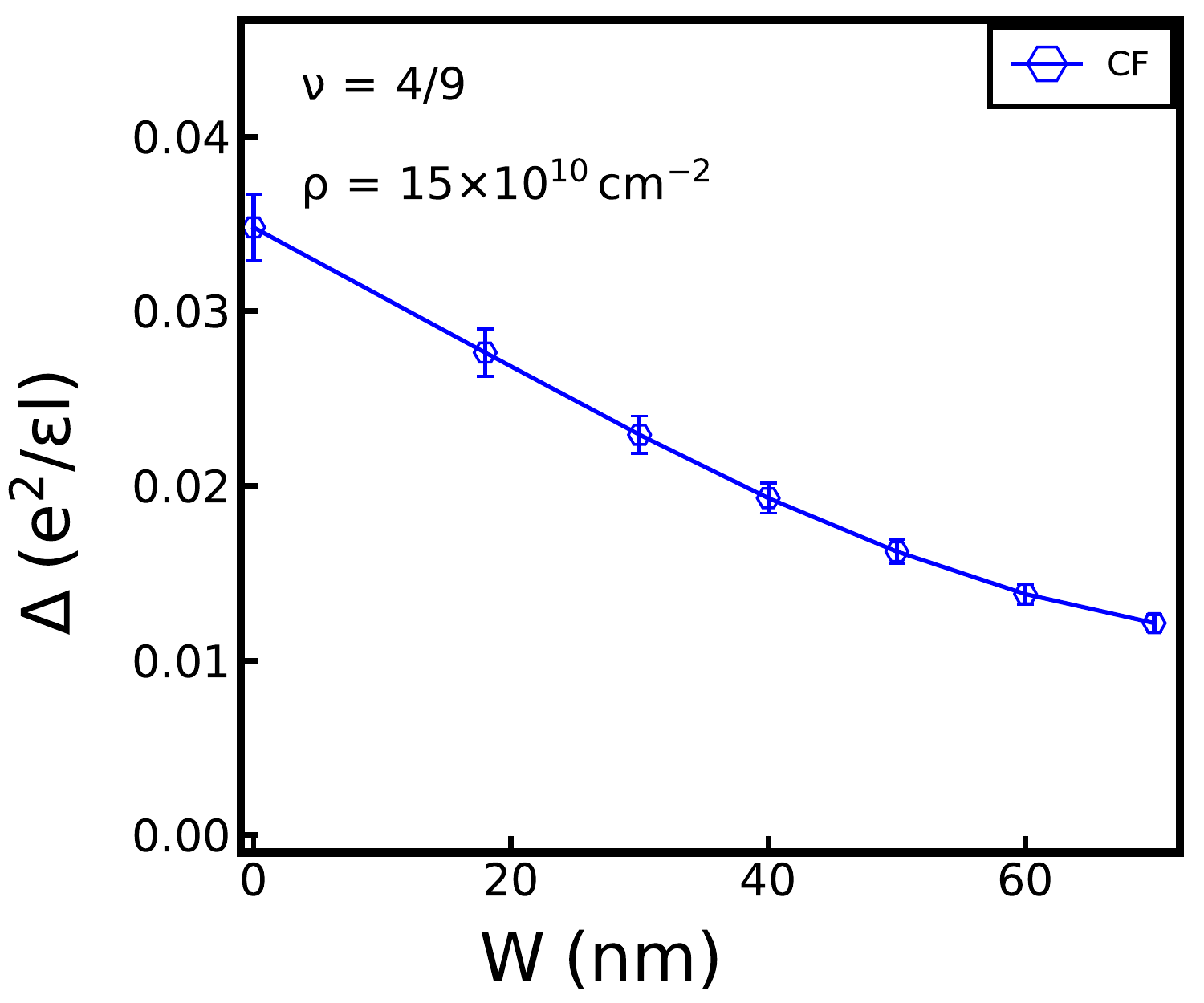}
	\includegraphics[width=0.32 \linewidth]{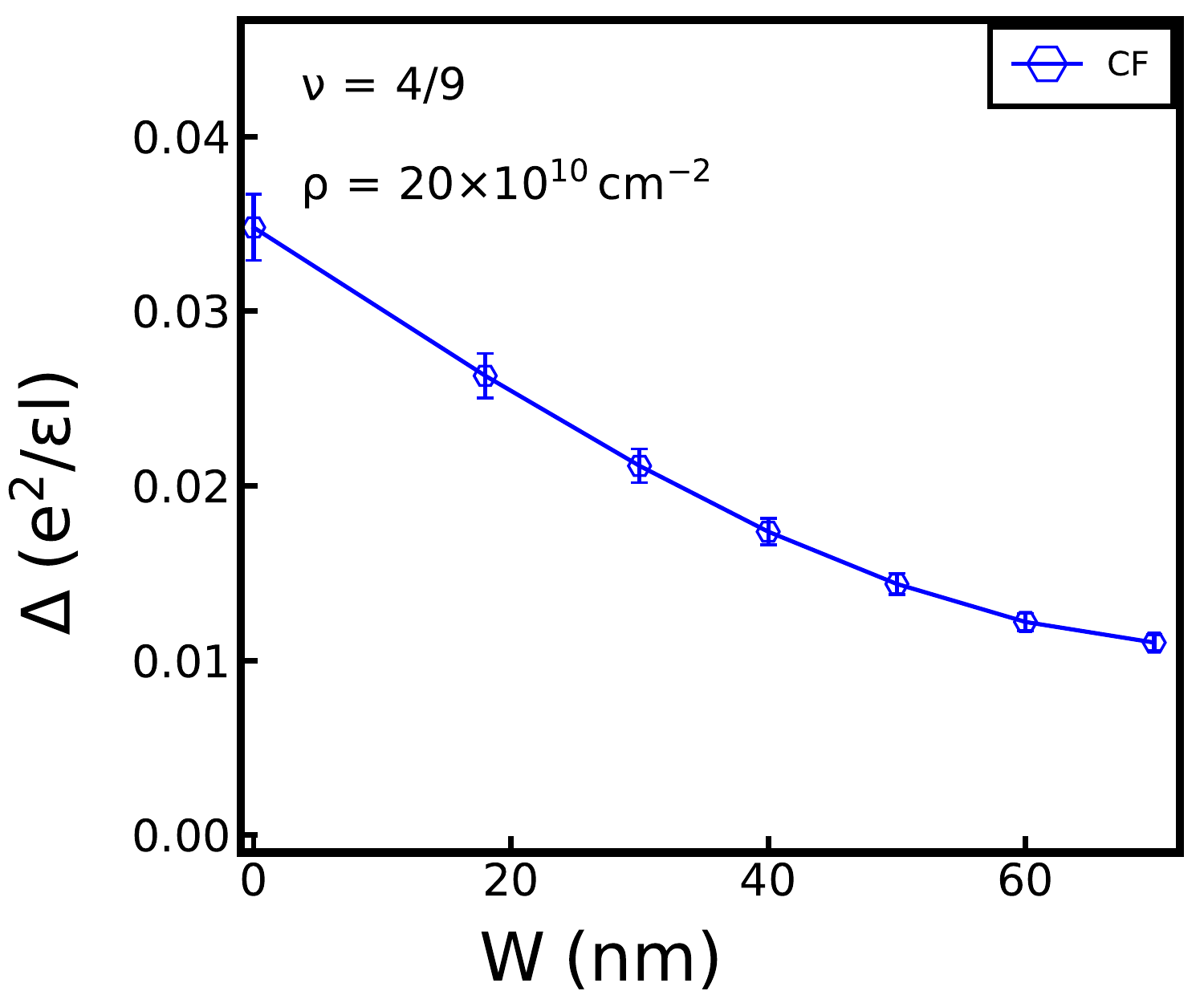}
	\includegraphics[width=0.32 \linewidth]{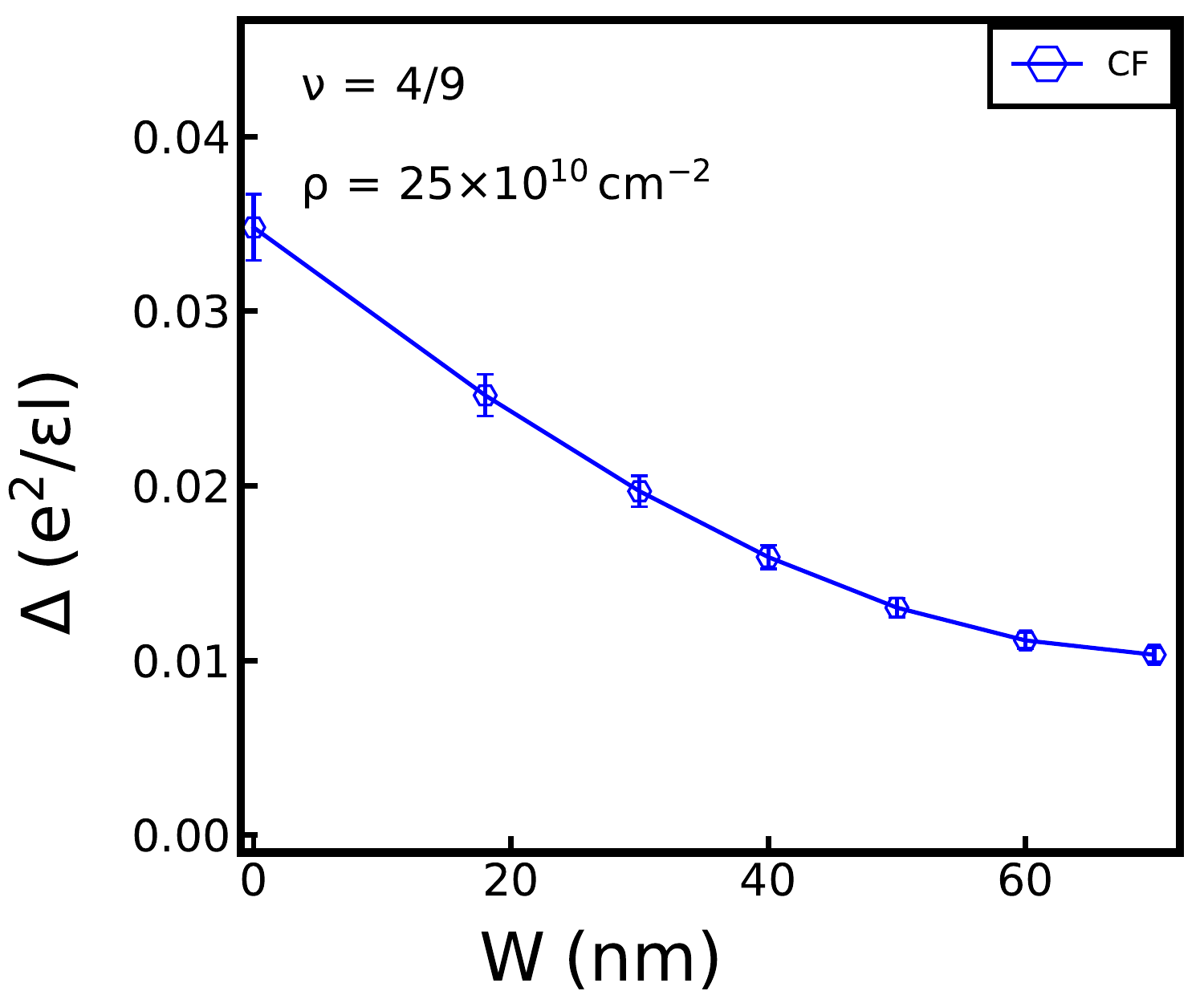}
	\includegraphics[width=0.32 \linewidth]{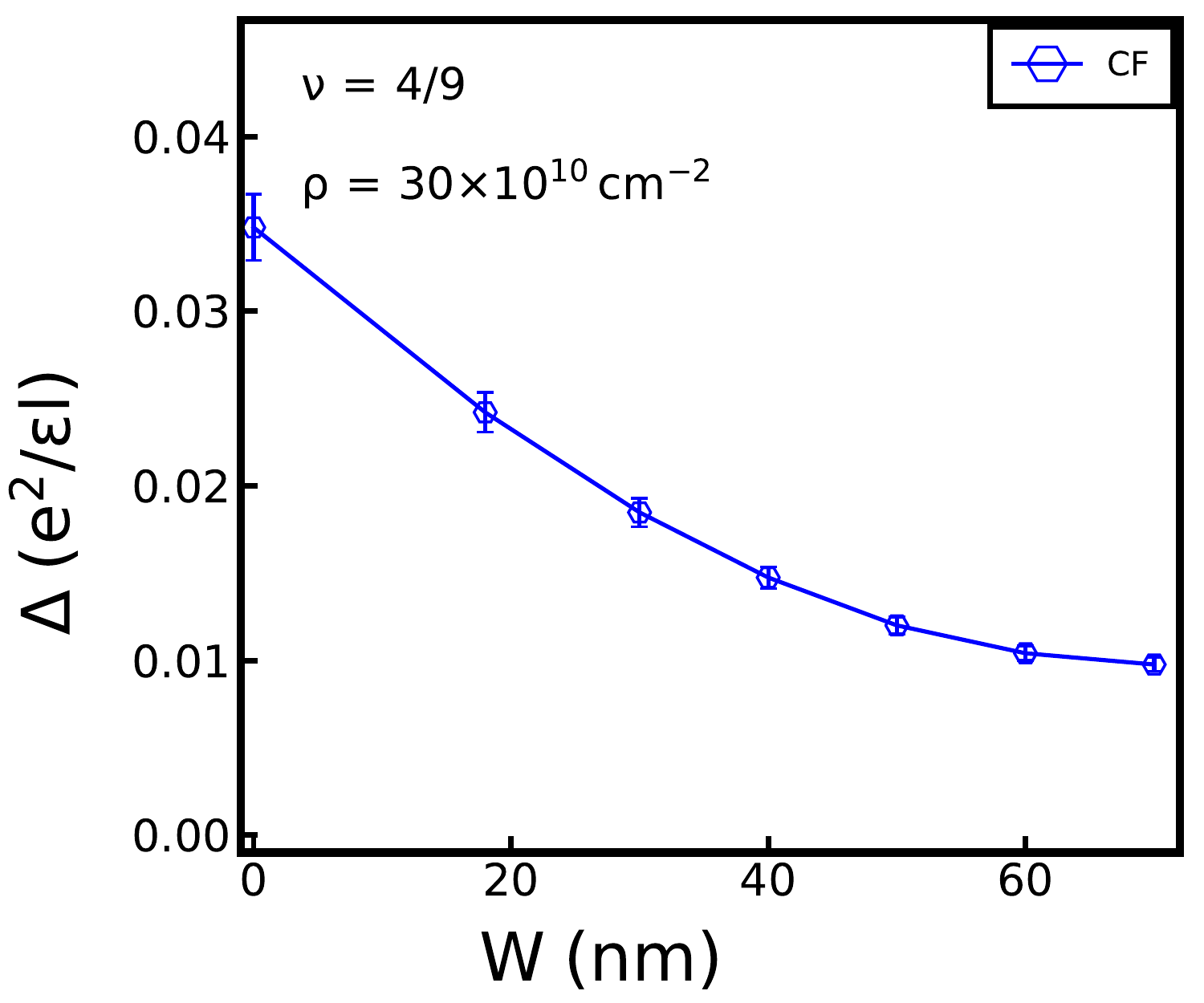}
	\caption{Transport gaps calculated by variational Monte Carlo (VMC) at $\nu=4/9$ for several widths and densities. }\label{X_fig:VMC_49_other}
\end{figure*}

\begin{figure*}[ht!]
	\includegraphics[width=0.32 \linewidth]{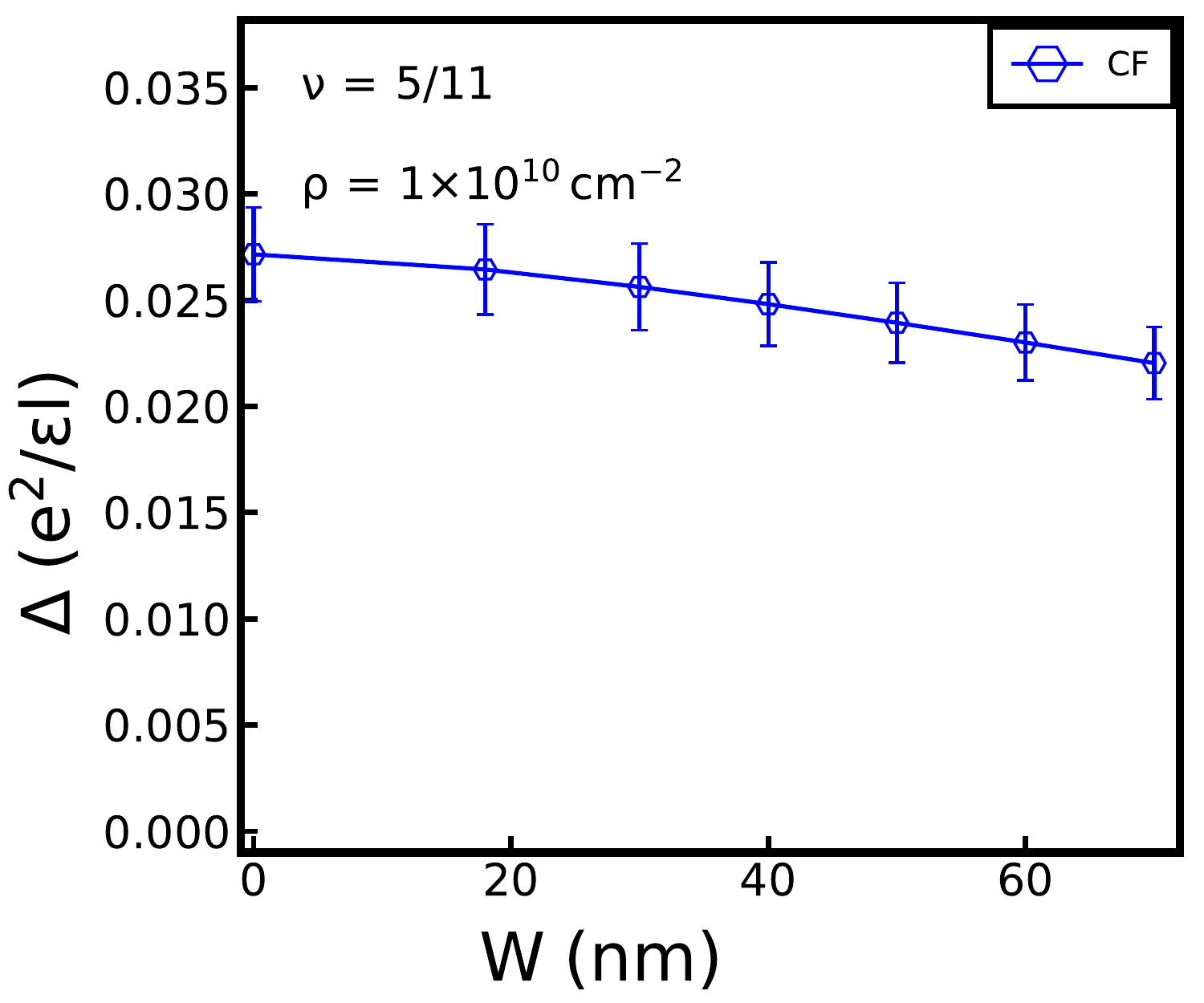}
	\includegraphics[width=0.32 \linewidth]{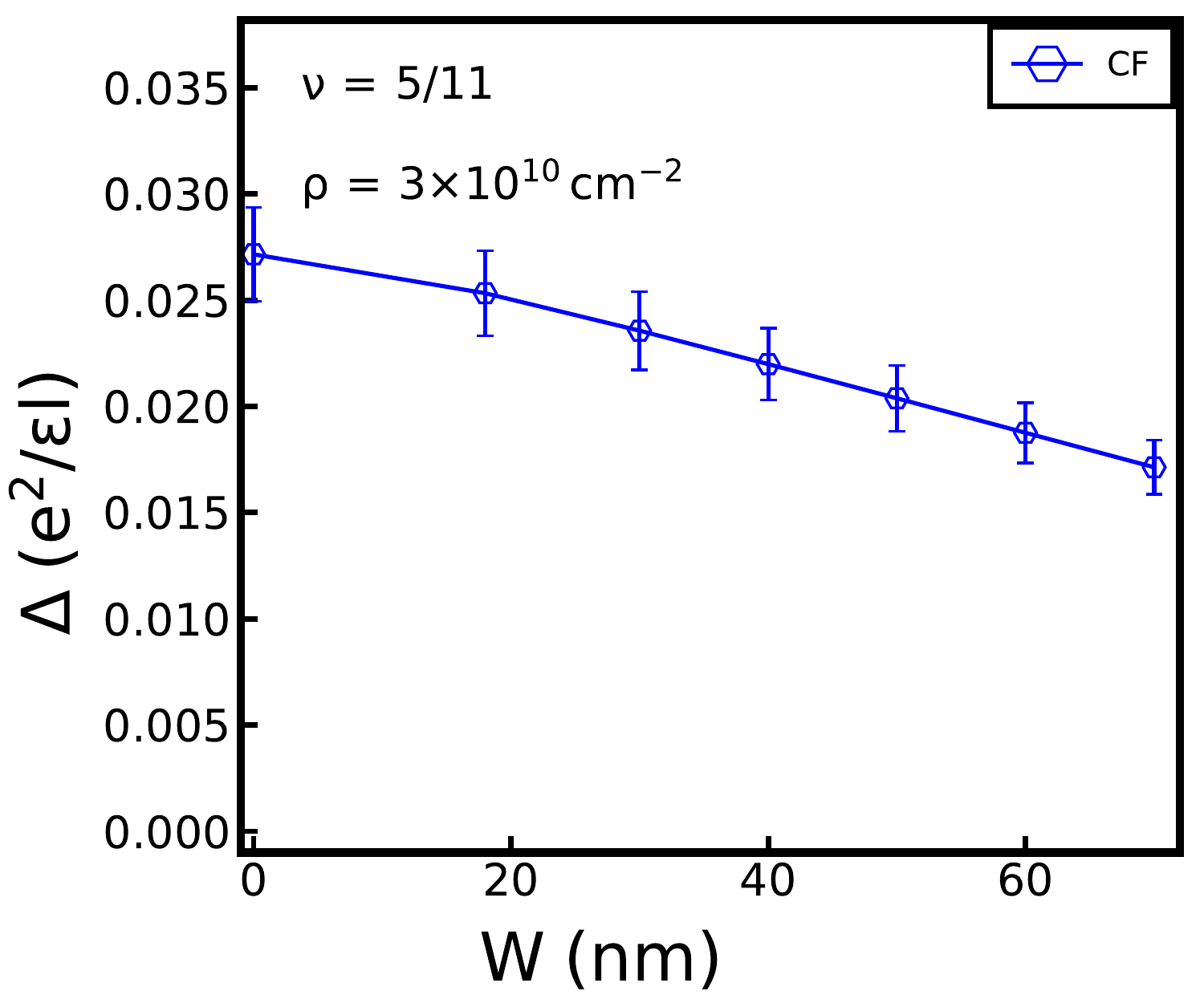}
	\includegraphics[width=0.32 \linewidth]{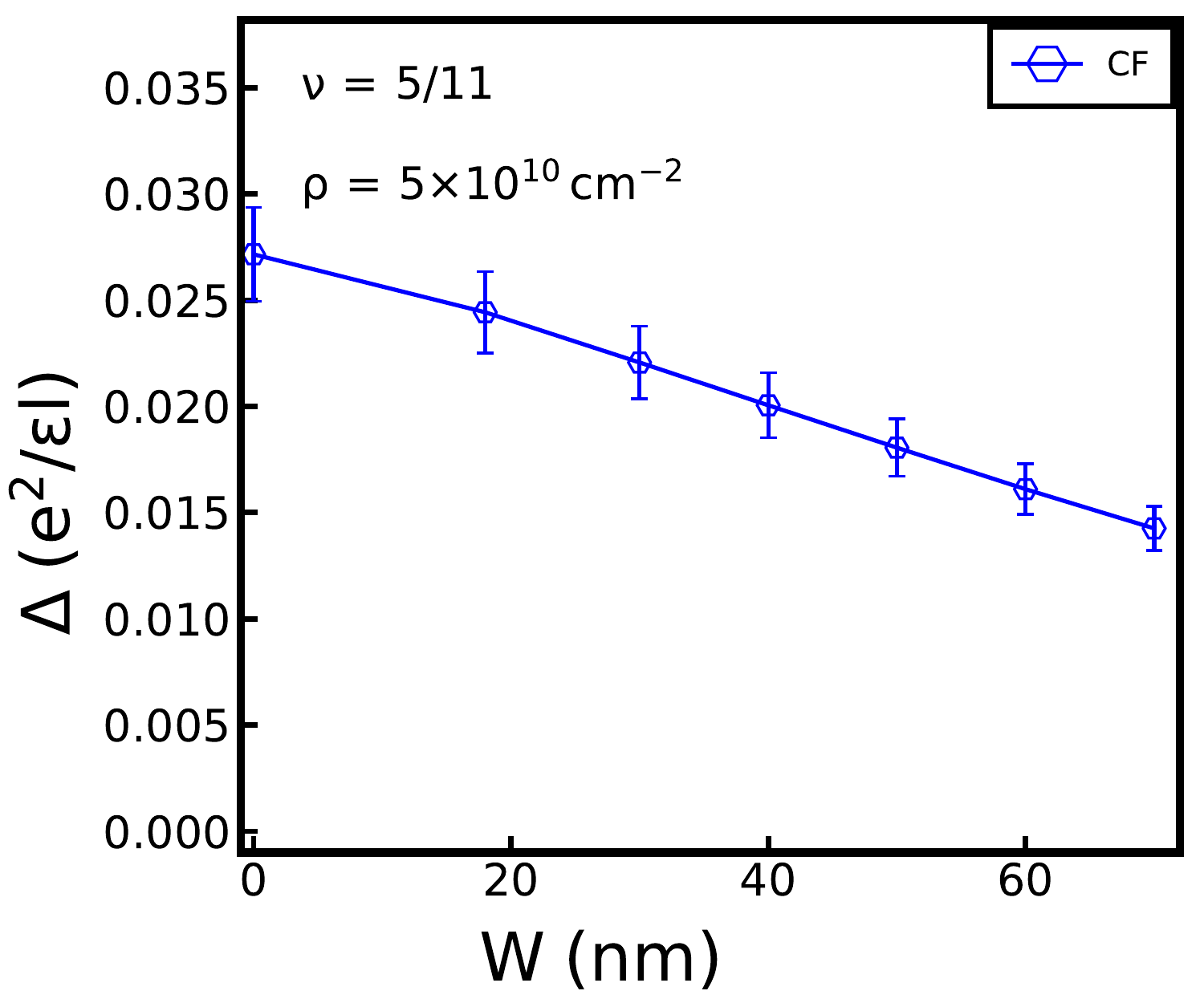}
	\includegraphics[width=0.32 \linewidth]{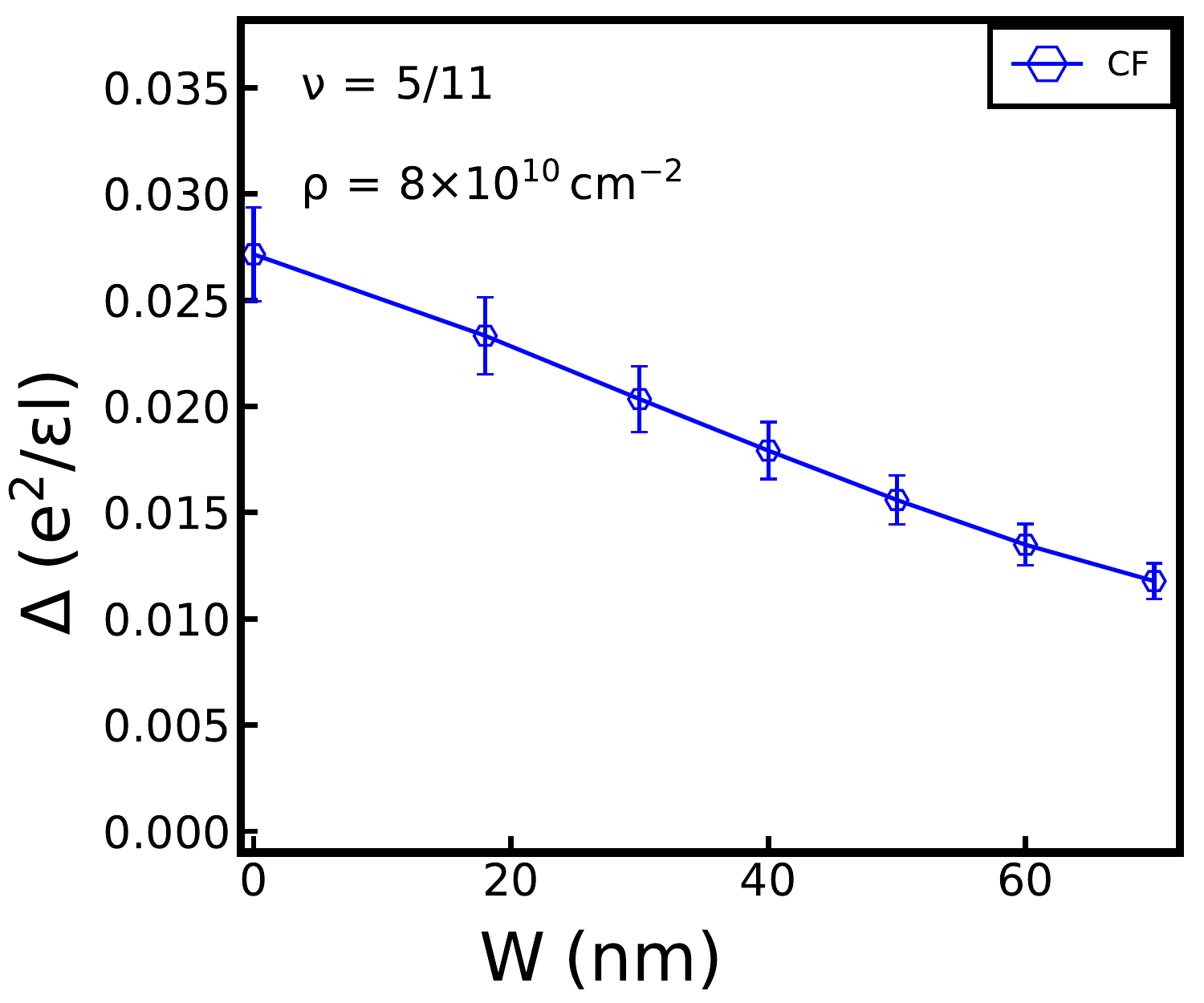}
	\includegraphics[width=0.32 \linewidth]{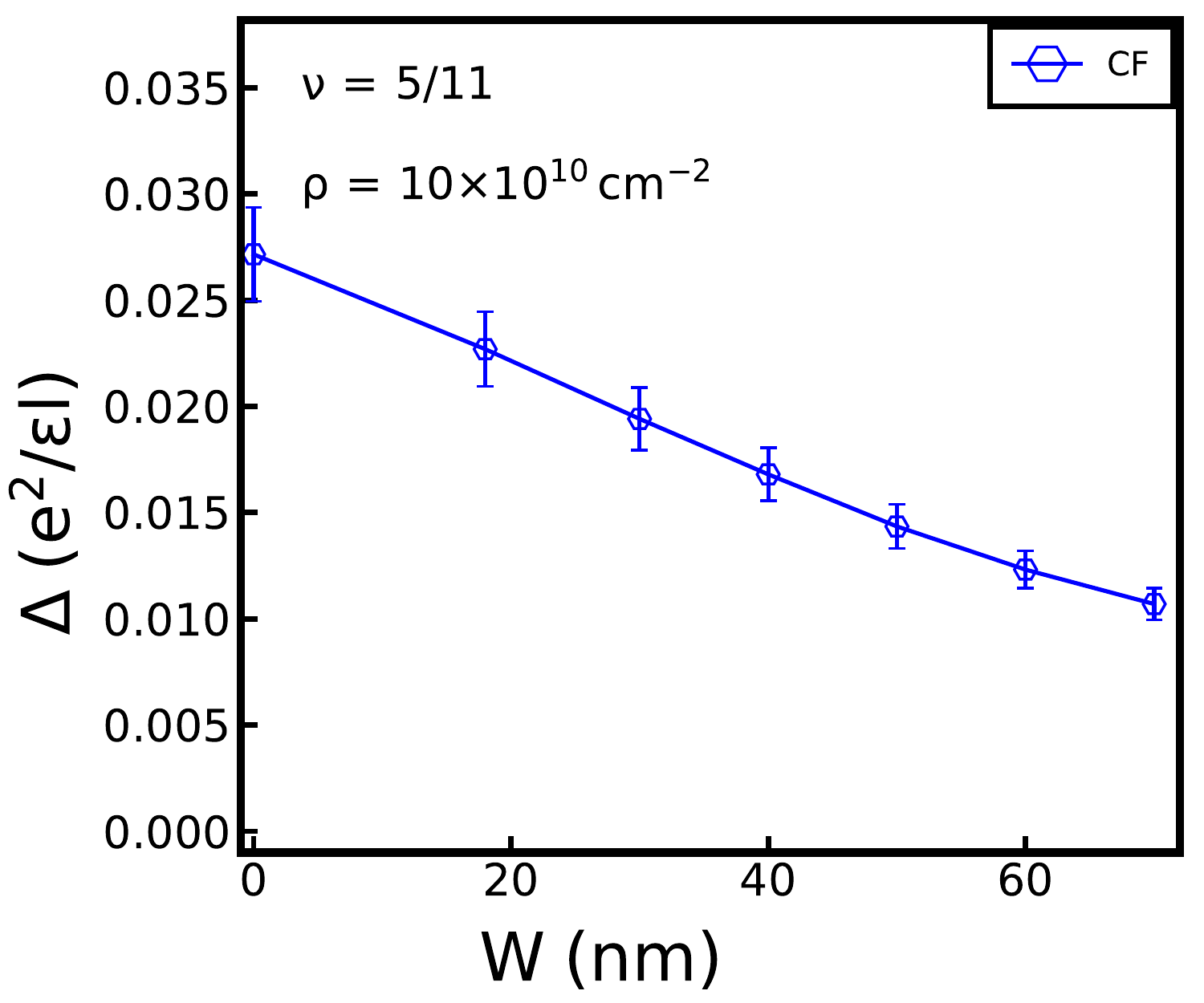}
	\includegraphics[width=0.32 \linewidth]{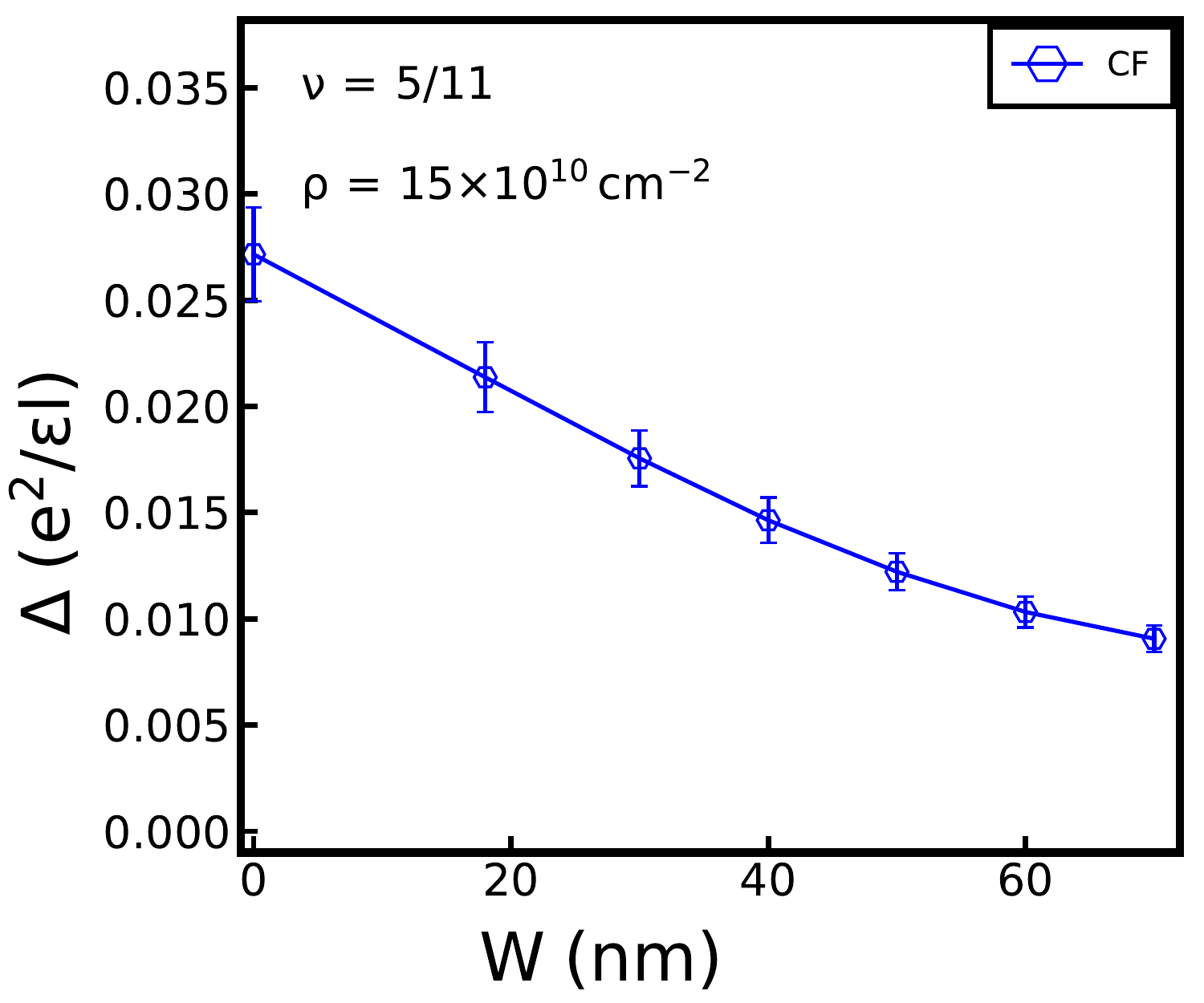}
	\includegraphics[width=0.32 \linewidth]{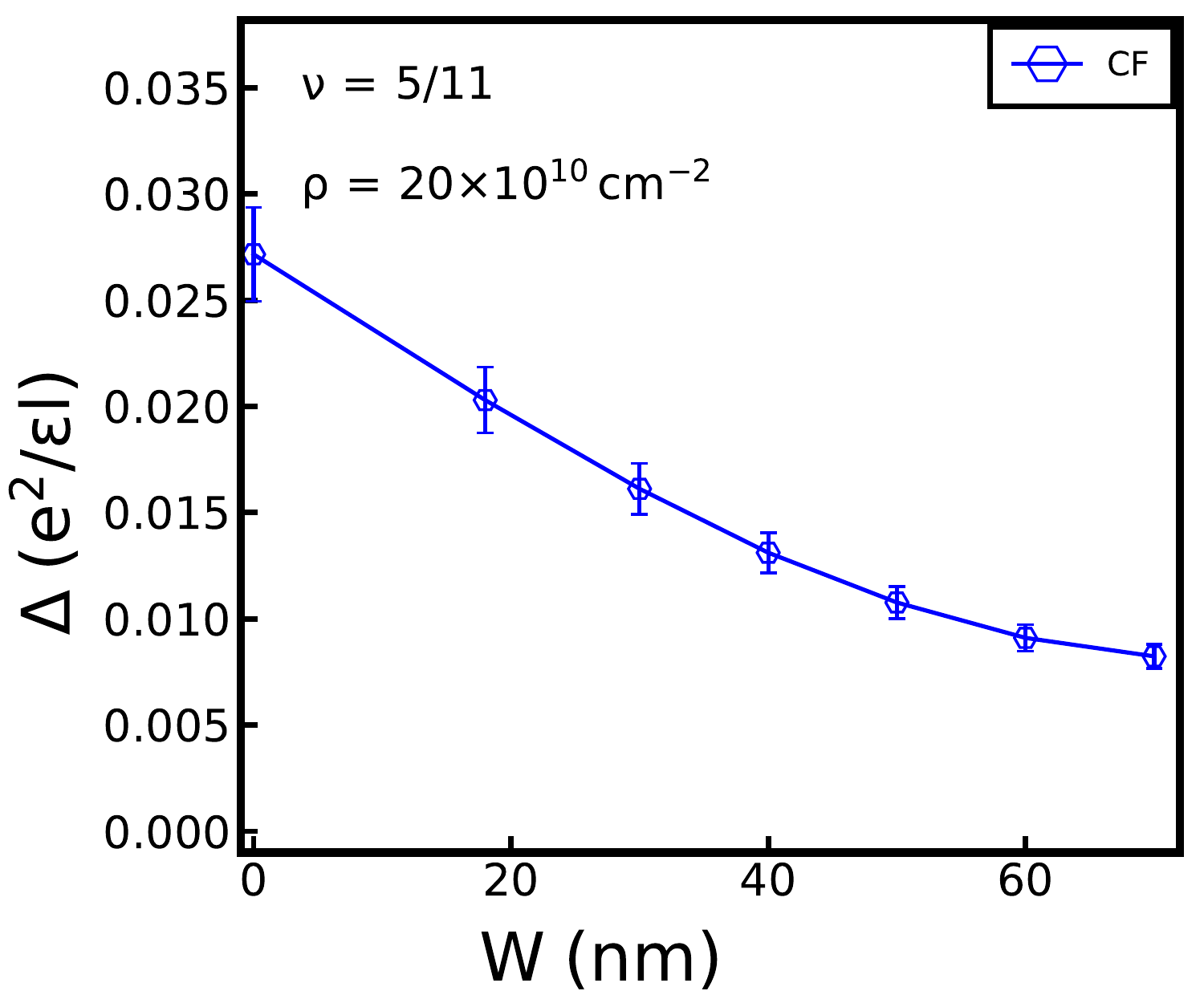}
	\includegraphics[width=0.32 \linewidth]{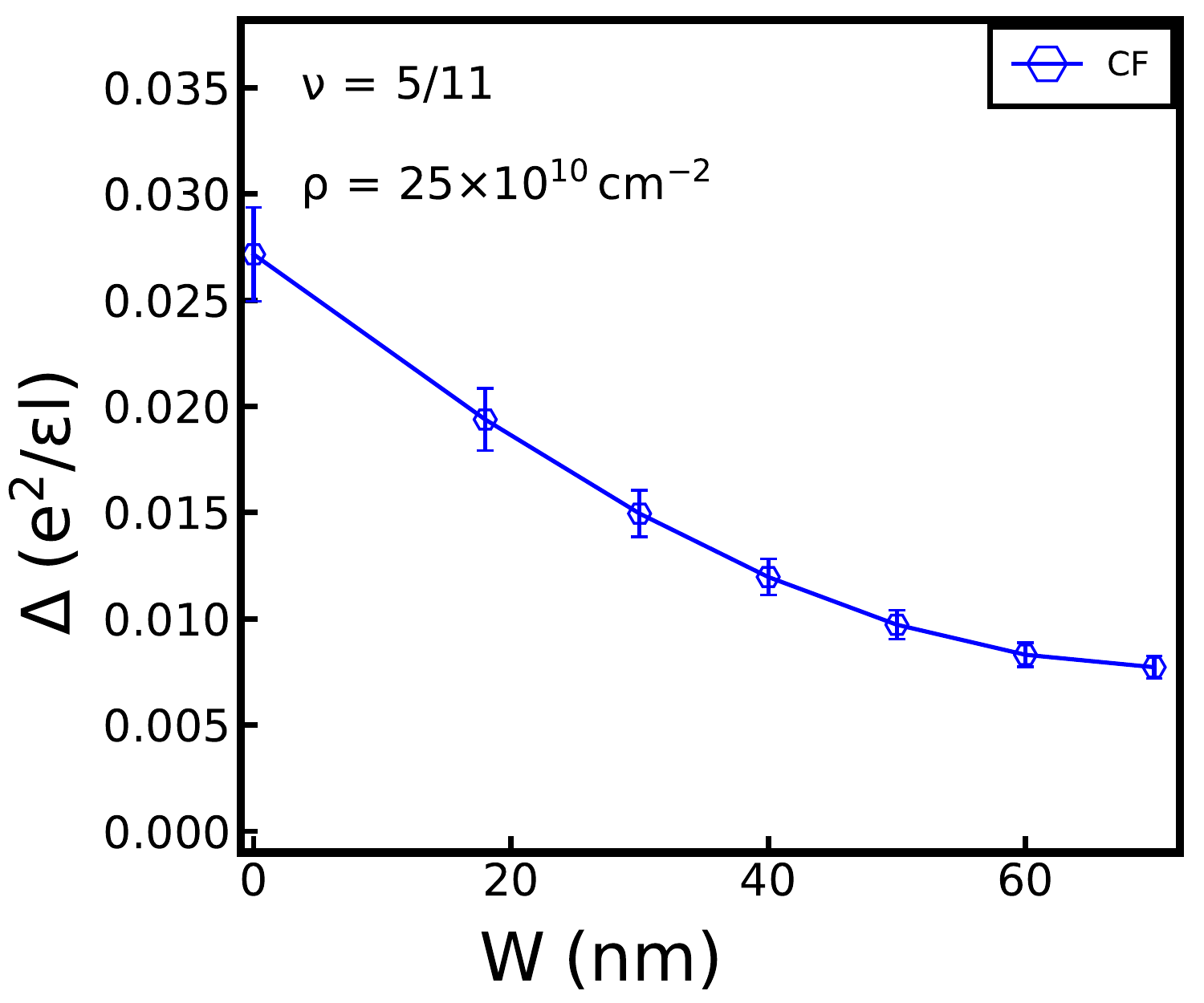}
	\includegraphics[width=0.32 \linewidth]{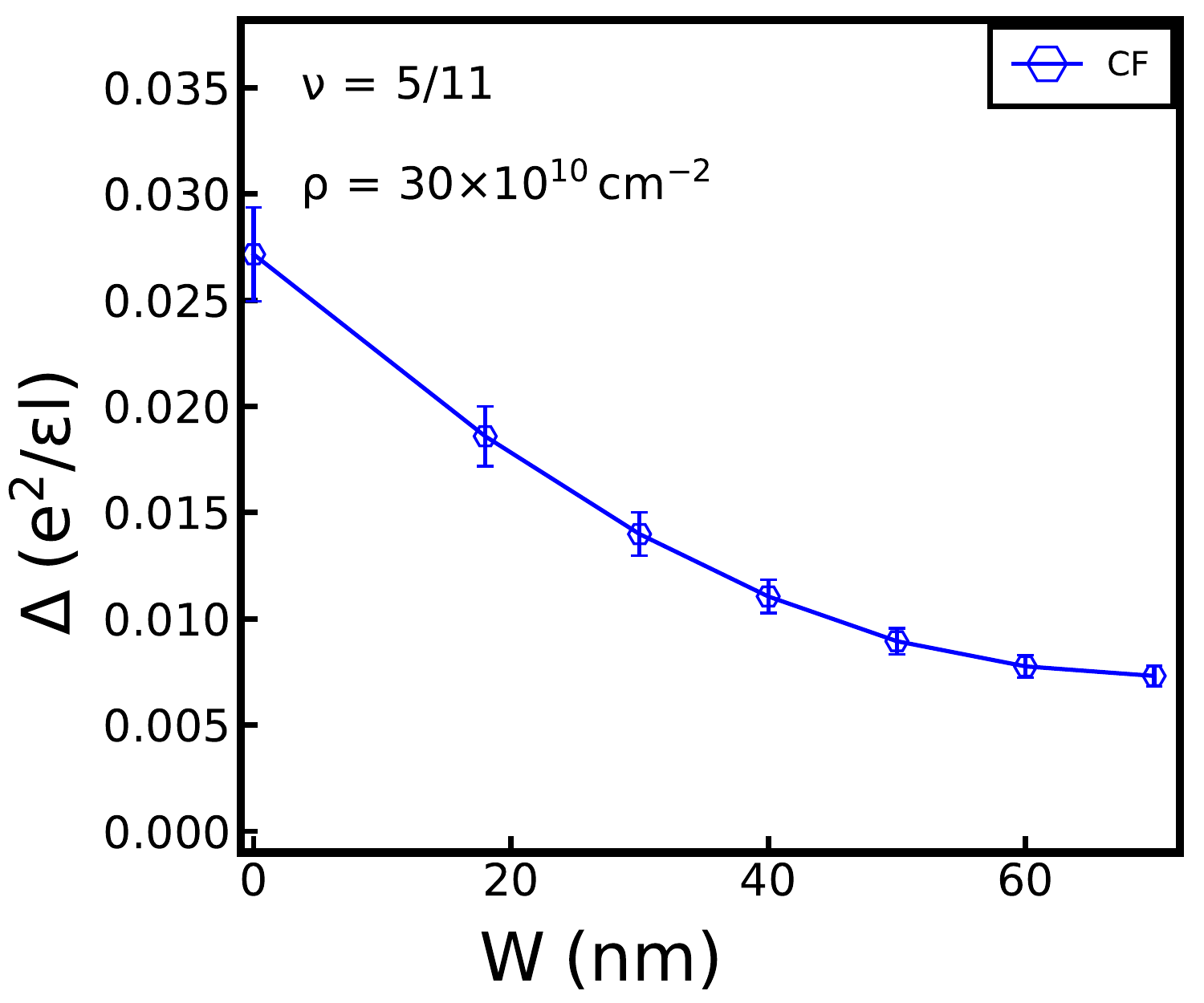}
	\caption{Transport gaps calculated by variational Monte Carlo (VMC) at $\nu=5/11$ for several widths and densities. }\label{X_fig:VMC_511_other}
\end{figure*}

\begin{figure*}[ht!]
\includegraphics[width=0.32 \linewidth]{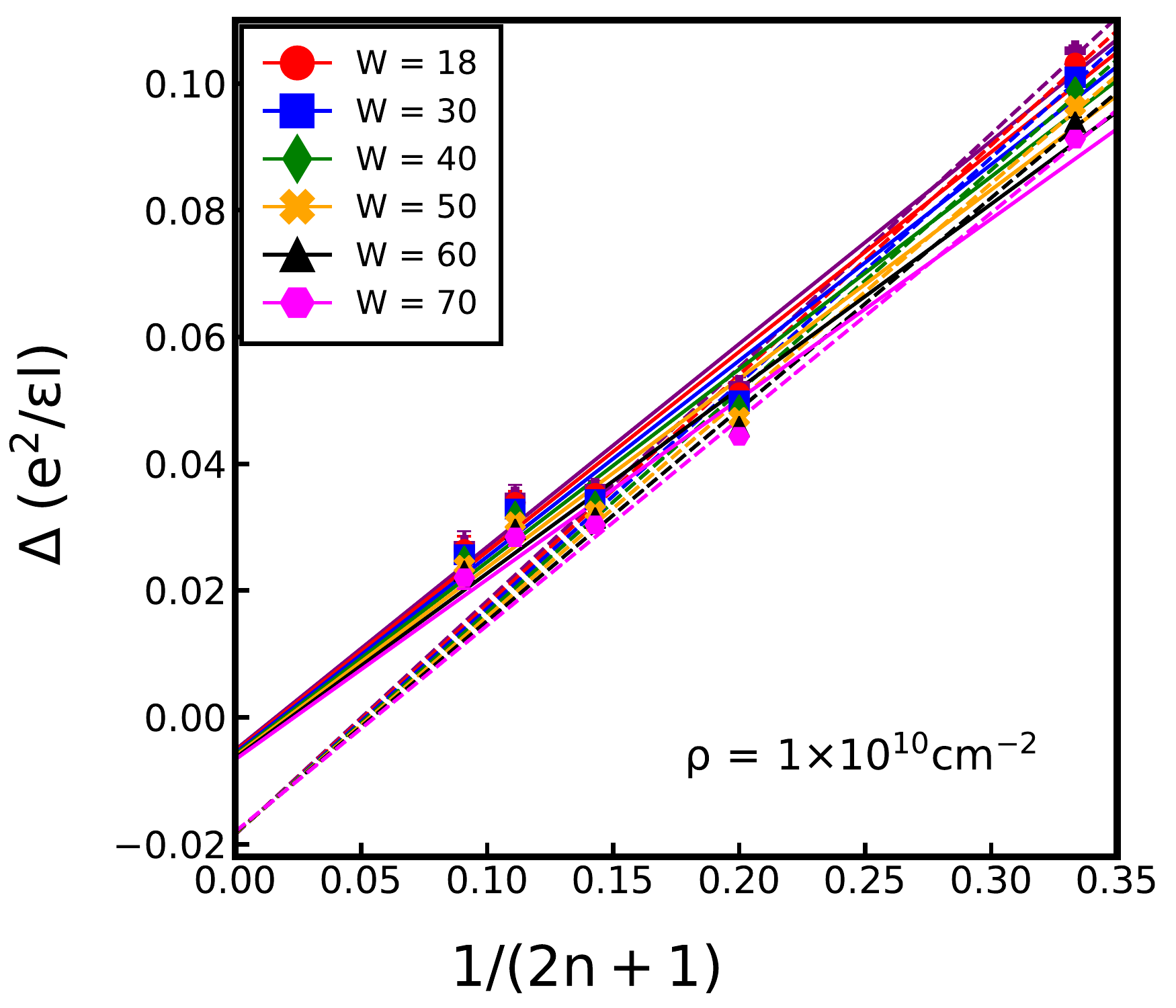}
\includegraphics[width=0.32 \linewidth]{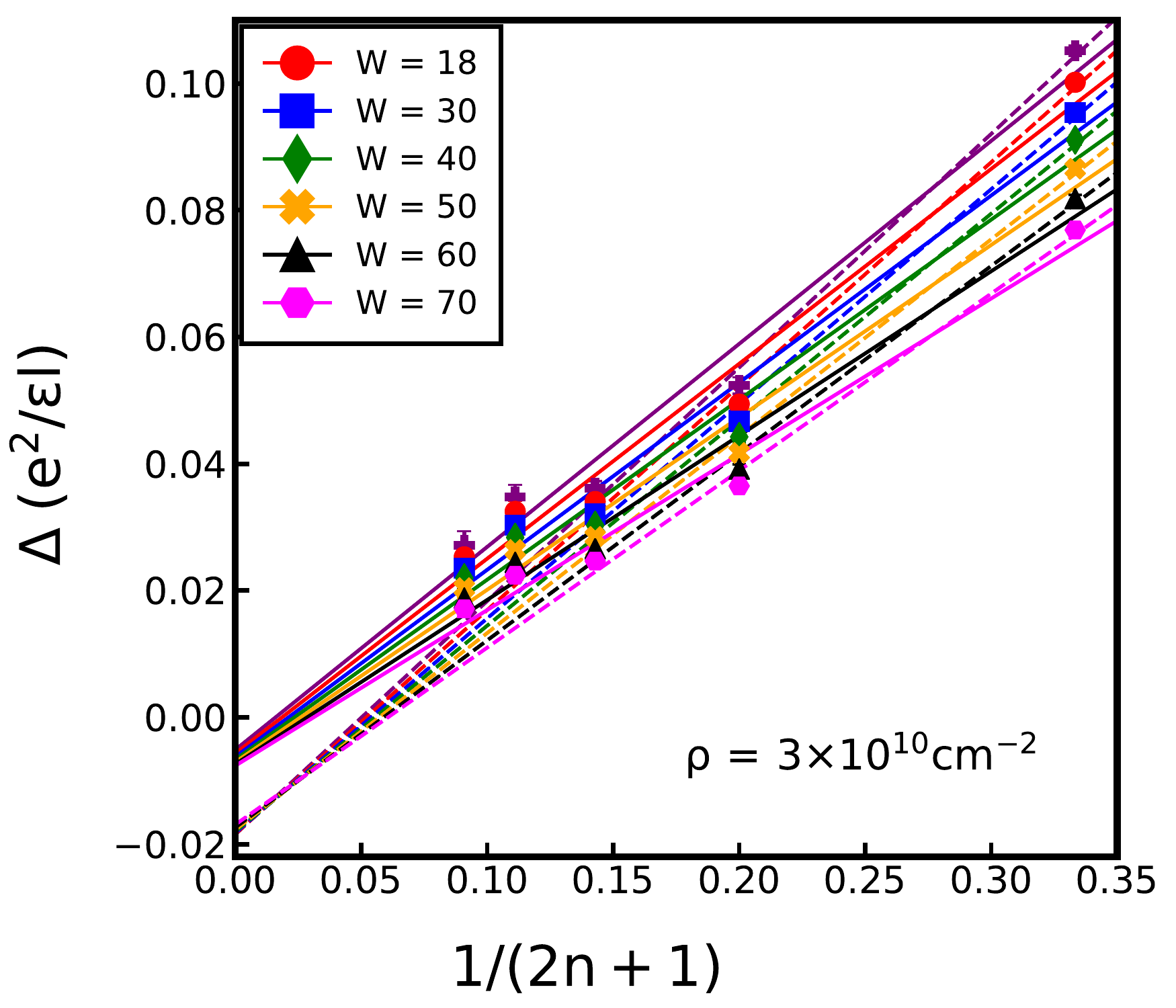}
\includegraphics[width=0.32 \linewidth]{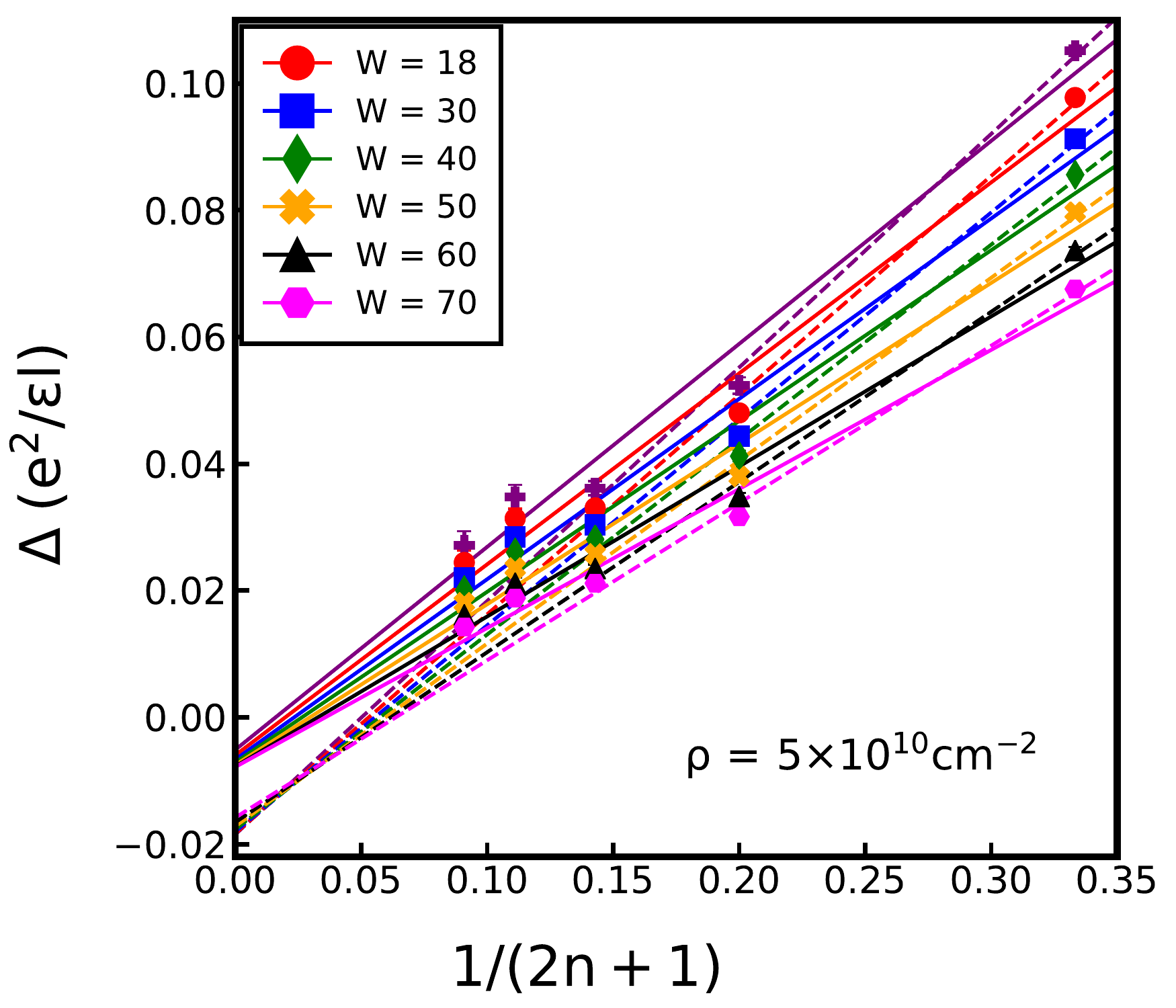}
\includegraphics[width=0.32 \linewidth]{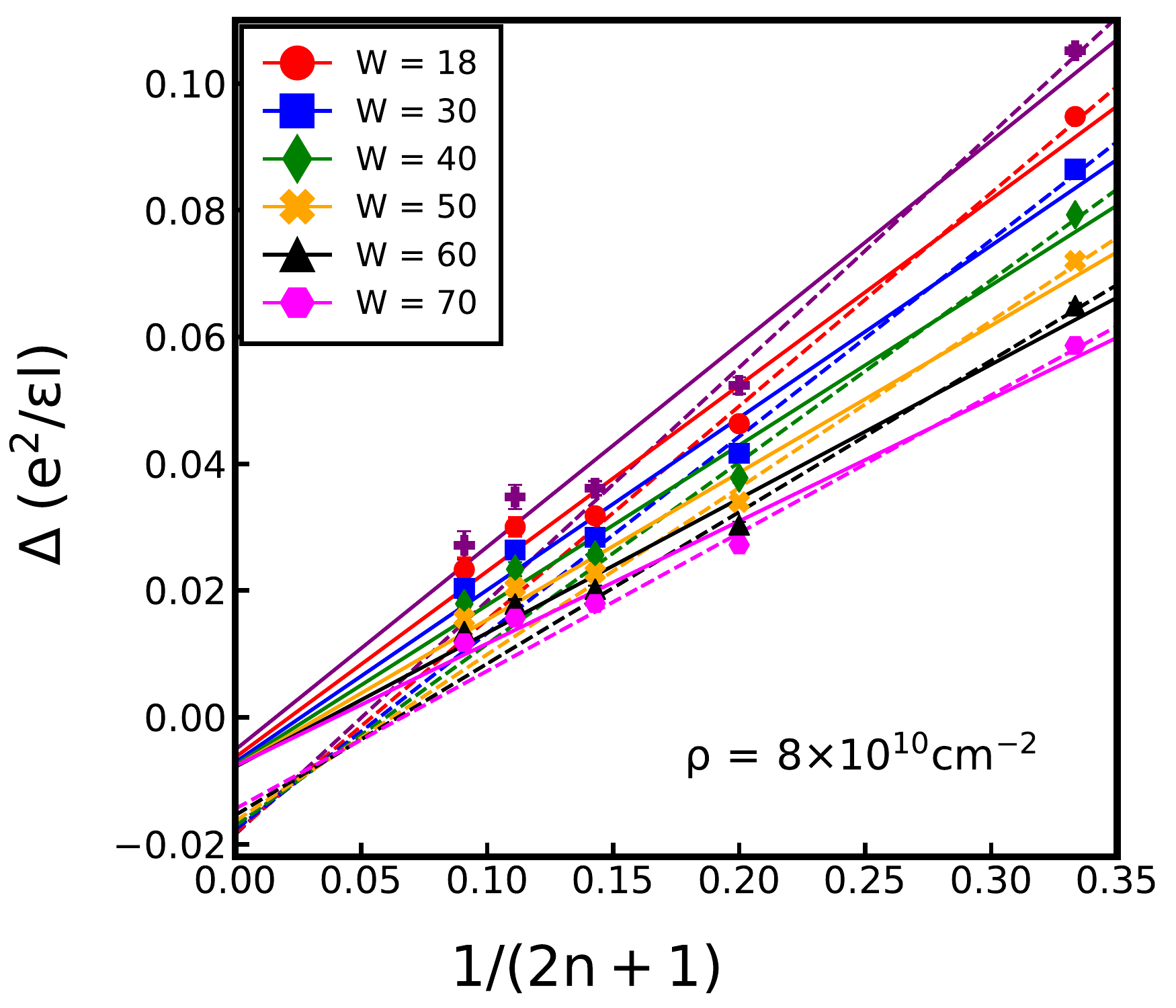}
\includegraphics[width=0.32 \linewidth]{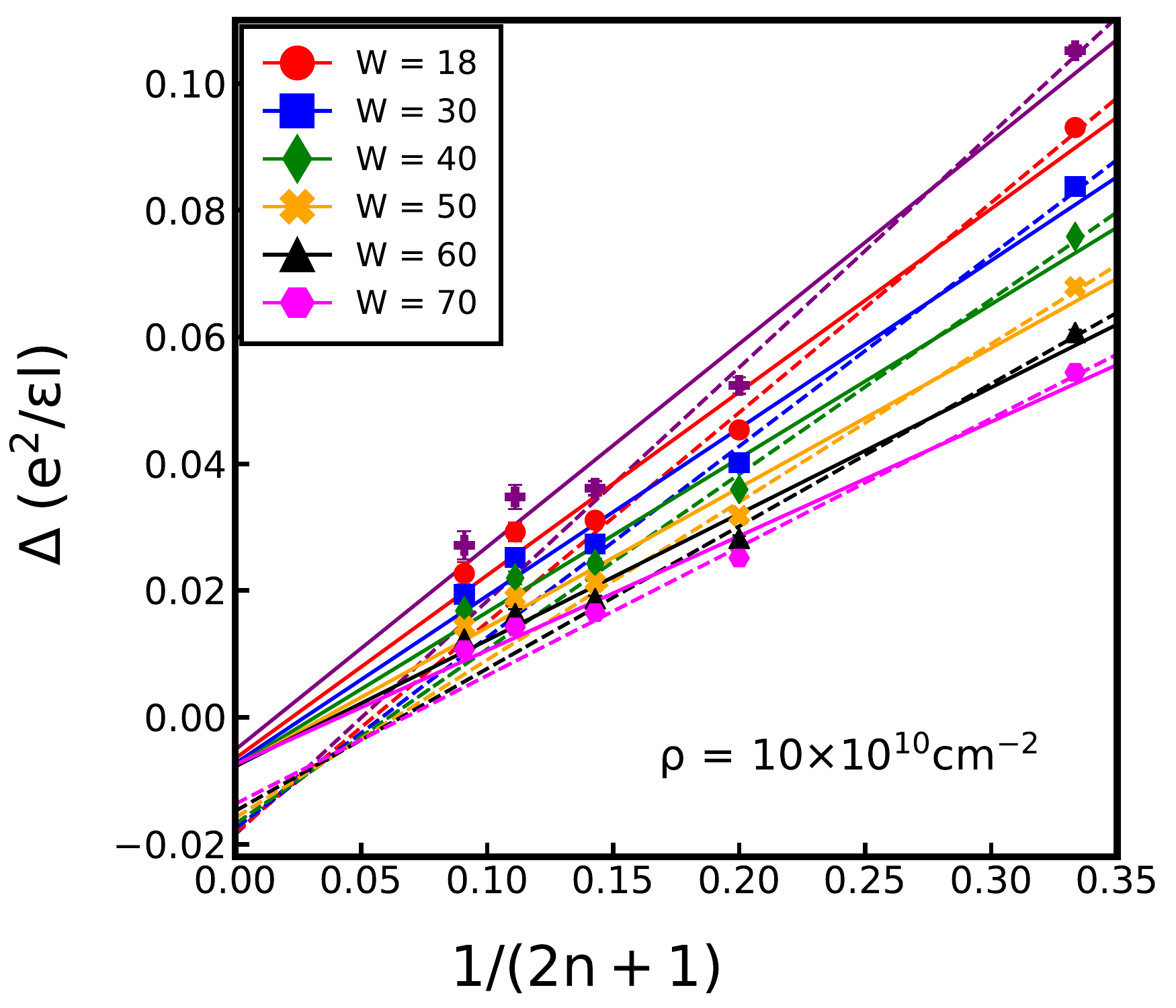}
\includegraphics[width=0.32 \linewidth]{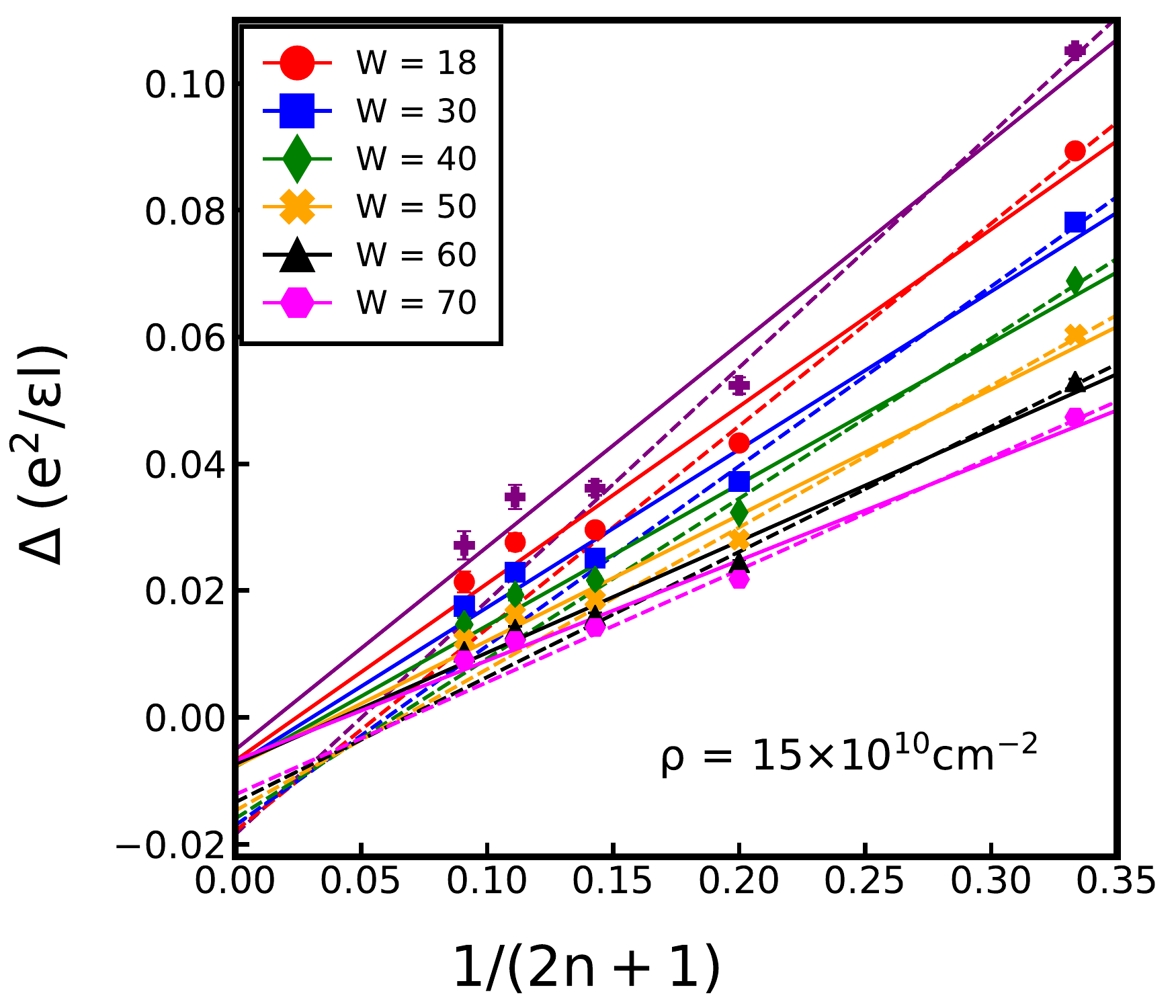}
\includegraphics[width=0.32 \linewidth]{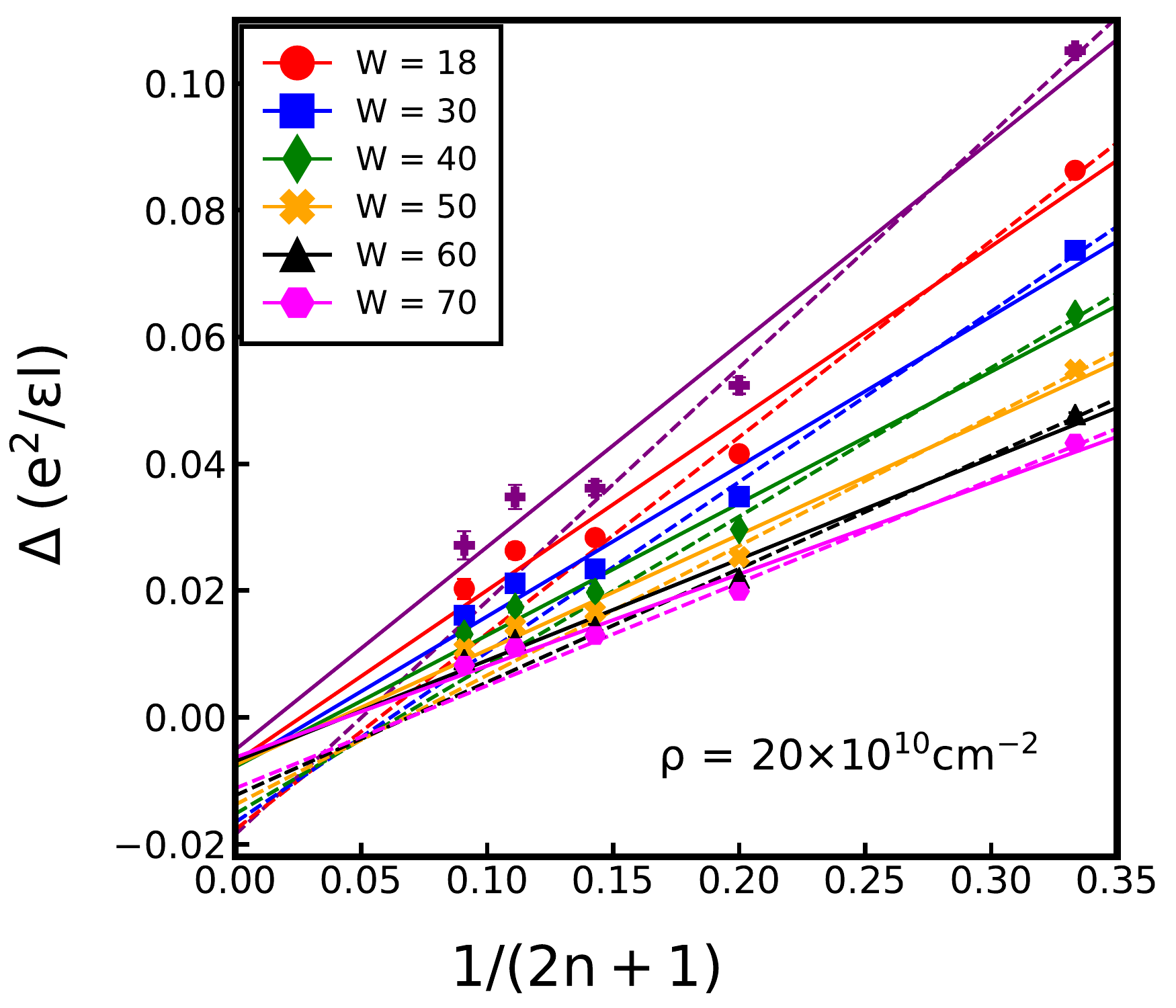}
\includegraphics[width=0.32 \linewidth]{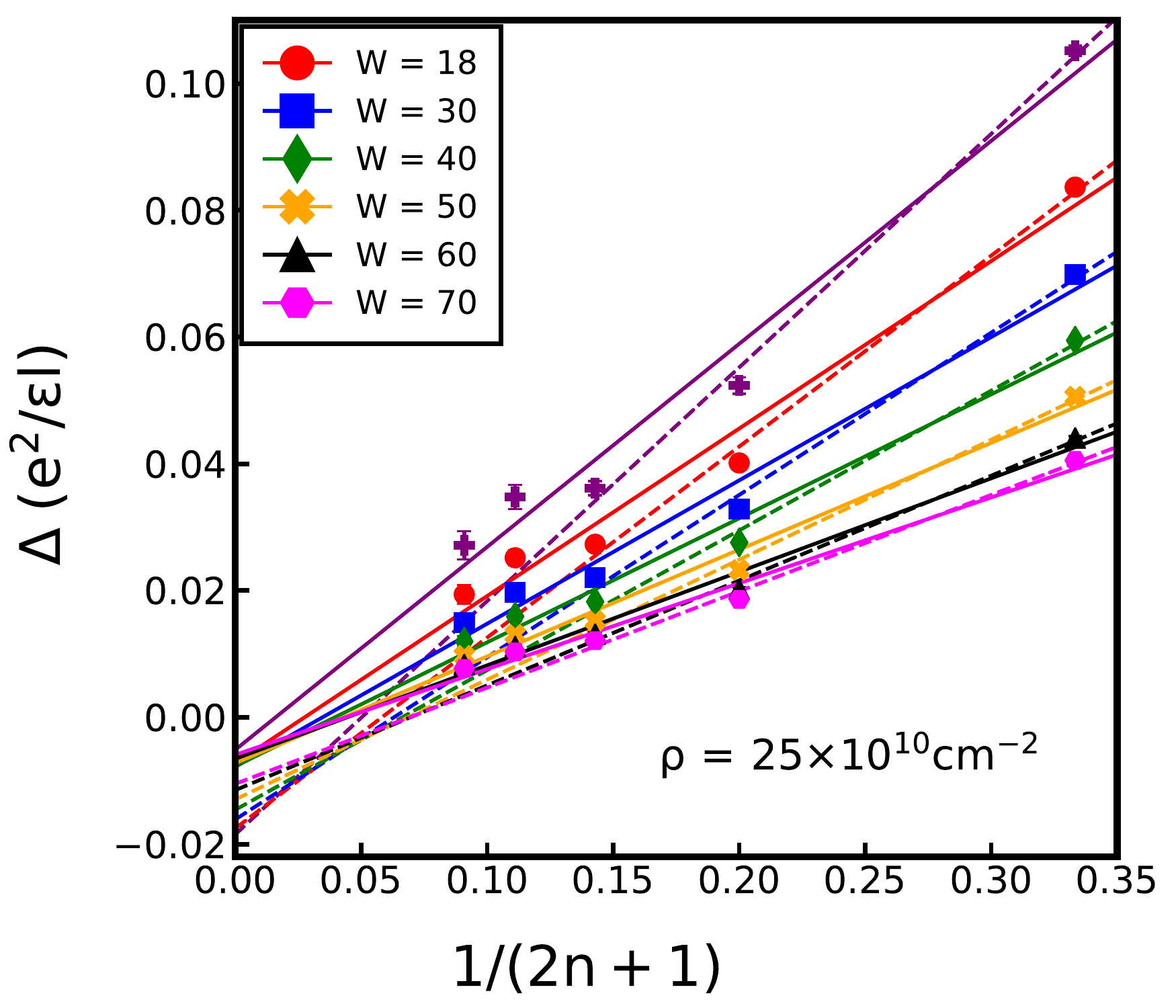}
\includegraphics[width=0.32 \linewidth]{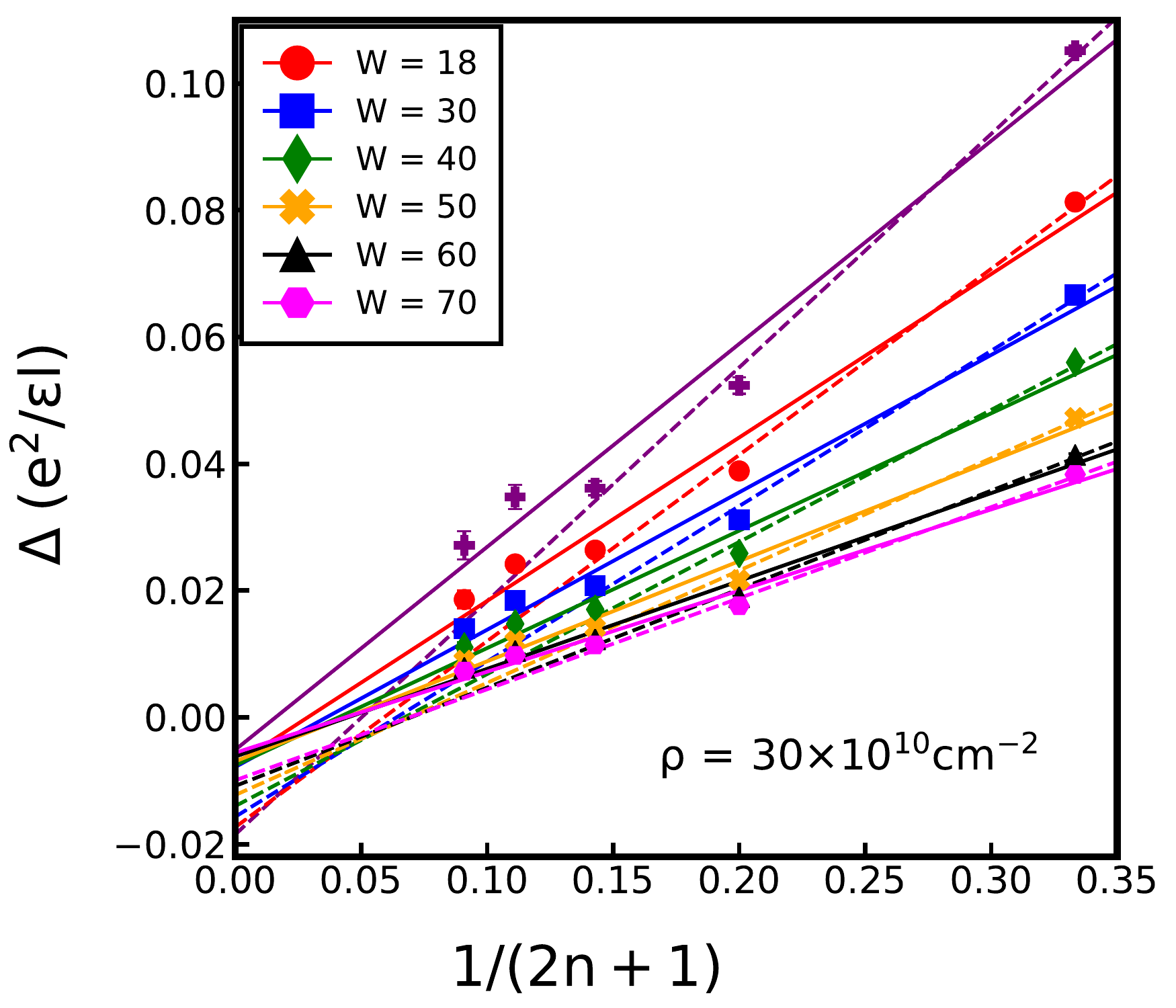}
\caption{Theoretical transport gap $\Delta$ as a function of $1/(2n+1)$ for $n/(2n+1)$ states in GaAs quantum wells at different widths $W$ with densities. The gap is calculated using the variational Monte Carlo (VMC) method. Different markers correspond to different well-widths and the density is indicated on each panel. The well-widths are labeled in units of nanometers. The dashed lines are fits using data points at $\nu=1/3, 2/5, 3/7$ only while the solid lines are fits using data points at all filling factors.}
\label{X_fig:Gamma_extrap_fig}
\end{figure*}

\begin{figure*}[ht!]
\includegraphics[width=0.32 \linewidth]{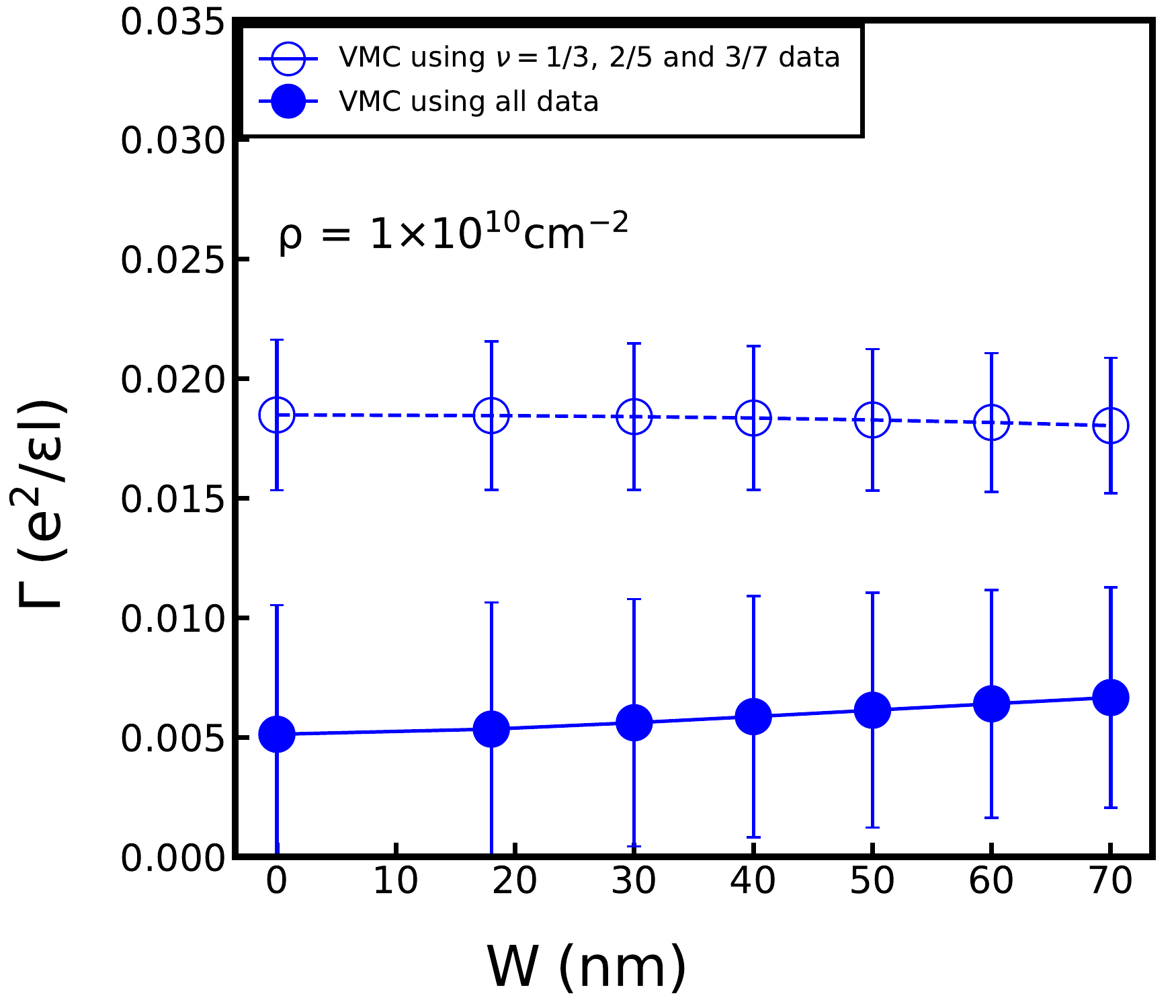}
\includegraphics[width=0.32 \linewidth]{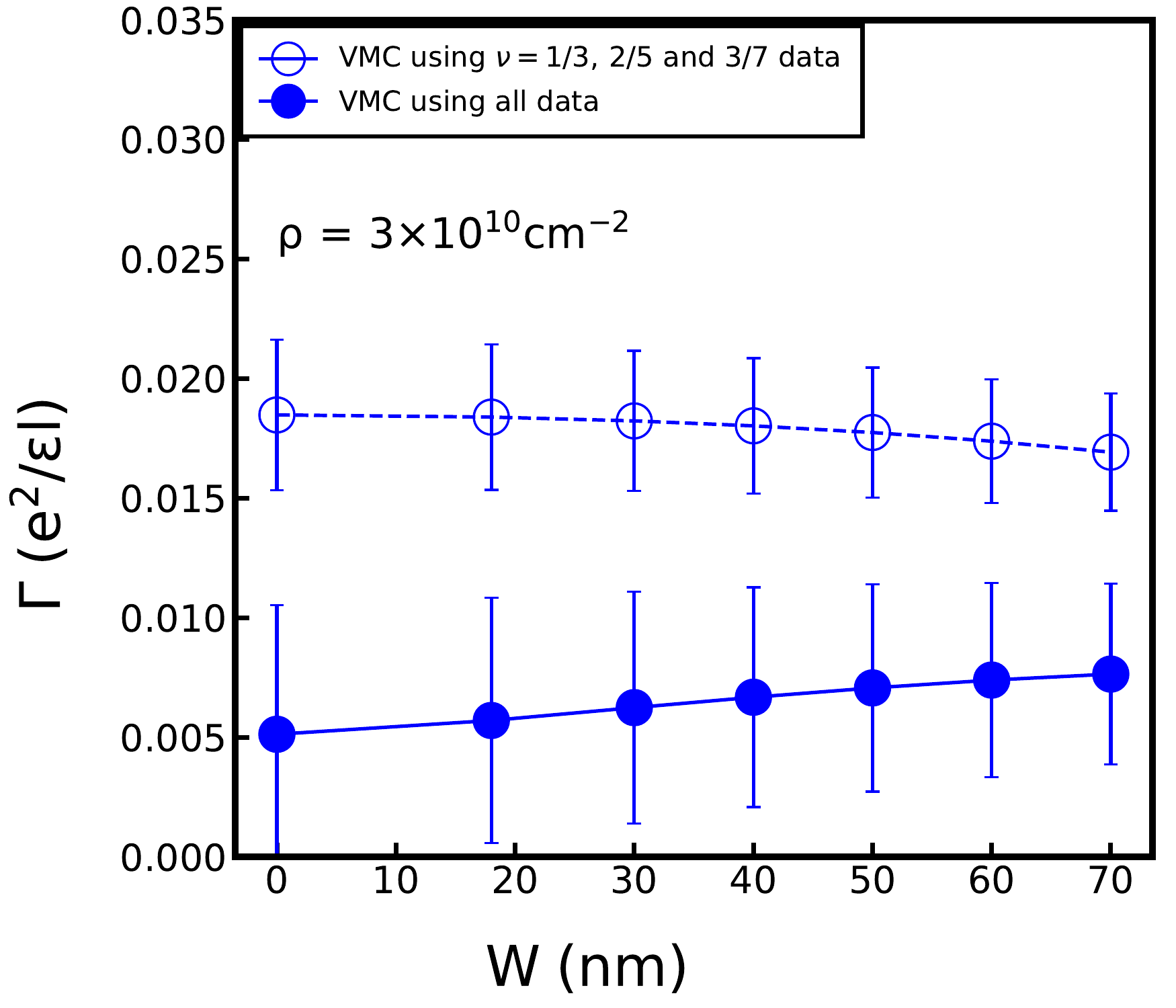}
\includegraphics[width=0.32 \linewidth]{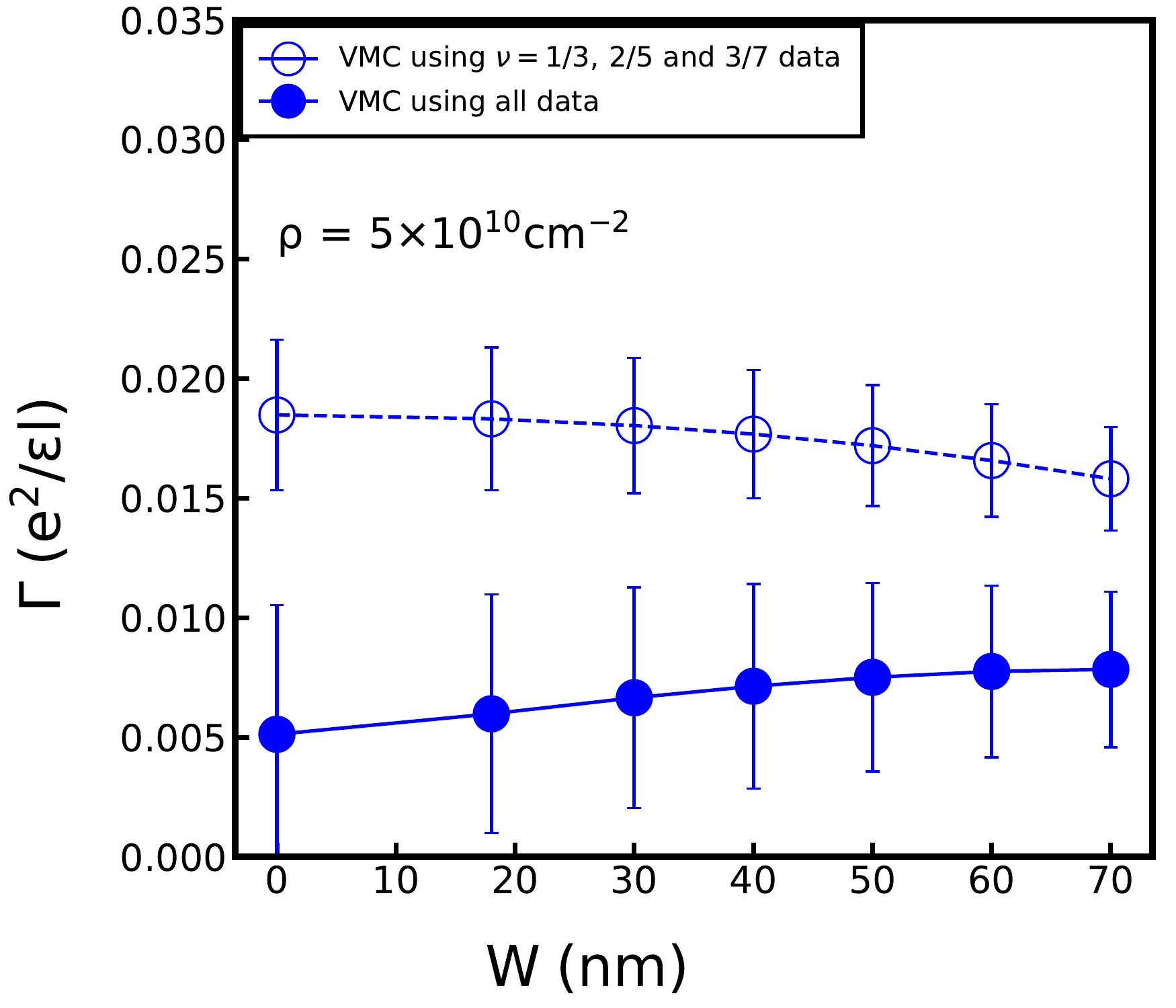}
\includegraphics[width=0.32 \linewidth]{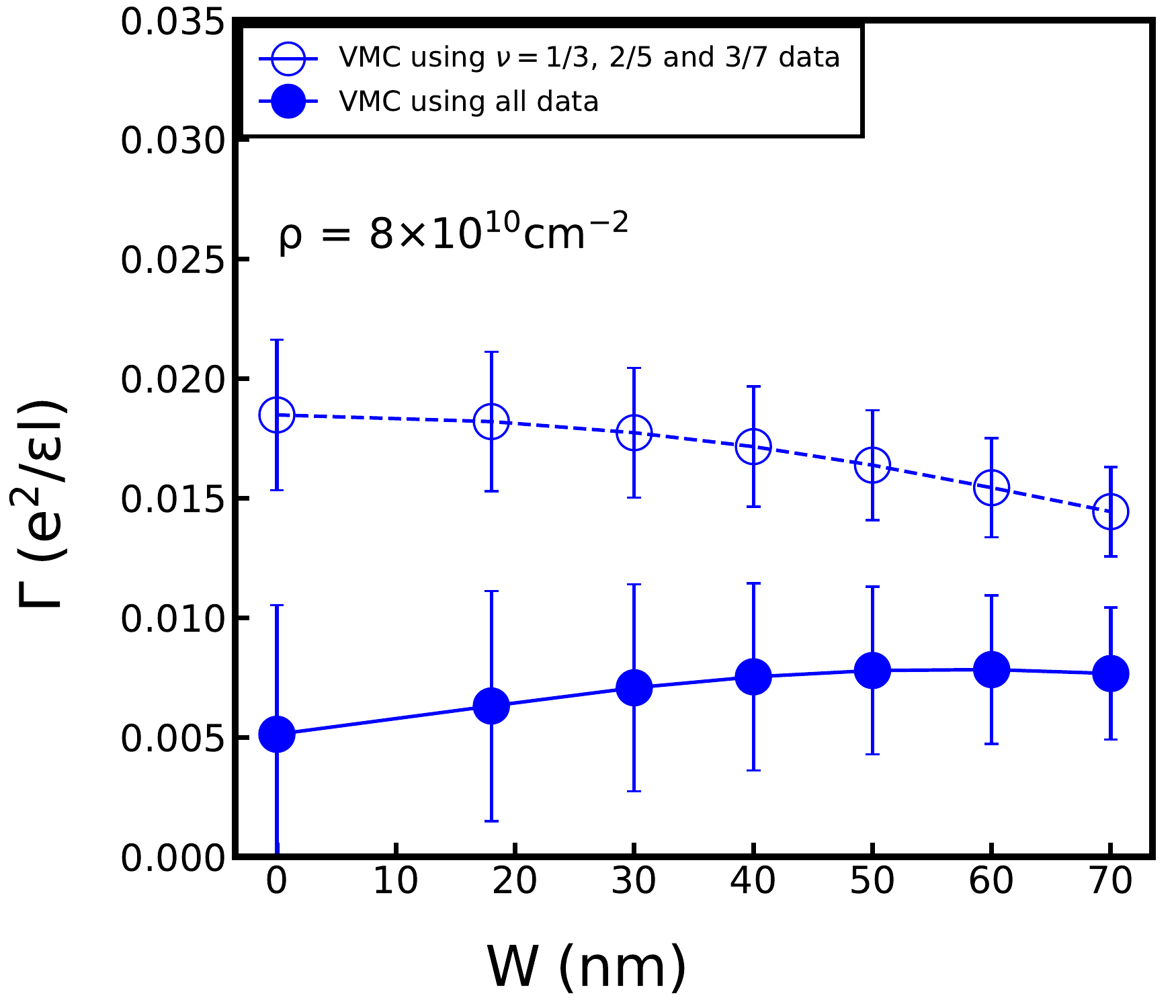}
\includegraphics[width=0.32 \linewidth]{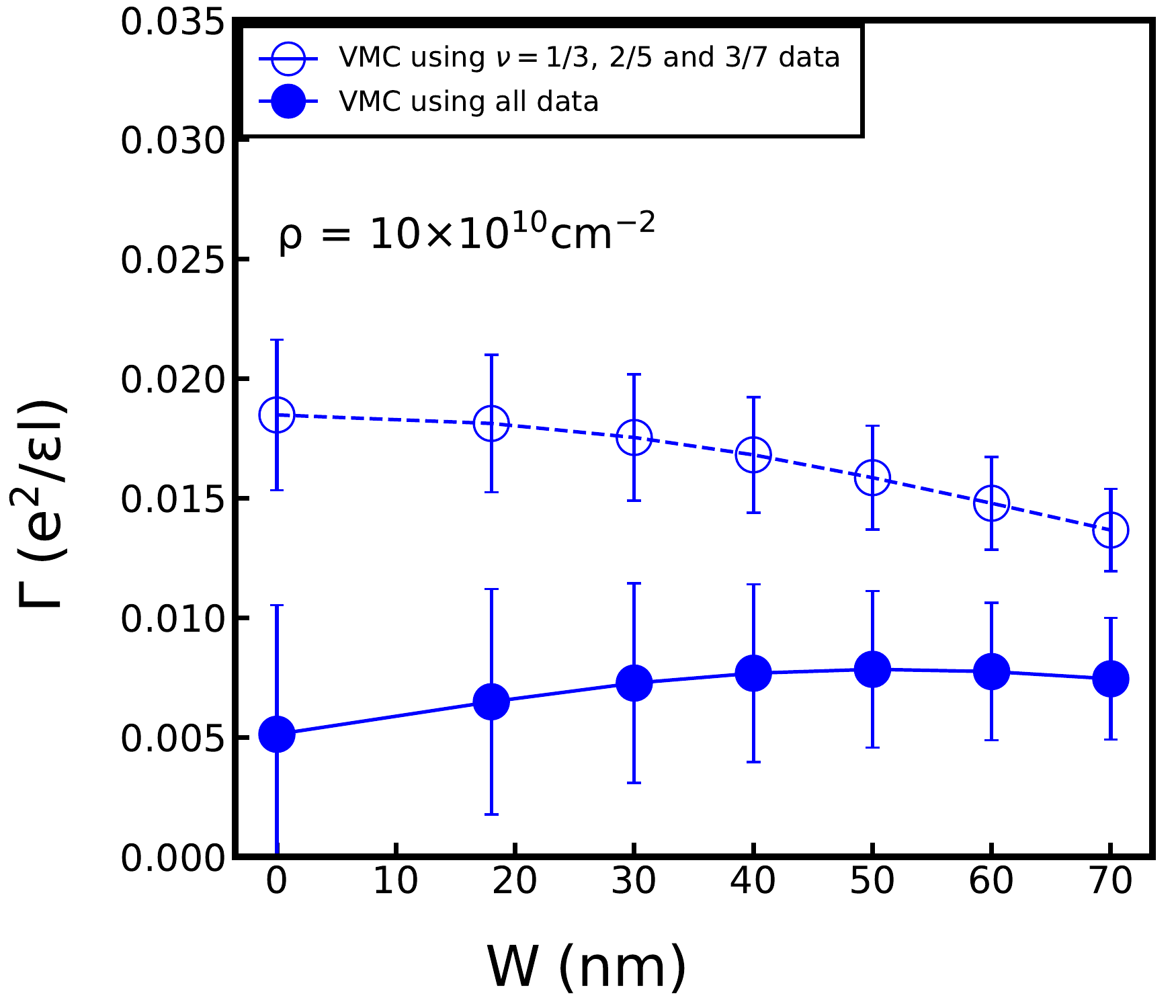}
\includegraphics[width=0.32 \linewidth]{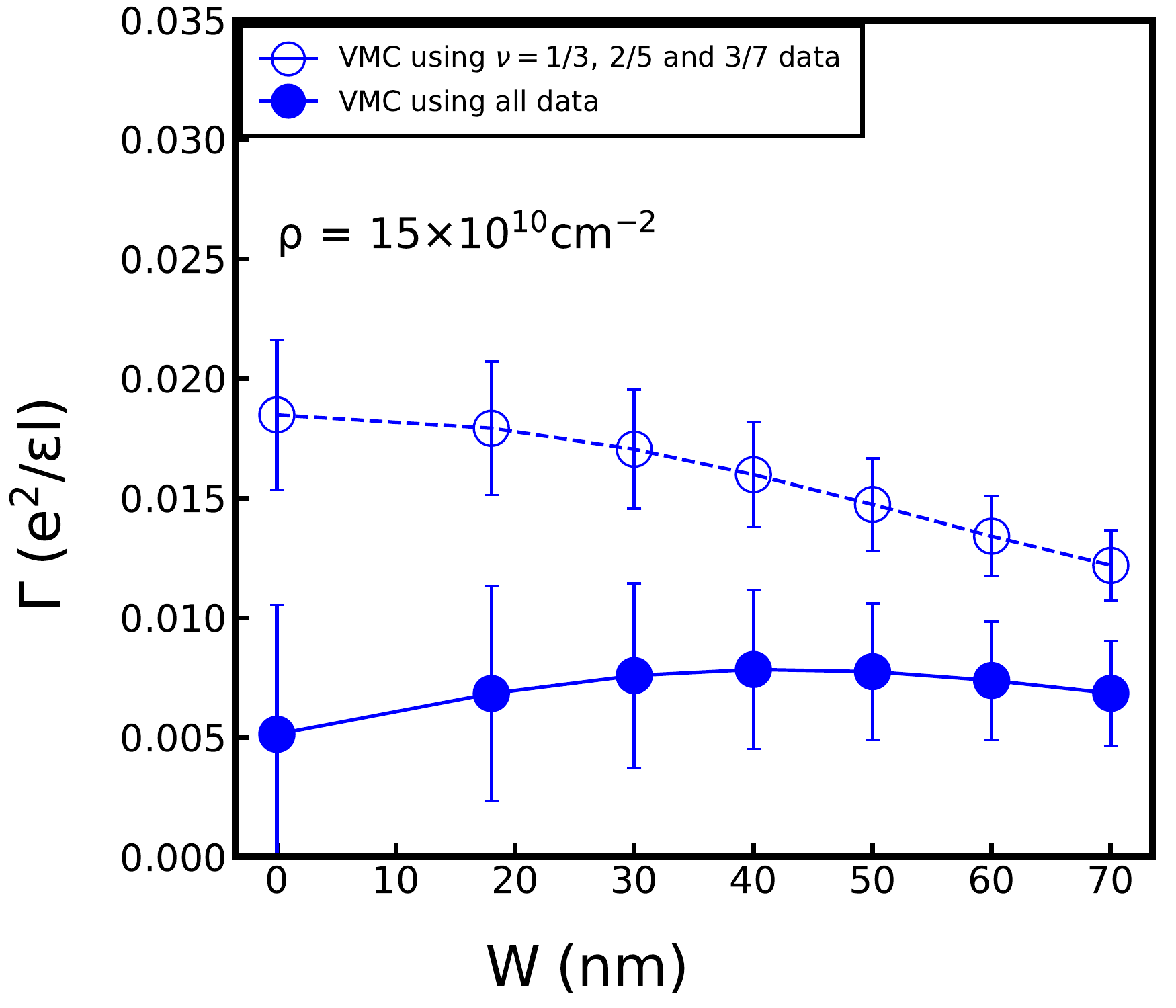}
\includegraphics[width=0.32 \linewidth]{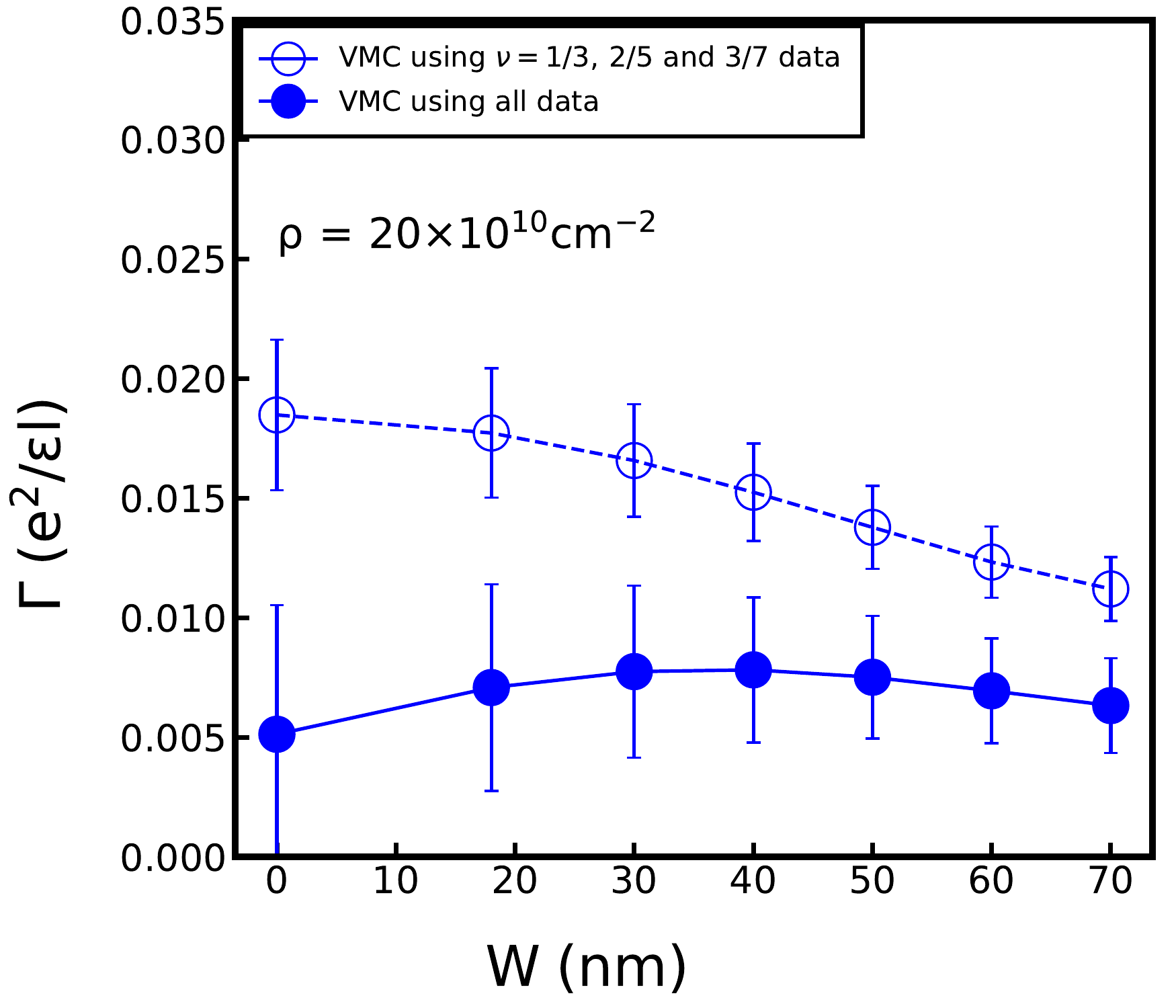}
\includegraphics[width=0.32 \linewidth]{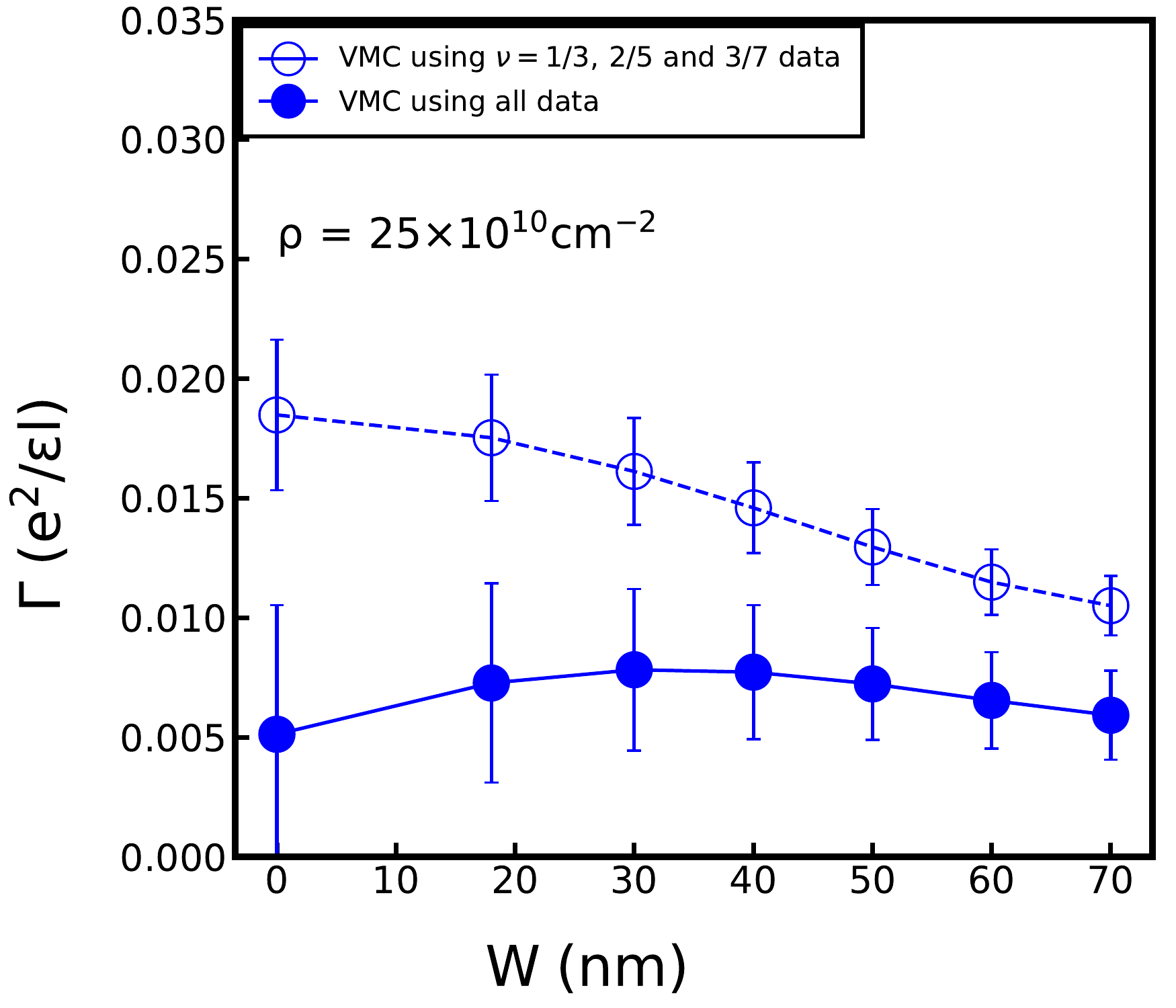}
\includegraphics[width=0.32 \linewidth]{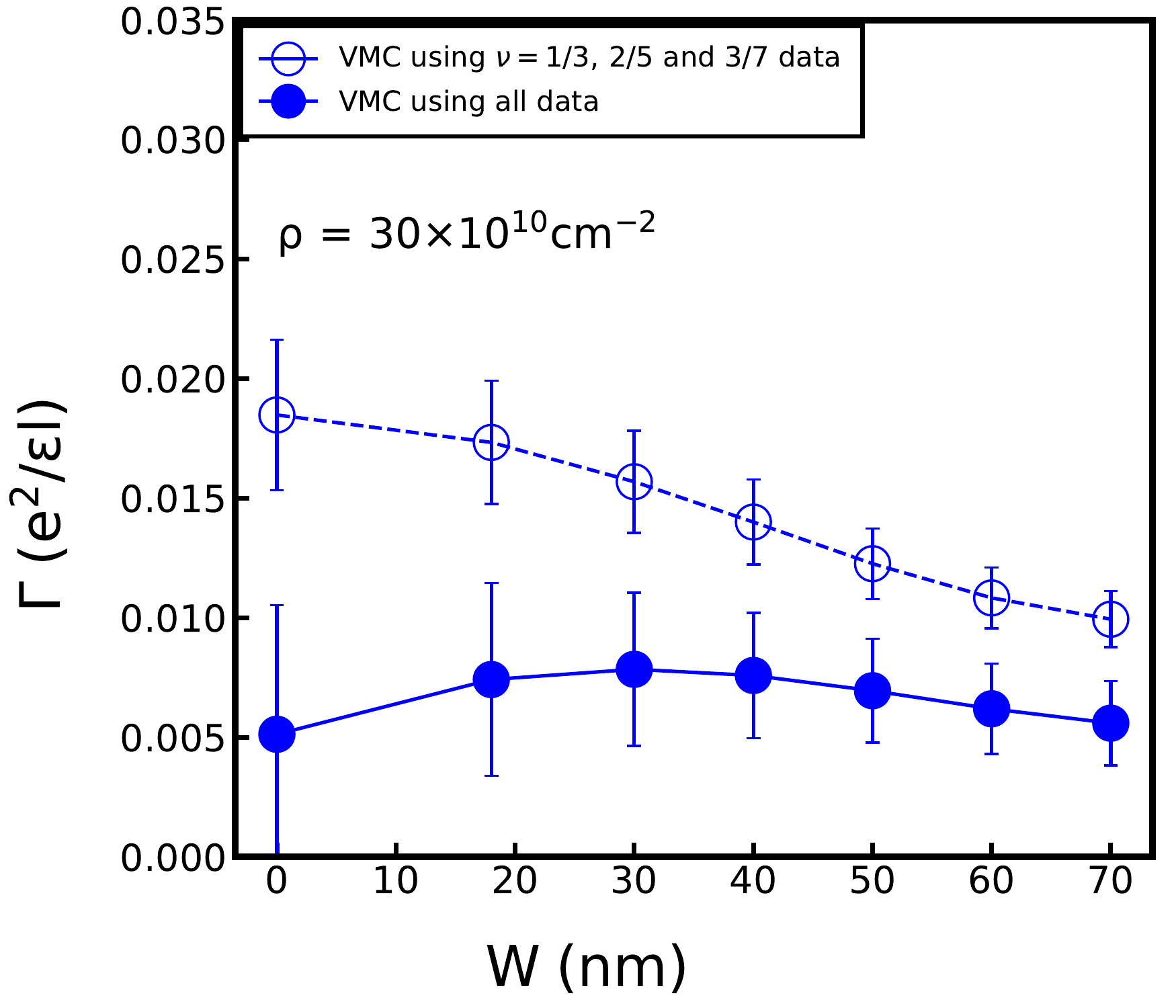}
\caption{$\Gamma$ estimated from VMC gaps as a function of the well-width $W$ for various densities. Each panel corresponds to a specific density shown. The solid markers are obtained by linear regression of data points at all filling factors ($\nu=1/3, 2/5, 3/7, 4/9$ and $5/11$), which correspond to the solid lines in Fig.~\ref{X_fig:Gamma_extrap_fig}, while the dashed markers are obtained by linear regression of gaps only for $\nu=1/3, 2/5$ and $3/7$, which correspond to the dashed lines in Fig.~\ref{X_fig:Gamma_extrap_fig}.}\label{X_fig:Gamma_thermo_fig}
\end{figure*}

\begin{figure*}[ht!]
	\includegraphics[width=0.32 \linewidth]{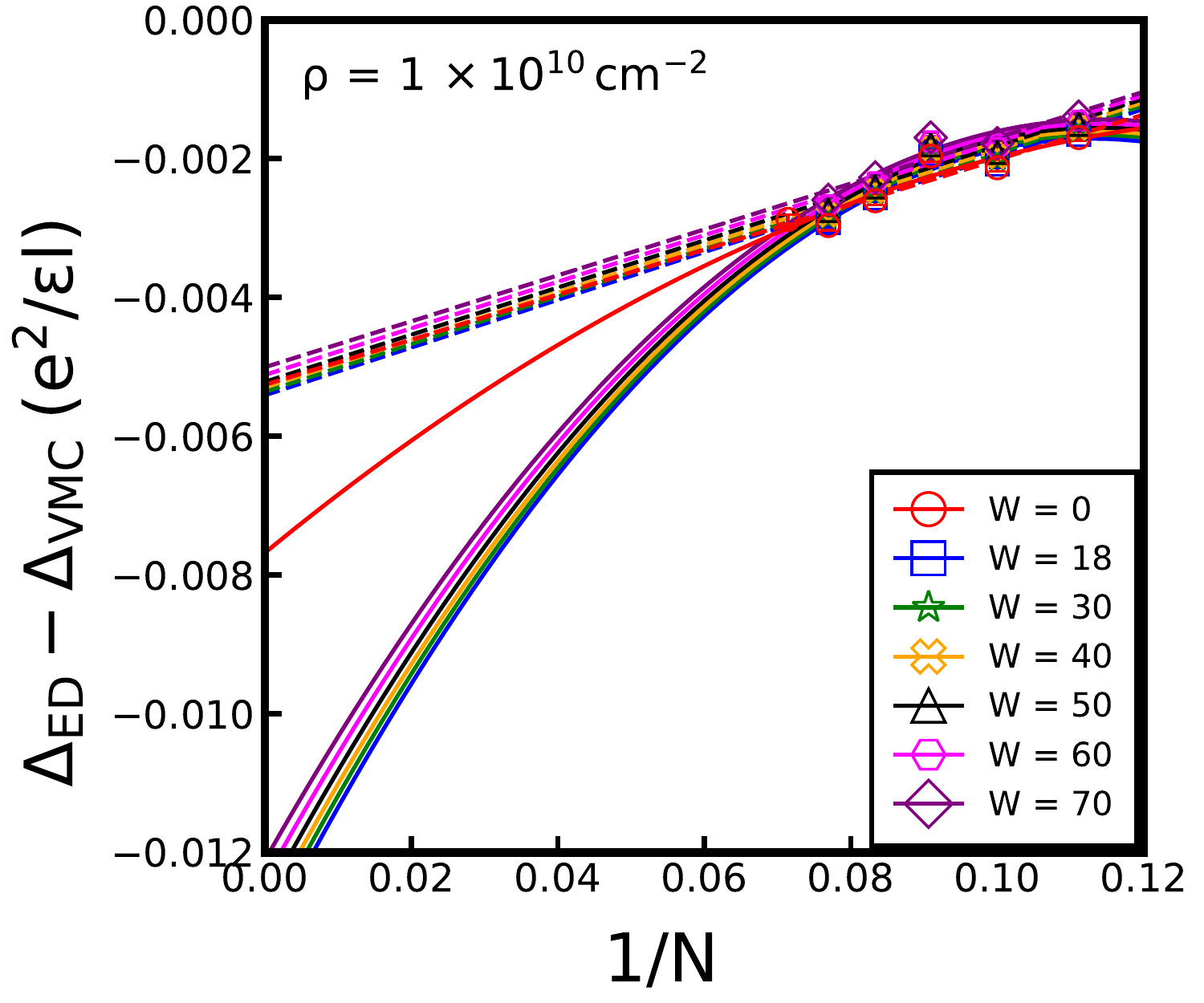}
	\includegraphics[width=0.32 \linewidth]{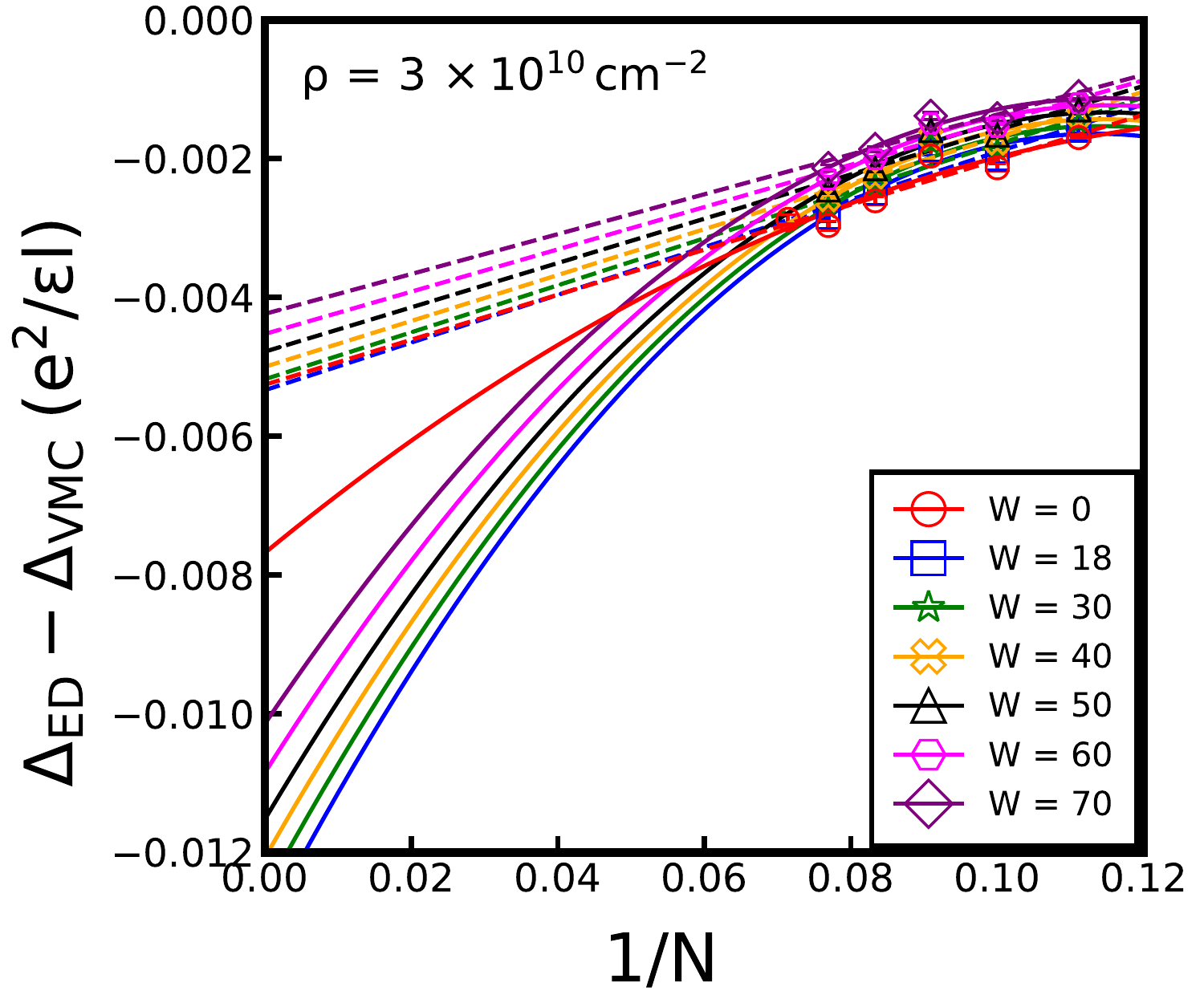}
	\includegraphics[width=0.32 \linewidth]{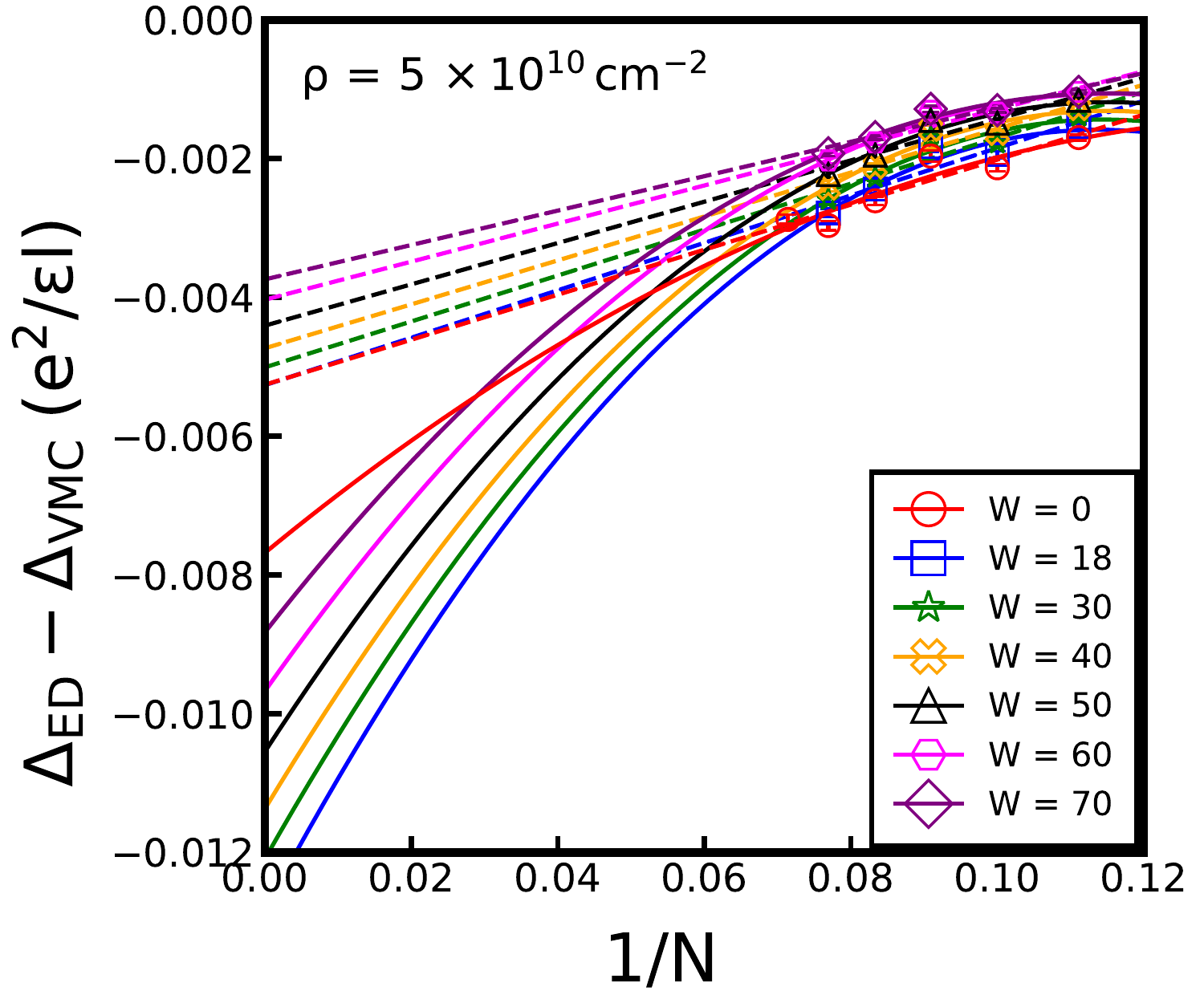}
	\includegraphics[width=0.32 \linewidth]{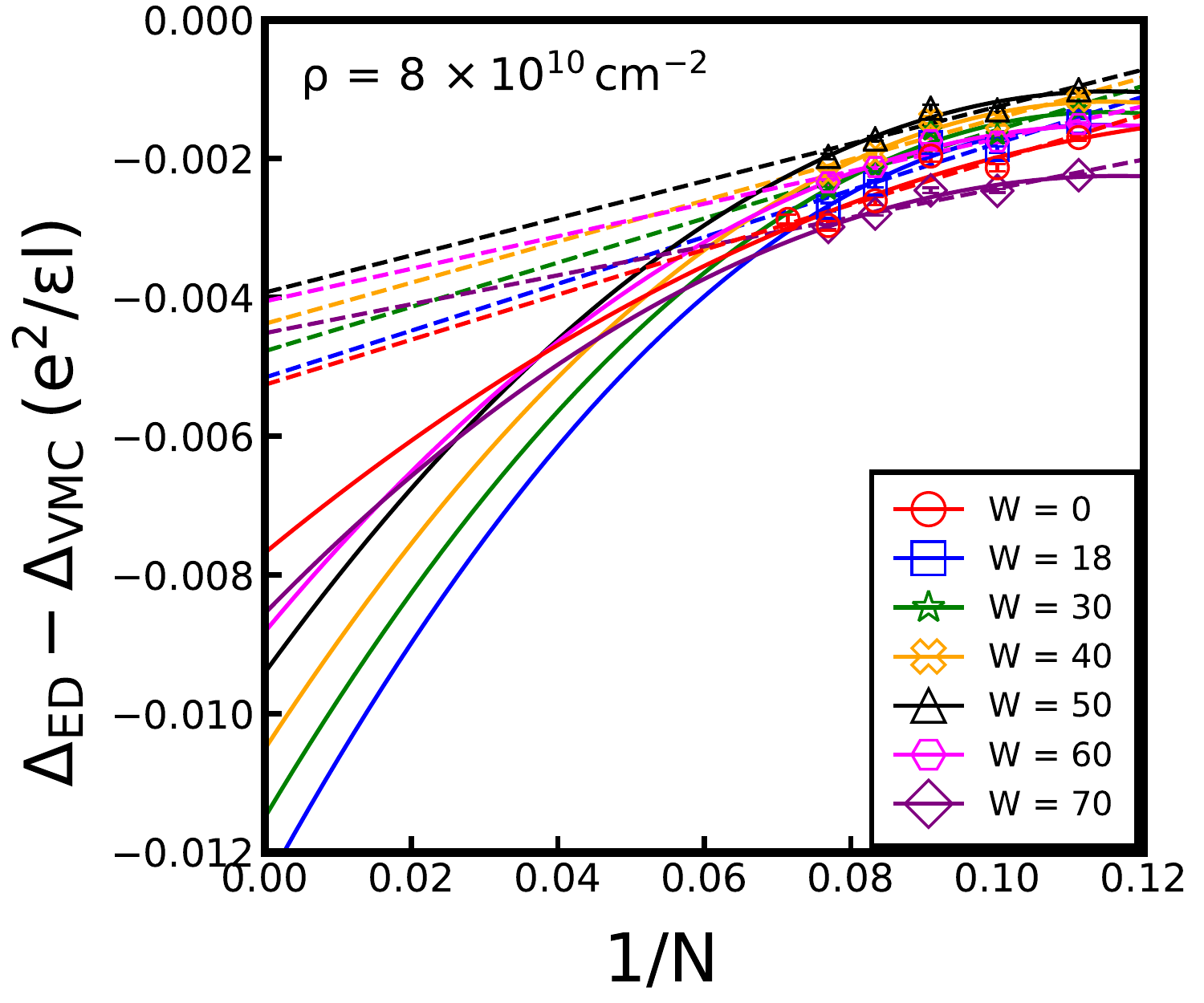}
	\includegraphics[width=0.32 \linewidth]{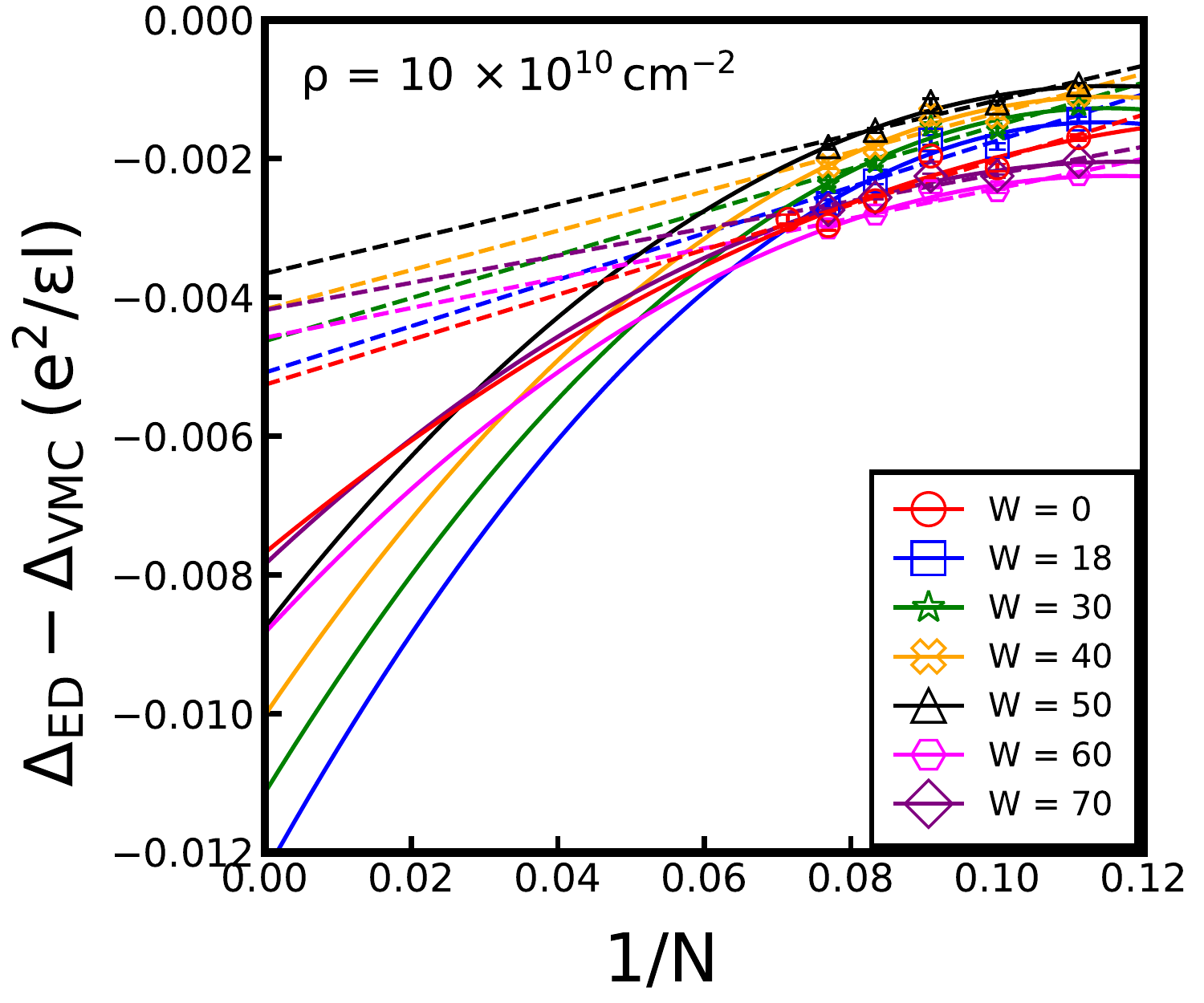}
	\includegraphics[width=0.32 \linewidth]{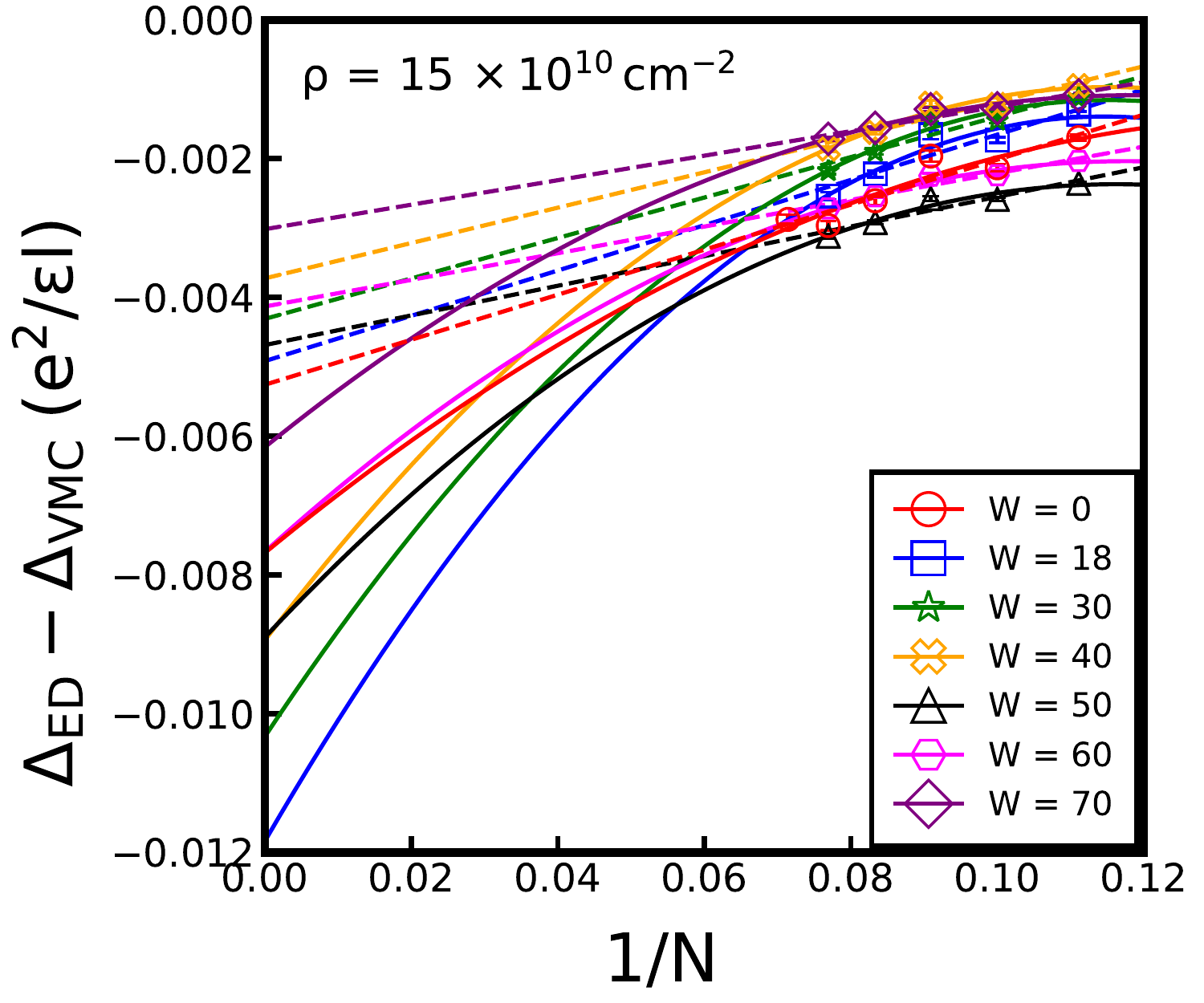}
	\includegraphics[width=0.32 \linewidth]{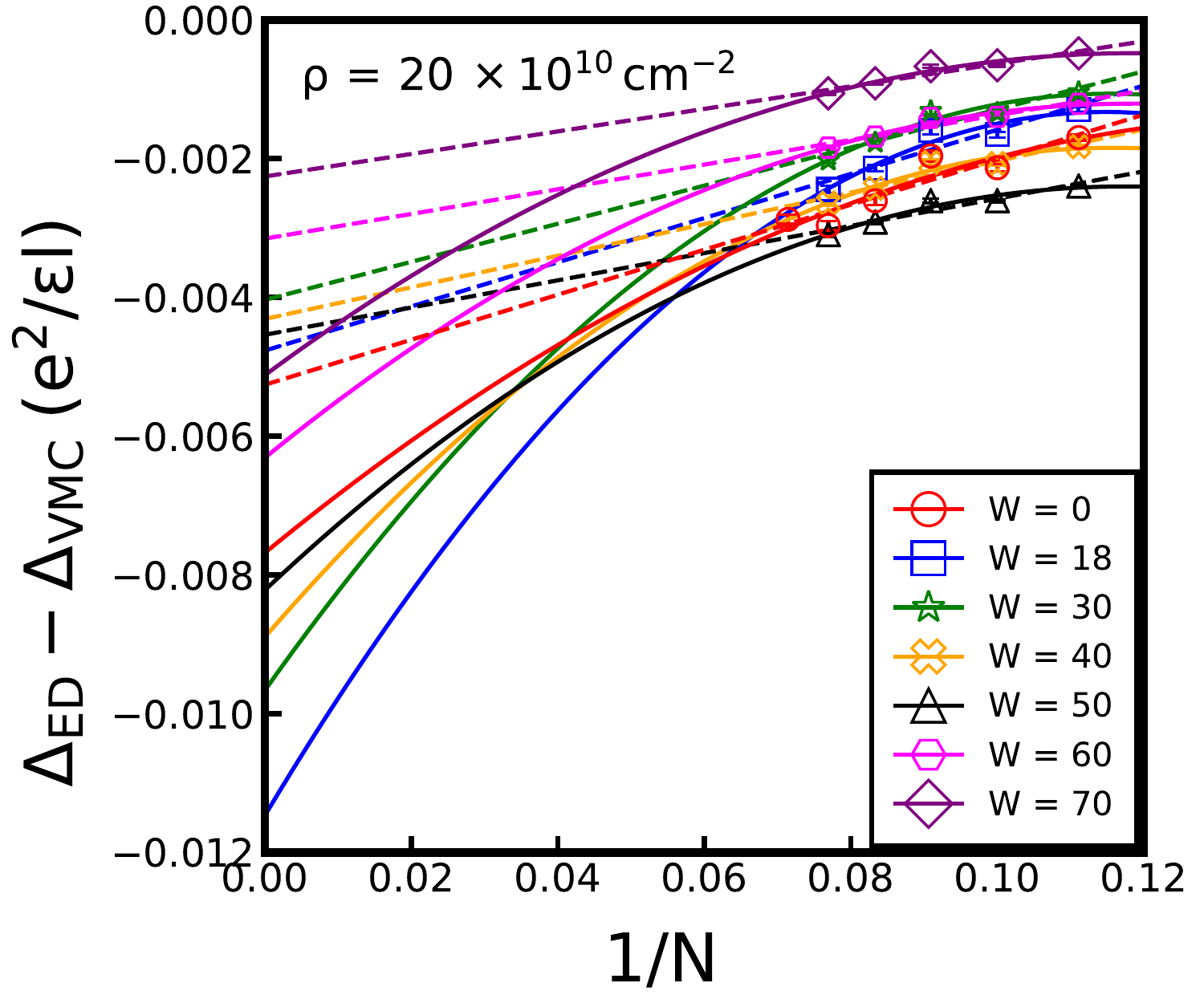}
	\includegraphics[width=0.32 \linewidth]{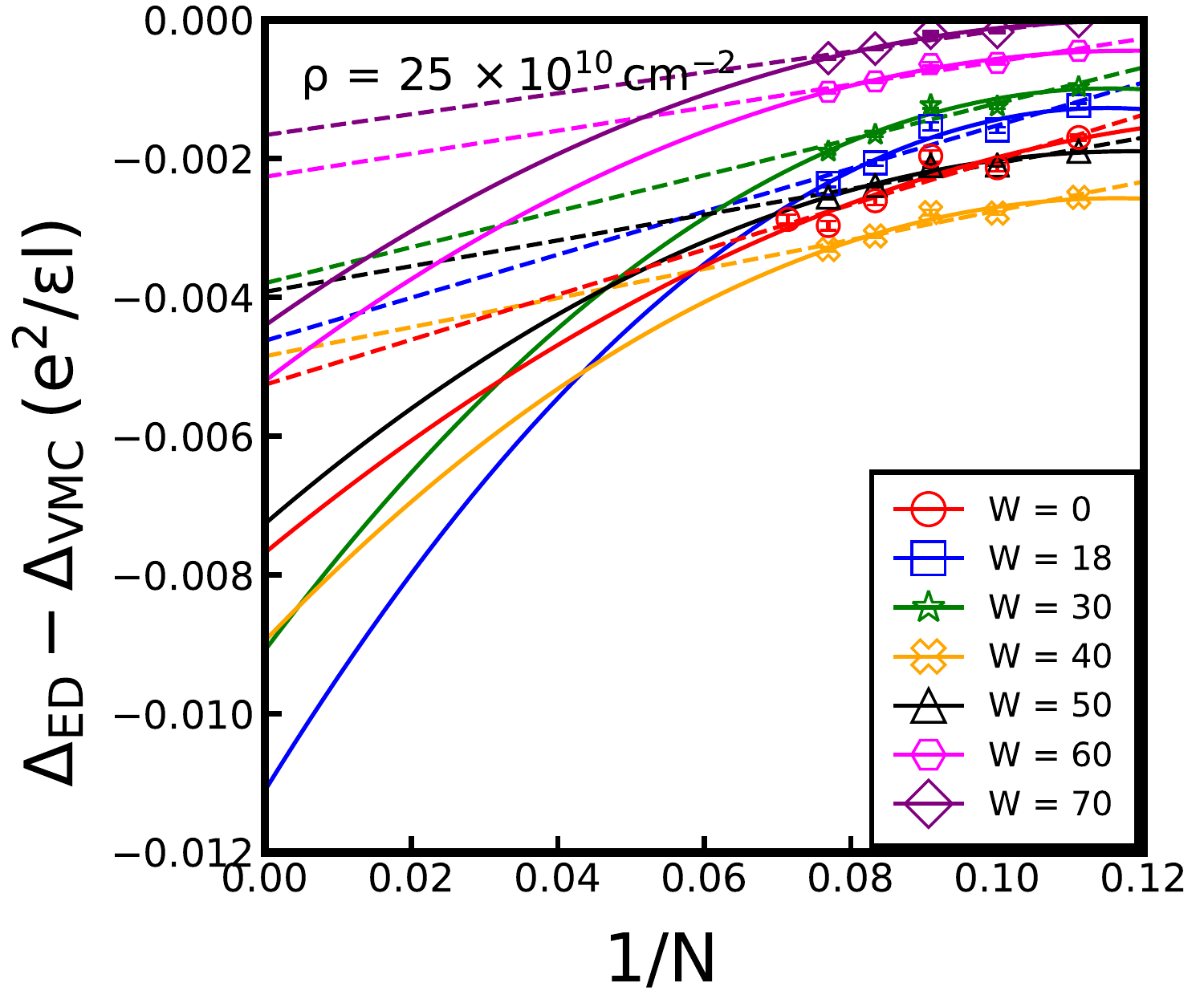}
	\includegraphics[width=0.32 \linewidth]{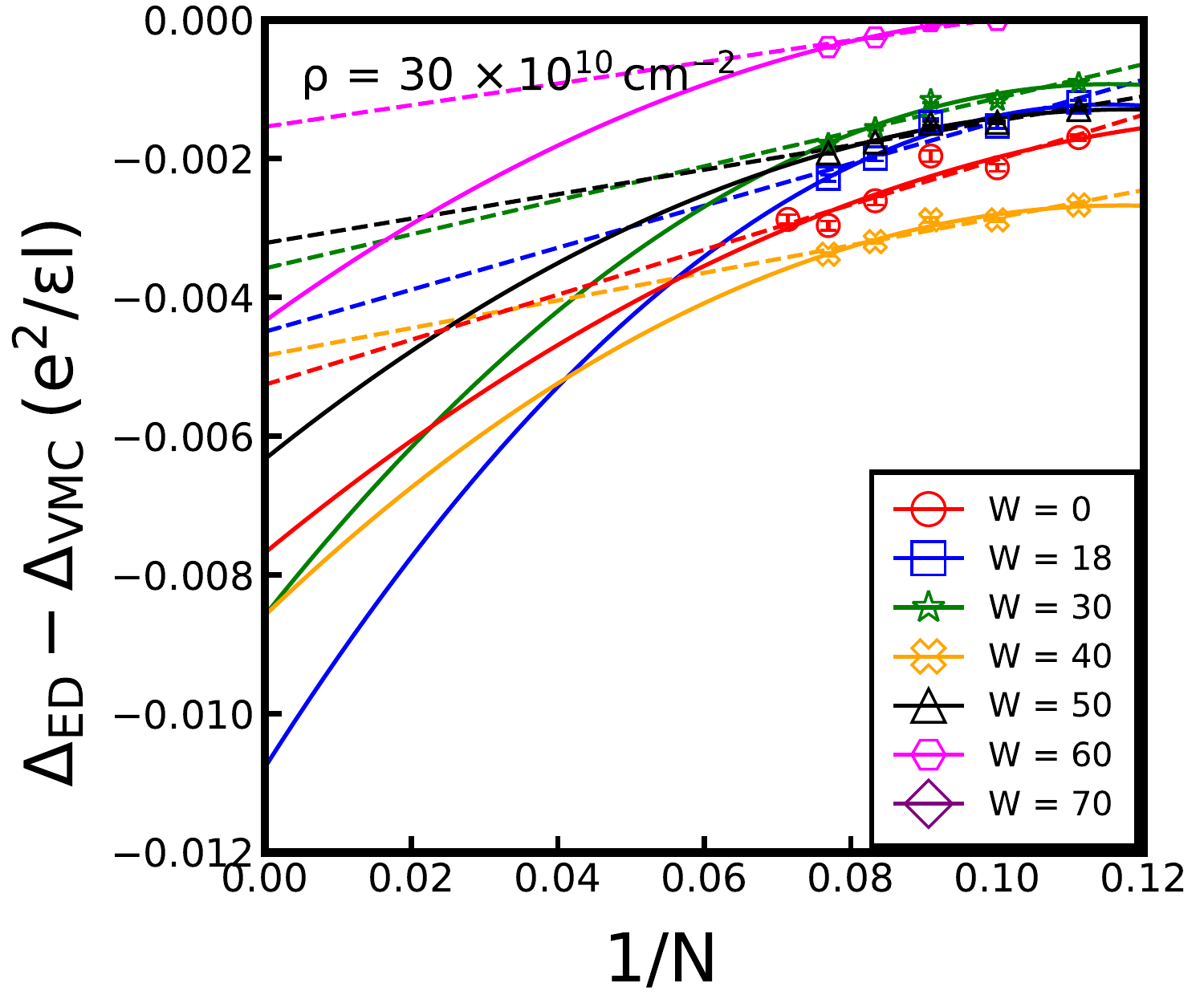}
	\caption{Thermodynamic extrapolation of the variational error, i.e., correction to the gap due to the difference between the variational wave function and the eigenstate obtained from exact diagonalization (ED) at $\nu=1/3$ for several widths and densities (labeled in plots). To calculate the ED energies, we obtain the two-particle pseudopotentials of the effective interaction (Eq. (4) of the main text).  The dashed lines are obtained from linear regressions while the solid lines are from quadratic fits in $1/N$ where $N$ is the number of particles. The ED corrections in the thermodynamic limit included in Fig. 1 of the main text are the mean values of the linearly and quadratically fitted values, and the difference between the linear and quadratic fits are used to estimate the uncertainties. The well-widths shown in the legends are in units of nanometers.}\label{X_fig:ED_correction_13}
\end{figure*}

\begin{figure*}[ht!]
	\includegraphics[width=0.32 \linewidth]{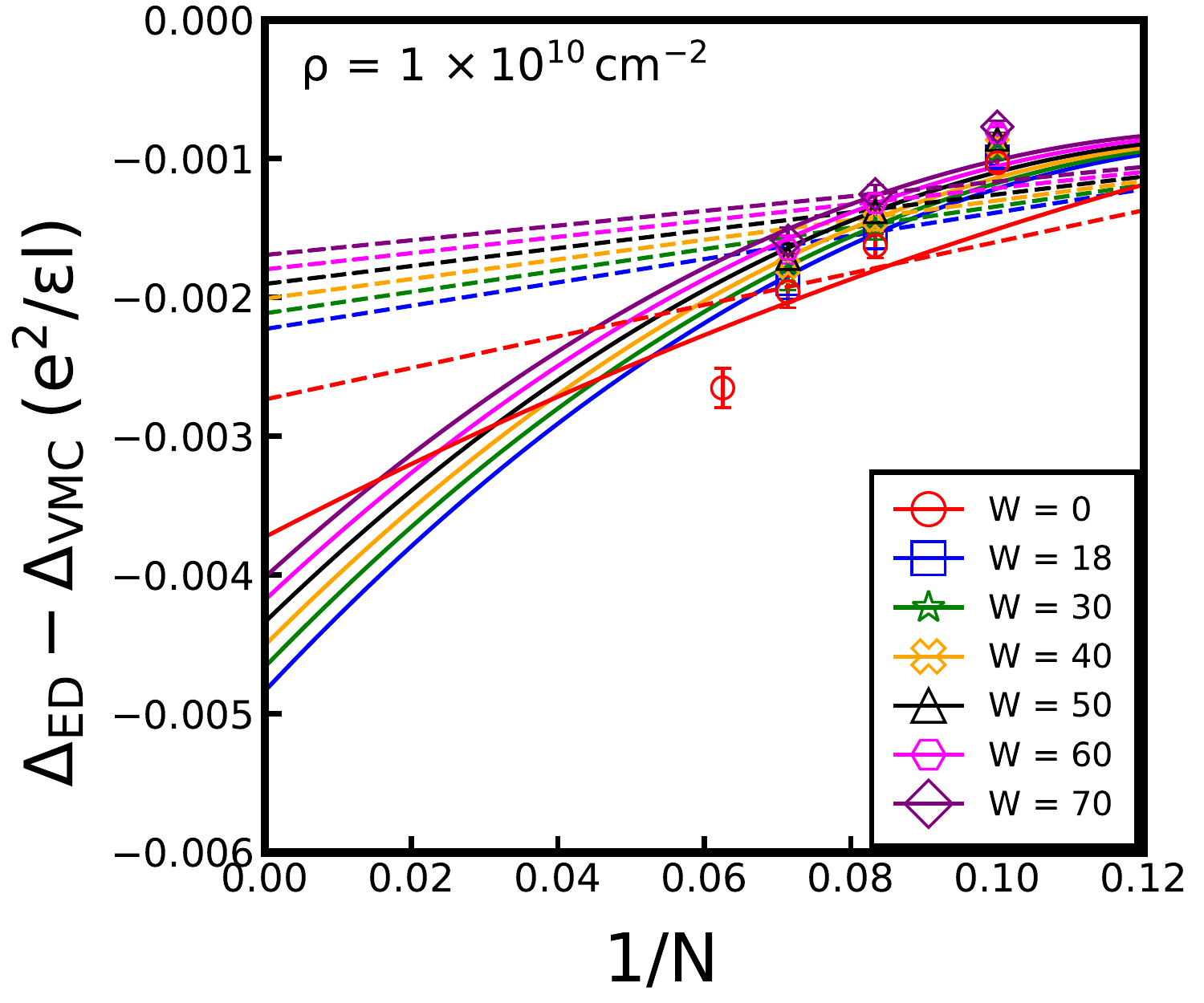}
	\includegraphics[width=0.32 \linewidth]{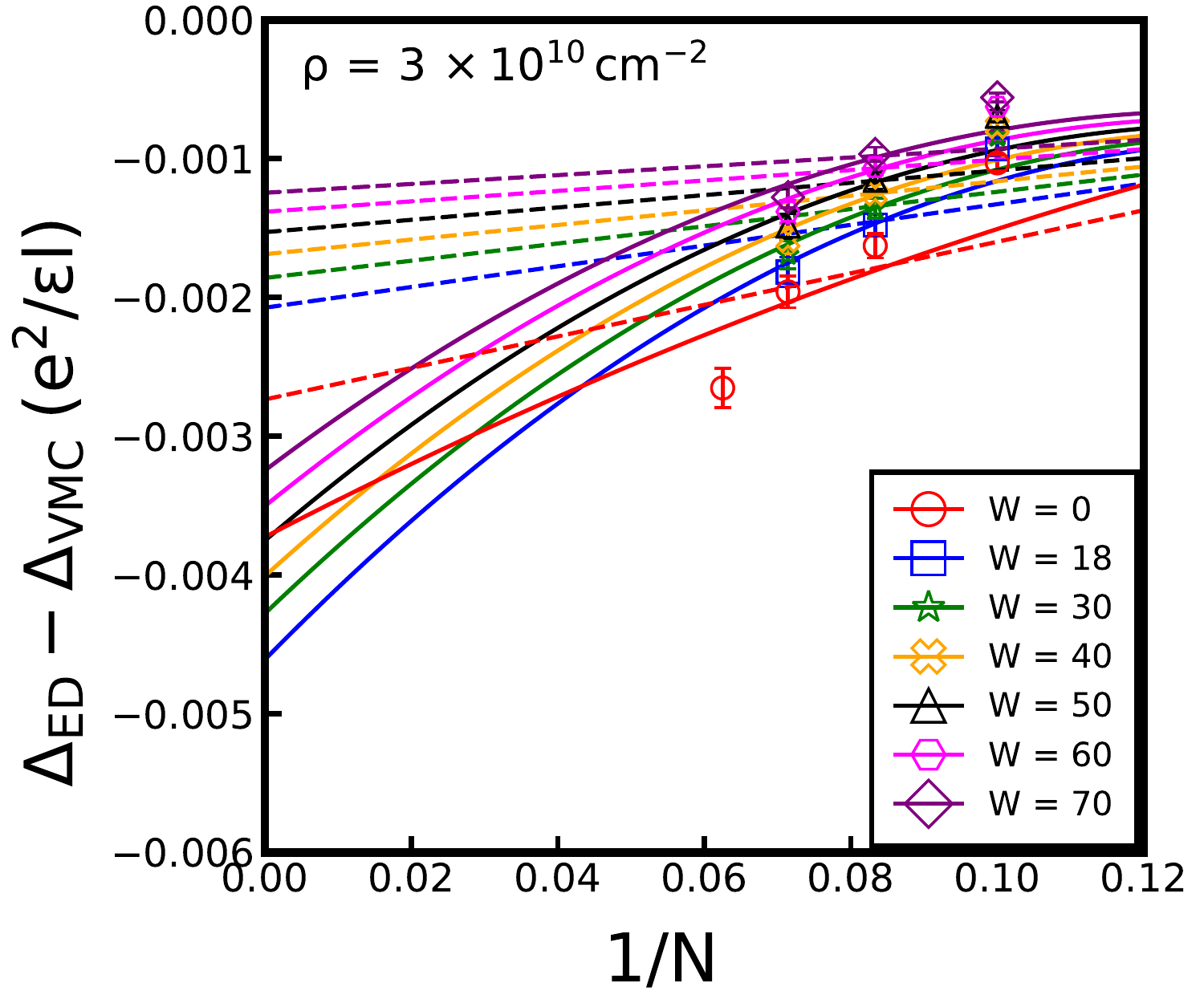}
	\includegraphics[width=0.32 \linewidth]{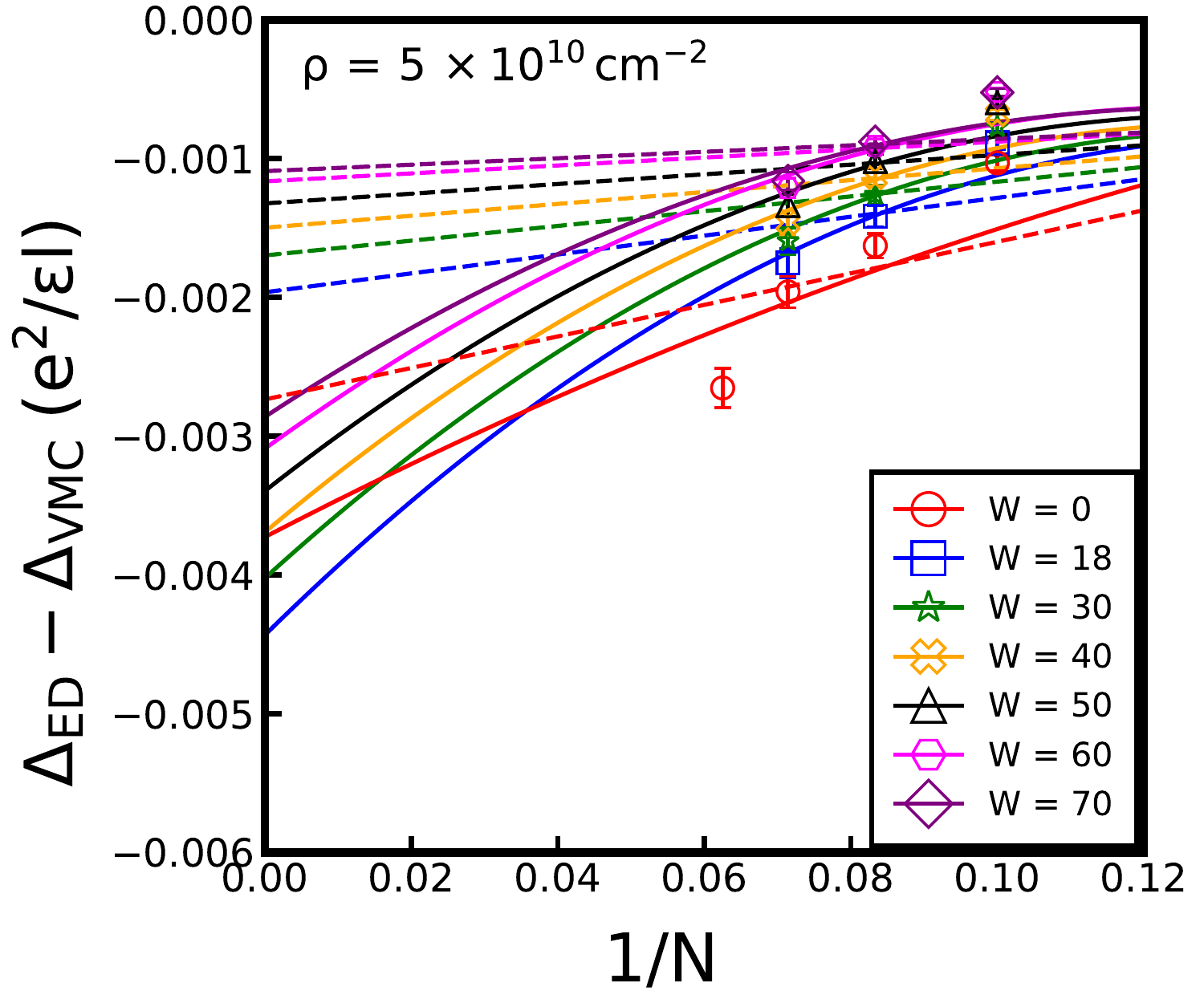}
	\includegraphics[width=0.32 \linewidth]{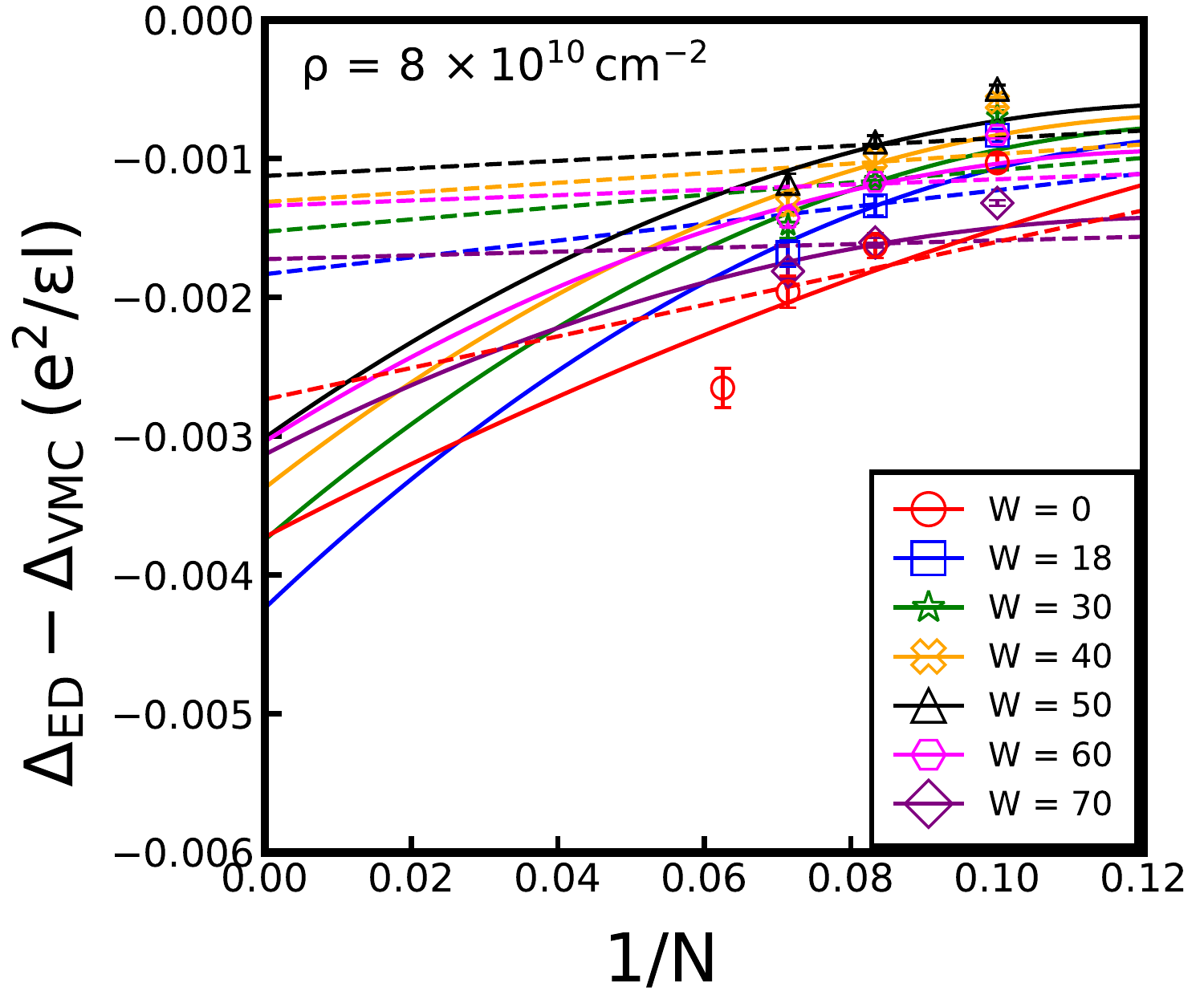}
	\includegraphics[width=0.32 \linewidth]{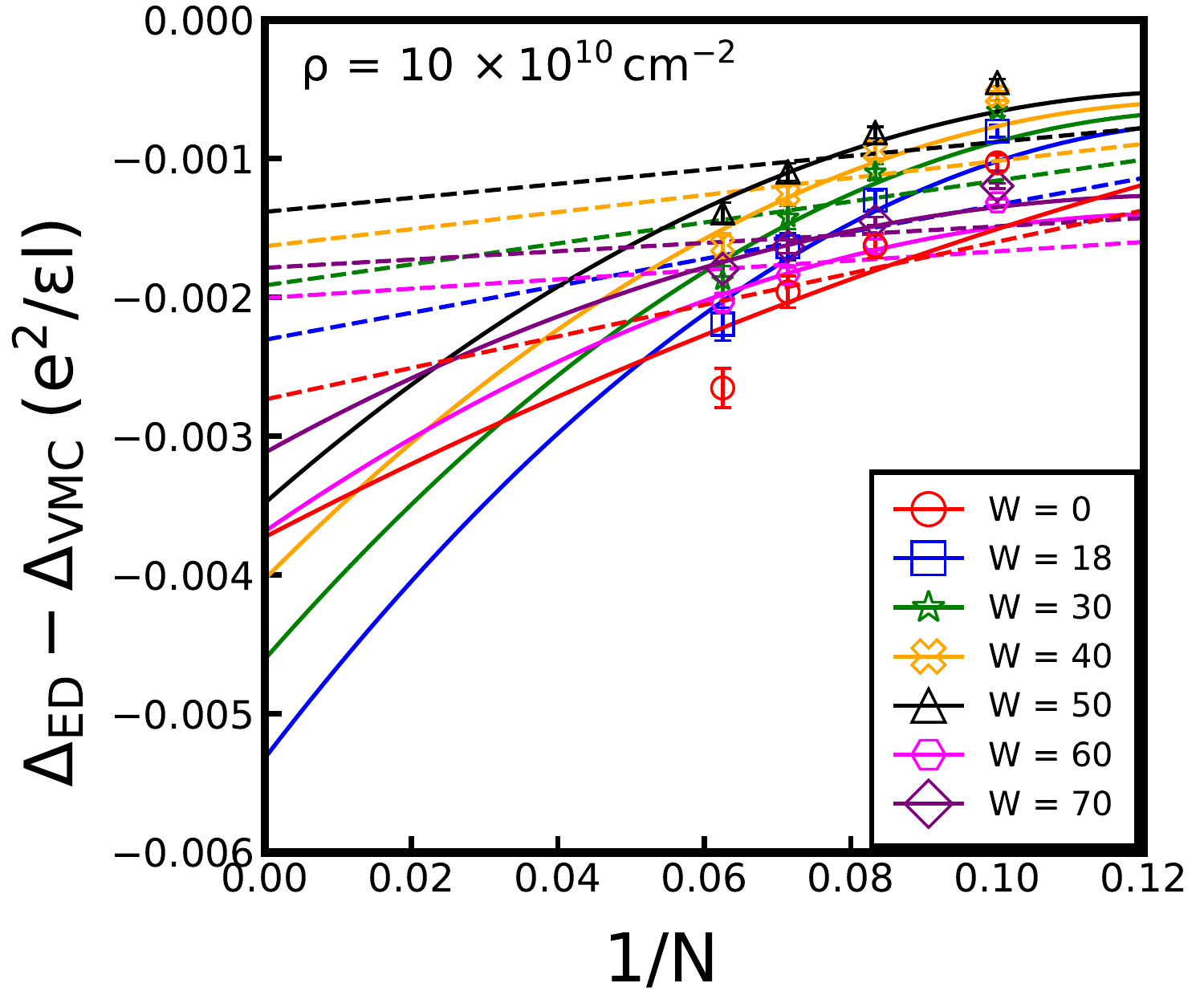}
	\includegraphics[width=0.32 \linewidth]{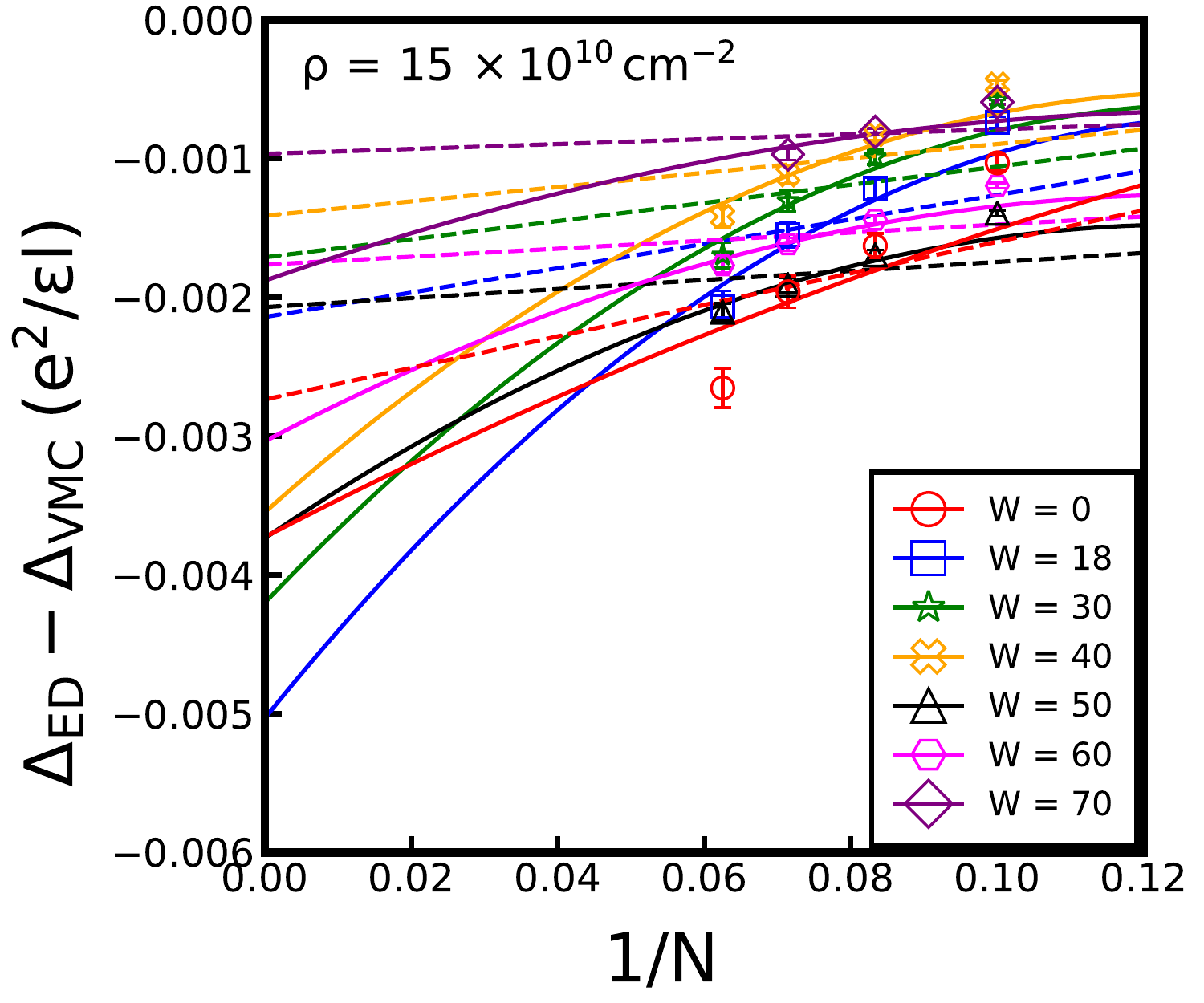}
	\includegraphics[width=0.32 \linewidth]{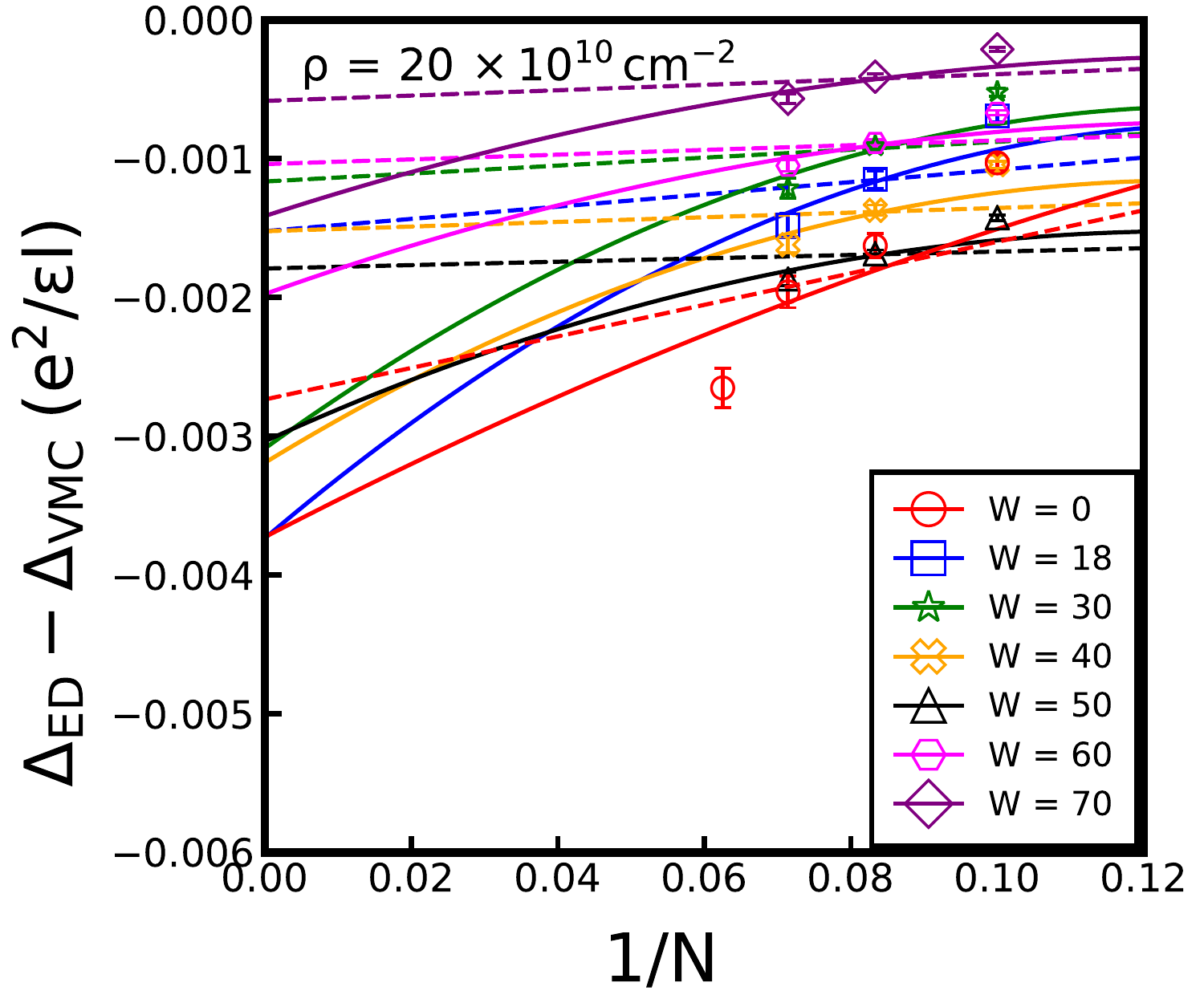}
	\includegraphics[width=0.32 \linewidth]{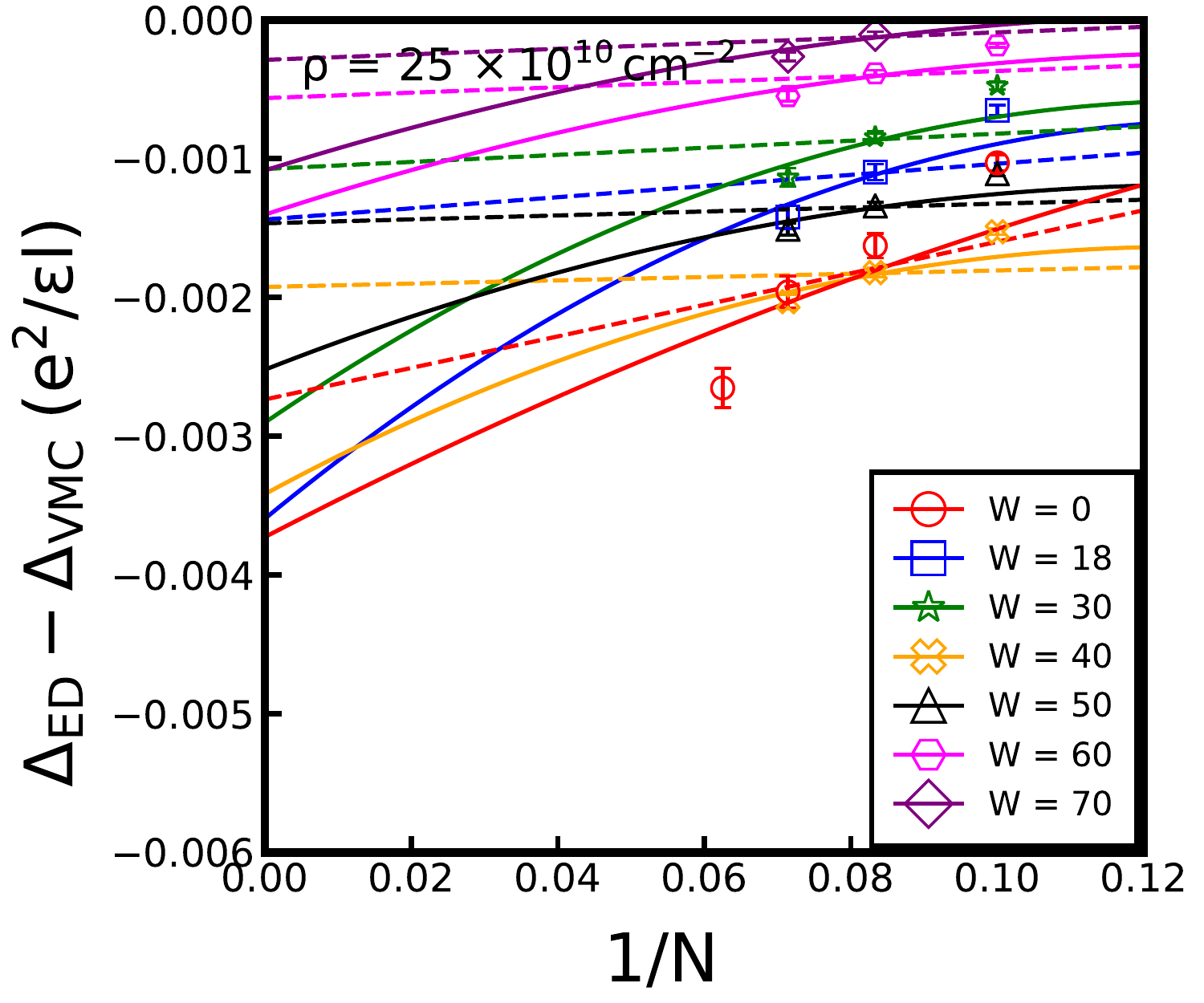}
	\includegraphics[width=0.32 \linewidth]{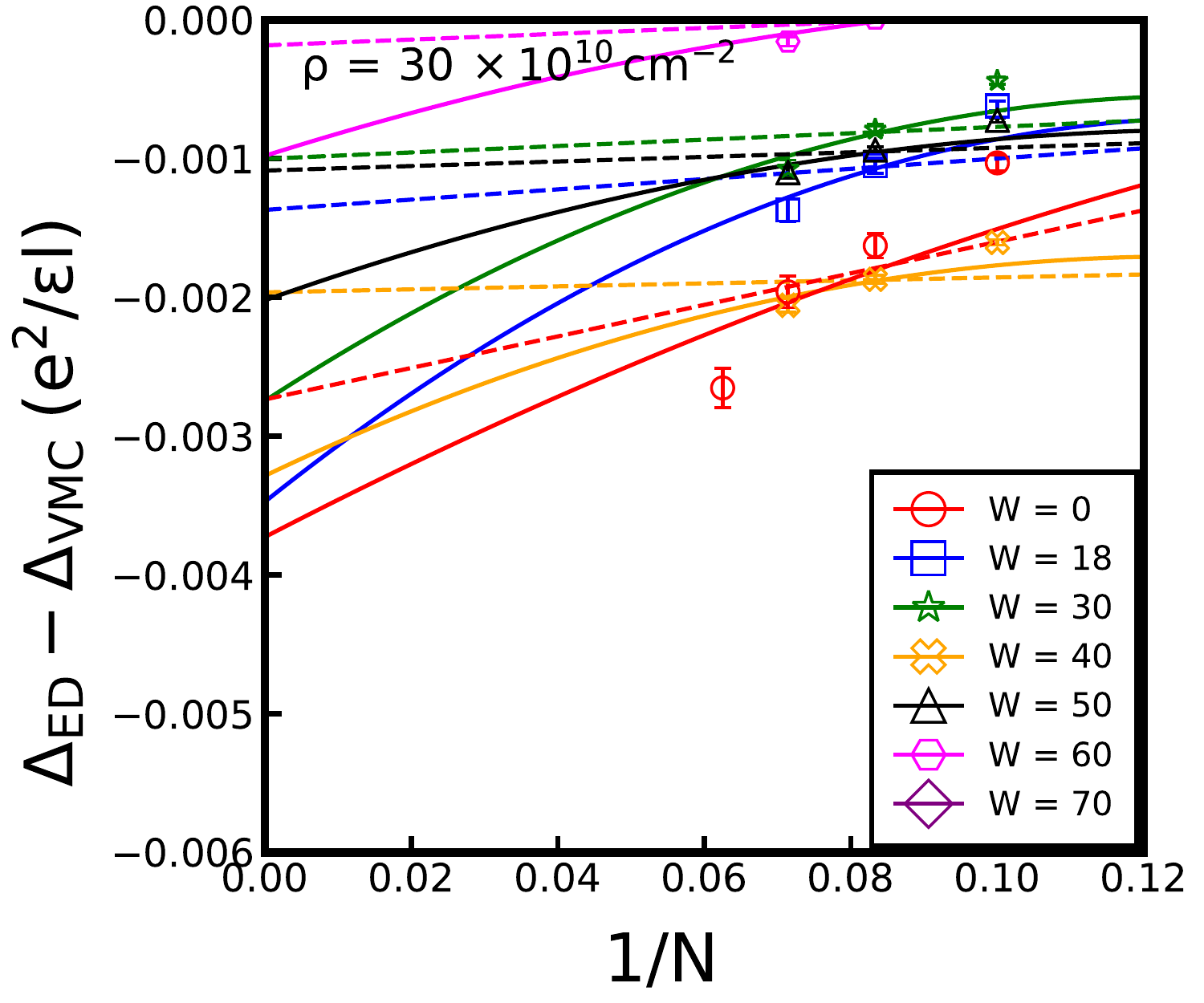}
	\caption{Thermodynamic extrapolation of the variational error, i.e., correction to the gap due to the difference between the variational wave function and the eigenstate obtained from exact diagonalization (ED) at $\nu=2/5$ for several widths and densities (labeled in plots). To calculate the ED energies, we obtain the two-particle pseudopotentials of the effective interaction (Eq. (4) of the main text).  The dashed lines are obtained from linear regressions while the solid lines are from quadratic fits in $1/N$ where $N$ is the number of particles. The ED corrections in the thermodynamic limit included in Fig. 1 of the main text are the mean values of the linearly and quadratically fitted values, and the difference between the linear and quadratic fits are used to estimate the uncertainties. The well-widths shown in the legends are in units of nanometers.}\label{X_fig:ED_correction_25}
\end{figure*}

\begin{figure*}[ht!]
	\includegraphics[width=0.32 \linewidth]{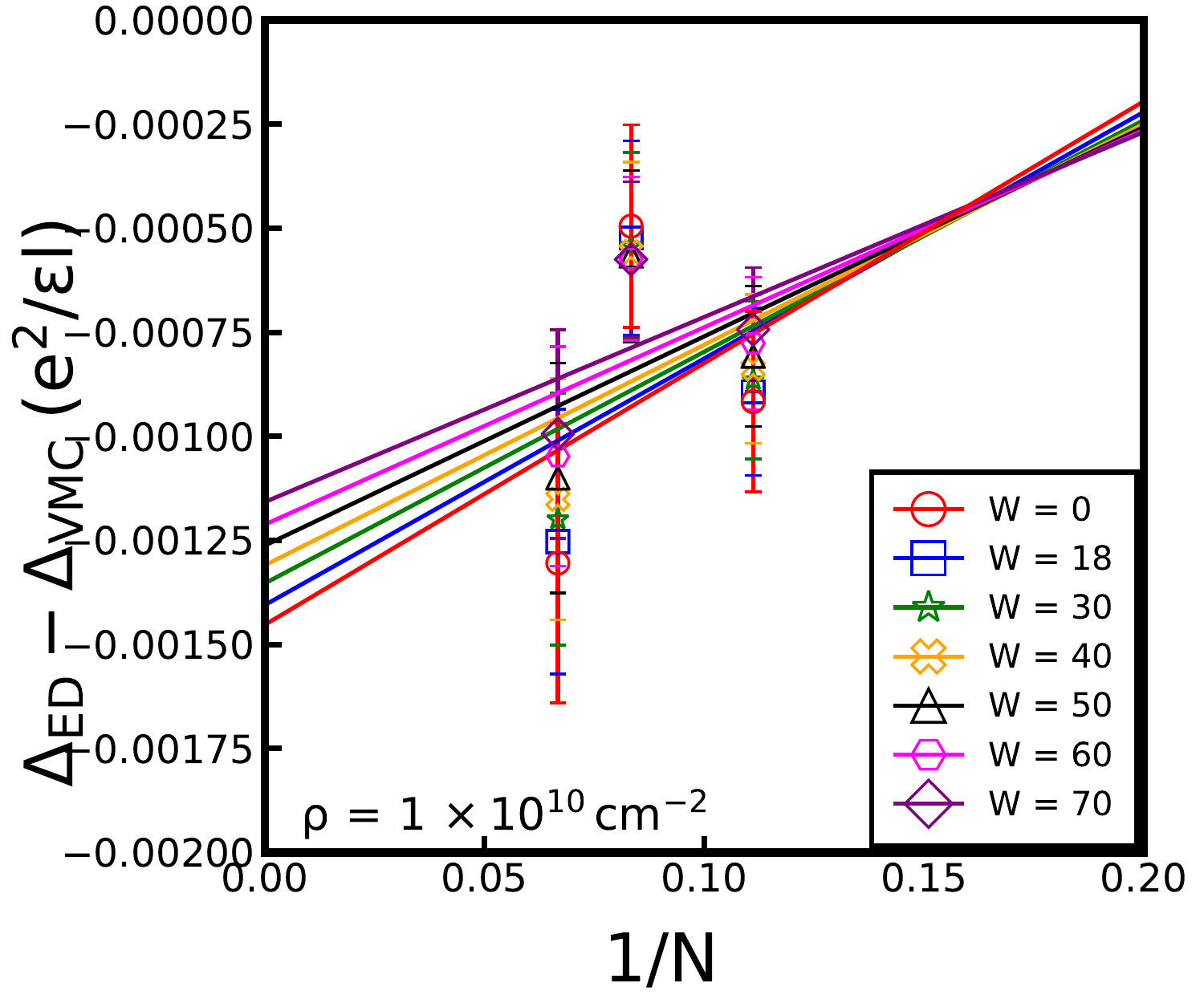}
	\includegraphics[width=0.32 \linewidth]{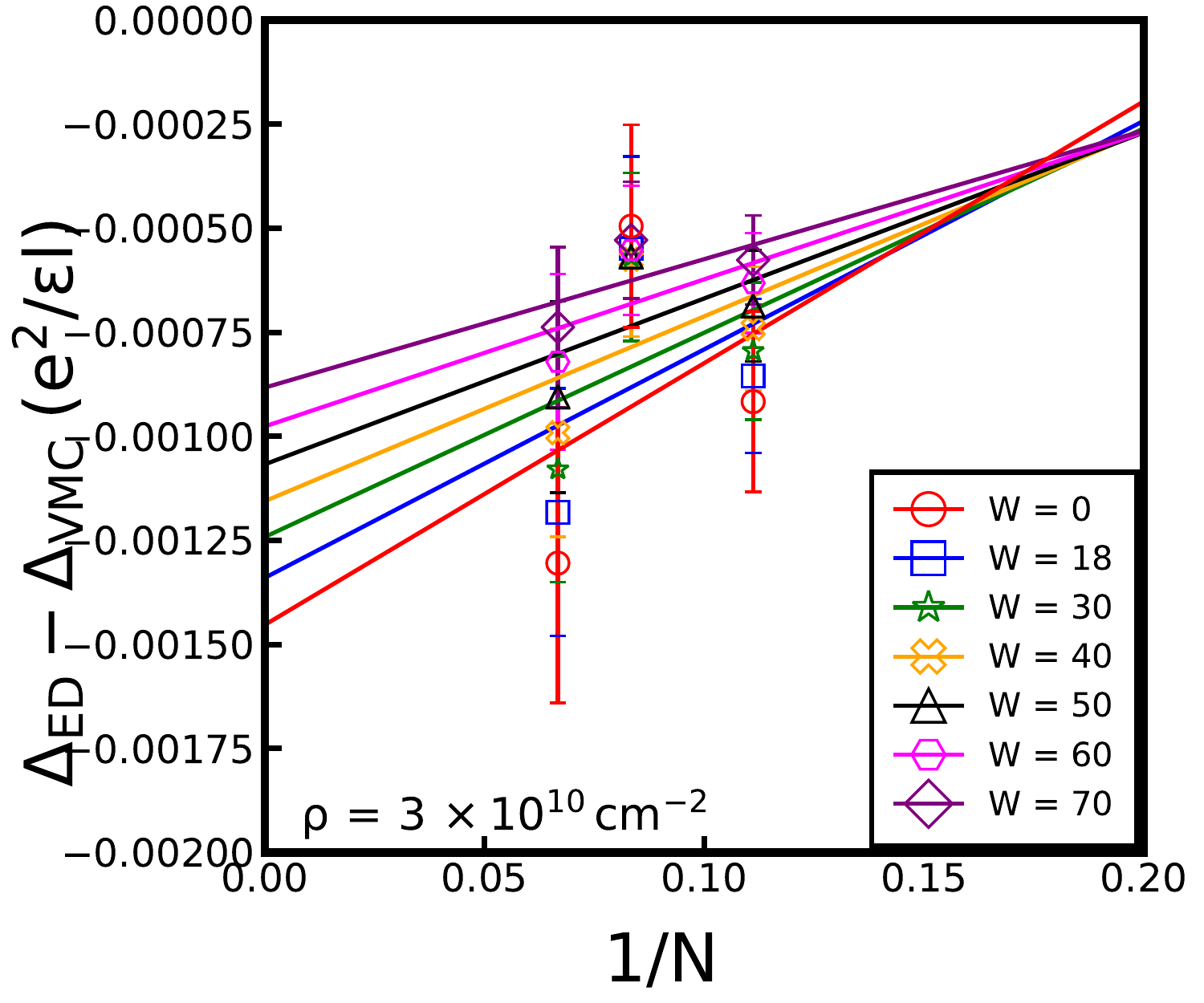}
	\includegraphics[width=0.32 \linewidth]{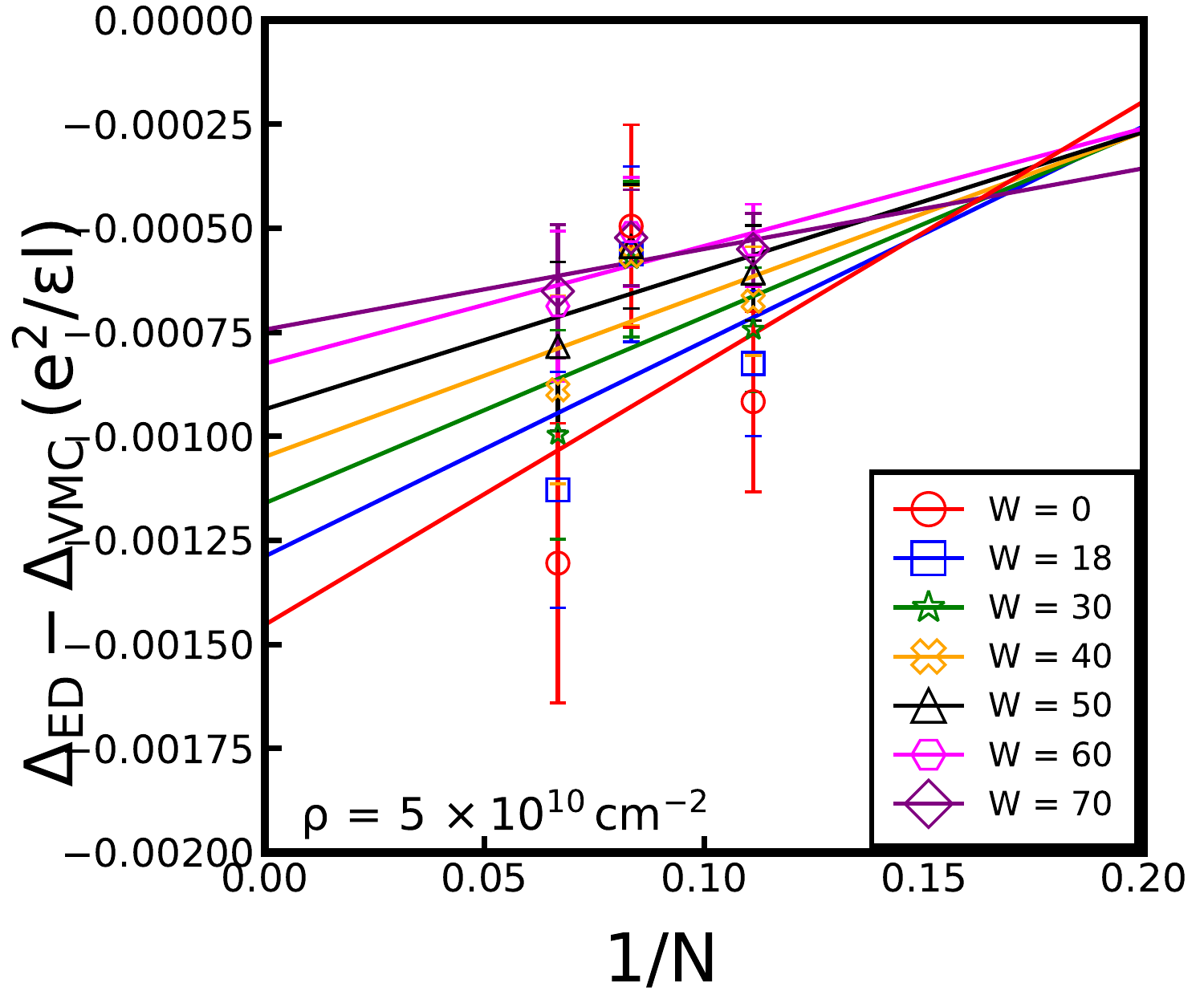}
	\includegraphics[width=0.32 \linewidth]{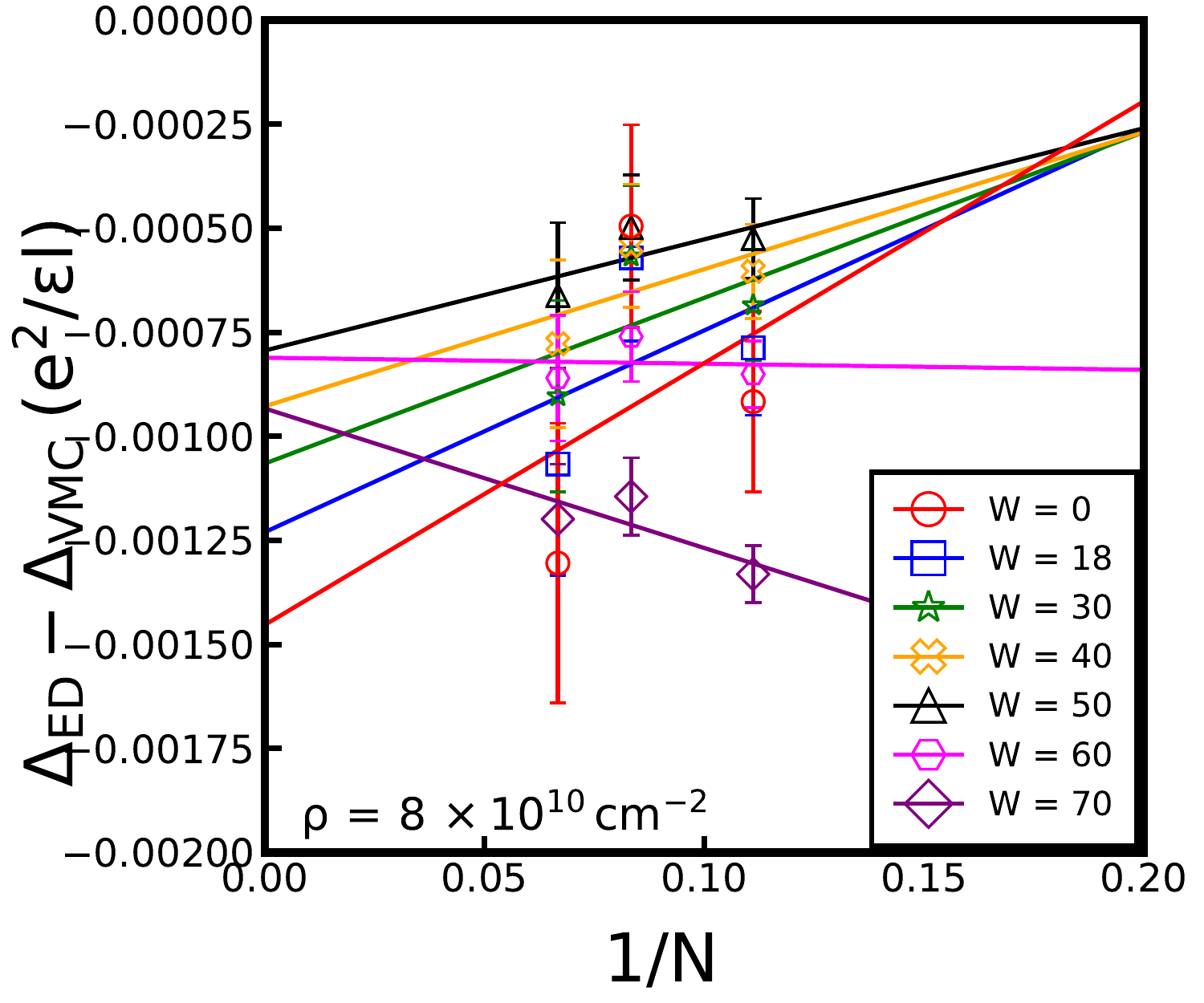}
	\includegraphics[width=0.32 \linewidth]{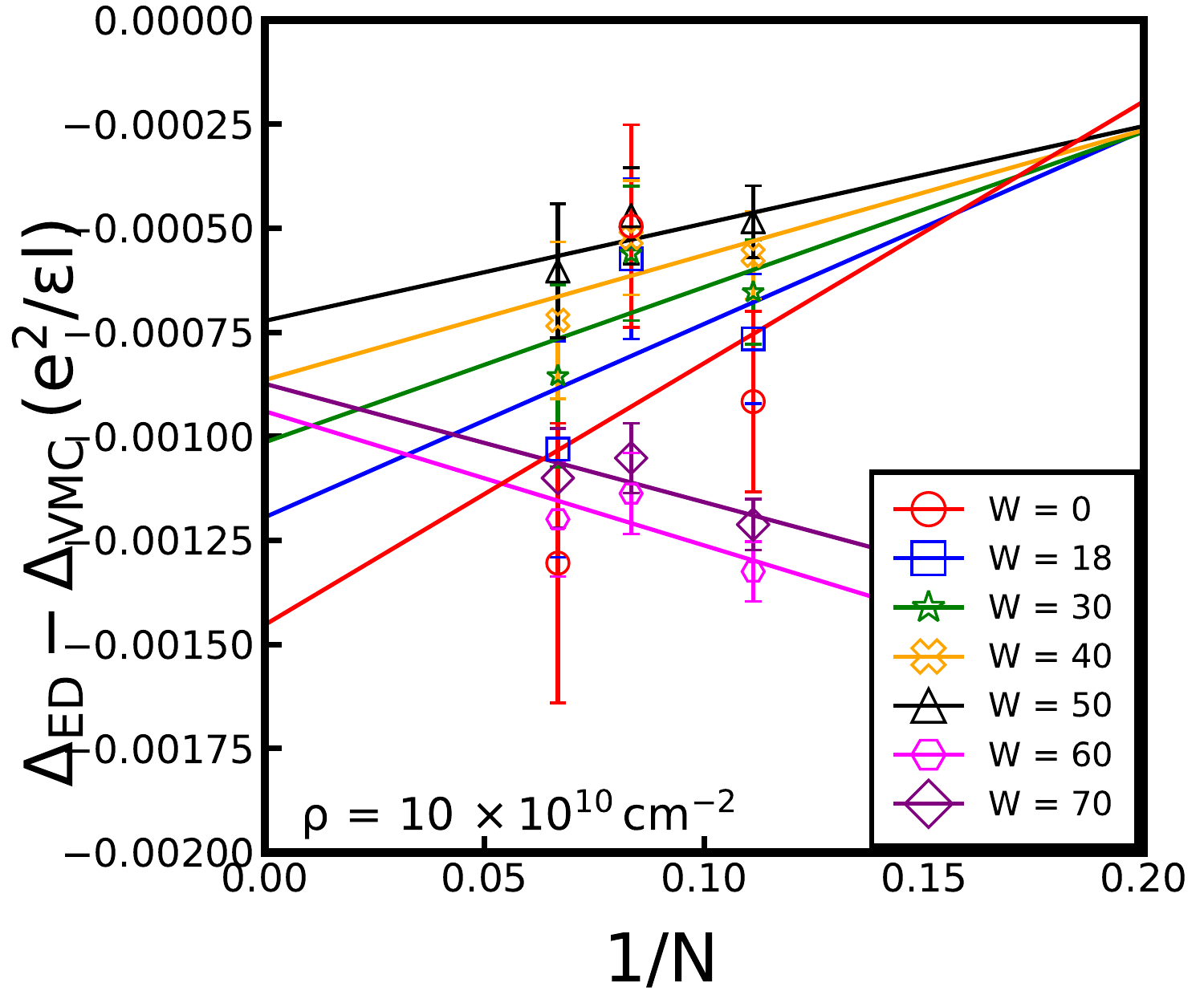}
	\includegraphics[width=0.32 \linewidth]{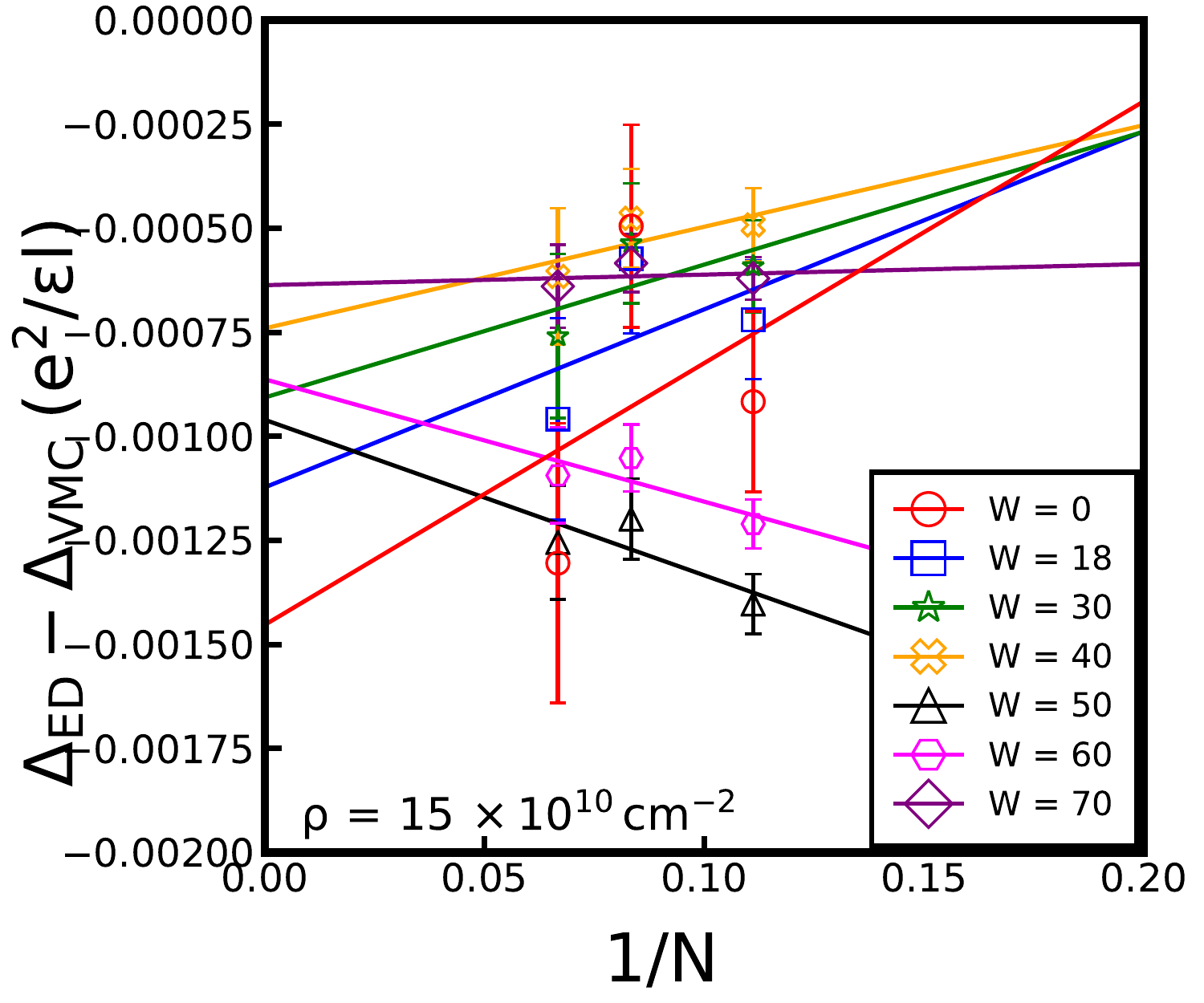}
	\includegraphics[width=0.32 \linewidth]{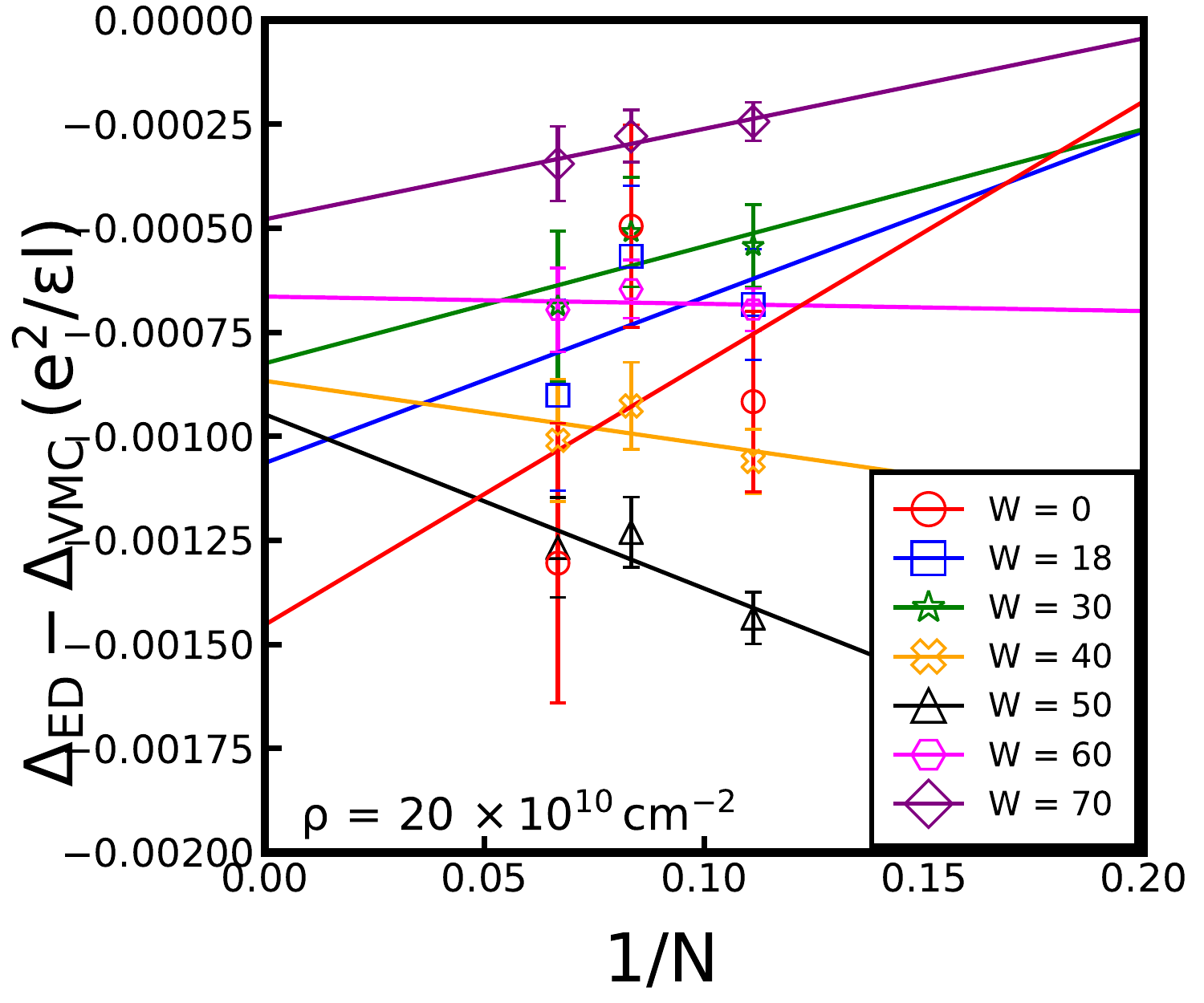}
	\includegraphics[width=0.32 \linewidth]{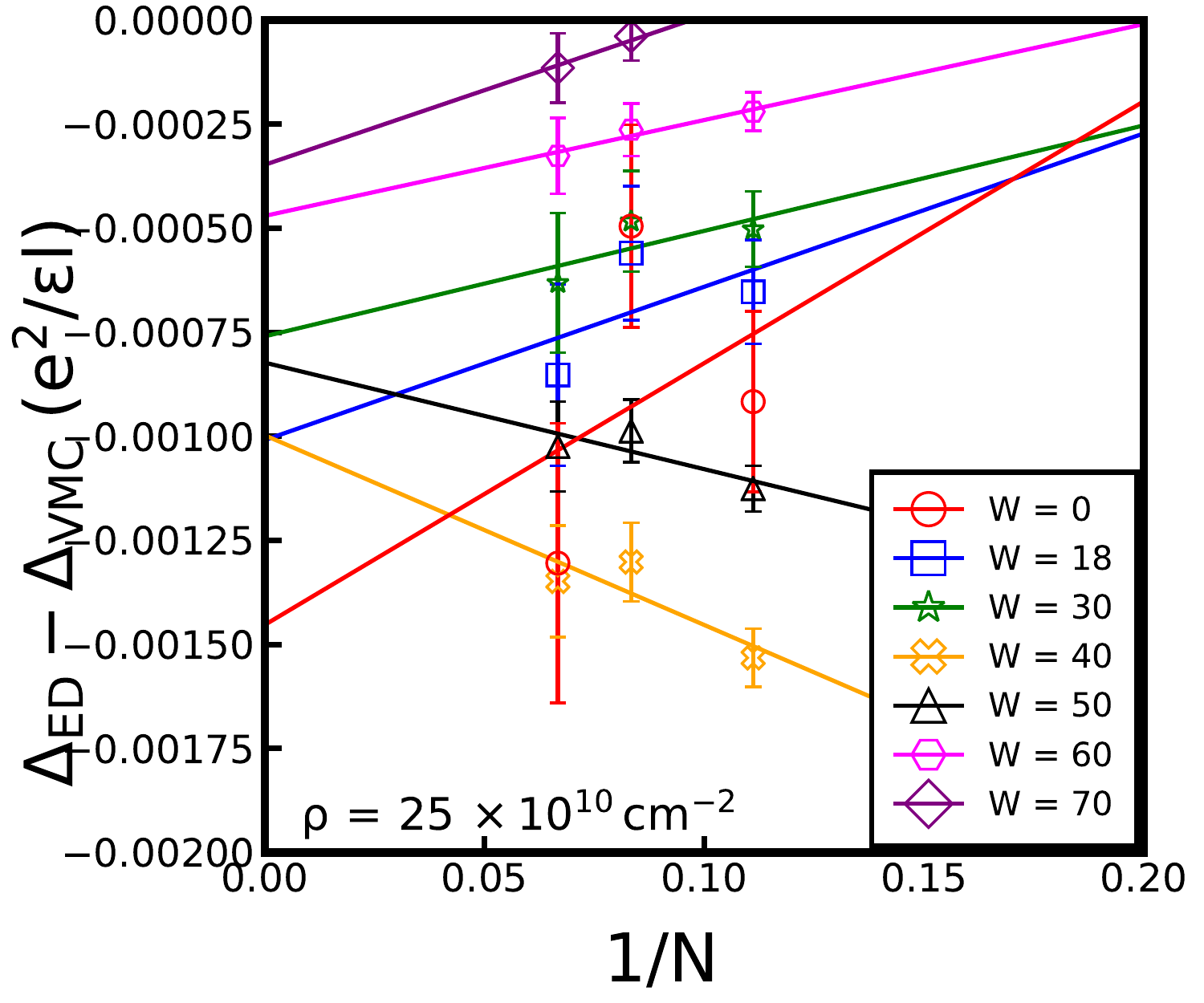}
	\includegraphics[width=0.32 \linewidth]{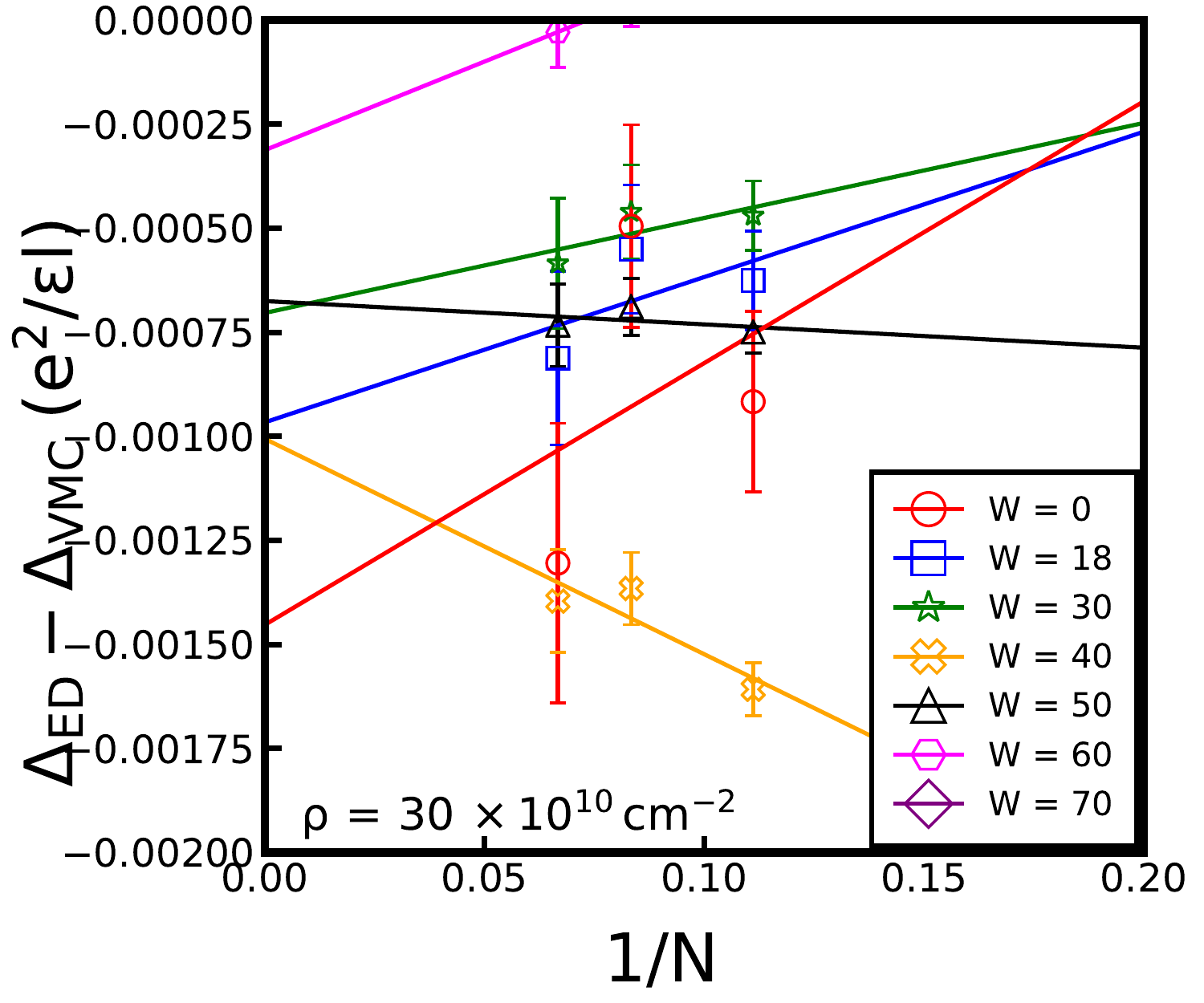}
	\caption{Thermodynamic extrapolation of the variational error, i.e., correction to the gap due to the difference between the variational wave function and the eigenstate obtained from exact diagonalization (ED) at $\nu=3/7$ for several widths and densities (labeled in plots). To calculate the ED energies, we obtain the two-particle pseudopotentials of the effective interaction (Eq.  (4) of the main text).  The dashed lines are obtained from linear regressions while the solid lines are from quadratic fits in $1/N$ where $N$ is the number of particles. The ED corrections in the thermodynamic limit included in Fig. 1 of the main text are the mean values of the linearly and quadratically fitted values, and the difference between the linear and quadratic fits are used to estimate the uncertainties. The well-widths shown in the legends are in units of nanometers.}\label{X_fig:ED_correction_37}
\end{figure*}

We give estimates of variational errors for the gaps of the $1/3, 2/5$ and $3/7$ FQHE states in Figs.~\ref{X_fig:ED_correction_13},\ref{X_fig:ED_correction_25},\ref{X_fig:ED_correction_37}. Correcting our VMC gaps for the variational error produces what we label as ``exact" gaps. These are closer to the experimental values than the VMC gaps.  The ``exact"  gaps are shown in Fig.~1 of the main text for the FQHE at $1/3$ and $2/5$, and in Fig.~\ref{X_fig:VMC_ED_37_n11} for the FQHE at $3/7$.

\subsection{Thermodynamic extrapolation of the transport gap using exact diagonalization (ED)}
\label{SM_subsec:ED}

\begin{figure*}[ht!]
	\includegraphics[width=0.32\linewidth]{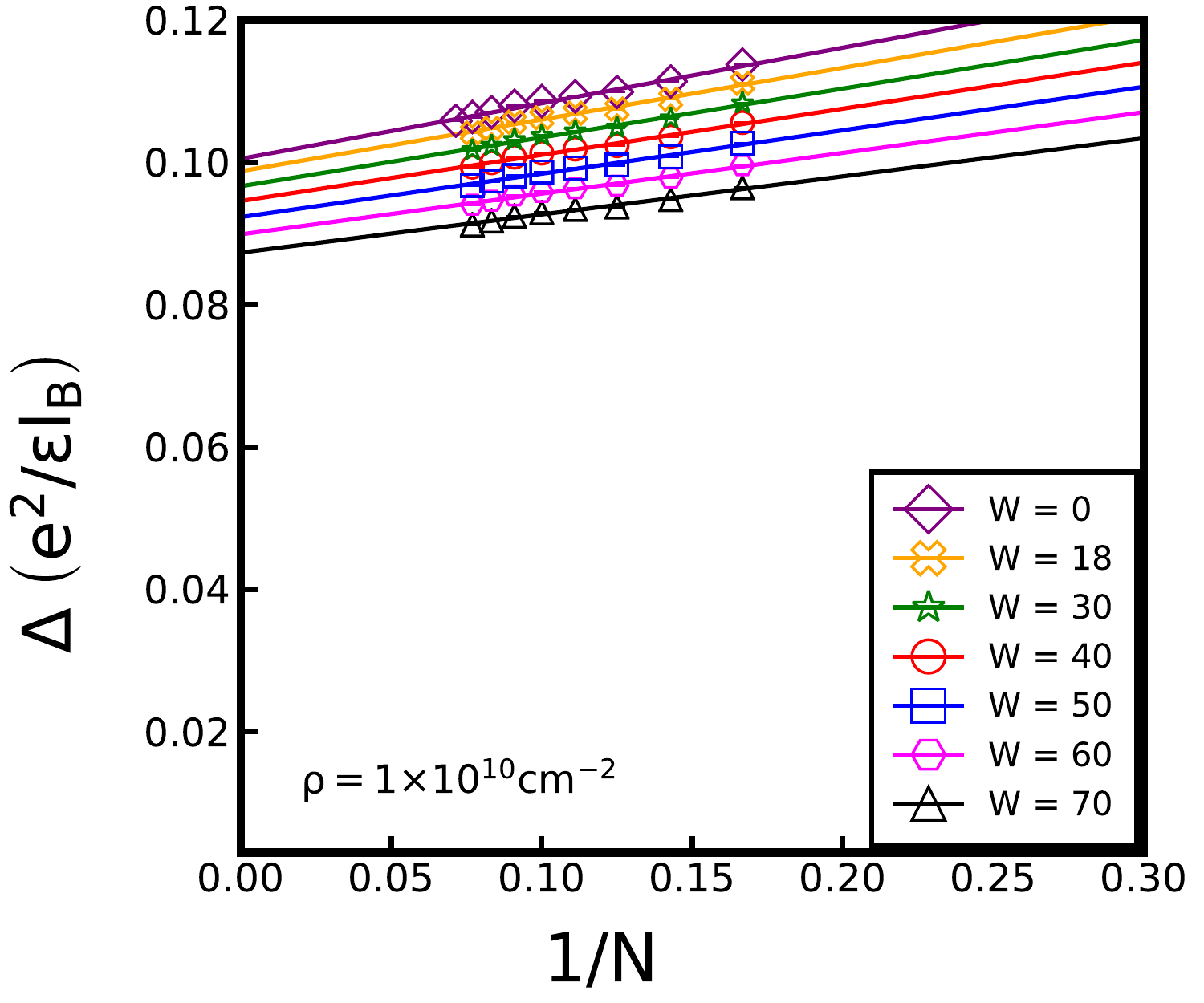}
	\includegraphics[width=0.32\linewidth]{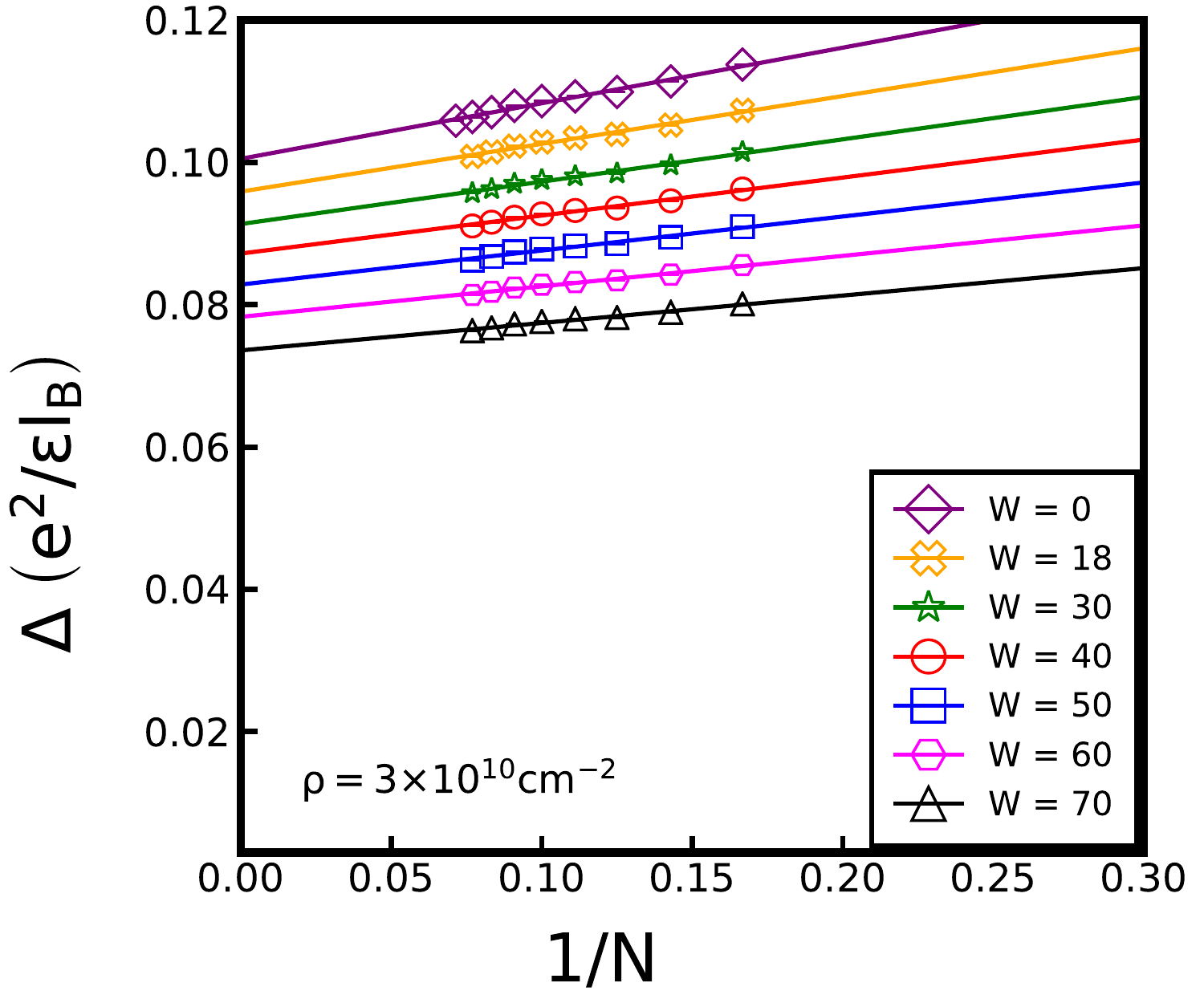}
	\includegraphics[width=0.32\linewidth]{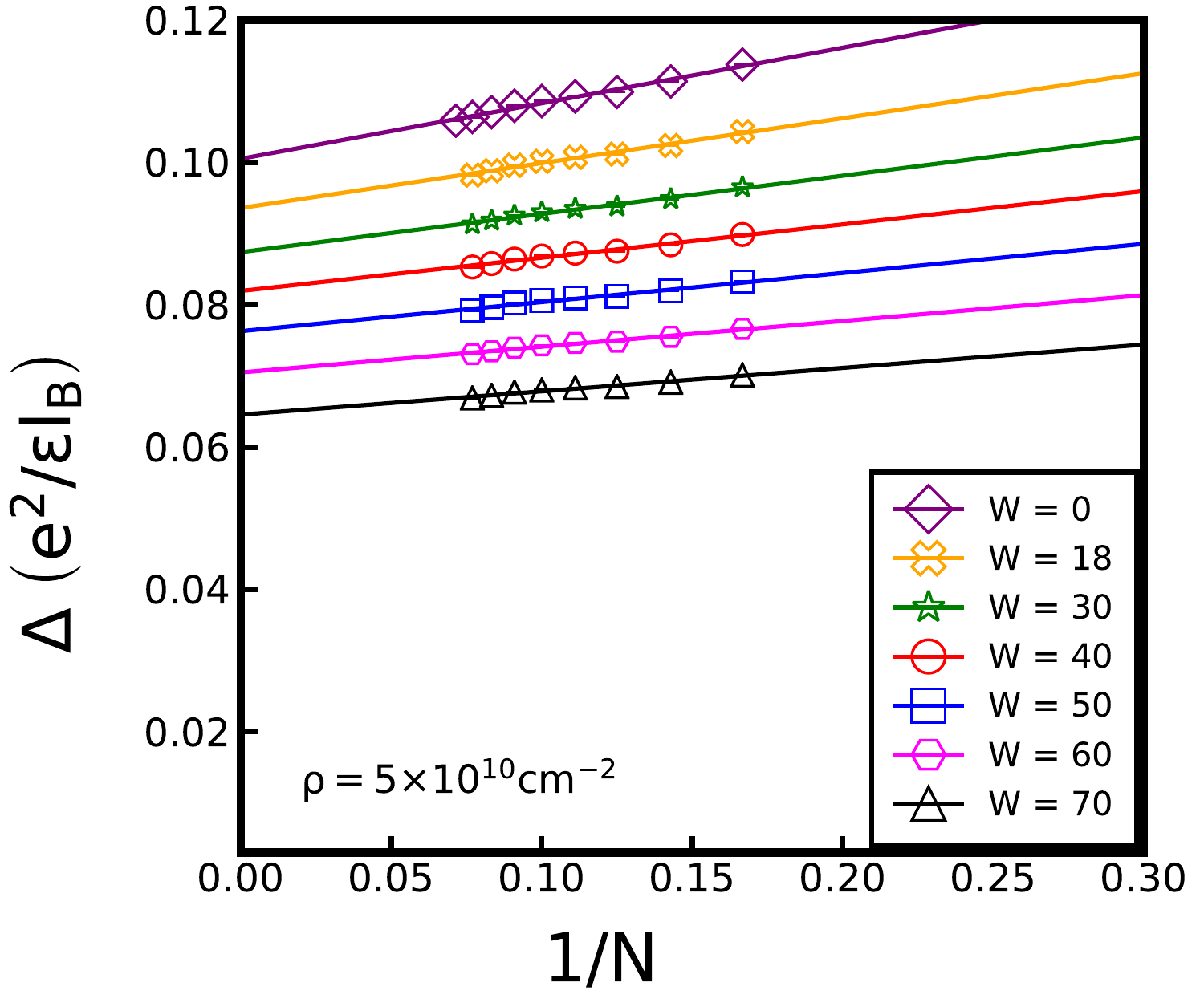}
	\includegraphics[width=0.32\linewidth]{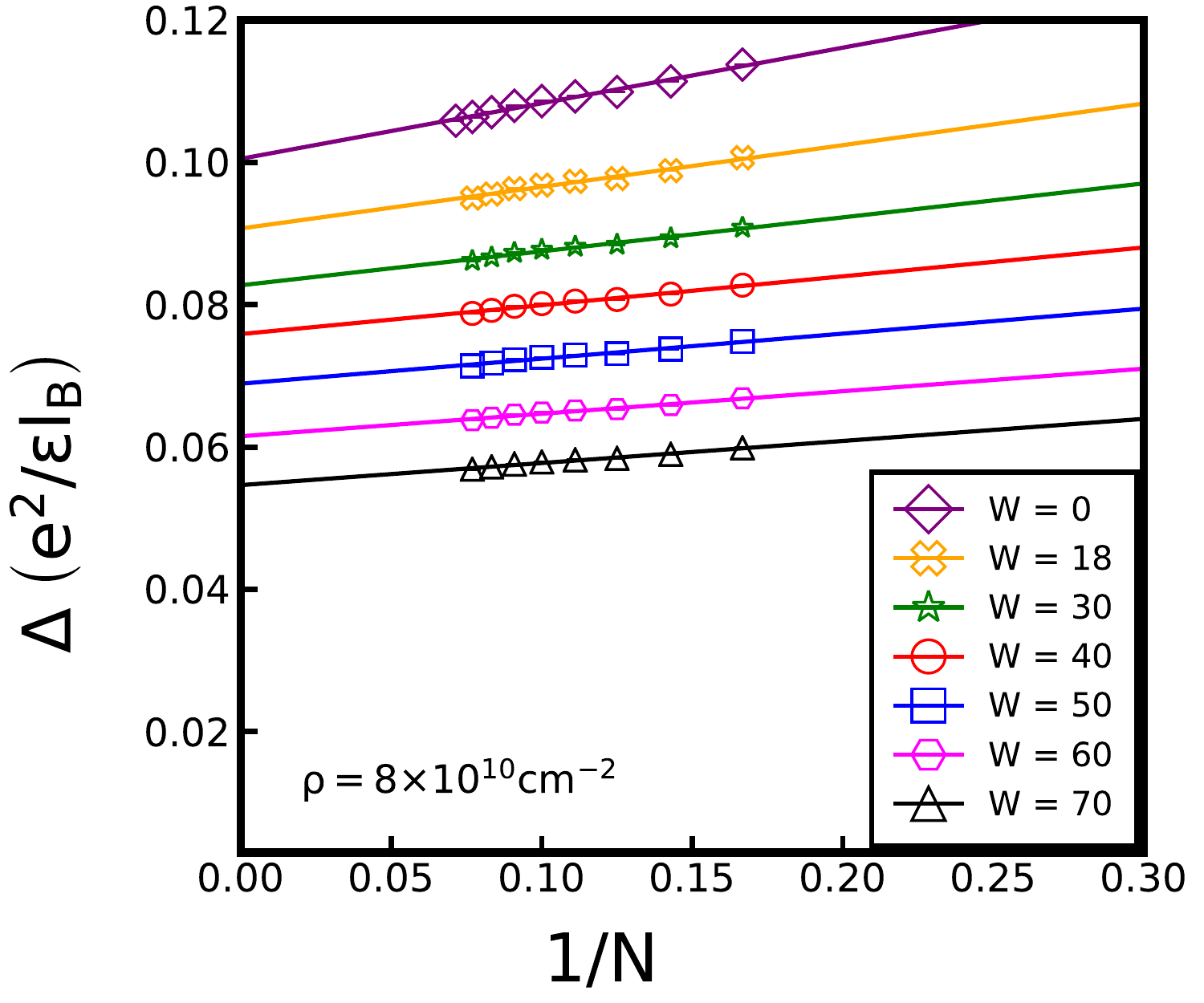}
	\includegraphics[width=0.32\linewidth]{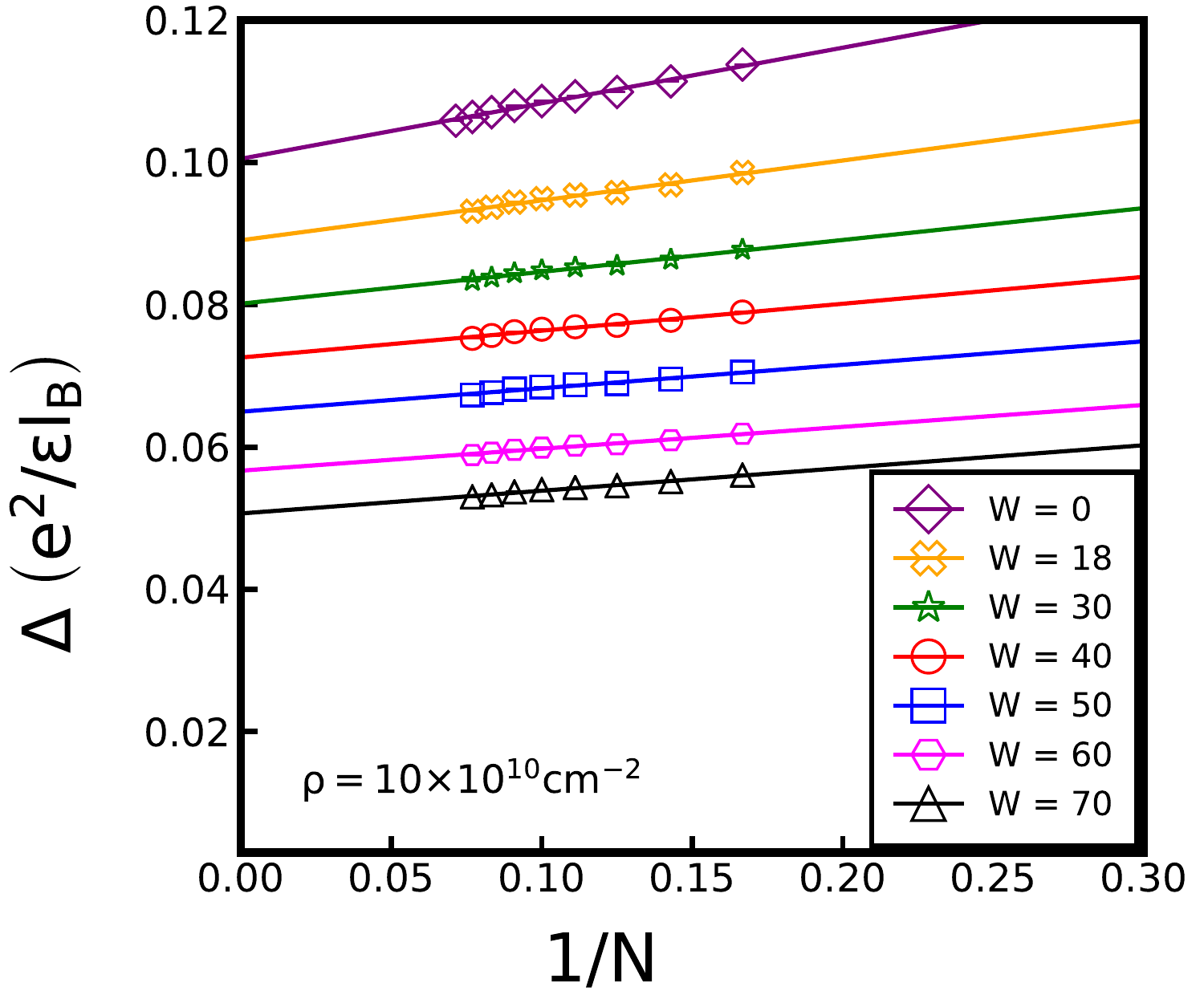}
	\includegraphics[width=0.32\linewidth]{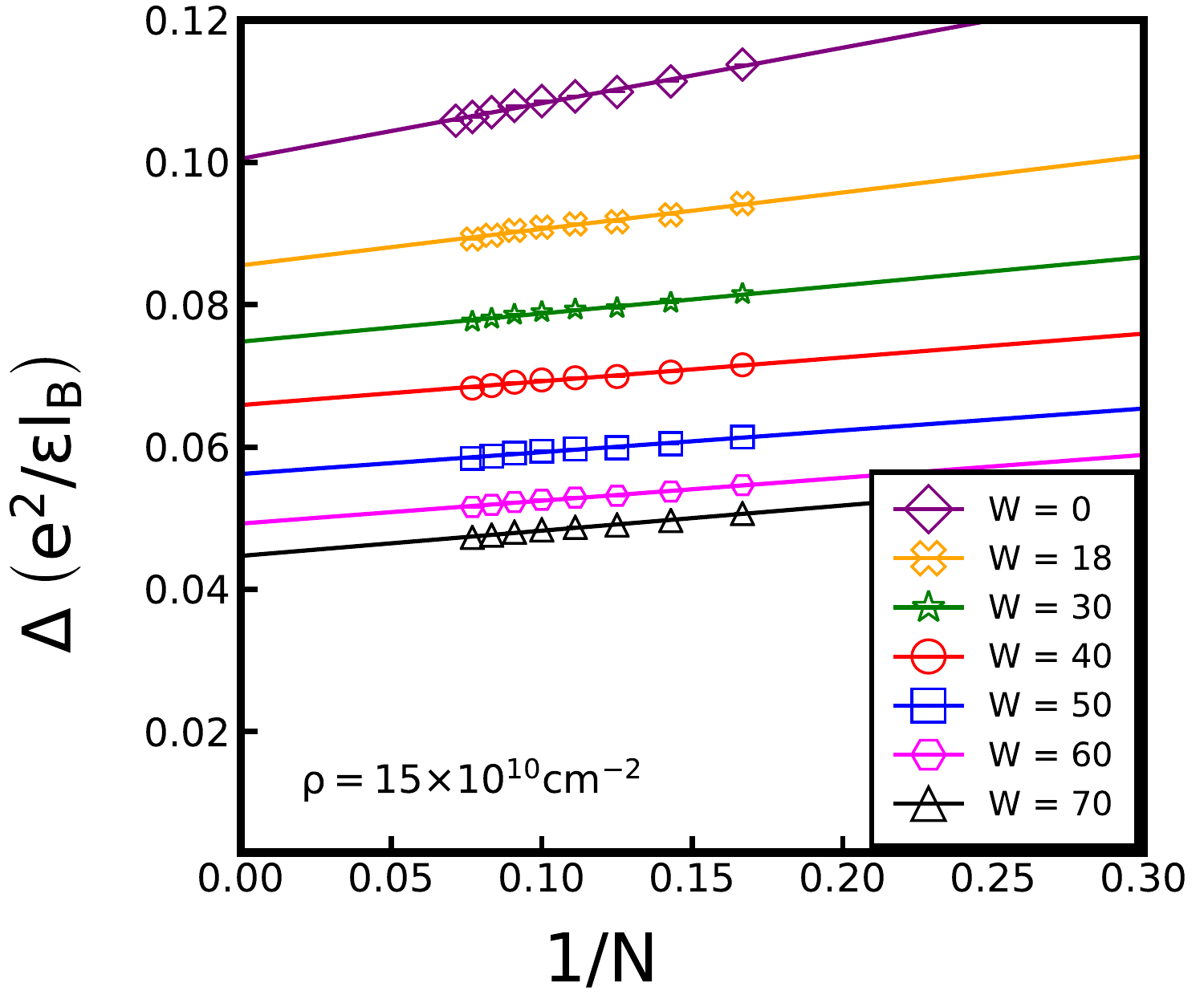}
	\includegraphics[width=0.32\linewidth]{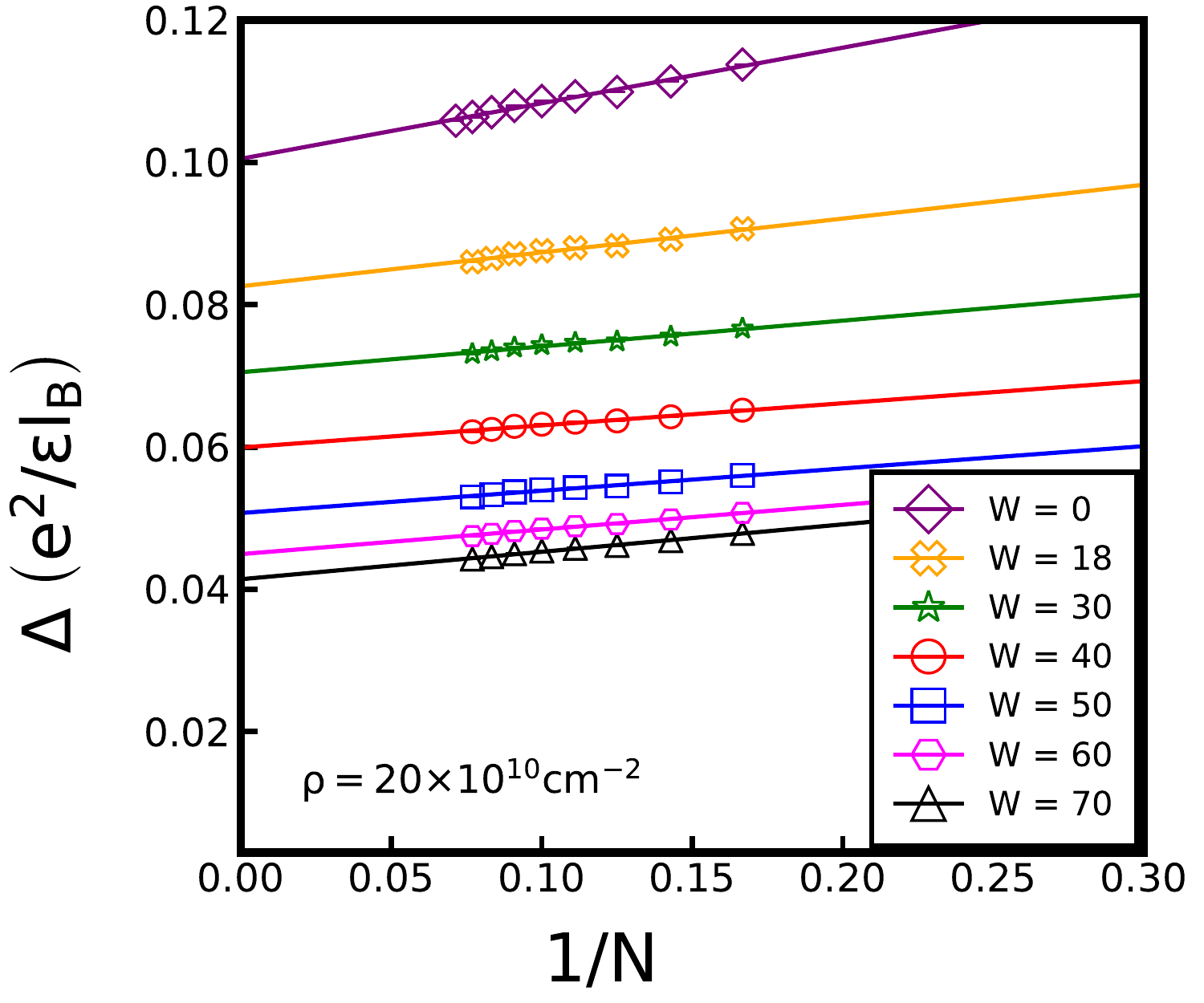}
	\includegraphics[width=0.32\linewidth]{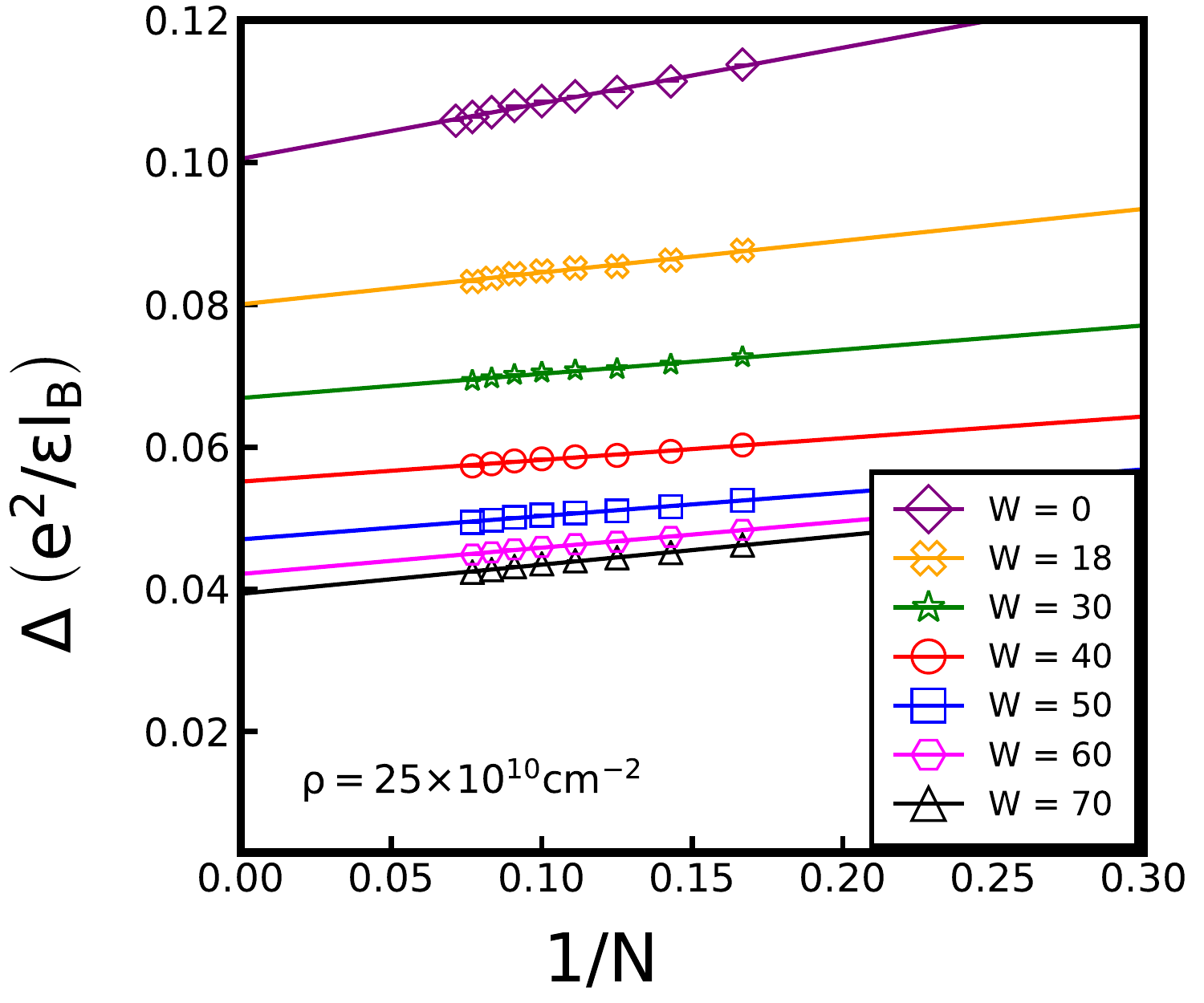}
	\includegraphics[width=0.32\linewidth]{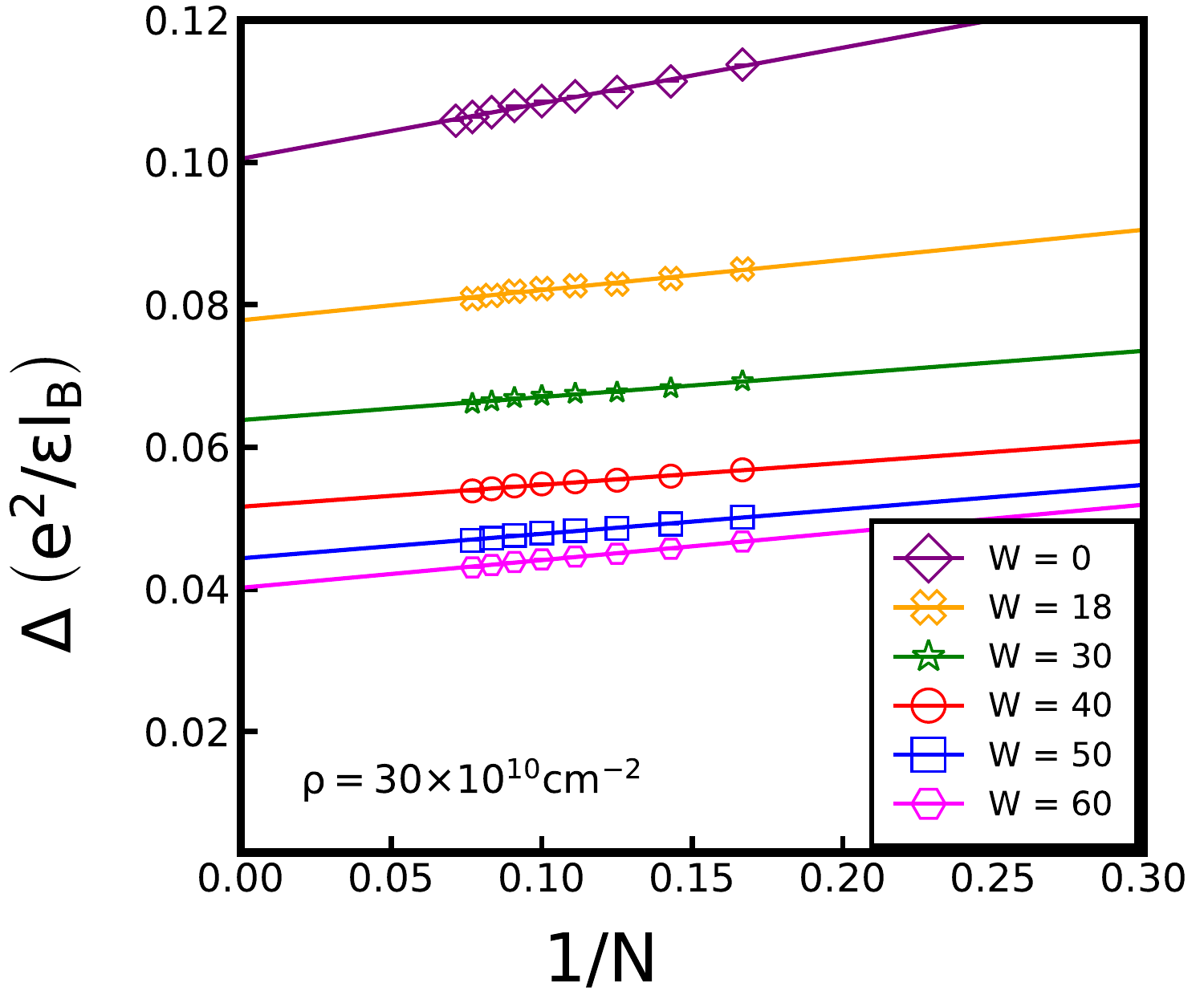}
	\caption{Thermodynamic extrapolation of the transport gap for $\nu=1/3$, calculated by the exact diagonalization (ED). Each plot presents the gaps at a specific density (labeled on the plot) with different markers labeling different well-widths in units of nanometers.}
	\label{X_fig_ED_extrap_13}
\end{figure*}

\begin{figure*}[ht!]
	\includegraphics[width=0.32\linewidth]{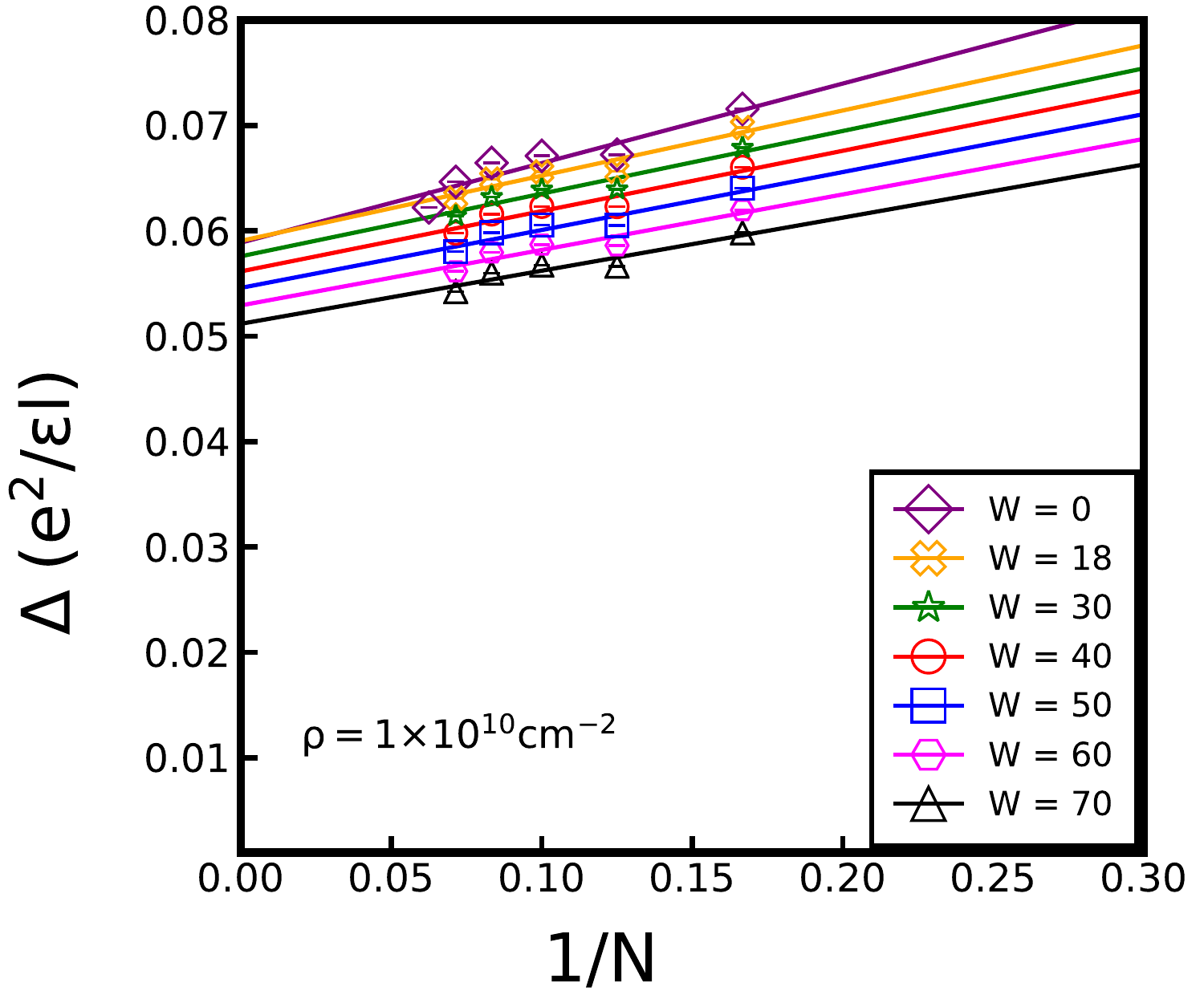}
	\includegraphics[width=0.32\linewidth]{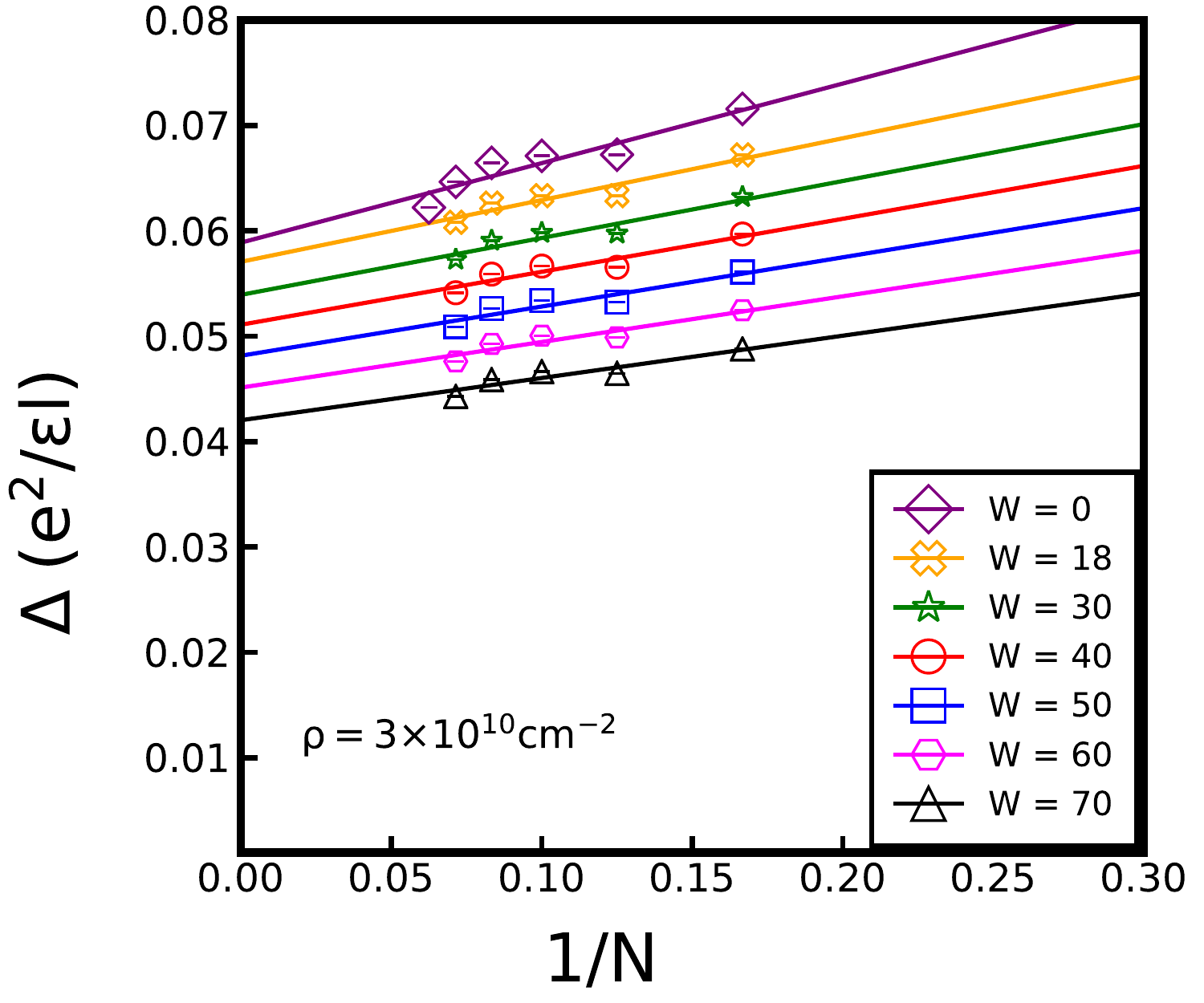}
	\includegraphics[width=0.32\linewidth]{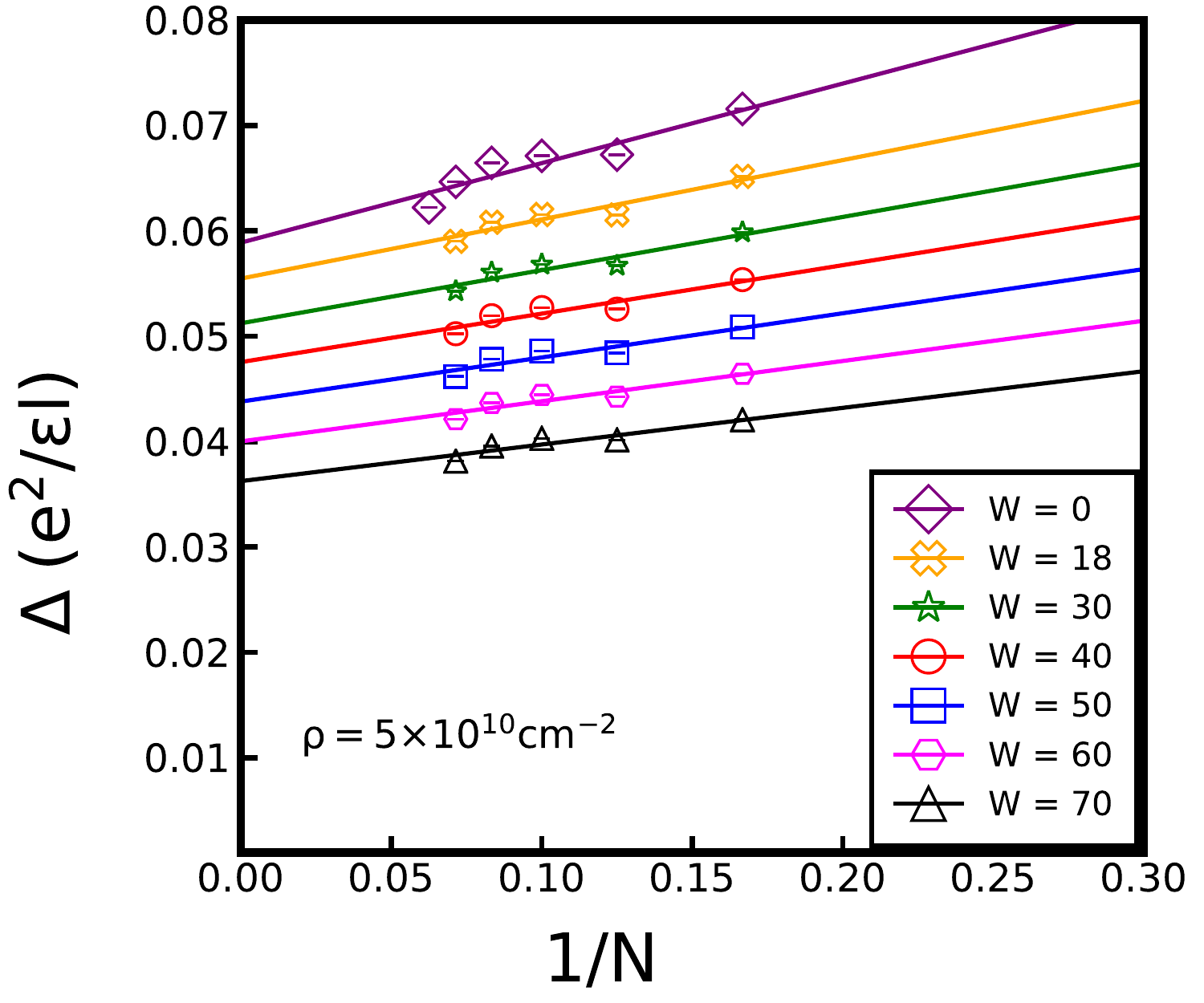}
	\includegraphics[width=0.32\linewidth]{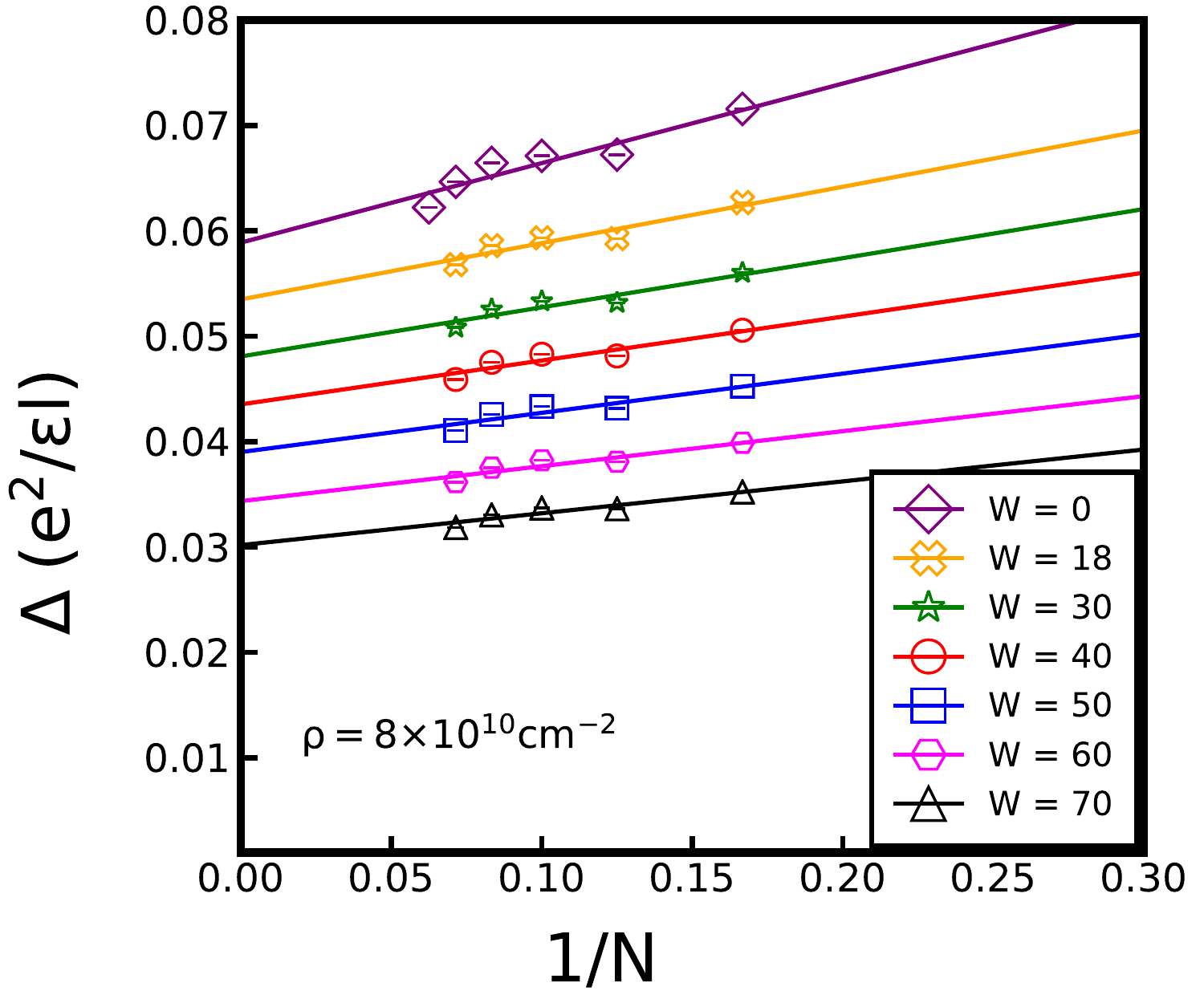}
	\includegraphics[width=0.32\linewidth]{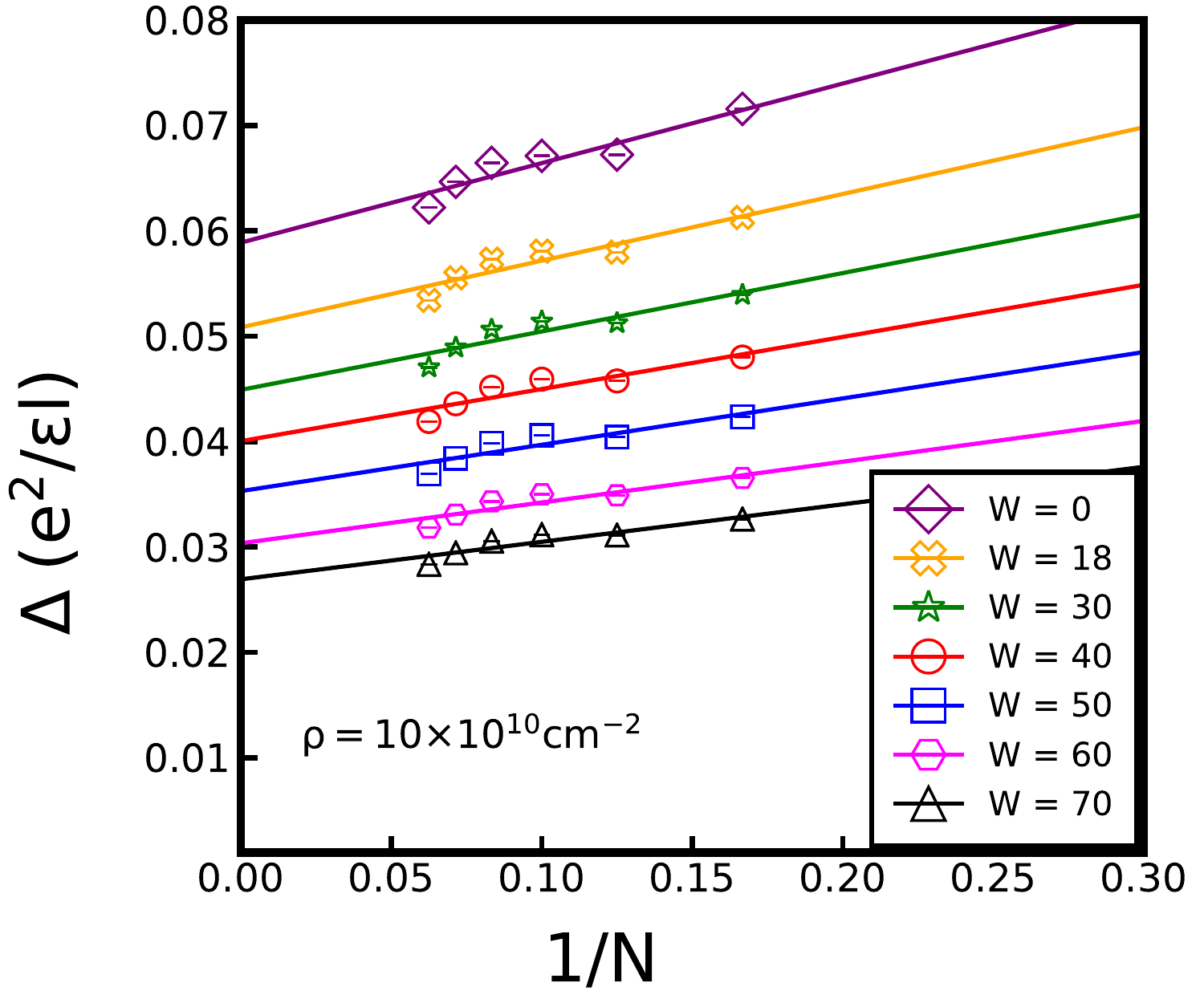}
	\includegraphics[width=0.32\linewidth]{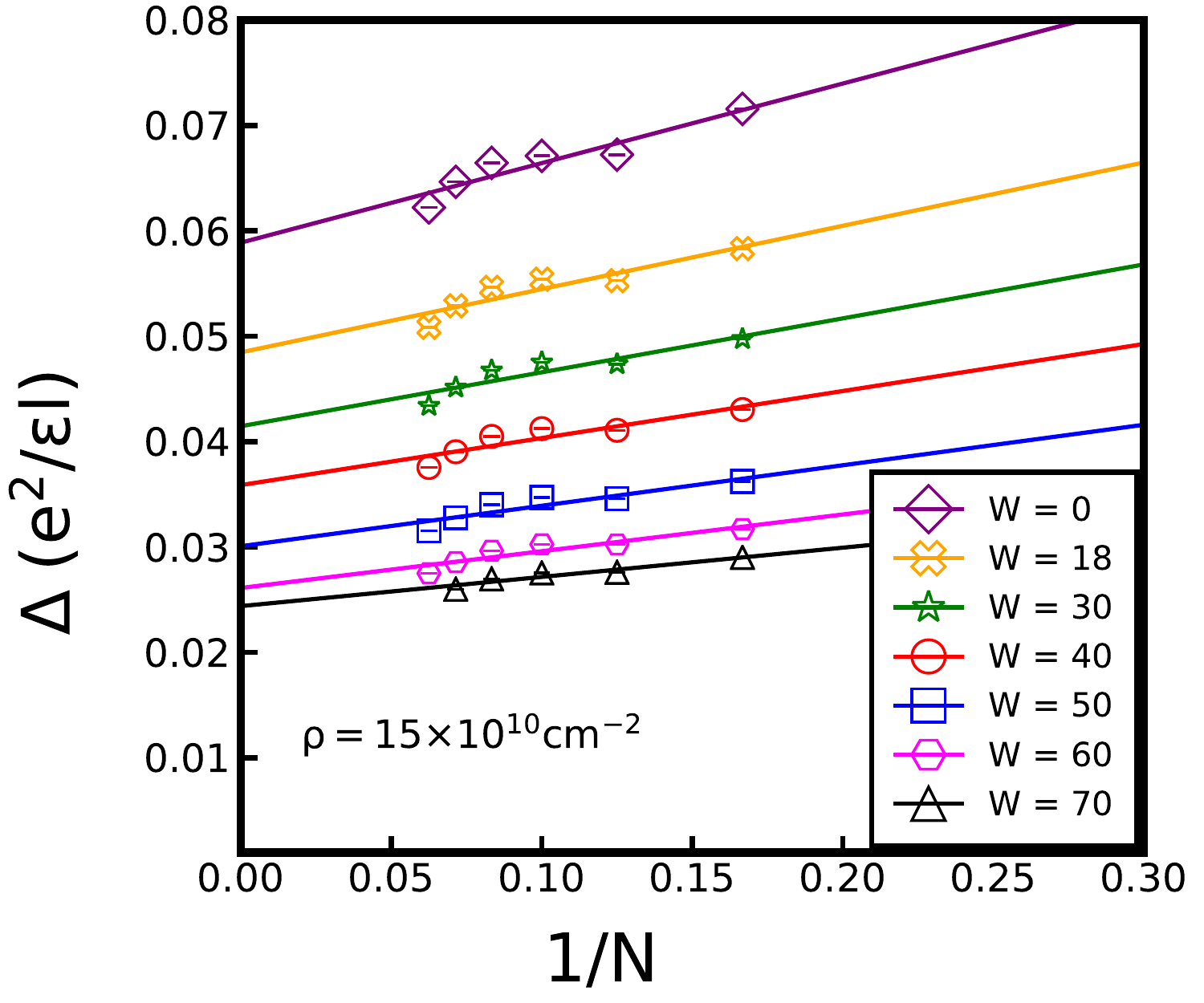}
	\includegraphics[width=0.32\linewidth]{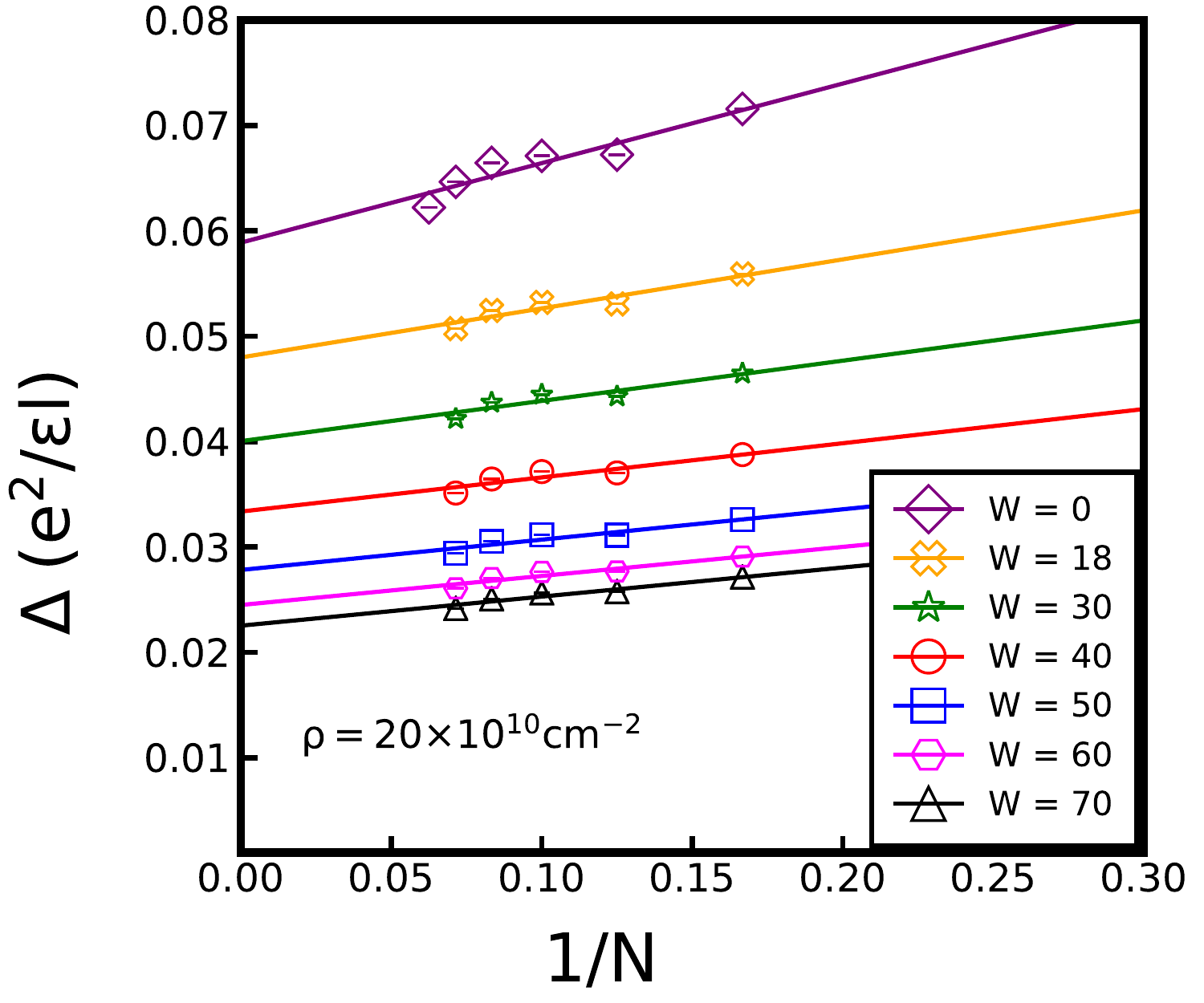}
	\includegraphics[width=0.32\linewidth]{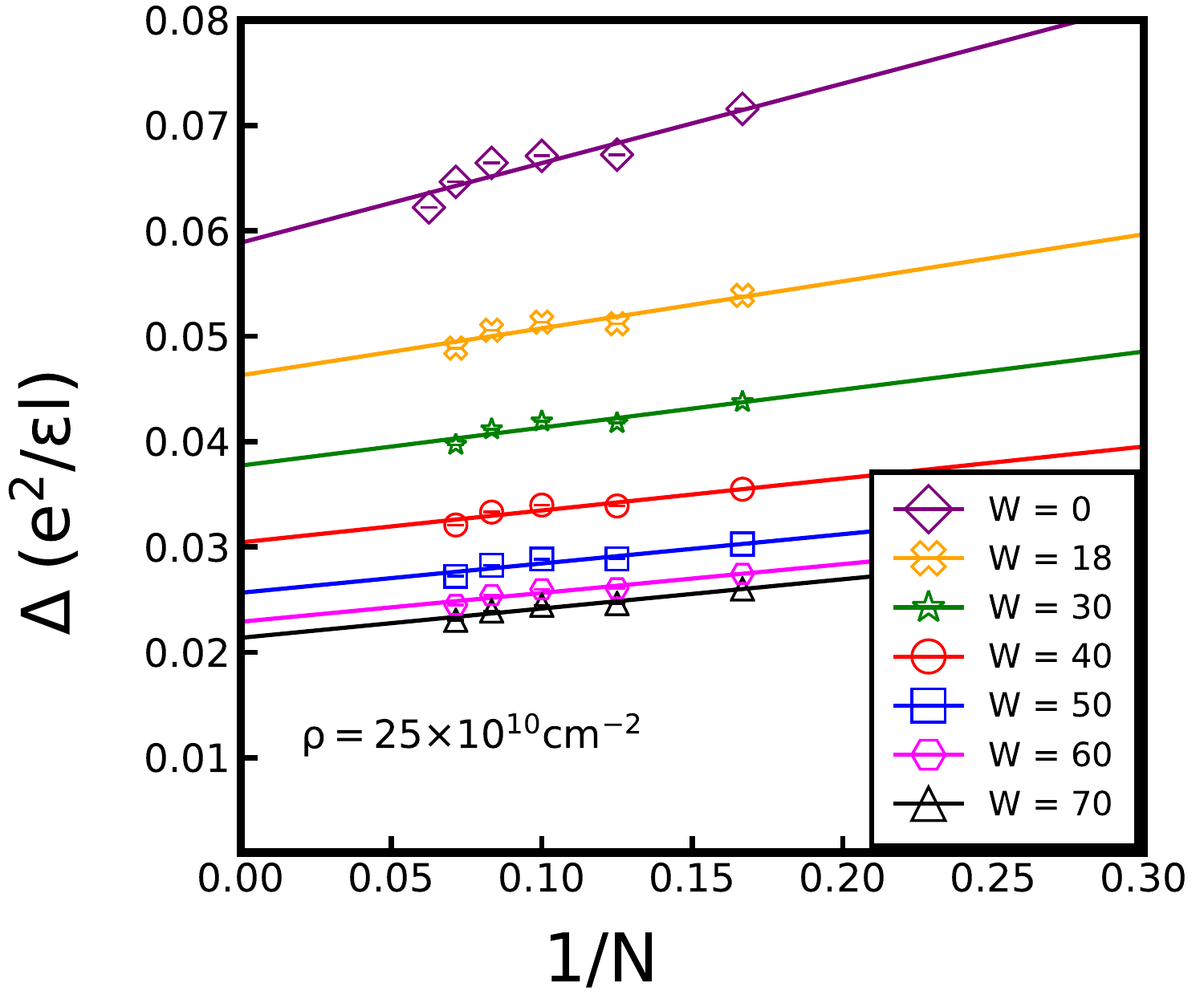}
	\includegraphics[width=0.32\linewidth]{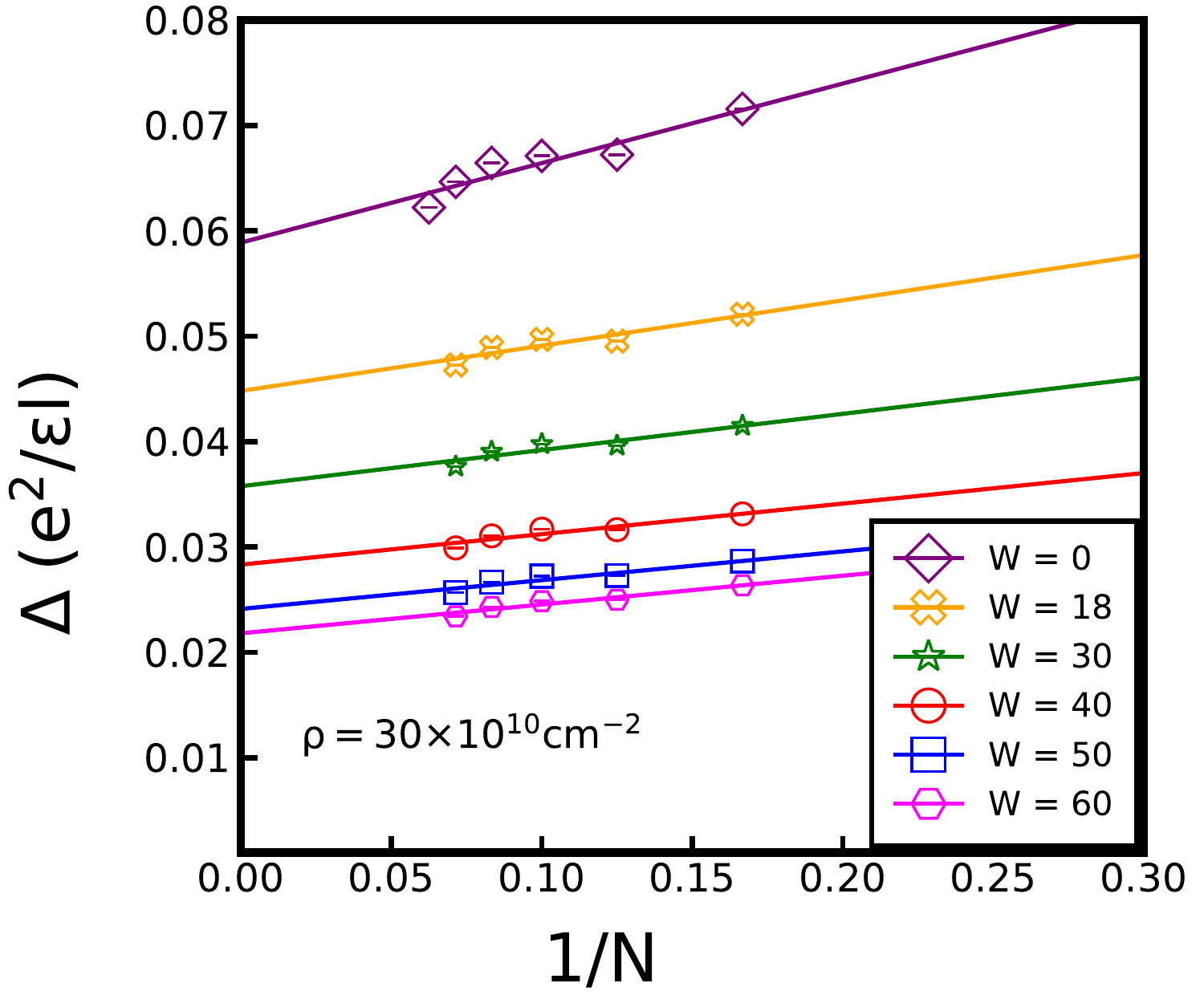}
	\caption{Thermodynamic extrapolation of the transport gap for $\nu=2/5$, calculated by the exact diagonalization (ED). Each plot presents the gaps at a specific density (labeled on the plot) with different markers labeling different well-widths in units of nanometers.}
	\label{X_fig_ED_extrap_25}
\end{figure*}

\begin{figure*}[ht!]
	\includegraphics[width=0.32\linewidth]{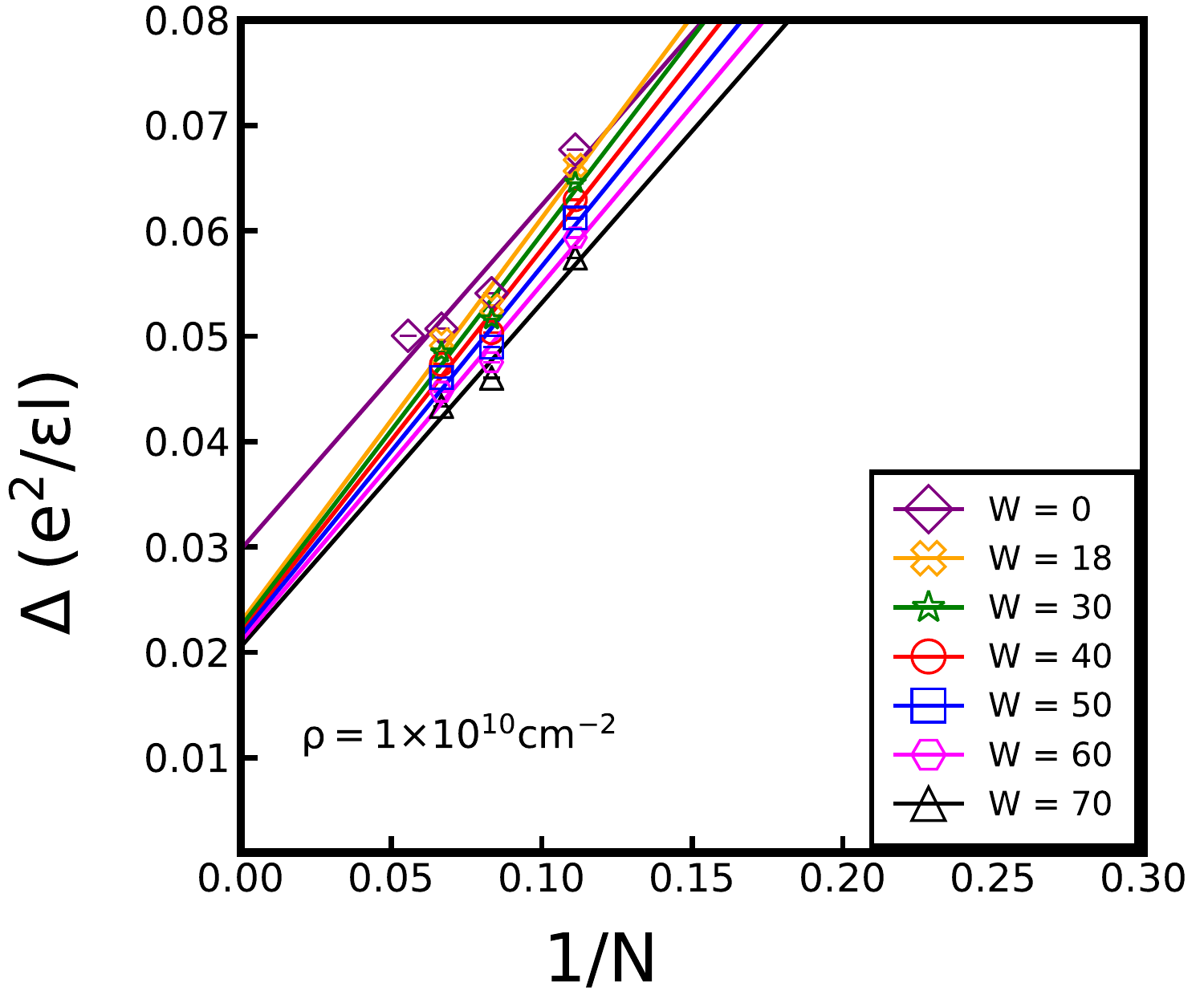}
	\includegraphics[width=0.32\linewidth]{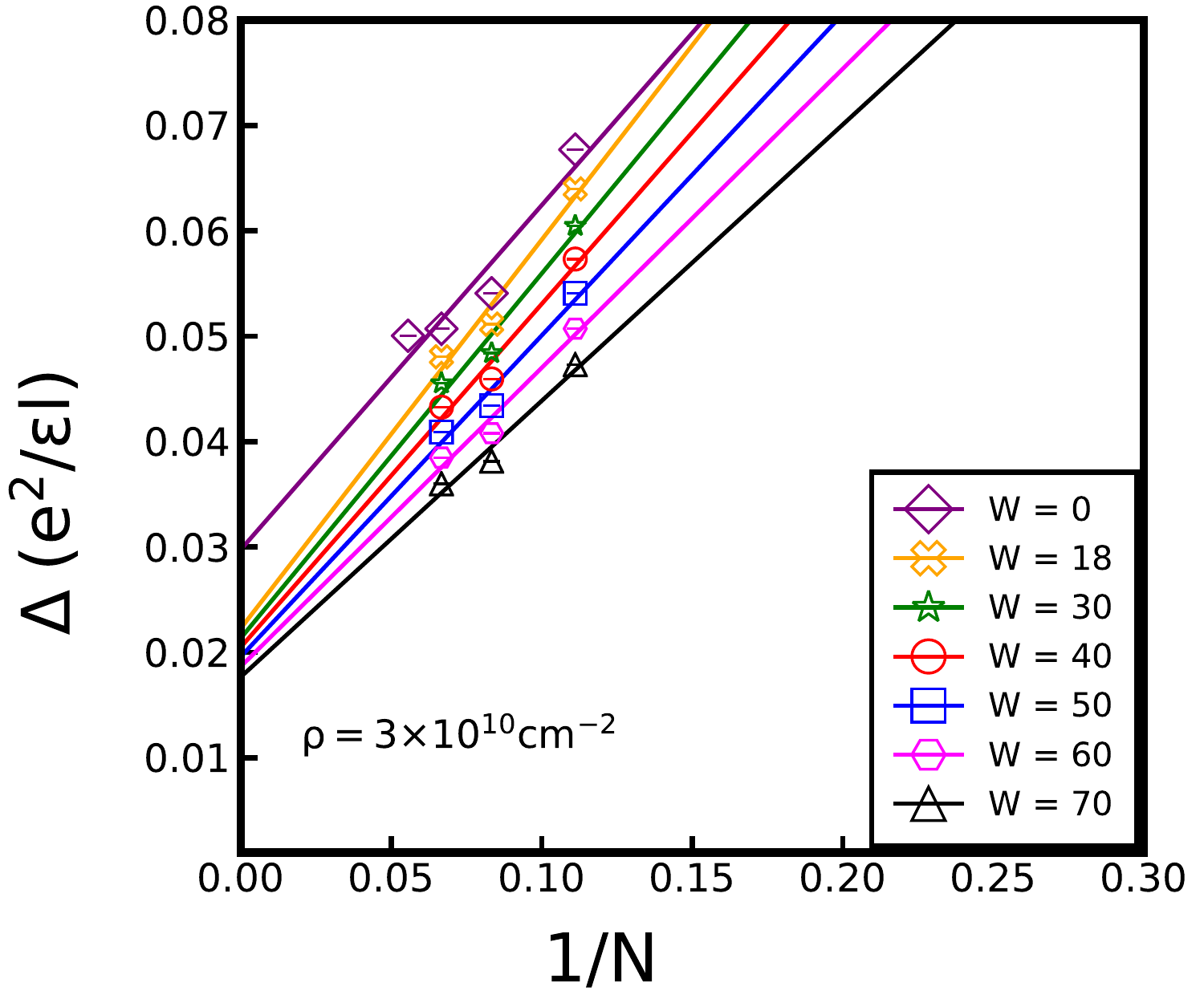}
	\includegraphics[width=0.32\linewidth]{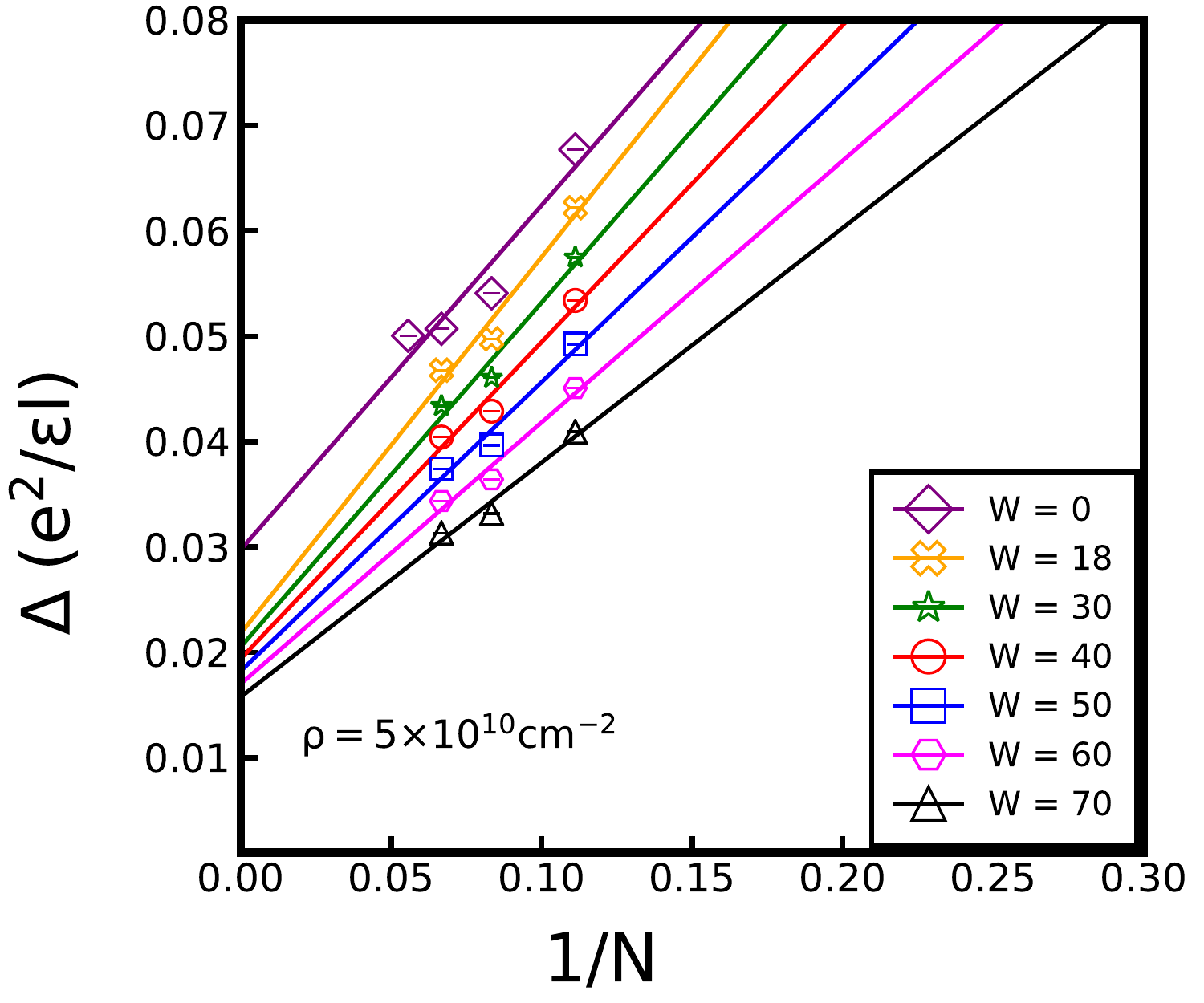}
	\includegraphics[width=0.32\linewidth]{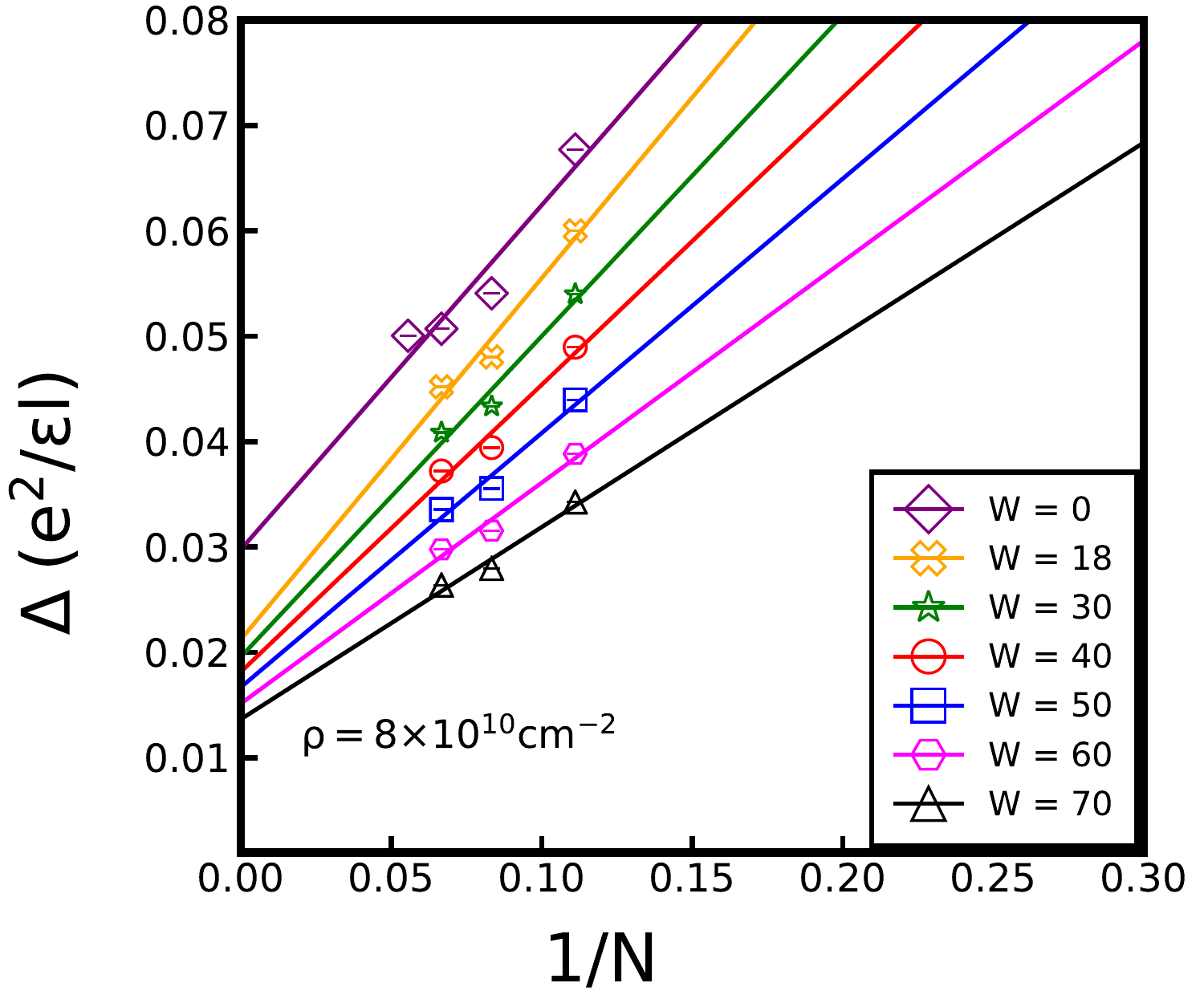}
	\includegraphics[width=0.32\linewidth]{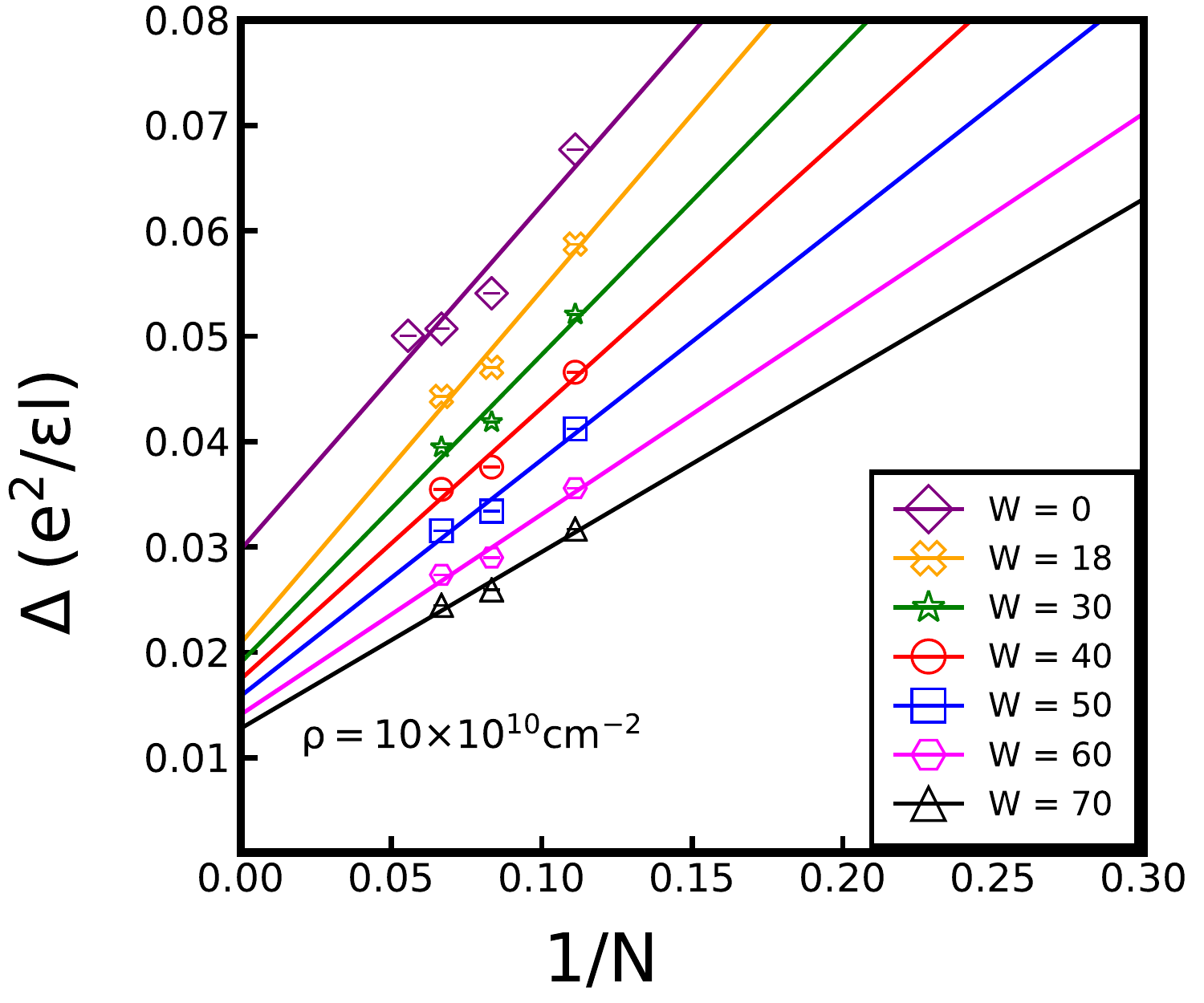}
	\includegraphics[width=0.32\linewidth]{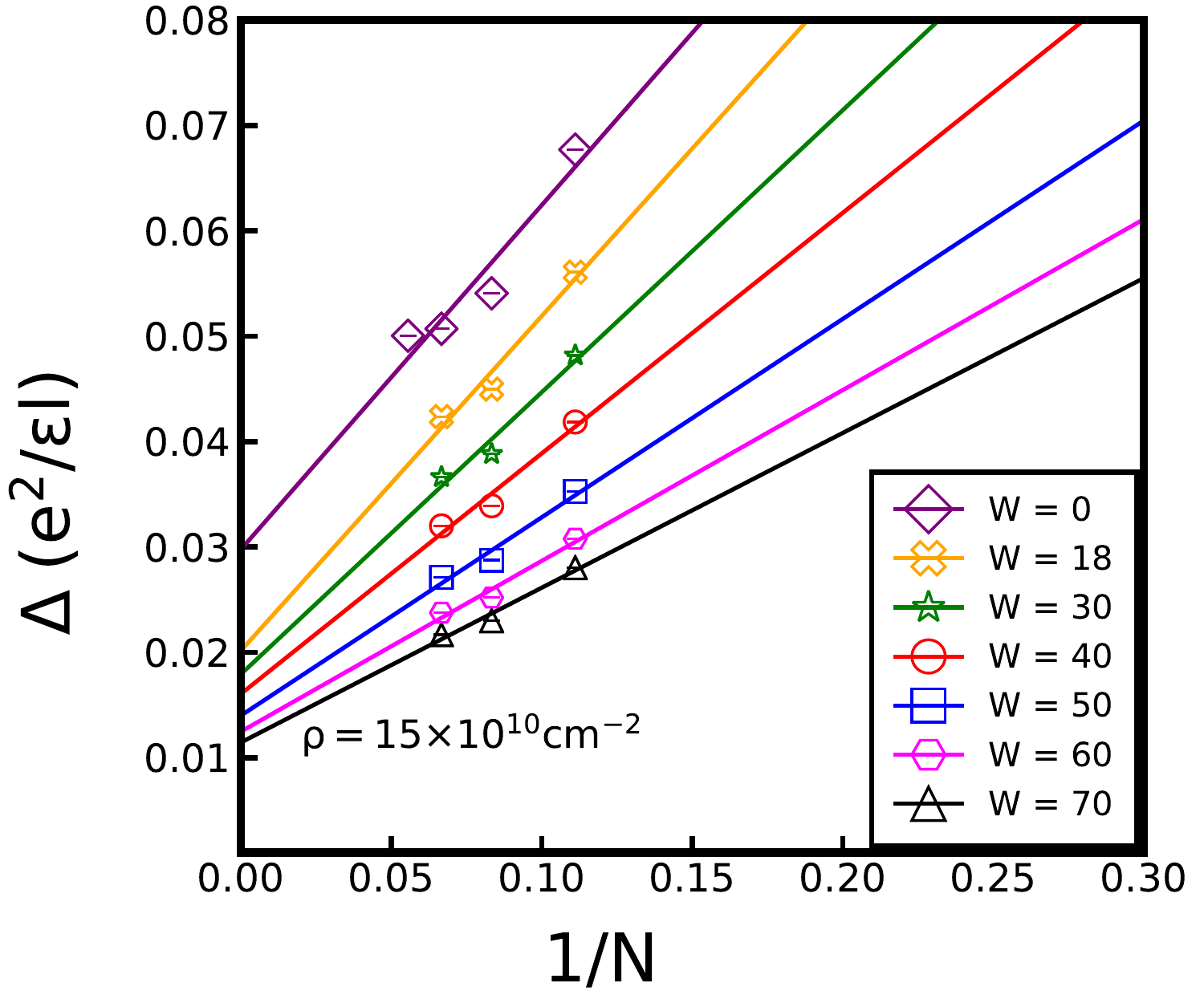}
	\includegraphics[width=0.32\linewidth]{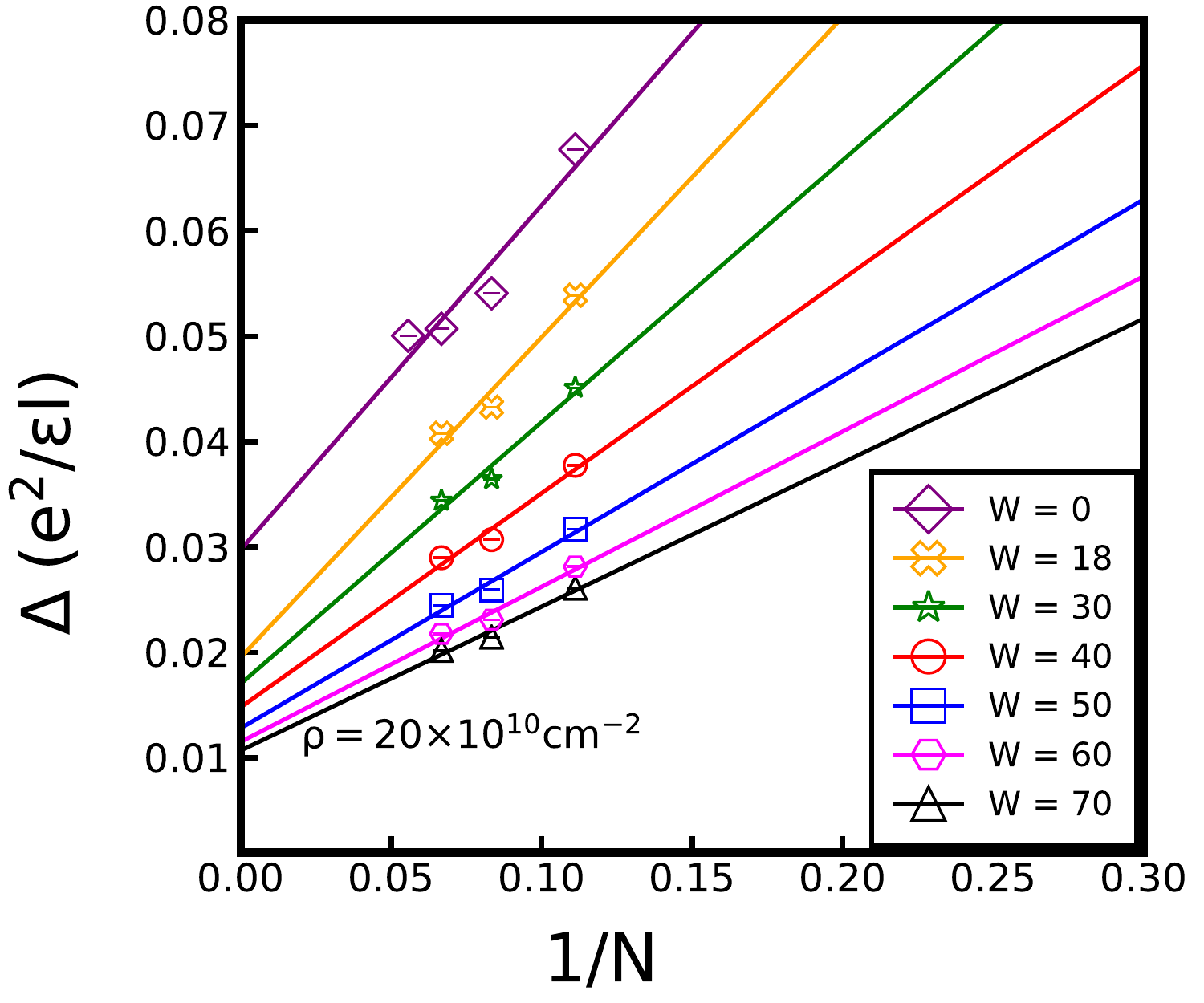}
	\includegraphics[width=0.32\linewidth]{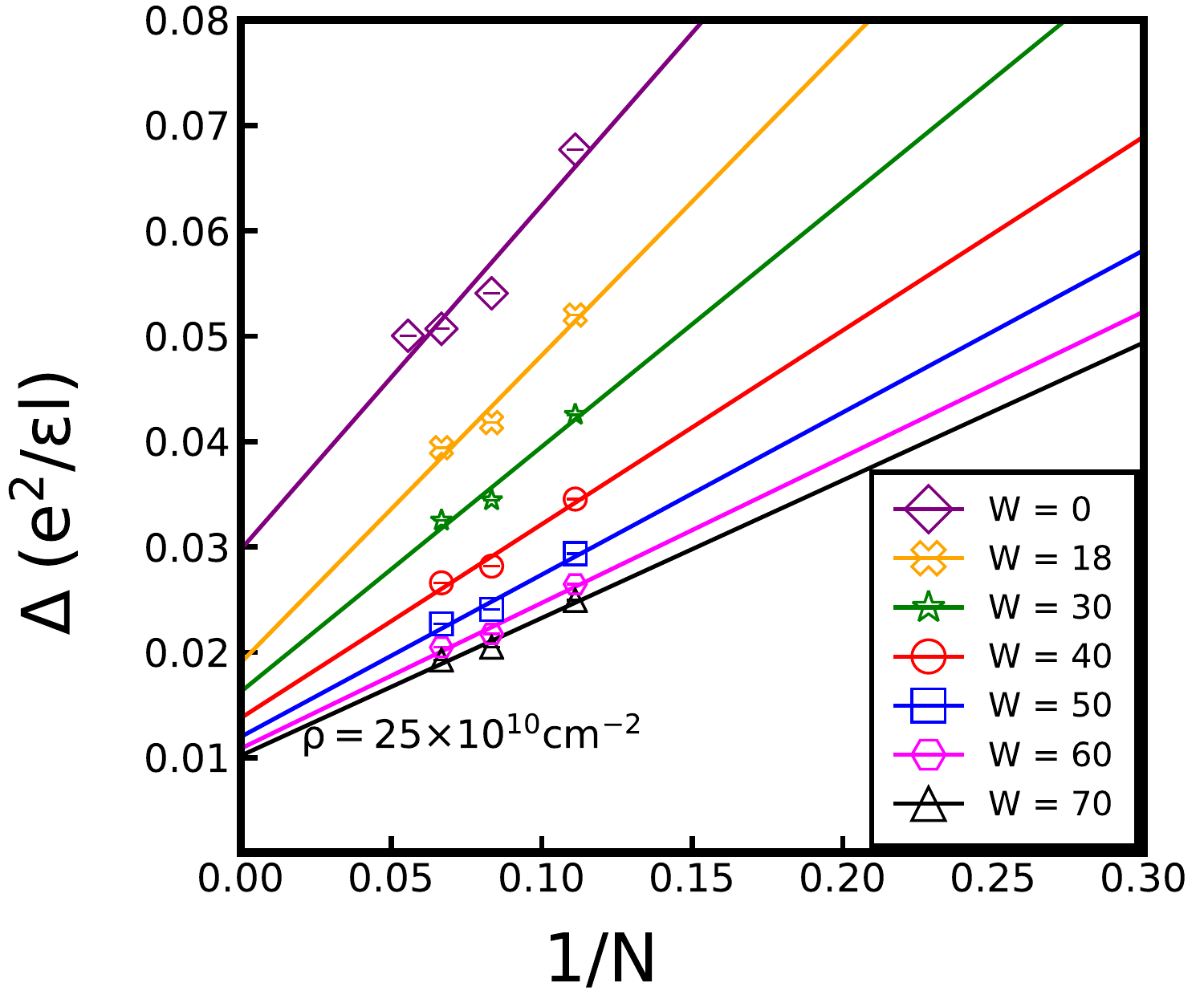}
	\includegraphics[width=0.32\linewidth]{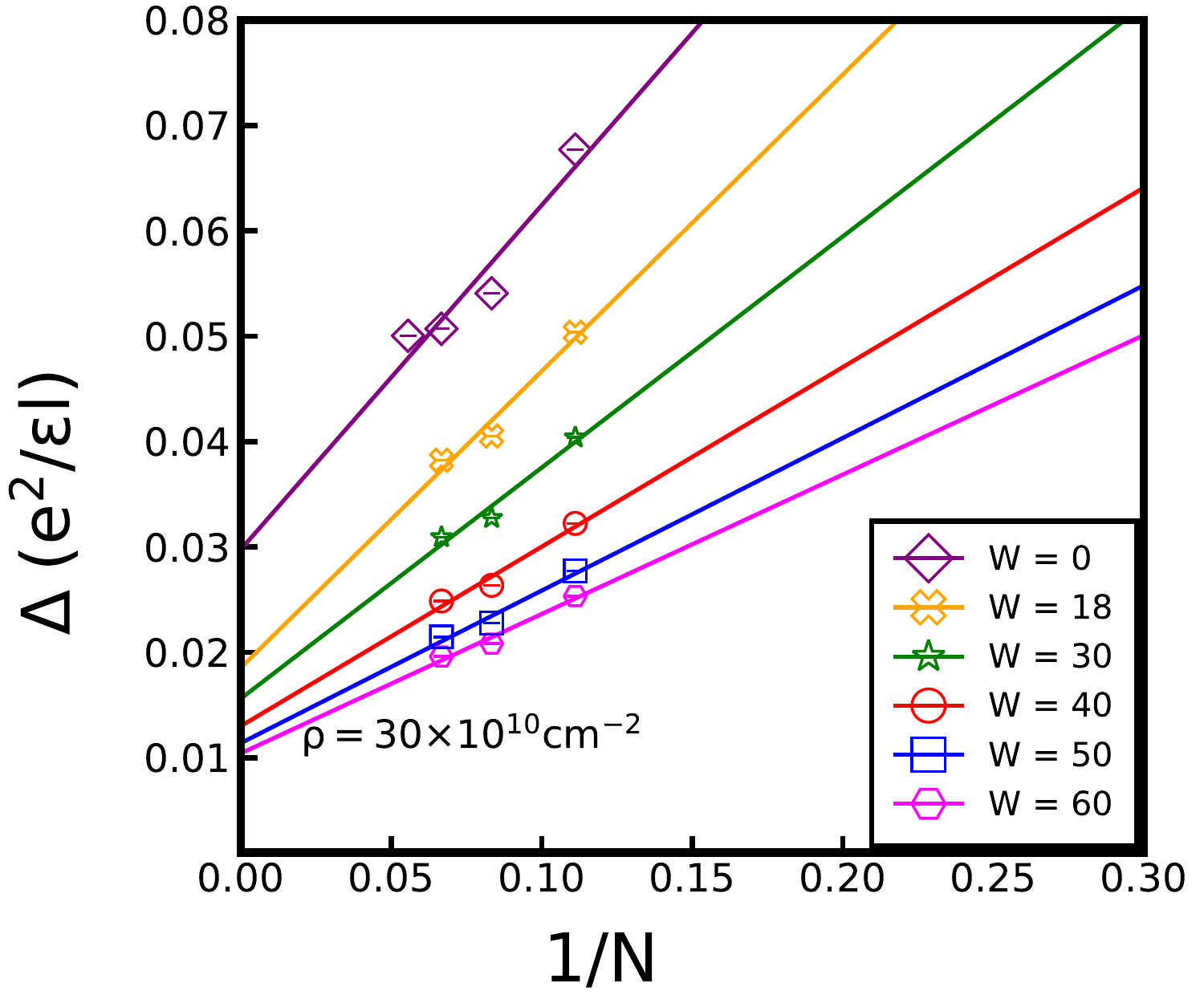}
	\caption{Thermodynamic extrapolation of the transport gap for $\nu=3/7$, calculated by the exact diagonalization (ED). Each plot presents the gaps at a specific density (labeled on the plot) with different markers labeling different well-widths in units of nanometers.}
	\label{X_fig_ED_extrap_37}
\end{figure*}

In this section, we show the thermodynamic extrapolations of the gaps calculated by the ED for the $1/3$, $2/5$ and $3/7$ FQHE. Fig.~\ref{X_fig_ED_extrap_13} shows the results for $1/3$ FQHE, Fig.~\ref{X_fig_ED_extrap_25} shows the results for $2/5$ FQHE, and Fig.~\ref{X_fig_ED_extrap_37} shows the results for $3/7$ FQHE. We note that the linear extrapolation for $1/3$ works relatively well, and it leads to the same value as the ``exact" energy we obtained in Fig.~1 of the main text within numerical error. However, one can see that for $\nu=2/5$, direct thermodynamic extrapolations of the ED gaps are no longer reliable, as the trend lines deviate from data substantially. For $\nu=3/7$, fits are even worse, as there are even fewer data points from the ED calculation. Based on this observation and the fact that the fits of $\Delta_\text{ED}-\Delta_\text{VMC}$ against $1/N$ are smoother, we have chosen to estimate $\lim\limits_{N\to \infty} (\Delta_\text{ED}-\Delta_\text{VMC})$ rather than directly evaluate $\lim\limits_{N\to \infty} \Delta_\text{ED}$.

Morf~\emph{et al.}~\cite{Morf02} also obtained gaps by directly extrapolating the ED results to the thermodynamic limit.  While our ``exact" gaps are very similar to theirs, there are some differences. For example, we find that the ED results are not very linear as a function for $1/N$, in particular for $\nu=2/5$ (see Appendix~\ref{SM_subsec:ED} for details).  One reason for the difference is that Morf~\emph{et al.}~\cite{Morf02} obtained the gap by constructing the quasihole and quasiparticle separately, which occur at monopole strengths that are different from that of the ground state. Also, the transverse distribution considered in their work is assumed to be a Gaussian distribution for square quantum well. We note that our ED data for zero-width systems are reproduced from Ref.~\cite{Balram13}.

\subsection{LL mixing: Fixed phase diffusion Monte Carlo method}
\label{SM_subsec: LLM}

In the main text, we treated the effects of LL mixing (LLM) in a perturbative approach. Here, we show results using the alternative method of fixed phase diffusion Monte Carlo (FPDMC). The general diffusion Monte Carlo (DMC) method is a standard quantum Monte Carlo method designed to solve the ground state of the many-body Schr\"odinger equation by stochastic method\cite{Foulkes01, Reynolds90, Mitas98}, and the FPDMC is a specific form of the DMC\cite{Ortiz93}. Essentially, the FPDMC fixes the phase of the wave function to be that of a well-defined trial wave function $\Psi_\text{T}$ (in our case $\Psi_\text{T}$ is chosen to be the lowest-Landau-level projected CF wave function) and searches for the lowest-energy state within this phase sector. The FPDMC algorithm revises the system from the initial state $\Psi_\text{T}$ to the target state $\Psi$, during which higher LLs are automatically included by optimizing the energy of the system. The strength of the LLM is controlled by $\kappa$. The details of the method can be found in earlier papers \cite{Ortiz93,Zhao18, Zhang16, Zhao20}. It is worth noting that in the limit $\kappa \to 0$, the FPDMC calculation reduces to the VMC calculation where the final state is fixed to be the trial wave function without any LLM.

\begin{figure*}[ht!]
\includegraphics[width=0.48 \linewidth]{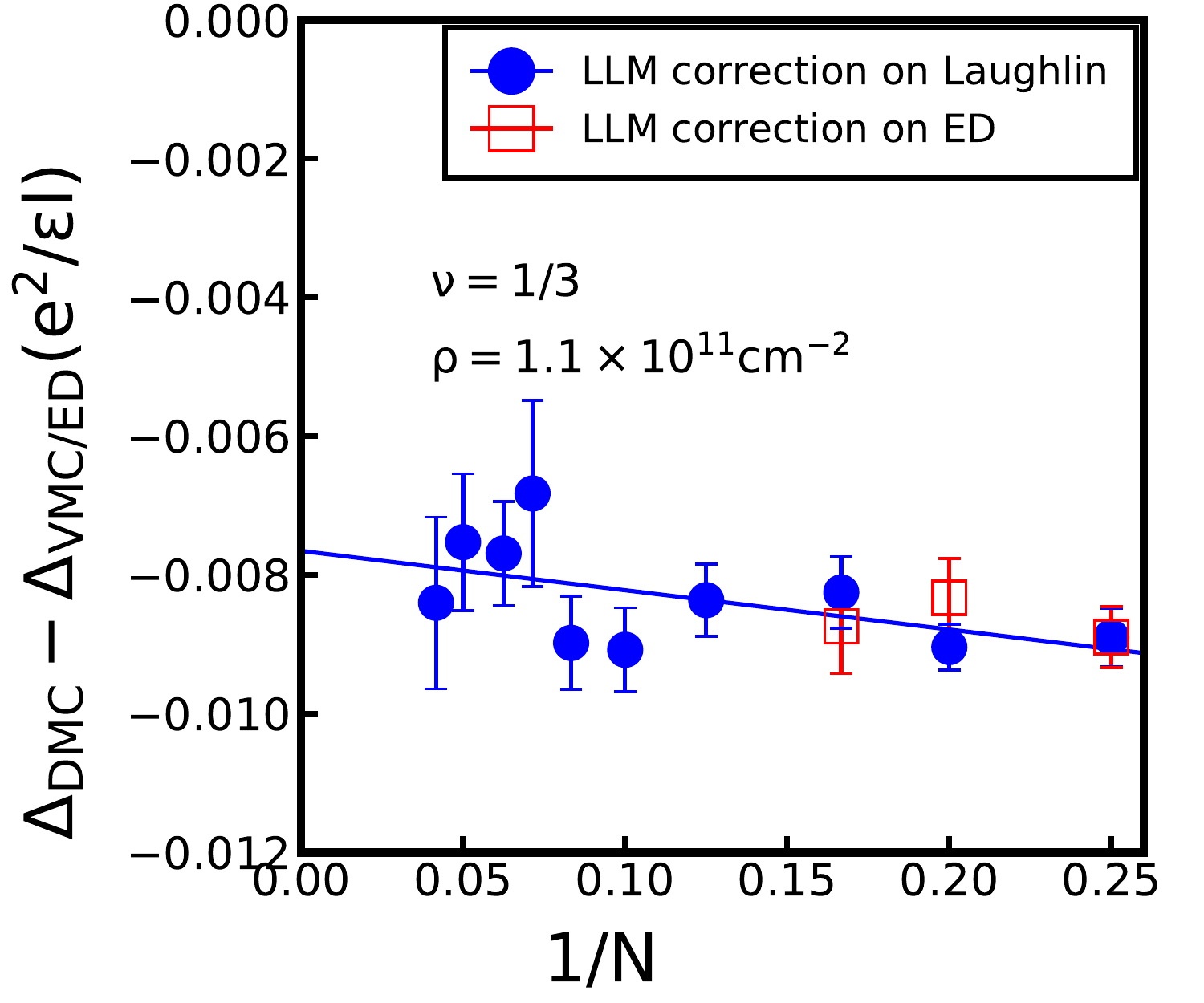}
\includegraphics[width=0.48 \linewidth]{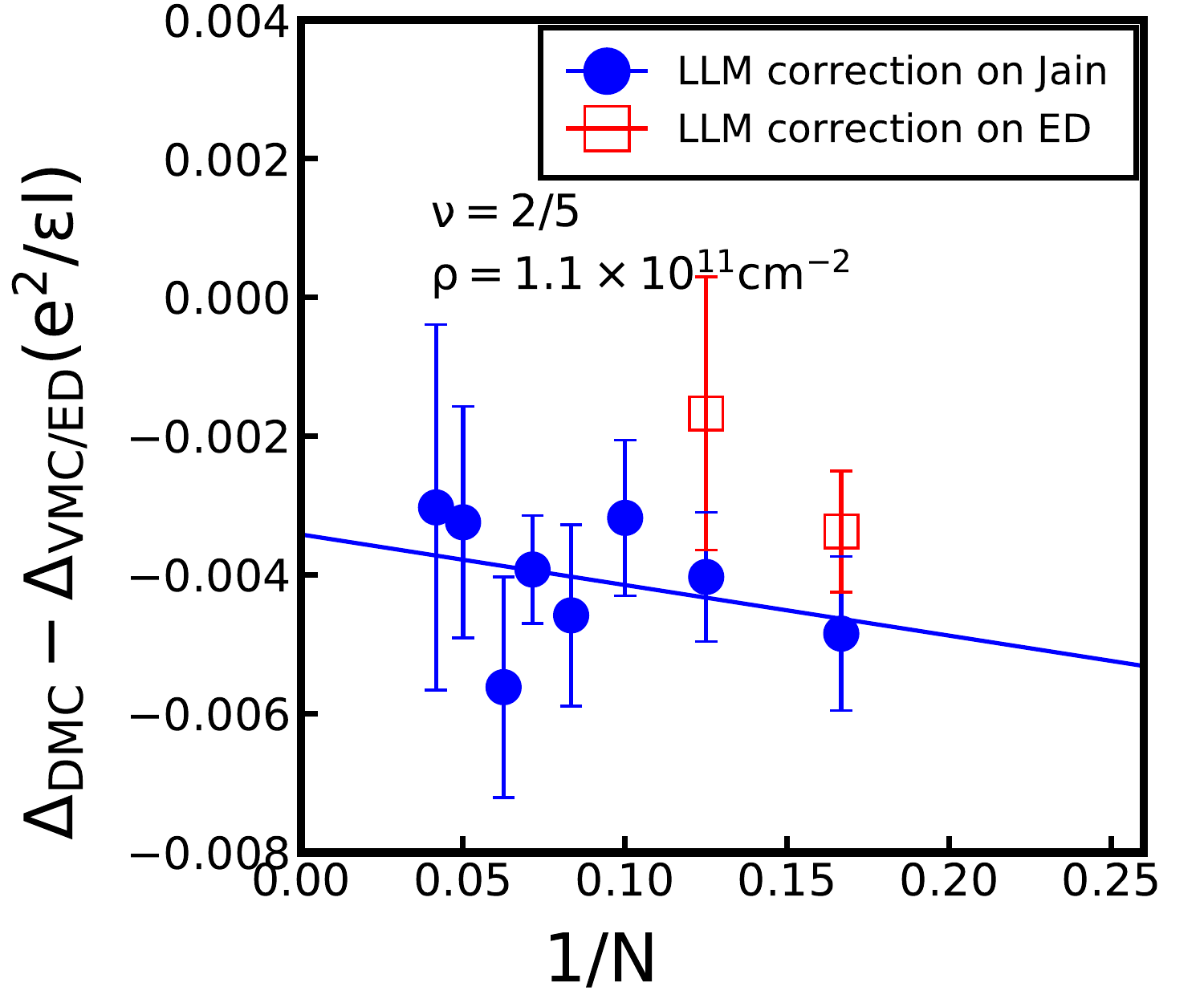}
\caption{Landau level mixing (LLM) corrections to the transport gaps calculated using fixed-phase diffusion Monte Carlo (FPDMC) at zero width and $\rho=1.1\times10^{11} \mathrm{cm}^{-2}$ for $\nu=1/3$ (left) and $2/5$ (right) FQHE. For the trial wave function of the FPDMC we have used the Laughlin / Jain trial wave functions (solid blue circles) and also the exact diagonalization (ED) wave functions (open red squares). 
}\label{LLM_Correction}
\end{figure*}

Fig.~\ref{LLM_Correction} shows our results for the LLM correction on the VMC gaps at both $\nu=1/3$ and $\nu=2/5$ at $\rho=1.1\times10^{11}\text{cm}^{-2}$ and zero width. To test how sensitive the FPDMC results are to the choice of the initial trial wave function, we have performed FPDMC using both the Laughlin/Jain states and the ED wave functions (for the latter, we can investigate relatively small systems). The results show that the LLM corrections over the CF wave function and the ED wave function are similar to each other, which is not surprising since the two wave functions have a very high overlap (for $N=12$, the overlaps between the CF wave functions and the exact wave functions in a large parameter region are $>98.7\%$ for $\nu=1/3$, $>99.5\%$ for $\nu=2/5$ and $>99.6\%$ for $\nu=3/7$, see Ref.~\cite{Zhao20}). The thermodynamic limit of the LLM correction is taken to be $\lim_{N\to \infty} (\Delta_\text{DMC}-\Delta_\text{VMC})$. The dashed green line in Fig.~1 of the main text is obtained with the  assumption that the percentage LLM corrections does not depend on the quantum well width; a  more extensive treatment for finite width quantum wells has not been performed given that the correction due to LLM is quite small. We note that we have not performed the FPDMC calculation for the $3/7$ FQHE, because the statistical uncertainty for this state is too large to obtain useful results. 

\section{Composite fermion mass}
\label{app: CF_mass}
For non-interacting electrons in the vacuum, the cyclotron energy is $\hbar \Omega_c=\hbar eB /m_{e}c$, where $\Omega_c$ is the cyclotron frequency and $m_e$ is the electron mass in the vacuum. By analogy, one interprets the transport gap as the CF cyclotron energy and defined the CF mass as 
\begin{equation}\label{X_eq:cf_mass_eq1}
\Delta = \hbar \omega_c^* = \hbar \frac{e B^*}{m^* c}=\frac{1}{2n+1}\frac{\hbar e B}{m^* c},
\end{equation}
where $\hbar \omega_c^*$ is the cyclotron energy of CFs.  Comparing Eq.~\eqref{X_eq:cf_mass_eq1} with the electron's cyclotron energy in the vacuum, one obtains the following relation:
\begin{equation}
\begin{aligned}\label{cf_mass_eq2}
\Delta& = \frac{1}{2n+1}\frac{m_e}{m^*} \frac{\hbar eB}{m_e c}\\
&=\frac{1}{2n+1}\frac{m_e}{m^*} \frac{m_b}{m_e}\frac{\hbar eB}{m_b c}\\
&=\frac{1}{2n+1}\frac{m_b}{m_e}\frac{m_e}{m^*} \frac{1}{\kappa} \frac{e^2}{\varepsilon l}\\
&=\frac{0.067}{2n+1}\frac{m_e}{m^*} \frac{1}{\kappa} \frac{e^2}{\varepsilon l},\\
\frac{m^*}{m_e}&=\frac{0.067}{\tilde{\Delta}}\frac{1}{\kappa}\frac{1}{2n+1},
\end{aligned}
\end{equation}
where $\tilde{\Delta}=\Delta/\left(e^2/\varepsilon l\right)$ is the activation gap measured in the Coulomb energy and we have incorporated the value of the band mass in the GaAs material.

\begin{figure*}
	\includegraphics[width=\columnwidth]{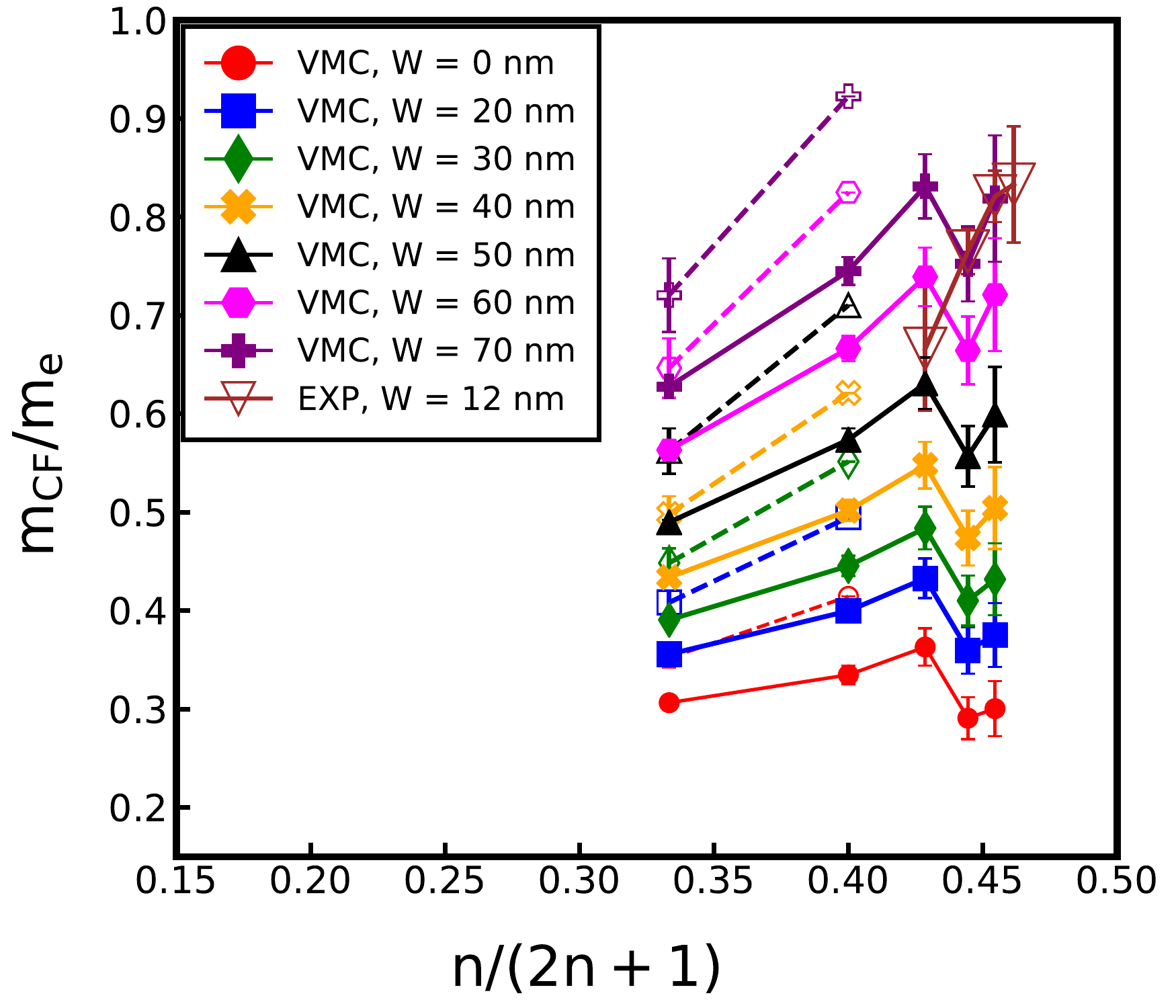}
	\caption{Comparison between the theoretically calculated and the experimentally measured values of the CF mass at density $\rho=1.1\times 10^{11}\text{cm}^{-2}$ for different quantum well-widths. For theoretical values, solid markers joined by solid lines are from the bare variational Monte Carlo (VMC) calculation, and the hollow makers joined by dashed lines are the theoretical values after including the corrections from the variational error and Landau level mixing (we only have these corrections for $\nu=1/3$ and $2/5$ states). The experimental data are taken from Ref.~\cite{Coleridge95}.}
	\label{CF_mass}
\end{figure*}

The CF mass for $\rho=1.1\times 10^{11} \text{cm}^{-2}$ is shown in Fig.~\ref{CF_mass}. In general, the CF mass calculated from the VMC gaps is roughly half of the CF mass measured in the experiment based on the Shubnikov-de Haas resistance oscillations\cite{Du93,Du94,Coleridge95}. When we include the corrections that account for the variational error and the LLM effect, we find that the CF mass increases by about $10\%$ to $20\%$, depending on parameters.

\section{Gaps in the first excited Landau level of graphene}
\label{SM_sec:graphene}

\begin{figure*}
\label{graphene_vmc_extrap}
	\includegraphics[width=0.32 \linewidth]{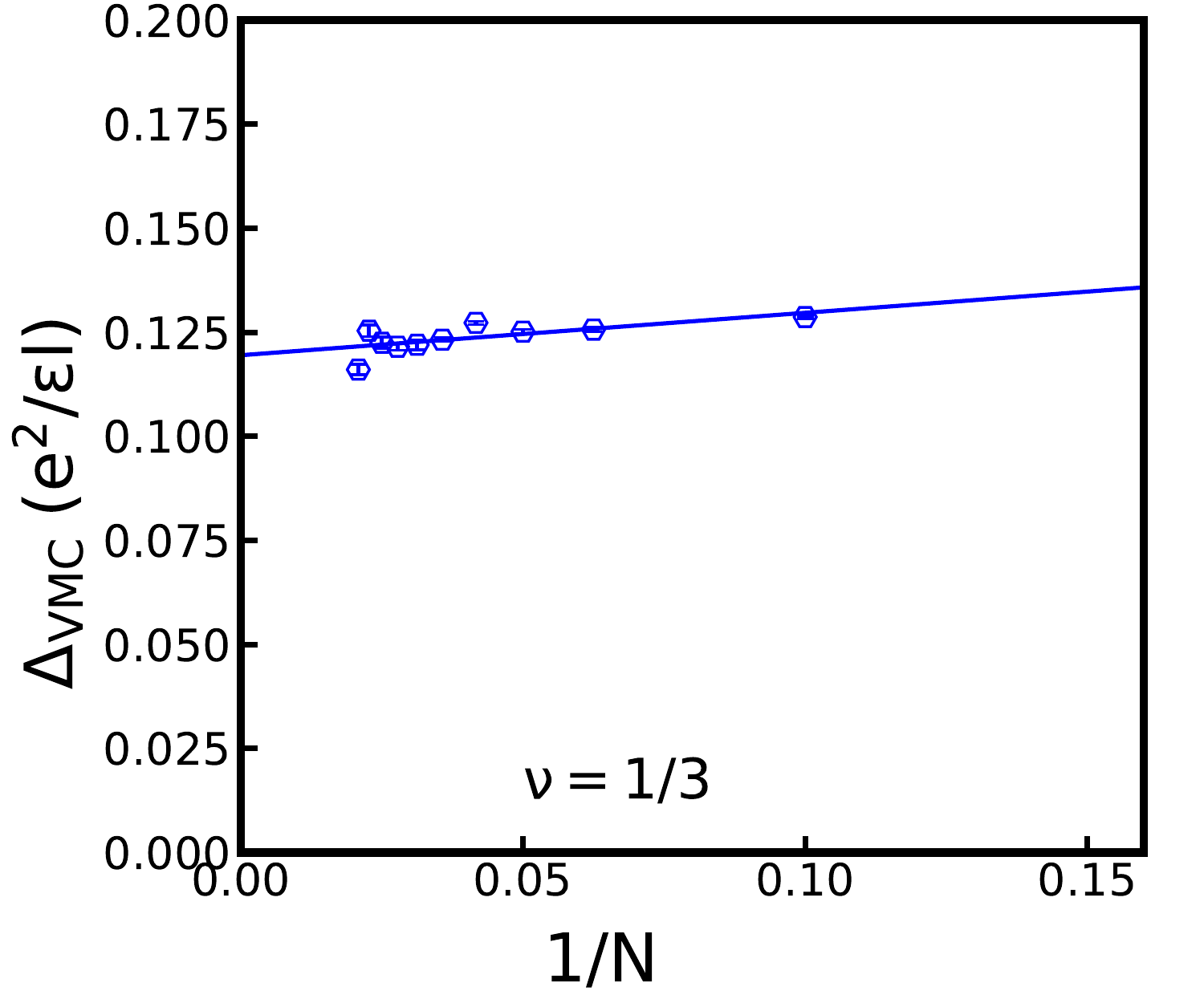}
	\includegraphics[width=0.32 \linewidth]{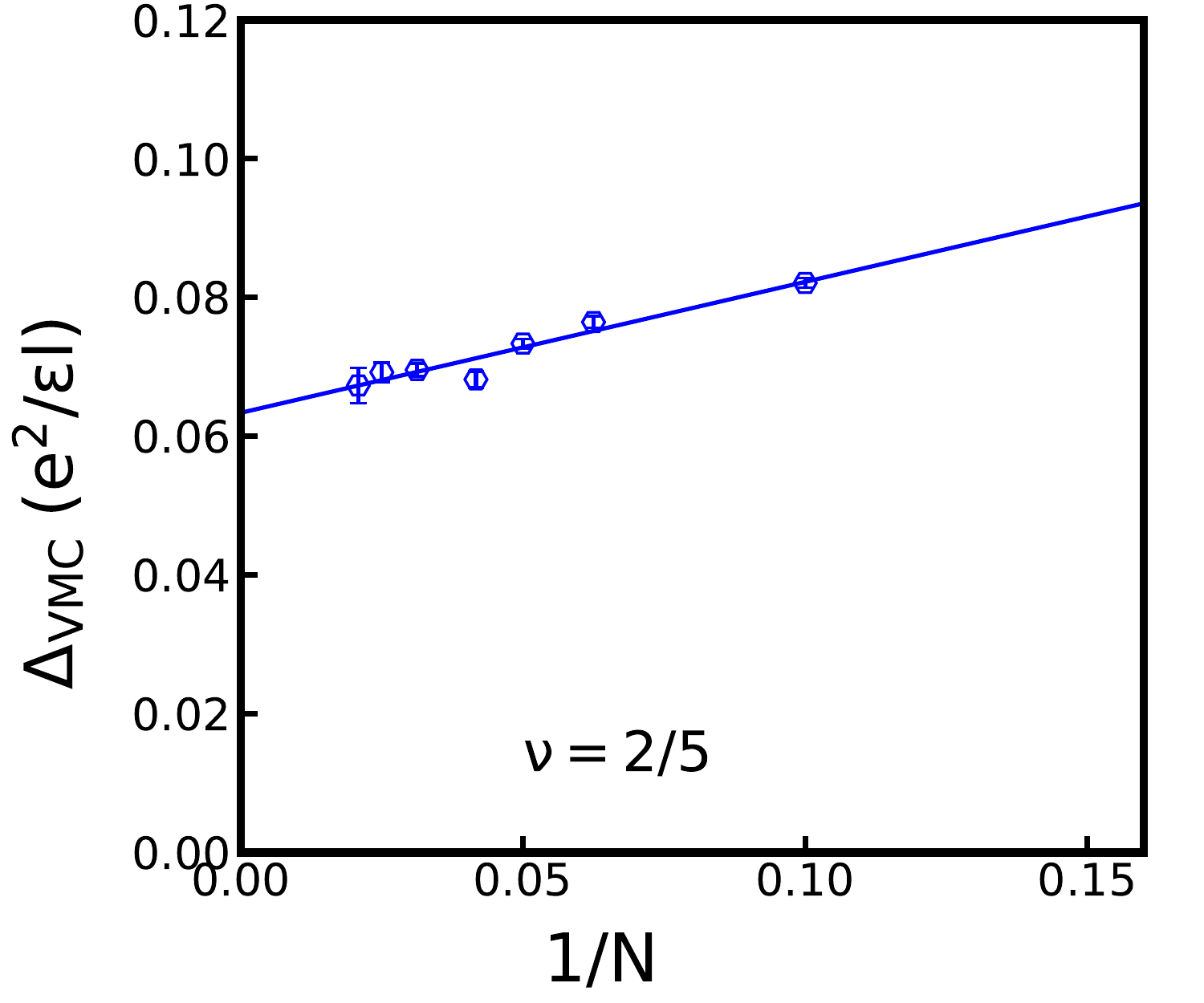}
	\includegraphics[width=0.32 \linewidth]{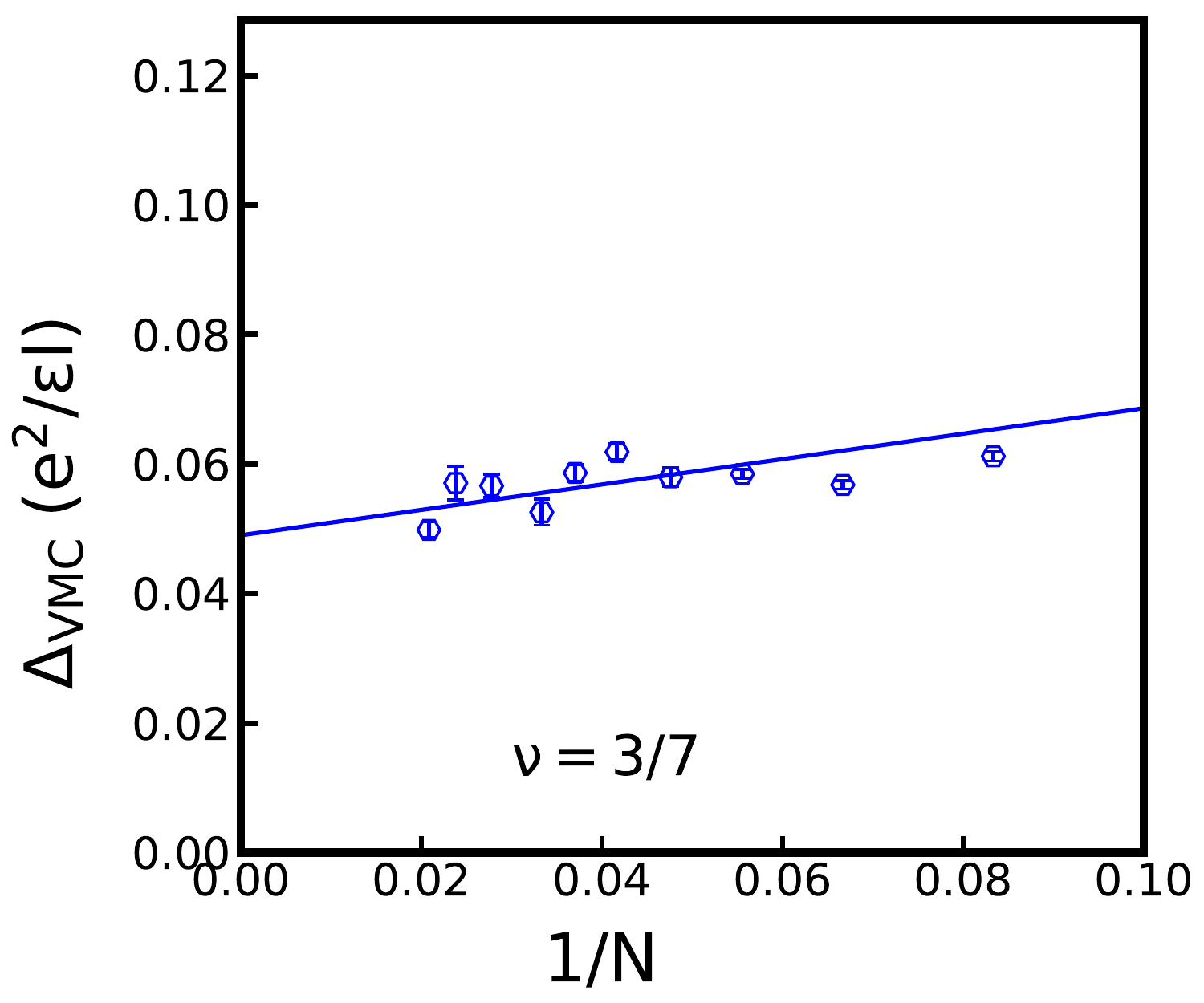}
	\includegraphics[width=0.32 \linewidth]{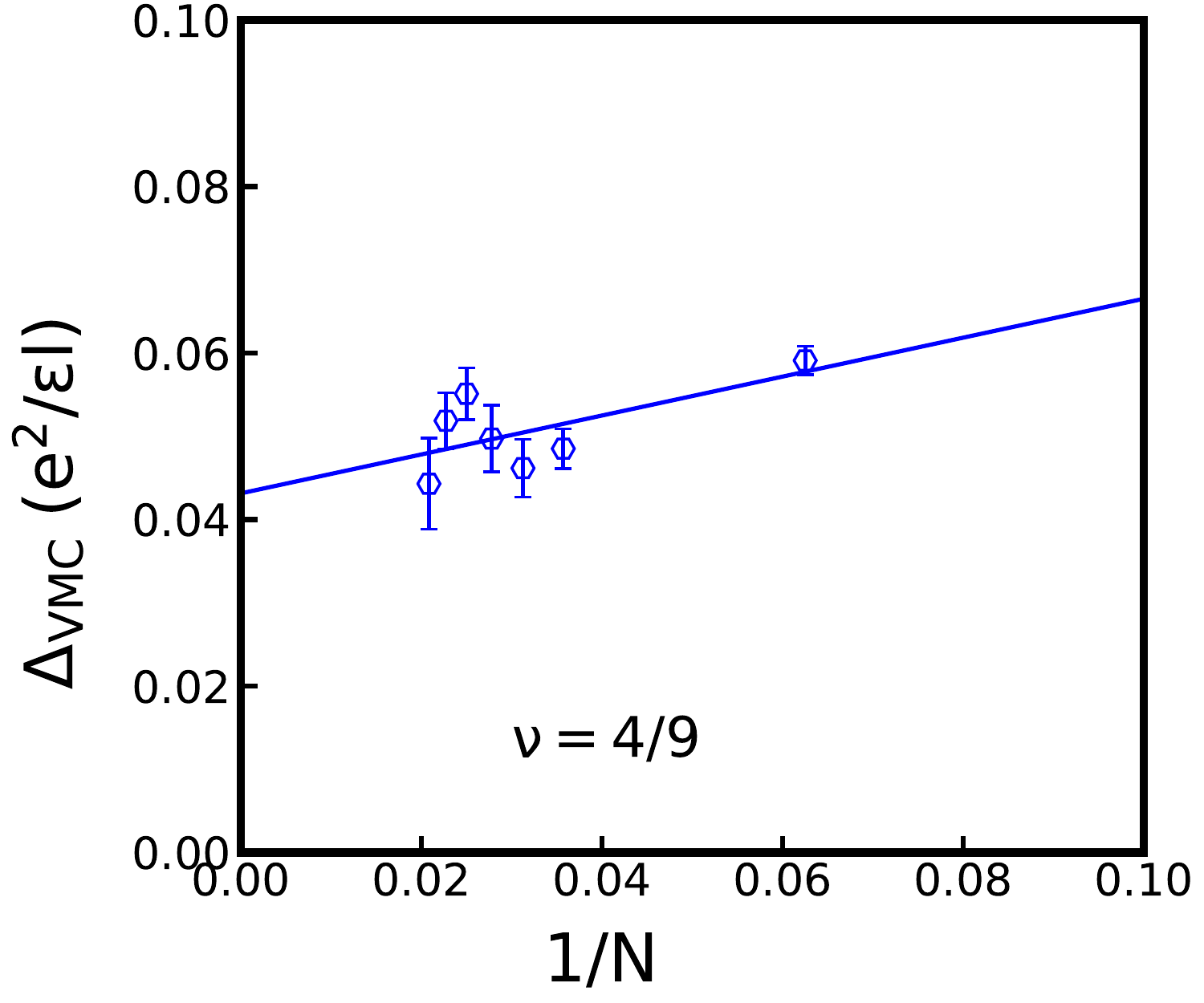}
	\includegraphics[width=0.32 \linewidth]{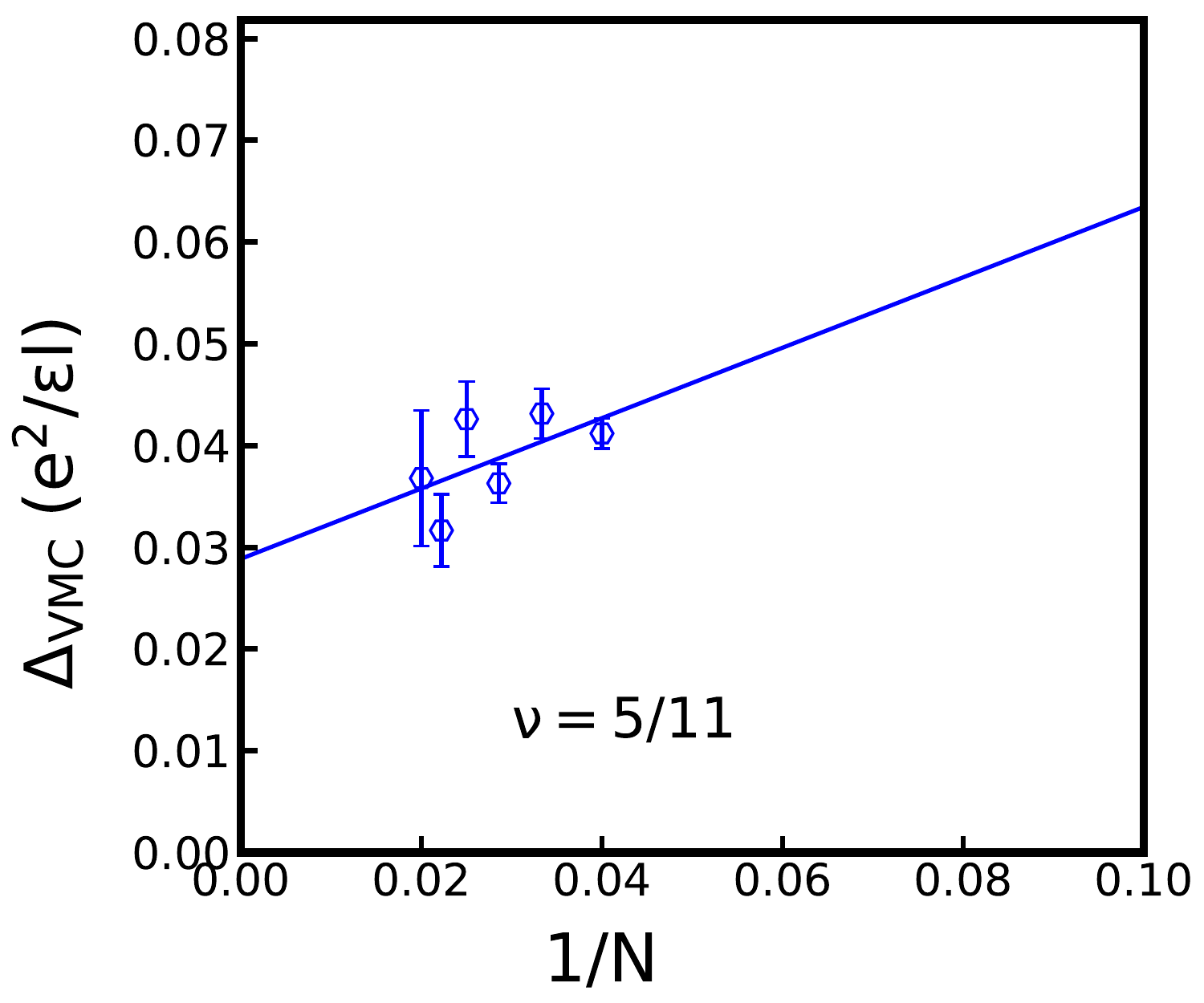}
	\caption{Thermodynamic extrapolation of the activation gaps for FQHE states at filling $1/3, 2/5, 3/7, 4/9, 5/11$ in the first excited Landau level of graphene.}
	\label{fig_graphene_extrap}
\end{figure*}

Our zero width results above apply to the zeroth LL of graphene to the extent LLM corrections can be neglected~\cite{Balram15a}.  The first excited LL of graphene, however, is different from the first excited LL in GaAs.  Earlier studies have shown that the CF theory gives an excellent account of the FQHE states in the first excited LL of graphene~\cite{Balram15c, Balram21b}. We have obtained the VMC activation gaps for the Jain  $n/(2n+1)$ FQHE states in this LL, using the effective interaction from Ref.~\cite{Balram15c}. Fig.~\ref{fig_graphene_extrap} shows the thermodynamic extrapolation of the activation gaps for FQHE states at fillings $1/3, 2/5, 3/7, 4/9$ and $5/11$. The results at $\nu=1/3, 2/5$ and $3/7$ are consistent with the previous results of Ref.~\cite{Balram16d} (We note that we have subtracted the interaction between the CF hole and the CF particle from the total gap, which results in values slightly higher than those reported in Ref.~\cite{Balram16d}.). We find that the statistical uncertainties of the thermodynamic values of the activation gaps for $4/9$ and $5/11$ states are very large because of the strong finite-size effects at these higher-order fractions.  Fig.~3 of the main text shows that these gaps are consistent with $\Gamma=0$. The experimental gaps, also shown in that figure, show a finite value of $\Gamma$; it is not known to what extent LLM and disorder contribute to it.

\bibliography{biblio_fqhe}

\pagebreak





\end{document}